\documentclass[11pt,a4paper]{article} 
\usepackage{jheppub}

\usepackage[T1]{fontenc}
\usepackage[normalem]{ulem}
\usepackage{amscd,amsmath,amssymb,amsfonts,xspace,mathrsfs,amsthm,bbm}
\usepackage{color}
\usepackage{mathtools}

\usepackage{extarrows}
\usepackage{braket}
\usepackage[dvipsnames]{xcolor}
\usepackage{url}

\usepackage{latexsym}
\usepackage{graphicx}
\usepackage{dsfont}
\usepackage{longtable}
\usepackage{enumitem}
\usepackage{mathdots}
\usepackage{float}
\usepackage[utf8]{inputenc}
\usepackage{multirow}
\usepackage{caption}
\usepackage{subcaption}
\usepackage{subcaption}
\usepackage{tikz-cd}
\usetikzlibrary{bending}
\usepackage{tikz,textcomp}
\usetikzlibrary{decorations.pathreplacing,decorations.pathmorphing,decorations.markings,calc,fadings,fit,shapes,arrows,positioning,chains,matrix}
\usepackage{cleveref}
\crefname{subsection}{subsection}{subsections}
\usepackage{bbold}
\usepackage{booktabs}  
\usepackage{array} 

\newtheorem{Thm*}{Theorem}

\newtheorem{thm-int}{Theorem}
\theoremstyle{definition}
\newtheorem{Def-s}[Thm]{Definition}

\newenvironment{enumerate*}{\begin{enumerate}}{\end{enumerate}}

\makeatletter
\newtheorem*{rep@theorem}{\rep@title}
\newcommand{\newreptheorem}[2]{%
\newenvironment{rep#1}[1]{%
 \def\rep@title{#2 \ref{##1}}%
 \begin{rep@theorem}}%
 {\end{rep@theorem}}}
\makeatother

\def\bea{\begin{eqnarray}}
\def\eea{\end{eqnarray}}
\def\be{\begin{equation}}
\def\ee{\end{equation}}
\def\ba{\begin{align}}
\def\ea{\end{align}}
\def\bse{\begin{subequations}}
\def\ese{\end{subequations}}

\def\det{\,{\rm det}\, }

\def\tr{\,{\rm tr}\, }

\def\Im{\,{\rm Im}\,}
\def\Re{\,{\rm Re}\,}

\newcommand{\etale}{\text{\'{e}t}}
\newcommand{\Frob}{\mathbf{Fr}}

\def\({\left(}
\def\){\right)}
\def\[{\left[}
\def\]{\right]}
\def\<{\left\langle}
\def\>{\right\rangle}

\newcommand{\cE}{\mathcal{E}}

\newcommand{\cL}{\mathcal{L}}

\newcommand{\R}{{\mathbb R}}
\newcommand{\Z}{{\mathbb Z}}

\newcommand{\F}{\mathbb{F}}
\newcommand{\M}{\mathcal{M}}

\newcommand{\abs}[1]{\left\lvert {#1}\right\rvert}

\newcommand{\N}{\mathbb{N}}
\renewcommand{\Z}{\mathbb{Z}}
\renewcommand{\R}{\mathbb{R}}
\newcommand{\C}{\mathbb{C}}
\newcommand{\Q}{\mathbb{Q}}

\newcommand{\ii}{\mathrm{i}}
\renewcommand{\P}{\mathbb{P}}
\renewcommand{\vec}[1]{\underline{#1}}

\newcommand{\dd}{\mathrm{d}}

\newcommand{\uz}{{\underline z}}

\renewcommand{\cal}{\mathcal}
 
\newcommand{\Mcs}{{\cal M}_{\text{cs}}(X)}

\newcommand{\Gr}{\mathrm{Gr}}

\numberwithin{equation}{section}

\title{Geometry and Arithmetic of Special Loci in the Moduli Spaces of Type II String Theory}

\author[a]{Paul Blesse,}
\author[a]{Janis Dücker,}
\author[a,b]{Albrecht Klemm}
\author[a]{and Julian F.\ Piribauer}
\affiliation[a]{Bethe Center for Theoretical Physics, Universit\"at Bonn, D-53115, Germany}
\affiliation[b]{Department of Mathematical and Physical Sciences, University of Sheffield, S3 7RH Sheffield, UK}
\abstract{We use Dwork's deformation method to calculate the Hasse-Weil Zeta function 
of multi-parameter families of Calabi-Yau three and fourfolds. This information is used to 
identify subslices of codimension one in the complex-structure moduli space, where the Hodge structure 
splits in particular ways and different type IIB flux vacua  emerge. We calculate the corresponding background fluxes and their potential that
drives the IIB string compactification to these subslices and analyse the properties of the corresponding physical vacua. We address 
the question whether the subslices correspond to fixed loci of symmetries acting on the original family and whether they
can be identified with consistent complex-structure moduli spaces of Picard-Fuchs systems with standard integral 
monodromy bases for fewer complex deformation parameters. We distinguish between supersymmetric vacua and singular 
subslices. In the latter case a standard geometrical basis can be expected if  a physical transition 
leads to a smooth type II vacuum. In many cases  the differential equations on the subslice are fulfilled by the restricted periods 
after adding an inhomogeneous term. This suggests that the resolution of the singularity provides 
three-chains and we indeed find that the corresponding integrals allow an integral expansion 
compatible with their interpretation as generating functions of disk instantons.}

\emailAdd{pblesse@uni-bonn.de}
\emailAdd{jduecker@uni-bonn.de}
\emailAdd{a.klemm@sheffield.ac.uk}
\emailAdd{piribauer@uni-bonn.de}

\begin{document}
\preprint{\begin{minipage}[t]{8cm}\begin{flushright} 
BONN-TH-2025-36 \\
      \end{flushright}\end{minipage}}

\setlength{\parskip}{0.2cm} 

\maketitle


\section{Introduction}
Moduli stabilisation in Type II string compactifications on complex families of Calabi-Yau  (CY) manifolds $X_{\underline z}$ by 
flux induced superpotentials $W$~\cite{Polchinski:1995sm,Michelson:1996pn,Gukov:1999ya,Mayr:2000hh,Curio:2000sc,Giddings:2001yu} yields vacuum solutions ${\underline z}_*$ in the complex moduli space of $X$  in which the  
Hodge structure in the fibres $X_{{\underline z}_*}$ over the loci ${\underline z}_*$ has special arithmetic 
properties. 
For example $N=2$ supersymmetric flux vacuum solutions require in the fibres the existence of a rank two lattice 
$\Lambda\subset H^3(X_{{\underline z}_*},\mathbb{Z})\cap 
(H^{2,1}(X_{{\underline z}_*})\oplus H^{1,2}(X_{{\underline z}_*}))$ so that $H^3(X_{{\underline z}_*})$ splits over $\mathbb{Q}$: 
$H^3(X_{{\underline z}_*},\mathbb{Q})=\Lambda_{\mathbb{Q}}\oplus \Lambda_\mathbb{Q}^\perp$. These loci have been 
described in~\cite{Moore:1998pn} and are called rank two attractors as they are fixed points of  the 
attractor flow mechanism for $N=2$ supersymmetric black holes with fixed electric and magnetic 
charges in $H_3(X_{\underline z},\mathbb{Z})$~\cite{Ferrara:1995ih,Ferrara:1996dd}. Identifying the charges with the 
flux quanta relates the $N=2$ vacuum solutions to the attractor loci as pointed out in~\cite{Curio:2000sc} and made more precise in~\cite{Kachru:2020abh,Bonisch:2022slo}. Here the absolute value of the flux superpotential is proportional to the 
central charge of the black hole and hence its entropy~\cite{Ferrara:1995ih,Ferrara:1996dd}.

The splitting of the Hodge structure over $\mathbb{Q}$ can be inferred  from the factorisation properties 
of the  rational Hasse-Weil Zeta function for sufficiently large lists of good primes. We refer to the
relevant polynomials that correspond to the middle cohomology of the Calabi-Yau space as the {\sl Euler factors}. By the Chinese 
remainder theorem (CRT) the further factorisation of the Euler factors also determines the locus of the rank two attractors for ${z}_*\in \mathbb{Q}$ 
as well as for $z_*\in\mathbb{Q}[\underline{\alpha}]$, at least for simple algebraic extensions $\mathbb{Q}[\underline{\alpha}]$ of $\mathbb{Q}$. 
Using this method the first three rank two attractor points, which are not fixed point loci 
under obvious geometrical symmetries (r2A-nG), were discovered 
in a one-parameter family $X_{z}$. They occur over a symmetric subslice in the moduli space of a complete intersection of two degree $d_1=d_2=(1,1,1,1,1)$ hypersurfaces in $(\mathbb{P}^1)^5$, see~\cite{Bonisch:2020qmm}, 
at the loci  $z_*\sim \{-\frac{1}{7},33 \pm 8 \sqrt{-17} \}$ in \cite{Candelas:2019llw}.  
The corresponding one-parameter Picard-Fuchs operator is labelled $4.3.1$ in the database~\cite{cycluster}\footnote{Here the first 
number stands for the order of the differential operator, the second for its degree in the $z$ variable and the last is an ordering number. 
Historically these operators were labelled by AESZ\#, with $\#=34$ for the 4.3.1 operator. The database~\cite{cycluster} contains the old and new labels.} of Calabi-Yau operators 
see \cite{vanstraten2017calabiyauoperators}, for a review. Geometrically the operator is also realised by a one-parameter family over a  
sub-slice of the moduli space of the Hulek-Verill CY 3-fold~\cite{hulek2005modularity}. Specific solutions of an inhomogeneous  extension of the operator 4.3.1 
compute the four-loop banana integral in two dimensions\footnote{These banana integrals play an important role in $\epsilon$ regularised generic 4d QFT~\cite{Bonisch:2021yfw}.}~\cite{Bonisch:2020qmm}. 
Further r2A-nG points have been found in simpler hypergeometric CY complete intersection families $X_{33}(1^6)$ at $z_*=-1/2^3 3^6$ and in $X_{43}(1^5,2)$ at $z_*=-1/2^4 3^3$ 
corresponding to the operators 4.1.4 and 4.1.11. A longer list of rank two attractor points can be 
found in Table 1 of \cite{Bonisch:2022slo}. 

As pointed out in \cite{Elmi2020,handle:20.500.11811/11048}, if a degree four Calabi-Yau operator is obtained from Hadamard product of two elliptic curves ${\cal C}^{(k)}_{t^{(k)}}$, an
exhaustive list of those which lead to CY operators of degree two in $z$ can be found in \cite{almkvist2023calabi}, and both curves admit an involution symmetry $t^{(k)}\mapsto 1/c^{(k)} t^{(k)}$ in their 
complex structure parameters $k=1,2$, then the CY operator has also  an involution symmetry $z\mapsto 1/(c^{(1)} c^{(2)}) z$. At its fixed point $z_*$ 
the $(1,1,1,1)$ Hodge structure can split as above into a motive of a weight four newform and a (-1)-Tate twist of a weight two newform 
$(1,0,0,1)\oplus (0,1,1,0)$  over $\mathbb{Q}$ yielding a rank two attractor or product of a weight 3 and a weight 2 motive $(1,0,1)\otimes (1,1)$. The physical 
significance of the latter possibility and some generalisations has not been spelled out. On the modular side the generalisations involve Bianchi- and Hilbert modular 
motives studied in \cite{handle:20.500.11811/11048}. We call rank two attractor loci, which correspond to fix loci of geometrical symmetries, rank two attractors 
of geometrical origin (r2A-G). 

The Euler factors contain much more detailed information about the Hodge structure 
than its splitting and in particular the values of the period matrix at the special fibres. For example at 
the rank two attractors $H^3(X_{{\underline z}_*},\mathbb{Q})=\Lambda_{\mathbb{Q}}\oplus \Lambda_\mathbb{Q}^\perp$ 
splits into two irreducible representations of the Galois group as detected by the splitting of the Euler 
factors, see Section \ref{sec:specialfibres}.
The coefficients of the factors determine modular forms and the complete 
period matrix in the $X_{z_*}$ fibre can be expressed up to $\mathbb{Q}$-factors in terms of four periods $\omega^{(k)}_\pm$ 
of these holomorphic Hecke newforms $f_k\in S_k(\Gamma_0(N_k))$ and four quasiperiods $\eta^{(k)}_\pm$ of the 
associated meromorphic Hecke eigenforms with vanishing residuum $f_k\in \mathbb{S}_k(\Gamma_0(N_k))$ of weight $k=2,4$ 
respectively~\cite{Bonisch:2022slo}. Remarkably the flux superpotential and the black hole entropy are expressible in 
terms of the periods of these modular forms. 

At the singular conifold fibres the degree of the Euler factors drops and 
the one-parameter models degenerate to rigid Calabi-Yau spaces and the period matrix can likewise be 
completely determined \cite{Bonisch:2025cax} in terms of periods of $f_4\in S_4(\Gamma_0(N))$, quasiperiods 
of $F_4\in \mathbb{S}_4(\Gamma_0(N))$ and periods of new types of meromorphic forms $g_{4,N}$ with 
non-vanishing residua as well as chain integrals, which are bounded by points of complex 
multiplications in the base of the Kuga-Sato variety. Interestingly the values of the periods, quasiperiods and open integrals at the  conifold nearest to the large radius point determine 
the asymptotic growth of the BPS indices at the latter~\cite{AKP}. The BPS indices contribute to the microscopic 
entropy of spinning black holes in five dimensions~\cite{Katz:1999xq}.  

Further factorisation of the Euler factors in special fibres are discussed in Section~\ref{sec:specialfibres}. In particular at points  
where a rational conformal field theory in the world sheet theory occurs, the Euler factors split, 
eventually over algebraic extensions of $\mathbb{Q}$, into linear factors with Gr\"ossencharacters, like 
for elliptic curves at points of complex multiplication. The $L$-function values for these characters involve 
$\Gamma$-functions at rational values.  For elliptic curves at complex multiplication 
that also follows from the Chowla-Selberg formula.  Together with the roots of the algebraic extension they determine the period matrix, 
which  in our examples can be alternatively computed by exactly Barnes integrals~\cite{Bonisch:2022slo}, corroborating the  
occurrence of these numbers. As in rational conformal theories for the Calabi-Yau worldsheet theories the correlation functions 
are determined by Feigin-Fuchs integrals with rational exponents, they are also  typically given in terms 
of $\Gamma$-functions values at rational values~\cite{Distler:1988ms}\cite{Fuchs:1989kz}. We therefore 
conjecture that rational conformal field theories RCFT can only occur if the Euler factors factorize  into linear
factors involving Gr\"o\ss encharacters. Similar observations have been made in~\cite{Jockers:2025fgv}.

In this paper we extend the arithmetic method of analysing special  properties of the Hodge 
structure in special fibres over higher co-dimensional subslices in the complex moduli space of multi 
parameter models of three- and fourfolds. The methods rely crucially on a fast algorithm to obtain the relevant Euler factors in the 
Hasse-Weil Zeta function. The latter is efficiently obtained by an implementation of Dwork's p-adic deformation 
of the periods at the MUM point, which was developed  for one-parameter families 
in~\cite{Candelas:2019llw}\cite{samol}\cite{Candelas:2021tqt}\cite{bonisch}\cite{Bonisch:2022mgw} and extended 
to multi-parameter Calabi-Yau  families in~\cite{Candelas:2024vzf}. It is strictly speaking 
still conjectural that it produces the Hasse-Weil zeta function obtained by the very tedious point counting 
on $X$ modulo $p$. However the numerical checks concerning period values~\cite{bonisch}\cite{Bonisch:2022mgw} and functional equations 
for the $L$-functions~\cite{Gegelia:2024vhd} are overwhelming. We improve first the method for the one-parameter models to include apparent singularities. This is necessary to provide lists of Euler factors for all one-parameter CY operators in the AESZ list \cite{cycluster}. This is important to identify one-parameter subfamilies and transitions in multi-parameter models as explained below, to provide further
checks of the modularity conjecture \cite{Gross2016OnTL}\cite{Gegelia:2024vhd}\cite{Blessemaster} and 
eventually to find new RCFT points in Calabi-Yau compactifications. In general the method to obtain the Euler factors for 
the full $m$-dimensional moduli space ${\cal M}_{\rm cs}$  by calculating the Frobenius traces using the 
p-adic deformation of the multivariate periods is too slow. We therefore restrict first to linear 
subspaces in ${\cal M}_{\rm cs}$ and derive a one-parameter differential operator of order $2m+2$ 
from which we evaluate the Euler factors by p-adic deformation of the latter. If the line crosses
a rank 2 attractor subslice, one expects a persistent factorisation. Choosing different lines 
and using the CRT one can reconstruct the subslice. In this way we found  r2A-nG subslices, 
which would be hard to find by other methods. Once a one-parameter subslice has been identified 
the persistent factor in the Euler factors  can be used to identify the one-parameter model based 
on its Euler factor. 

The method also detects consistent one-parameter subslices eventually followed by geometric 
transitions~\cite{Greene:1995hu}\cite{Klemm:1996kv}\cite{Morrison:1996pp}\cite{Candelas_a_Font_1998}. We provide 
a new way of studying them using the A-discriminants of Gelfand, Kapranov and Zelevinsky~\cite{gelfand1994discriminants}\cite{MR1264417}. 
We focus on the ones related to one faces and two faces of the reflexive polyhedron for 3-folds, 
but the technique works similar for three faces of 4-fold polyhedra.  The idea 
is to fulfill vanishing of the  A-determinant by setting the deformation parameters of the 
face polynomial  to special symmetric values. In particular if the Calabi-Yau manifold in question 
has a fibration structure whose base is geometrically described by the face, which can 
be read from the toric data as explained in \cite{Avram:1996pj}, we find then quite systematically 
that we can describe the lower dimensional model as a Hadamard product of the periods of the  
base geometry and the ones of the fibre geometry.

\section{Calabi--Yau period geometry and integral period bases}\label{sec:periodgeometry}
Due to the local Torelli theorem, periods locally parametrize the moduli space of Calabi-Yau $n$-folds $X$ 
and their geometry in an {\sl integral basis} is essential to study its special subslices. In this section we briefly recall this period geometry. 
Subsection \ref{sec:biliniears} introduces three important concepts: Bilinear relations, the Gauss-Manin connection and Griffiths transversality. These lead to special geometry on CY 3-folds and higher dimensional generalisations, discussed in detail in \cite{Ducker:2025wfl}, which can serve as 
a general reference. In subsection~\ref{sec:integralbases} we give explicit formulas for the {\sl integral  basis}, which is symplectic for $n$ odd ($n=3$) and symmetric for $n$ even ($n=4$), at  points of maximal unipotent monodromy (MUM). On the mirror manifold $\hat X$ the MUM point corresponds to 
large radius geometries and the integral basis can be inferred often from the standard $\hat \Gamma(T{\hat X})$ formalism, 
in terms of the topological data of $\hat X$. Here we restrict our self to conventional geometric 
realisations of the A-model geometries. A-model geometries $\hat X$
with non-trivial B-field, torsion  and non-commutative resolutions~\cite{Schimannek:2021pau} are 
important to understand the most general subslices and transitions~\cite{Katz:2022lyl}\cite{Katz:2023zan}\cite{Schimannek:2025cok}, generalisations 
of the $\hat \Gamma(T{\hat X})$ formalism and hence the different integral bases that can be associated to a  
Picard-Fuchs differential ideal on the $B$-model side. It should explain systematically the adhoc 
modifications of the $\hat \Gamma(T{\hat X})$ class formulas in the search of integral 
monodromy bases\footnote{Some of them have been already explained in \cite{Katz:2022lyl}\cite{Katz:2023zan}\cite{Schimannek:2025cok}.} in~\cite{arXiv:2505.07685}. From the arithmetic point of view these geometries  
will be discussed in \cite{BDKPT}. Subsection \ref{sec:toricambients} describes shortly the class of CY in toric ambient spaces, highlighting 
some formulas for the periods that are used in the subsequent analysis.            

\subsection{Bilinears and period geometry}
\label{sec:biliniears} 
Each Calabi--Yau manifold $X$ is endowed with two pairings that contain information about the complex structure and yield physical data such as D-brane charges.
The \emph{intersection form} in homology is the bilinear product
\begin{equation}\label{eq:Sigmahat}
    \hat{\Sigma}:\ H_n(X,\Z) \times H_n(X,\Z)\longrightarrow \Z
\end{equation}
given by the intersection numbers of $n$-cycles.
The second important pairing is the \emph{period matrix}
\begin{equation}\label{eq:periods}
    \mathbf{\Pi}:\ H_n(X,\Z)\times H^n(X,\Q)\longrightarrow \C\,,
\end{equation}
which consists of the integrals of cohomology classes in $H^n(X,\Q)$ over elements in $H_n(X,\Z)$.
We denote a topological basis for $H_n(X,\Z)$ by $\{\Gamma_i\}_{1\leq i\leq b_n}$ and denote the matrix coefficients of \cref{eq:Sigmahat} by $\hat{\Sigma}_{ij}=\Gamma_i\cap \Gamma_j$\,.
The intersection in homology induces a dual pairing in cohomology (the cup product), whose representation in the dual basis $\{\gamma^i\}_{1\leq i\leq b_n}$ with ${\int_{\Gamma_i}\gamma^j=\delta_i^j}$ is given by $\Sigma_{ij} = (\hat{\Sigma}^{-1})_{ij}$\,.
The unique $(n,0)$-form of $X$ is denoted by $\Omega$ and can be expanded as
\begin{equation}
    \Omega = \gamma^i\Pi_i\,,\quad \vec{\Pi} = \int_{\vec{\Gamma}}\Omega\,.
\end{equation}
The vector $\vec{\Pi}$ is referred to as \emph{period vector}\,.
The Kähler potential for the complex-structure moduli space can then be expressed as
\begin{equation}\label{eq:Kcs}
    e^{-K} = \ii^{n}\int_X\overline{\Omega}\wedge\Omega = \ii^n\vec{\Pi}^\dagger\Sigma\,\vec{\Pi}>0\,.
\end{equation}
\paragraph{}
Changes in the complex structure of $X$ generate a family of Calabi--Yau manifolds $\mathcal{X}$ defined over its moduli space with a projection $\pi:\ \mathcal{X}\longrightarrow \Mcs$. 
For families defined as complete intersections of $r$ hypersurfaces in toric ambient spaces, convenient\footnote{This is due to faithfulness of the parametrisation and the large complex structure limit being positioned at $\vec{z}=0$,  allowing for an identification with the Kähler moduli space of the mirror.} coordinates for $\Mcs$ are given by Batyrev coordinates \cite{Batyrev:1993oya}, see \cref{eq:batyrevz}.
The fibres $\pi^{-1}(\vec{z})\eqqcolon X_{\vec{z}}$ have $(n,0)$-forms $\Omega(\vec{z})$, which, together with the periods, vary smoothly with $\vec{z}$ along generic loci.
This leads to pure variations of Hodge structure \cite{MR2393625,deligne1997local}.
\paragraph{}
For certain values of $\vec{z}$, the fibres $X_{\vec{z}}$ become singular. 
The collection of these points form the discriminant $\Delta\subset \Mcs$ consisting of irreducible components $\Delta_i$ of codimension one. 
For toric models, the components can be obtained from the underlying polytope~\cite{gelfand1994discriminants}, see \cref{eq:disclambda}. 
While the topological basis $\{\Gamma_i\}_{i}$ is locally invariant under changes in the complex structure, it undergoes monodromy transformations when transported around singular loci.
In the next subsection, we will review how mirror symmetry yields such a basis with integral monodromy transformations.
\paragraph{}
The middle cohomology has a filtration $F^0\subset\ldots \subset F^n=H^n(X_{\vec{z}},\C)$ with $F^i= \bigoplus_{j=0}^i H^{n-j,j}(X_{\vec{z}})$.
These become locally free constant sheaves $\mathcal{F}^p$ when passing to families ~\cite{Voisin_2002}.
The periods are flat sections of the Gauss--Manin connection, which acts on above filtrants as
\begin{equation}
    \nabla:\ \mathcal{F}^n\longrightarrow \mathcal{F}^n\otimes \Omega_{\Mcs}^1\,.
\end{equation}
Here, $\Omega_{\Mcs}^1$ denotes the bundle of one-forms on the complex-structure moduli space.
Griffiths transversality restricts the image of $\mathcal{F}^p$ to the next filtrant
\begin{equation}
    \label{eq:Griffithstransversality}
        \nabla \mathcal{F}^p \subset \mathcal{F}^{p+1}\otimes \Omega_{\Mcs}^1\,.
\end{equation}
The filtration implies a differential equation for the period vector
\begin{equation}\label{eq:GMAi}
    (\partial_{z^i}-A_i)\,\vec{\Pi} =0\,,
\end{equation}
with matrices $A_i\in \Q(\vec{z})$\,. 

As a consequence, only the $n$-th derivative of $\Omega(\vec{z})$ has non-trivial cup-product with $\Omega(\vec{z})$
\begin{equation}\label{eq:Griffithsexpl}
        \int_X\partial_\mathcal{I} \Omega\wedge \Omega = \begin{cases}
        0 & \text{if }\lvert \mathcal{I}\rvert <n\,,\\
        C_{i_1,\ldots ,i_n}(\vec{z}) & \text{if }\lvert \mathcal{I}\rvert =n\,.
        \end{cases}
    \end{equation}
where the \emph{Yukawa couplings} $C_{i_1,\ldots ,i_n}$ are rational functions in $\vec{z}$\,.
For $n>3$, these decompose into triple couplings \cite{Greene:1993vm,Mayr:1996sh,Klemm:1996ts}.
\paragraph{}
The first order system of the Gauss--Manin connection \eqref{eq:GMAi} can be recast into scalar form known as Picard--Fuchs differential ideal (PFDI). 
It is generated by a set of Fuchsian operators $\{\mathcal{L}^{(k_i)}_i(\vec{z})\}_i$ of orders $k_i$ that has a $b_n$-dimensional solution space containing the periods.
The minimal number of generators is model dependent; for one-parameter models, there is a single operator of order $b_n$\,.
Any single period determines the PFDI and thus the remaining periods. Explicit formulas for the periods and the $\{\mathcal{L}^{(k_i)}_i(\vec{z})\}_i$  
and hints of their derivation in the class of CY in toric ambient spaces, can be found in \cref{sec:toricambients}. 
The solutions to the PFDI form a complex vector space and their precise linear combinations giving the integral period vector $\vec{\Pi}(\vec{z})$ follow in particular 
for this class from mirror symmetry as will be made explicit in the next section \ref{sec:integralbases}.
\paragraph{}
The form $\Omega$ and thus the periods are subject to a Kähler gauge symmetry $\Omega\mapsto e^f\Omega$ for holomorphic functions $f\equiv f(\vec{z})$\,. 
The Kähler potential \eqref{eq:Kcs} transforms as $K\mapsto K-f-\bar{f}$.
One defines
\begin{align}
    \begin{split}\label{eq:covariantD}
        D_i:\ H^{n,0}(X,\C)&\rightarrow H^{n-1,1}(X,\C)\,,\\
        \Omega  &\mapsto \partial_i\Omega  + (\partial_i K)\Omega\,,
    \end{split}
    \end{align}
such that $D_i\Omega$ transforms covariantly\footnote{One can also define a connection mapping $H^{n-k,k}(X,\C)\rightarrow H^{n-k-1,k+1}(X,\C)$ that is covariant under  
K\"ahler transformations and reparametrizations on ${\cal M}_{\rm cs}$, see \cite{MR3965409} for a review.}.
Since physical quantities are invariant under Kähler gauge 
transformations, they typically involve covariant derivatives 
with compensating factors such as $e^{K/2}$\,, see for example \cref{eq:scalarpot}.

\subsection{Mirror symmetry and integral period bases}
\label{sec:integralbases} 
Mirror symmetry relates the complex-structure moduli space of a Calabi--Yau family $\mathcal{X}$ with the complexified Kähler moduli space of its mirror partner $\hat{\mathcal{X}}$\,.
The identification is local and valid only in the large complex structure and large volume limit.
If a mirror pair exists, the Hodge numbers satisfy $h^{n-1,1}(X)=h^{1,1}(\hat{X})$ and $h^{1,1}(X)=h^{n-1,1}(\hat{X})$.
We choose the point of large complex structure to be at $\vec{z}=0$. At this point, called MUM point, the monodromy is maximally unipotent. For example in Batyrev construction 
MUM points are at $\vec z $ in the coordinates \eqref{eq:batyrevz}.  
For a basis of curves $\{\mathcal{C}^1,\ldots ,\,\mathcal{C}^{h^{1,1}}\}$ on the mirror manifold $\hat{X}$ together with its complexified Kähler form $\omega_\C = B-\ii \omega$, the complexified Kähler moduli space is parametrized by
\begin{equation}
    t^k = \frac{1}{(2\pi\ii)^2\alpha'}\int_{\mathcal{C}^{k}}\omega_\C\,.
\end{equation}
The volume of the curve $\mathcal{C}^k$ is given by $v^k=\frac{1}{4\pi^2\alpha'}\Im t^k$.
On the original family $\hat{X}$ near the MUM point $\vec{z}=0$, the torus period is denoted by $X^0$. 
There are additional $m=h^{n-1,1}$ solutions to the PFDI with single-logarithmic singularities \cite{deligne1997local} at $\vec{z}=0$, called $X^i$ for $1\leq i\leq m$\,.
By a local Torelli theorem, these $m+1$ periods form homogeneous coordinates for the moduli space of complex structure.
The mirror maps
    \begin{equation}\label{eq:mirrormap}
        t^i(\vec{z}) = \frac{X^i(\vec{z})}{X^0(\vec{z})} = \frac{\log(z^i)}{2\pi\ii}+\mathcal{O}(z)\,.
    \end{equation}
are inhomogeneous coordinates and are identified with the Kähler moduli of the mirror family above. 
By construction, the large complex-structure limit $z^i\rightarrow 0$ corresponds to the large volume limit $t^i\rightarrow \ii\,\infty$\,.
\paragraph{}
The period integrals in \cref{eq:periods} compute central charges of A-branes in type IIB on $X$. 
Homological mirror symmetry \cite{Kontsevich:1994dn} equates them to the central charges of B-branes in type IIA on its mirror $\hat{X}$ in the large volume/complex-structure limit.
Identifying the B-branes with structure sheaves $\mathcal{E}$ on $\hat{X}$ and on iterated intersections of divisors $D_i$, $1\leq i\leq h^{1,1}(\hat{X})=m$, one obtains elements of an asymptotic integral period basis via~\cite{iritani2009integral,MR2683208}
\begin{equation}\label{eq:Piasy}
    \Pi_{\mathcal{E}}^{\text{asy}} = \int_{\hat{X}}e^{\sum_i t^iJ_i}\,\hat{\Gamma}(T\hat{X})\,\text{ch}(\mathcal{E})\,.
\end{equation}
A basis change is necessary to bring the intersection form (given by the Hirzebruch--Riemann--Roch pairing) into block-anti-diagonal form. 
We refer to the literature for a more detailed account of the computation of B-brane charges and suitable basis changes~\cite{Gerhardus:2016iot,MR1319280,libgober1998chern,CaboBizet:2014ovf,Ducker:2025wfl,Piribauer2025}.
We choose a normalization of the Frobenius bases as
\begin{equation}
    \label{eq:frobbasis}
        \vec{\varpi}^ {(n=3)}(\vec{z}) ~ {\footnotesize \begin{pmatrix}
        1\\
        \log z^1\\
        \vdots\\
        \log z^m\\
        \frac{1}{2!}c_{mij}\log z^i\,\log z^j\\
        \vdots \\
        \frac{1}{2!}c_{1ij}\log z^i\,\log z^j\\
        \frac{1}{3!}c_{ijk}\log z^i\,\log z^j\,\log z^k\\
        \end{pmatrix}},\ 
        \vec{\varpi}^ {(n=4)}(\vec{z}) ~ {\footnotesize\begin{pmatrix}
        1\\
        \log z^1\\
        \vdots\\
        \log z^m\\
        \frac{1}{2!}c_{\vec{1} ij}\log z^i\,\log z^j\\
        \vdots \\
        \frac{1}{2!}c_{\vec{m} ij}\log z^i\,\log z^j\\
        \frac{1}{3!}c_{mijk}\log z^i\,\log z^j\,\log z^k\\
        \vdots\\
        \frac{1}{3!}c_{1ijk}\log z^i\,\log z^j\,\log z^k\\
        \frac{1}{4!}c_{ijkl}\log z^i\,\log z^j\,\log z^k\,\log z^l\\
        \end{pmatrix}}\,,
    \end{equation}
    where $c_{i_1,\ldots ,i_n}=D_{i_1}\cap \ldots \cap D_{i_n}$ denote the classical intersection numbers. 
    These are rotated into an integral basis $\vec{\Pi} = T_{0}\vec{\varpi}$ by transition matrices $T_{0}$ depending on topological data of the mirror $\hat{X}$.
    \paragraph{}
    For $n=3$, one finds
    \begin{equation}
    \label{eq:Tmum3}
        T_{0} = \begin{pmatrix}
        1 & 0 & 0 & 0 \\
        0 & \frac{\mathbbm{1}}{2\pi\ii} & 0 & 0\\
        -\frac{c_2\cdot \vec{\tilde D}^T}{24} & \mathbf{A} & \frac{\mathbbm{1}}{(2\pi\ii)^2} & 0\\
        \frac{\chi\,\zeta(3)}{(2\pi\ii)^3} & -\frac{c_2\cdot \vec{D}}{24(2\pi\ii)} & 0 & \frac{1}{(2\pi\ii)^3}
        \end{pmatrix}
    \end{equation}
    with $\vec{\tilde D} = (D_m,\ldots ,D_1)$\,.
    The entries of the matrix $\mathbf{A}$ are defined by ${A_{ij} = \frac{1}{2}\int_{\hat{X}} i_*c_1(D_i)\wedge J_j}$. 
    Integer shifts reflect the Sp$(4,\Z)$ symmetry of the period lattice and leave the intersection form invariant.
    The complex-structure moduli space of threefolds is \emph{special} Kähler, implying the existence of a prepotential
    \begin{equation}\label{eq:prepotential}
        \mathcal{F} = \frac{c_{ijk}}{3!}t^it^jt^k - \frac{A_{ij}}{2!}t^it^j - \frac{c_2\cdot D_i}{24}t^i - \frac{\chi\,\zeta(3)}{2(2\pi\ii)^3} + \mathcal{F}_\text{inst}(\vec{q})
    \end{equation}
    with instanton corrections given by
    \begin{equation}\label{eq:instantons}
        \mathcal{F}_\text{inst}(\vec{q}) = \frac{1}{(2\pi\ii)^3}\sum_{\vec{l}>0} n_{\vec{l}}\,\mathrm{Li}_3\left(\vec{q}^{\vec{l}}\right).
    \end{equation}
    The period vector can then be written as
    \begin{equation}\label{eq:PiasCY3}
        \vec{\Pi} = X^0\left( 1, t^1, \ldots , t^{h^{2,1}} ,\partial_{t^{h^{2,1}}}\mathcal{F}, \ldots ,\partial_{t^1}\mathcal{F},2\mathcal{F} - t^i\partial_i \mathcal{F}\right)
    \end{equation}
    and the Yukawa couplings \eqref{eq:Griffithsexpl} in the mirror coordinates $\vec{t}$ is given by \begin{equation}
        C_{ijk}=\partial_i\partial_j\partial_k\mathcal{F}= c_{ijk}+ \sum_{\vec{l}>0}n_{\vec{l}}\, l_i l_j l_k\frac{\vec{q}^{\vec{l}}}{1-\vec{q}^{\vec{l}}}\,.
    \end{equation}
    \paragraph{}
    The case $n=4$ is more involved due to the basis choice for $H_4$ in terms of intersections of toric divisors.
    We denote a certain basis by $H_\alpha = D_{\alpha_1}\cap D_{\alpha_2}$, $1\leq \alpha\leq h^{2,2}(X)$, where we defined implicitly a map $H_4(\hat{X})\rightarrow H_2(\hat{X})\times H_2(\hat{X})$\,.
    To obtain integral pairing in the cohomology, one must scale the K-theory classes of certain sheaves with a matrix $S$.
    We find that a rotation of the periods in \cref{eq:Piasy} rendering the intersection form block-anti-diagonal given by
    \begin{equation}\label{eq:Tmum4}
        T_{\text{MUM}} = S\begin{pmatrix}
        1 & 0 & 0 & 0 &0\\
        0 & \frac{\mathbbm{1}}{2\pi\ii} & 0 & 0 & 0\\
        -\frac{c_2\cdot H_{\alpha}}{24} & -\frac{c_{ii\alpha}}{2(2\pi\ii)} & \frac{\mathbbm{1}}{(2\pi\ii)^2} & 0 & 0\\
        
        \frac{c_2\cdot \tilde{D}_i^2}{48} + \frac{\zeta(3)\,c_3\cdot \tilde{D}_i}{(2\pi\ii)^3} & \frac{\tilde{c}_{ijjj}/6+c_2\cdot \tilde{D}_i D_j/24}{2\pi\ii} + R_{ij} & -\frac{\mathbf{P}_{ii,\alpha}}{2(2\pi\ii)^2} & \frac{\mathbbm{1}}{(2\pi\ii)^3}& 0 \\
        
        \frac{7c_2^2-4c_4}{5760}-1 & \frac{\zeta(3)\,c_3\cdot D_i}{(2\pi\ii)^4} & \mathfrak{c}^{ij\alpha}\,c_2\cdot D_i\cdot D_j & 0 & \frac{1}{(2\pi\ii)^4}
        \end{pmatrix}S\,.
    \end{equation}
    Here, $\tilde{c}_{i,j,j,j}=c_{m-i+1,j,j,j}$ and $\mathfrak{c}^{ij\alpha}$ are inverses of the classical triple couplings given by ${c_{ij\alpha}=D_i\cap D_j\cap H_\alpha}$.
    They are defined implicitly by the condition $\mathfrak{c}^{ij\alpha}c_{ij\beta} = \delta^\alpha_\beta$\,.
    The components of the matrix $\mathbf{R}$ in \cref{eq:Tmum4} read
    \begin{align}
        R_{m-i+1,j} = \begin{cases}
        \frac{1}{12} D_i\cdot D_j\cdot \left(c_2-3 D_i\cdot D_j+2 D_i^2+2 D_j^2\right)&,\text{for }i>j\,,\\
        \frac{1}{24} D_i^2\cdot \left(c_2+D_i^2\right)&,\text{for }i=j\,,\\
        0&,\text{else.}
        \end{cases}
    \end{align}
    and $\mathbf{P}_{\alpha,ii}$ expresses $D_{m-i+1}\cap D_{m-i+1}$ in the basis generated by $H_\alpha$\,.
    The four-point couplings are reducible into triple couplings $C_{ijkl}=C_{ij\alpha}C_{kl}^{\ \ \alpha}$. 
    The triple couplings are again corrected by genus-zero curve counts
    \begin{equation}\label{eq:invariantsijalpha}
        C_{ij}^{\ \ \alpha} = c_{ij}^{\ \ \alpha} + \sum_{\vec{l}>0}n_{\vec{l}}^{\alpha}\, l_i l_j \frac{\vec{q}^{\vec{l}}}{1-\vec{q}^{\vec{l}}}\,.
    \end{equation}
    \paragraph{}
    Many identities among the periods are apparent only in a basis with block-anti-diagonal intersection form. This includes the splitting of the period matrix into a unipotent and semisimple component \cite{Ducker:2025wfl} useful to obtain $\epsilon$-factorized differential equations in the computations of Feynman integrals \cite{Henn:2013pwa}, see also \cite{Gorges:2023zgv,Duhr:2025lbz}.
    Depending on the dimension of the manifold, the integral period basis described above has intersection form
    \begin{equation}\label{eq:intformscohom}
        \Sigma_{(n=3)}=\left(\begin{array}{cccccc}
             &&&&&1\\
             &&&&\iddots&\\
             &&&1&&\\
             &&-1&&&\\
             &\iddots &&&&\\
             -1&&&&&\\
        \end{array}\right),\quad \Sigma_{(n=4)}=\left(\begin{array}{ccccccccc}
             &&&&&&&&1\\
             &&&&&&&-1&\\
             &&&&&&\iddots&&\\
             &&&&&-1&&&\\
             &&&&\sigma &&&&\\
             &&&-1&&&&&\\
             &&\iddots &&&&&&\\
             &-1&&&&&&&\\
              1&&&&&&&&\\
        \end{array}\right)
    \end{equation}
    with $\sigma\in\text{Mat}_{h^{2,2}\times h^{2,2}}(\Z)$ the intersection form of the chosen basis $\{H_\alpha\}_\alpha$ for $H_4(\hat{X},\Z)$\,.
\subsection{Hypersurfaces and complete intersections in toric ambient spaces}\label{sec:toricambients}
We consider toric spaces described by a reflexive lattice polytope $\Delta$ with a triangulation $\Sigma$ that is star, fine and regular, for which the topological data used in the previous section for the construction of an integral period basis can be obtained straightforwardly. 
Reflexivity implies that the polar dual of $\Delta$ defined by
\begin{equation}\label{eq:polardual}
    \Delta^\circ = \left\{x\in \Z^{n+1}\otimes \R\,\big|\, \braket{x | y}\geq -1\ \forall y\in\Delta\right\},
\end{equation}
is again a lattice polytope.
Batyrev showed in \cite{Batyrev:1993oya} that generic sections of the anti-canonical bundle ($\Delta_{-\mathcal{K}}=\Delta^\circ$)
\begin{equation}\label{eq:defpolyDelta}
    P_{\Delta}(\vec{a},\vec{x}) = \sum_{\nu^\circ\in\Delta^\circ}a_{\nu^\circ} \prod_{\rho\in\Sigma(1)} x_{\rho}^{\braket{\rho | \nu^\circ}+1}\,,
\end{equation}
have smooth Calabi--Yau manifolds as vanishing locus. 
The mirror family is obtained by exchanging $\Delta$ for $\Delta^\circ$ and vice versa.
Moreover, their Hodge numbers can be computed from the polytopes alone
\begin{equation}\label{eq:h11Bat}
    h^{n-1,1}(\mathcal{X}_\Delta) = l(\Delta) - (n+2) -\sum_{\text{codim}(\theta)=1}\hat{l}(\theta) + \sum_{\text{codim}(\theta)=2}\hat{l}(\theta)\cdot \hat{l}(\theta^*)\,,
\end{equation}
where mirror symmetry implies
\begin{equation}
    h^{1,1}(\mathcal{X}_\Delta)=h^{n-1,1}(\mathcal{X}_{\Delta^\circ})\,.
\end{equation}
The generalization to complete intersections of $r>1$ hypersurfaces is due to Batyrev--Borisov~\cite{MR1463173}.
If the $n+r$ dimensional polytope has a Minkowski decomposition
\begin{equation}\label{eq:minkowskisum}
    \Delta = \Delta_1+\ldots +\Delta_r\,,
\end{equation}
the intersection of the vanishing loci of generic sections $\P_{\Delta_1}$ yields again Calabi--Yau manifolds.
This decomposition is also called a numerically effective (NEF) partition.

The Mori cone of the ambient space is identified with the cone of strictly convex, piecewise linear functions.
We will denote the generators of the cone by
\begin{equation}
    l^{(i)} = \left(-l^{(i)}_{01},\ldots ,-l^{(i)}_{0r} \,;\ l^{(i)}_1,\ldots ,l^{(i)}_n\right),
\end{equation}
where the entries $-l^{(i)}_{0j}$ correspond to the origin of $\Delta_j$\,.
These define the Batyrev coordinates for the complex-structure moduli space
\begin{equation}\label{eq:batyrevz}
    z^i = \frac{\prod_{j=1}^{\abs{\Delta}}a_j^{l^{(i)}_j}}{\prod_{k=1}^{r}(-a_{0k})^{l^{(i)}_{0k}} }
\end{equation}
with the property of $\vec{z}=0$ being a MUM point at large complex structure.
There can be further MUM points, which are mapped to large volume limits of different phases (different NEF cones) under mirror symmetry. 

The algebraic torus inside a toric ambient space contains an $n$-torus, whose periods can be given in terms of the Mori-cone generators $l$ above by \cite{MR1316509,MR1319280}
\begin{equation}\label{eq:fundamentalpiclosed}
        \varpi_0(\vec{z}) = \sum_{\vec{n}\in\N_0^m} \frac{\prod_{i=1}^r\Gamma(1+\sum_{j=1}^ml^{(j)}_{0i}n_{j})}{\prod_{i=1}^\delta \Gamma(1+\sum_{j=1}^ml^{(j)}_{i}n_{j})}\prod_{j=1}^m (z^j)^{n_j}\,,
    \end{equation}
where $\delta$ is the number of points not being origins in \cref{eq:minkowskisum}.
The polytope contains information about the singular loci of the family. 
Consider the extended polytope $\bar{\Delta}$ with points $\{(e_i,v_i)\ |\ v_i\in \Delta_i\,,\ 1\leq i\leq r\}$\,. 
For each face of $\bar{\Sigma}$ with linear dependencies among its points (this includes $\bar{\Delta}$), one collects the kernel in a matrix with entries $s_{ij}$\,.
Each such face yields an so called A-discriminant component parametrised by $(\lambda_1:\ldots :\lambda_r)\in \P^{r-1}$ satisfying the condition~\cite{gelfand1994discriminants}, see also~\cite{Aspinwall:2017loy},
\begin{equation}\label{eq:disclambda}
    z^i =  \sum_{j=1}^{\abs{\bar{\Sigma}}} \left(\sum_{k=1}^r s_{kj}\lambda_k\right)^{s_{ij}}\,.
\end{equation}
By eliminating $\lambda$, one obtains polynomials in $\vec{z}$ whose vanishing locus is the discriminant component.

\section{Special loci in type IIB compactifications}
The complex-structure moduli spaces of Calabi--Yau threefolds contain loci of particular interest from both a mathematical and physical perspective.
We differentiate between splittings of the Hodge structure of smooth fibres and limiting mixed Hodge structures at singularities of the Calabi--Yau family.
Physically, supersymmetric flux vacua in type IIB compactifications are an example of the former, while singularities can signal transitions to different families via black hole condensation.
Both concepts are reviewed in the following two subsections, where we will furthermore discuss the extension to its non-perturbative description in F-theory. 
\subsection{Flux vacua}
Compactification of string theory yields massless scalar fields in the four-dimensional theory, which must be stabilized by a scalar potential.
One way to obtain such a potential is to use an O$3/$O$7$-orientifold projection to break supersymmetry to $N=1$~\cite{Grimm:2004uq}. 
In the setup we use, $h^{2,1}=h_-^{2,1}$ and $h^{1,1}=h^{1,1}_+$ removes $B_2$ and $C_2$ from the spectrum and leaves $h^{2,1}$ complex-structure moduli $z_i$, the axio-dilaton $\tau$ together with $h^{1,1}$ Kähler moduli
\begin{equation}
    T_\alpha = -h_\alpha + \frac{\ii}{2}c_{\alpha\beta\gamma} v^\beta v^\gamma
\end{equation}
with $h_\alpha$ and $v^\alpha$ coming from the reduction of the metric and $C_4$, respectively. 
Their kinematic terms in the $N=1$ Lagrangian are given by $g_{i\bar{\jmath}}\,\partial_\mu\phi^i\partial^\mu\phi^{\bar{\jmath}}$ with $g_{i\bar{\jmath}}=\partial_{\phi^i}\partial_{\bar{\phi}^{\bar{\jmath}}}\,K$ the Kähler metrics with Kähler potentials
\begin{equation}
    K_{\text{K.s.}} = -2\log\left(c_{\alpha\beta\gamma} v^\alpha v^\beta v^\gamma\right)\,,\quad K_\tau = -\log\left(-\ii(\tau - \overline{\tau})\right)
\end{equation}
and $K_{\text{c.s.}}$ given in terms of the periods in \cref{eq:Kcs}. 
The total Kähler potential is denoted by $K_{\text{tot}} = K_{\text{c.s.}} + K_{\text{K.s.}} + K_\tau$\,.
\paragraph{}
After integrating-out the auxiliary fields $F^i$ of the superfields, the $N=1$ theory obtains a scalar potential 
\begin{equation}\label{eq:scalarpot}
    V = e^{K_\text{tot}} \left(\sum_{A,\overline{B}} g^{A\overline{B}} D_AW\,D_{\overline{B}}W-3\abs{W}^2\right),
\end{equation}
where the Gukov--Vafa--Witten superpotential~\cite{Gukov:1999ya,Taylor:1999ii} is given by
\begin{equation}
    W(\vec{z}) = \int G_3 \wedge \Omega(\vec{z}).
\end{equation}
Here, $G_3=F_3-\tau\,H_3$ and the summations in \cref{eq:scalarpot} run over all moduli fields.
The covariant derivatives are given by $D_A = \partial_A + \partial_A K_{\text{tot}}$\,.
The supersymmetry variations include the superpotential $W$ together with its derivatives and unbroken supersymmetry in a Minkowski background $(\Lambda =0)$ requires
\begin{equation}\label{eq:vacuacond}
    W=\partial_{z^A}W=0\,,\quad \forall\,A\,.
\end{equation}
This implies $G_3\in H^{2,1}(X)$ and, together with a reality condition and a Dirac--Zwanziger quantisation of the field strengths,
\begin{equation}
    F_3\,,\, H_3\in (H^{2,1}\oplus H^{1,2})\cap H^3(X,\Z)\,.
\end{equation}
The two elements $F_3$ and $H_3$ generate a rank-two lattice in the middle cohomology.
In the one-parameter case with $\dim H^3=4$, fibres with these properties are called rank two attractor points.
They are of special interest in the arithmetic analysis of rank-four motives due to their modular properties. 
We will discuss them in more detail in \cref{sec:arithmeticmethods}.
\paragraph{}
In M/F-theory compactifications on fourfolds, the scalar potential comes from the field strength $G_4$ of the 3-form field of M-theory.
The supersymmetric flux vacua conditions \eqref{eq:vacuacond} then imply
\begin{equation}
    G_4\in H^{2,2}_\text{prim}(X)\text{\quad and\quad} G_4+\frac{c_2}{2}\in H^4(X,\Z)\,.
\end{equation}
For three-dimensional $N=2$ supergravity obtained, for example, from M-theory on a CY fourfold, the coefficients of $\abs{W}^2$ is $-4$ instead of $-3$\,.
\paragraph{}
Background values of the field strengths demand the inclusion of brane source terms in the action. 
Integration of the altered equations of motions imply the tadpole conditions
\begin{align}
    n=3:&\quad 0=\frac{1}{2\kappa_{10}^2T_{\text{D}3}}\int_X F_3\wedge H_3 + Q_3=0\,,\\
    n=4:&\quad 0=\frac{1}{4\kappa_{11}^2T_{\text{M}2}}\int_X G_4\wedge G_4 + N_{\text{M}2} = \frac{\chi}{24}\,.
\end{align}
These include the gravitational constants $\kappa_D$, the brane tensions $T_{\text{brane}}$, the D$3$-brane charge $Q_3$ and number of M$2$-branes $N_{\text{M}2}$ and the Euler number of the fourfold $\chi$\,.
\subsection{Topological transitions and black hole condensation}\label{sec:bhc}
As the complex-structure moduli approach a conifold locus, a number of three-spheres shrink. 
The singularity can be resolved by gluing in two-spheres, which satisfy the same relations in homology as their $S^3$-counterparts.
Such a transition changes the Hodge numbers of the manifold and thus connects the moduli spaces of two separate families.
In the seminal article \cite{Greene:1995hu}, the authors showed that additional hypermultiplets appear in the string theory, that remove some of the vectormultiplets via the Higgs effect.
We will review this process and the implications on B-model quantities in the following.
\paragraph{}
Transitions due to different desingularizations of conic singularities were known to mathematicians since the 1980s \cite{CLEMENS1983107,Friedman} and Reid's conjecture \cite{Reid} that all Calabi--Yau families are connected by such transitions is still open. 
For complete intersections in products of projective spaces, the statement has been proven in \cite{Green:1988bp}.
Locally, the two desingularisations can be understood as \cite{Candelas:1989js}
\begin{equation}
    \begin{array}{ccc}
        \sum_{i=1}^4 w_i^2=0 & \hspace{1cm}\longrightarrow &\hspace{1cm} \sum_{i=1}^4 w_i^2=\epsilon\,, \\
        \rotatebox{90}{=} & & \\
        X\,Y-U\,V =\det \begin{pmatrix}
        X&U\\V&Y
    \end{pmatrix} =0 & \hspace{1cm}\longrightarrow &\hspace{1cm}\begin{pmatrix}
        X&U\\V&Y
    \end{pmatrix}
    \begin{pmatrix}
        \lambda_1\\ 
        \lambda_2
    \end{pmatrix}=0\,.
    \end{array}
\end{equation}
Here $(\lambda_1:\lambda_2)\in \P^1\cong S^2$ parametrises the small deformation and $\epsilon$ a smoothing of the singularity.
\paragraph{}
Physically, the singularity is cured by a condensation of extremal black holes \cite{Greene:1995hu}.
Strominger showed in \cite{Strominger:1995cz} that a D$3$-brane supported on such an $S^3$ yields a BPS black hole whose mass 
$M= e^{\frac{K}{2}}\abs{\int_{S^3}\Omega}$  shrinks with its $\Omega$-volume.
These black holes correspond to hypermultiplets in the four-dimensional theory and descend into the effective theory as their masses decrease.
\paragraph{}
In four-dimensional type IIB compactifications, the number of vectormultiplets is $h^{2,1}(X)$.
Besides the universal hypermultiplet, there are additional $h^{1,1}(X)$ hypermultiplets that are neutral under the vectors. 
In contrast, the hypermultiplet descending from the shrinking D$3$-brane is charged under the U$(1)$ vector coming from the shrinking cycle. 
Giving a vacuum expectation value to the hypermultiplet, the Higgs effect gives a mass to the vectormultiplet and removes it from the effective theory.
The singularity in the Coulomb branch is resolved into a mixed (Coulomb+Higgs) branch.
In the more general case of $k$ shrinking three-cycles subject to $r$ homological relations, each of the cycles introduces a hypermultiplet that gives its vectormultiplet a mass.
However, since there are only $k-r$ vectors to be ``Higgsed'', the other $r$ hypermultiplets remain in the spectrum of the theory. 
In the example discussed in \cite{Greene:1995hu}, there are 16 three-spheres that vanish at a conifold locus of the quintic with $(h^{2,1},h^{1,1})=(101,1)$\,.
These satisfy a single relation and the transition leads to $h^{2,1}=101-(16-1)=86$ and $h^{1,1}=1+1=2$ of the hypersurfaces $X_{22211}$\,.
\paragraph{}
In type IIA, a similar phenomenon includes the wrapping of D$2$-branes around $k$ shrinking two-spheres.
The U$(1)^{h^{1,1}+1}$ gauge group of the four-dimensional theory is enhanced to U$(1)^{h^{1,1}+1-k}\times G$, where intersections of the shrinking cycles yields the Cartan matrix of~$G$~\cite{Katz:1996ht,Klemm:1996kv}.
A dual perspective in heterotic string theory was given in~\cite{Candelas_a_Font_1998}.
At the intersection of singular loci, localised fundamental matter appears \cite{Katz:1996fh,Katz:1996xe}.
In the study of type IIB on the mirror family $X$, the shrinking of curves is represented by a vanishing of the Kähler moduli $t_i$ and leads via a resummation of instanton corrections to the prepotential of a model with $h^{1,1}(\tilde{X})=h^{1,1}(X)-k$\,.
For example, the intersection ring follows immediately with $R(\tilde{X})=R(X)|_{t_1=\ldots =t_k=0}$ and the Euler characteristic is corrected by the degree-zero instantons in $t_i$, $1\leq i\leq k$\,,
\begin{equation}
    \chi(\tilde{X}) = \chi(X) - 2 \sum_{\vec{l}\in \N^k_0}n_{\vec{l},\vec{0}}\,,
\end{equation}
where we used Li$_3(1)=\zeta(3)$ in \cref{eq:prepotential,eq:instantons}.
\paragraph{}
For hypersurfaces in toric ambient spaces, loci of transitions can sometimes\footnote{See the remarks in \cref{sec:remarks}.} be deduced from the polytope of the anti-canonical divisor:
moduli parametrising deformations whose points lie on a single edge are given by binomial coefficients at the transition locus. 
For each shrinking cycle in $H_{2}(\hat{X})$, one obtains two vanishing periods of $H_3(X)$\,.
The periods of such models can be written as Hadamard products of a K3 and a zero-fold.
We identify new transitions at conifold degenerations where three-spheres shrink. 
Torically, they arise at suitable values of moduli corresponding to points inside two-dimensional faces of the polytope. 
In the case of a single shrinking three-cycle, its dual cycle rips open and the remaining periods satisfy an inhomogeneous differential equation. 
The transition induces an integral period structure on the one-parameter model.
Explicit examples of both types of transitions are discussed in \cref{sec:sctrans,sec:contrans}.
A general analysis together with a cataloguing of transitions is given in \cref{sec:gentrans}.
\paragraph{}
Transitions also appear in higher-dimensional Calabi--Yau families. 
In the context of M- and F-theory compactifications, we study such special singular loci of fourfolds in \cref{sec:fourfolds}.
The results for strong-coupling and conifold transitions translates directly to CY3- and K3-fibred fourfolds with deformation points on edges and two-dimensional faces, respectively.
The precise generalisation together with examples can be found in \cref{sec:CY4fibr12}.
The shrinking $S^4$ along conifold loci can be used to construct a supersymmetric flux vacuum \cite{Grimm:2009ef,Cota:2017aal,Ducker:2025wfl}\,.
On the other hand, one can describe the one-parameter model satisfying the flux vacuum conditions again as a Hadamard product, where one factor is given by the fibre and the other originates from the base geometry, see \cref{sec:CY4fibr3}.

\section{Arithmetic methods}\label{sec:arithmeticmethods}
One problem in the study of CY manifolds is to find rank two attractor points in the complex structure moduli space. For a given point $z_{*}\in \M_{cs}$, this is equivalent to the existence of two vanishing periods,
\begin{align}
    f^T \Sigma \Pi(z_*)=0\,,\quad h^T \Sigma \Pi(z_*)&=0\,,
\end{align}
with integer coefficient vectors $f$ and $h$, which correspond to the chosen fluxes. On the level of the cohomology this implies the existence of a rank two lattice inside
\begin{equation}
    H^3(X_z(\C),\Q) \cap \left(H^{2,1}(X_z(\C)) \oplus H^{1,2}(X_z(\C) )\right).
\end{equation}
Apart from a few candidate choices for $z_*$, one has to check in principle all $z\in \Q$ to decide whether or not a family has attractor points or not. An elegant solution to this problem was provided in \cite{Candelas:2019llw}. The key idea is to treat the CY $X$ as a manifold over finite fields $\F_p$ with elements $\{ 0, 1, \dots , p-1\}$. It is clear that $X$ over one fixed $\F_p$ contains much less information than $X/\C$ but one recovers more and more information from reducing $X$ over many $\F_p$. Using the \emph{Chinese remainder theorem} (CRT), one can reconstruct a rational number (and elements in extensions of $\Q$) after sufficiently many reductions. The upshot is that we translate the problem of having one field $k$ with infinitely many elements to infinitely many fields $\F_p$ with finitely many elements. In practice, we compute the relevant quantities up to some prime number $p$ and the CRT returns a candidate $z_*\in \overline{\Q}$ that can be explicitly checked to allow for orthogonal $f$ and $g$. In the following, we will apply this arithmetic approach in a similar fashion to find splittings in the cohomology of multi-parameter threefolds and fourfolds. 

\subsection{Preliminaries}
The observation that multiple cohomology theories produce equivalent information is formalized in the theory of motives. The motive associated to an algebraic variety has realizations in all Weil cohomologies. It is well known that the comparison between the Betti cohomology and the de Rham cohomology produces the classical periods of $X$ and we can apply techniques of Hodge theory and the variation of Hodge structure. Grothendieck introduced Étale cohomology for an algebraic variety $X/k$ with coefficients in the $\ell$-adic numbers $\Q_{\ell}$, where  $\ell \neq \text{char}(k)$, originally to prove the Weil conjectures (see below). The Étale cohomology $H_{\etale}(X/\overline{k},\Q_{\ell})$ comes with a continuous action of $\text{Gal}(\overline{k}/k)$ which is generated for finite fields $k=\F_{p}$ by the Frobenius automorphism $x\mapsto x^{p}$. Note that fixed points of this map are exactly the points on $X_p\equiv X/\F_{p}$ and we can apply the Lefschetz fixed point  theorem to obtain 
\begin{equation}
\label{eq:lefschetz_fix_pt}
   \#X_p(\mathbb{F}_{p^n}) = \sum_{i=0}^{2n} (-1)^i \text{Tr}((\text{Fr}_p^*)^n|H^i_{\etale}(X/\overline{\F}_p,\Q_{\ell})).
\end{equation}
Here $\text{Fr}_p$ represents the inverse of $\text{Frob}_p$, also called the \emph{geometric Frobenius}. The above equation shows that we can compute arithmetic information (i.e. point counts over finite fields) from cohomological data. In the end we will not use eq.\eqref{eq:lefschetz_fix_pt} directly, but rather rephrase the statement in other, more computable, cohomology theories and arrange the point counts in a generating function 
\begin{align}
     Z_p(X,T)&\equiv \exp\left(\sum_{n=1}^{\infty}\#X_p(\mathbb{F}_{p^n})\frac{T^n}{n}\right)\\
     &\overset{(\ref{eq:lefschetz_fix_pt})}{=}\prod_{i=0}^{2n} \text{det}\left( 1-T \text{Fr}_p^*|H^i_{\etale}(X/\overline{\F}_p,\Q_{\ell})\right)^{(-1)^{i+1}}.
\end{align}
It was conjectured in \cite{weil1949numbers} that $Z_p(X,T)$ has some remarkable properties.
We follow \cite{goncharov2019weilconjecturesexposition} closely in stating these so-called \emph{Weil conjectures} for $Z_p(X,T)$ with smooth projective $X$:
\begin{itemize}
    \item[(I)] \textbf{Rationality}: $Z_p(X, T)$ is a rational function of $T$. Moreover, we have
    \begin{equation}
         Z_p(X, T) = \frac{P_1^{(p)}(X, T)P_3^{(p)}(X, T) \cdots P^{(p)}_{2n-1}(X,T)}{P^{(p)}_0(X, T)P^{(p)}_2(X, T) \cdots P^{(p)}_{2n}(X, T)}
    \end{equation}
    where $P^{(p)}_0(X, T) = 1 - T$, $P^{(p)}_{2n}(X, T) = 1 - p^n T$, and each $P^{(p)}_i(X, T)$ is an integral polynomial.

    \item[(II)] \textbf{Functional equation}: $Z_p(X, T)$ satisfies the functional equation
    \begin{equation}
    \label{eq:weil_func}
        Z_p(X, p^{-n} T^{-1}) =\epsilon_p p^{n \chi/2} T^\chi Z_p(X, T),
    \end{equation}
    where $\chi = \sum_i (-1)^i b_i$ for $b_i = \deg P^{(p)}_i(X,T)$ and $\epsilon_p=\pm1$ with $\epsilon_p=1$ for odd n.

    \item[(III)] \textbf{Betti numbers}: If $X$ lifts to a variety $X_1$ in characteristic 0, then $b_i$ are the (real) Betti numbers of $X_1$ considered as a variety over $\mathbb{C}$.

    \item[(IV)] \textbf{Riemann hypothesis}: For $1 \leq i \leq 2n - 1$, $P^{(p)}_i(T) = \prod_{j=1}^{b_i} (1 - \lambda_{i,j} T)$, where $\lambda_{i,j}$ are algebraic integers of absolute value $p^{i/2}$.
\end{itemize}
The Weil conjectures impose strong constraints on $Z_p(X,T)$. For threefolds and fourfolds the local zeta functions are constrained to take the form
\begin{align}
    &Z^{(n=3)}_p \left(X, T \right) = \frac{P^{(p)}_3(X, T)}{(1 - T)(1 - pT)^{h^{1,1}}(1 - p^2 T)^{h^{1,1}}(1 - p^3 T)}\,,\\
    &Z^{(n=4)}_p \left(X, T \right) = \frac{P_3^{(p)}(X,T)P_5^{(p)}(X,T)}{(1 - T)(1 - pT)^{h^{1,1}}P^{(p)}_4(X, T)(1 - p^3 T)^{h^{1,1}}(1 - p^4 T)}\,,
\end{align}
reducing in the threefold case its computation to only the polynomial $P^{(p)}_3(X, T)$ corresponding to the middle cohomology. For fourfolds, while in general $\deg P_3^{(p)}(X,T)=\deg P_5^{(p)}(X,T)=b_3\neq 0$, the interesting arithmetic information is captured by the piece $P_4^{(p)}(X,T)$ and will be our object of interest.

More generally, each factor $P_i^{(p)}(X,T)$ can be interpreted as the piece emerging from the respective cohomology groups $H^i(X)$ and their arithmetic information over all primes is encoded in the \emph{L-function}, 
\begin{align}
    L_i(X,s) &= \sum_{n=1}^{\infty} \frac{a^{(i)}_n}{n^s}\\
    &=\prod_{p\text{ prime}} P_i^{(p)}(X,T) \\
    &= \left(\prod_{p\text{ bad}} P_i^{(p)}(X,T)\right) \left(\prod_{p\text{ good}} P_i^{(p)}(X,T)\right) ,
\end{align}
that is convergent for $\Re(s)\gg 0$.
Above, a prime is of \emph{bad reduction} if $X_p$ is singular and otherwise it is called a \emph{good prime}. In the simplest case, where all $a_n=1$, one recovers the Riemann $\zeta$-function. Euler showed that it can be written as a product over all primes and therefore the $P_i^{(p)}(X,T)$ are called \emph{Euler factors}. The function $L_i(X,s)$ is an example of a motivic $L$-function. It can be completed by supplementing a factor $\Gamma_{\infty}(s)$, determined by the Hodge numbers of the underlying cohomology group, that can be thought of as coming from the infinite prime,
\begin{equation}
    \Lambda(s) = \Gamma_{\infty}(s) L_i(X,s)\,.
\end{equation}
Conjecturally, $\Lambda(s)$ fulfils a functional equation
\begin{equation}
    \Lambda(s) = \epsilon \Lambda(k-s)\,,
\end{equation}
for an integer $k$ and $\epsilon = \pm 1$. Specific examples are
\begin{align}
    \text{Riemann Zeta:}\quad  \Lambda_{\zeta}(s)&= \pi^{-s/2}\Gamma\left(\frac{s}{2}\right)\zeta(s)\,,\\
    \text{Elliptic Curve:} \quad\Lambda_{\cE}(s)&= \left(\frac{N^{s/2}}{(2\pi)^{s}}\right) \Gamma \left(s\right) L(\cE,s)\,,\\
    (1,1,1,1)\footnotemark : \quad \Lambda_{X}(s) &= \left( \frac{N}{4\pi}\right)^{s/2} \Gamma\left(\frac{s-1}{2}\right) \Gamma\left(\frac{s}{2}\right) \Gamma\left(\frac{s}{2}\right) \Gamma\left(\frac{s+1}{2}\right) L(X,s)\,.
\end{align}
\footnotetext{The completion of the $L$-function is determined by the Hodge vector of the underlying cohomology group. The example $(1,1,1,1)$ includes the case of one-parameter CY threefolds.}
We denote the Gamma function by $\Gamma(s)$ and the conductor of the Galois representation by $N$. Concretely, the conductor is given by
\begin{equation}
    N = \prod_{p\text{ bad}} p^{\delta_p}\,.
\end{equation}
In the case of elliptic curves, there exists a formula \cite{Ogg} for the exponents $\delta_p$, but for higher dimensional manifolds, much less is known.
The respective functional equations then read
\begin{align}
    \Lambda_{\zeta}(s)&=\Lambda_{\zeta}(1-s)\,,\\
    \Lambda_{\cE}(s)&= \epsilon \Lambda_{\cE}(2-s)\,,\\
    \Lambda_{X}(s) &= \epsilon \Lambda_{X}(4-s)\,.
\end{align}

The functional equation serves as a highly non-trivial test for the calculated Euler factors and the predicted conductor which was investigated in \cite{Gegelia:2024vhd}. In the spirit of the Langlands program, the motivic $L$-function associated to the Galois representation of an algebraic variety $X$ should match the $L$-function of an automorphic representation such that the conductor $N$ matches the level of the automorphic form.


\subsection{The deformation method}
For families of Calabi--Yau manifolds a powerful method, called the Dwork deformation  method \cite{deformation_zeta_1962,articleLauder}, has been developed to compute $P^{(p)}_n(X, T)$ from the periods of the holomorphic $(n,0)$-form of the Calabi--Yau \cite{Candelas:2021tqt,Candelas:2024vzf}, which we now briefly summarize\footnote{For some parts we closely follow the exposition given for the threefold case in \cite{Candelas:2024vzf}, adapted to our conventions and generalized to cover also the multi-parameter fourfold case. For more details, in particular more detailed derivations, we refer the reader to the aforementioned work and references therein.}. The idea is to consider the Frobenius action on the moduli $\uz \mapsto \uz^p=(z_1^p,\dots,z_{r}^p)$ and its induced action on the Hodge bundle of the family. In terms of the matrix $\Frob_p(\uz)$ representing the latter, the characteristic polynomial $P^{(p)}_n(X_\uz, T)$ is computed via:
\begin{align}
\label{eq:charpol}
    P^{(p)}_n\left( X_\uz , T \right)=\text{det}(1-T\Frob_p(\uz))\vert_{\uz=\text{Teich}(\underline{z})}\,,
\end{align}
where $\text{Teich}(\uz)\in \mathbb{Z}_p$ denotes the component-wise Teichmüller lift of $\uz$ with $\text{Teich}(\uz) =  \text{Teich}(\uz^p)$. From a compatibility condition between $\Frob_p(\uz)$ with the Gauss--Manin connection Dwork derived a differential equation, whose solutions take the form
\begin{align}
\label{eq:frob}
   \Frob_p(\uz)=\boldsymbol{\Pi}(\uz^p)^{-1}\mathbf{V}_0\boldsymbol{\Pi}(\uz)\,.
\end{align}
Here $\boldsymbol{\Pi}(\uz)$ denotes the period matrix in a basis of sections of the Hodge bundle. A canonical choice for the bases yields
\begin{align}
  \boldsymbol{\Pi}^{(n=3)}(\uz)&\equiv \left(\underline{\varpi}(\uz),\,\theta_i\underline{\varpi}(\uz),\,\mathfrak{c}^{ijk}\theta_j\theta_k\underline{\varpi}(\uz),\,\mathfrak{c}^{ijk}\theta_i\theta_j\theta_k\underline{\varpi}(\uz)\right),\\
 \boldsymbol{\Pi}^{(n=4)}(\uz)&\equiv \left(\underline{\varpi}(\uz),\,\theta_i\underline{\varpi}(\uz),\,\mathfrak{c}^{ij\alpha}\theta_i\theta_j\underline{\varpi}(\uz),\,\mathfrak{c}^{ijkl}\theta_j\theta_k\theta_l\underline{\varpi}(\uz),\,\mathfrak{c}^{ijkl}\theta_i\theta_j\theta_k\theta_l\underline{\varpi}(\uz)\right),  
\end{align}
where we introduced the logarithmic derivatives $\theta_i=z_i\frac{\partial}{\partial z_i}$ and the inverse couplings, implicitly defined via $\mathfrak{c}^{ijk}c_{ijl}=\delta^k_l$, $\mathfrak{c}^{ij\alpha}c_{ij\beta}=\delta^\alpha_\beta$ and $\mathfrak{c}^{ijkl}c_{ijkm}=\delta^l_m$.
The Frobenius basis $\underline{\varpi}(z)$ around the MUM point $\uz=0$ is as in \eqref{eq:frobbasis}. The moduli independent matrix $\mathbf{V}_0$ is conjecturally obtained by keeping in the MUM point transition matrix $T_0$ only the diagonal terms and terms proportional to $\zeta(3)$, i.e. omitting terms not needed for global rationality of the monodromies, and replacing $\zeta(3)$ by the $p$-adic zeta value $\zeta_p(3)$ and $1/2\pi\ii$ by $p$. In the threefold case this gives\footnote{We suppress in the $V_0$ matrices the conjectural possibility of an overall sign factor $\varepsilon_p=\pm 1$, which for our applications, namely observing factorizations and identifying sub-Hodge structures, will not be important and henceforth assumed to be $\varepsilon_p=1$.}
\begin{align}
    \label{eq:V0matrix}
   \mathbf{V}^{(n=3)}_0=\begin{pmatrix}
1 & 0 & 0 & 0 \\
0 & p\mathds{1}_{m\times m} & 0 & 0 \\
0 & 0 & p^2\mathds{1}_{m\times m} & 0 \\
\chi\,\zeta_p(3)\, p^3& 0 & 0 & p^3
\end{pmatrix},
\end{align}
and in the fourfold case we have
\begin{align}
    \label{eq:4foldV0matrix}
    \mathbf{V}^{(n=4)}_{0} =
\begin{pmatrix}
1 & 0 & 0 & 0 & 0 \\
0 & p\,\mathds{1}_{m\times m} & 0 & 0 & 0 \\
0 & 0 & p^{2}\,\mathds{1}_{h_{2,2}^{\mathrm{hor}}\times h_{2,2}^{\mathrm{hor}}} & 0 & 0 \\c_3\cdot \underline{D}^T\,\zeta_{p}(3)\,p^{3} & 0 & 0 & p^{3}\,\mathds{1}_{m\times m} & 0 \\
0 & c_3\cdot \underline{D}\,\zeta_{p}(3)\,p^{4} & 0 & 0 & p^{4}
\end{pmatrix}.
\end{align}
We wrote $h_{2,2}^{\text{hor}}$, since via the above recipe we only compute the factor $P^{(p)}_{4,\text{hor}}(X,T)$ corresponding to the \emph{horizontal subspace} of the middle cohomology, i.e. the part generated by derivatives of $\Omega(\uz)$. In particular, as discussed in \cite{Jockers:2023zzi}, we tacitly assume that the Frobenius has a well-defined action on $H^{4}_{\text{hor}}(X)$. 

Using \eqref{eq:charpol} we can express the $P_n^{(p)}(X,T)$ in terms of traces of the Frobenius action\footnote{from now on we will suppress in the notation the dependence on the fibre $\uz$.}:
\begin{align}
\label{eq:Frobexpansion}
    P_n^{(p)}(X,T) &= \det(1-T\mathbf{Fr}_p)\\
    &=\exp\left(\tr(\log\left(1-T\mathbf{Fr}_p)\right)\right)\\
    &=\exp\left(-\sum_{n=1}^{\infty}\tr(\Frob_p^n)\frac{T^n}{n}\right)\\
    &=1 + \tr ( \Frob_p ) \,T +  \frac{\tr ( \Frob_p)^2 - \tr (\Frob^2_p) }{2}\,T^2 + \dots\,.
\end{align}
We do not need to calculate all Frobenius traces up to $\tr(\Frob_p^{b_n})$ since the functional equation \eqref{eq:weil_func} for $Z_p(X,T)$ translates to a functional equation for $P^{(p)}_n(X,T)$:
\begin{align}
\label{eq:Pnfunctionalequation}
    P^{(p)}_{n}(X,T)= \epsilon_p\, p^{n\,b_n/2}T^{b_n}P^{(p)}_{n}(X,p^{-n}T^{-1})\,.
\end{align}
Writing
\begin{align}
    P^{(p)}_{n}(X,T)=1+\sum_{i=1}^{b_n}a^{(p)}_iT^i\,,
\end{align}
we then find by comparing coefficients on both sides of \eqref{eq:Pnfunctionalequation} the relations
\begin{align}
\label{eq:coeffrelations}
    a^{(p)}_{b_n-i}=\epsilon_p\,p^{n\left(b_n/2-i\right)}a^{(p)}_i\,, \quad a^{(p)}_{b_n}=\epsilon_p\,p^{n\,b_n/2}\,.
\end{align}
From the Riemann hypothesis we further have 
\begin{align}
P^{(p)}_{n}(X,T)&=\prod_{j=1}^{b_n} (1 - \lambda_{n,j} T)\\
                &=\sum_{i=0}^{b_n} (-1)^i e_i(\lambda_{n,1},\dots,\lambda_{n,b_n}) T^i\,,
\end{align}
where $e_i$ is the $i$-th elementary symmetric polynomial. Since $\abs{\lambda_{n,j}}=p^{n/2}$, we find the following bounds on the coefficients $a^{(p)}_i$:
\begin{align}
    \vert a^{(p)}_i \vert\leq \binom{b_n}{i}\,p^{\frac{n\,i}{2}}\,.
\end{align}
It then follows that, to compute the coefficient $a^{(p)}_i$ for a certain prime, it is enough to calculate the relevant Frobenius traces in \eqref{eq:Frobexpansion} to a finite $p$-adic accuracy $p^{n_{\text{acc}}}$, with $n_{\text{acc}}$ chosen as the smallest integer such that $\binom{b_n}{i}\,p^{n\,i/2}<p^{n_{\text{acc}}}$. Having calculated $a^{(p)}_i\mod p^{n_{\text{acc}}}$, the exact value is obtained through the centered lift. Due to the relations \eqref{eq:coeffrelations} we need only compute $a^{(p)}_1,\dots, a^{(p)}_{\lfloor{b_n/2\rfloor}}$ to determine, up to $\epsilon_p$, all coefficients of $P^{(p)}_{n}(X,T)$. The value of $\epsilon_p=\pm1$ can be deduced by calculating also $a^{(p)}_{\lfloor{b_n/2\rfloor}+1}$, to the same $p$-adic accuracy as $a^{(p)}_{\lfloor{b_n/2\rfloor}}$, and then demanding $a^{(p)}_{\lfloor{b_n/2\rfloor}+1}=\epsilon_p\, p^{n(b_n/2-\lceil{b_n/2\rceil}+1)}a^{(p)}_{\lceil{b_n/2\rceil}-1}$ to hold to that precision.

In particular for $n=3$ and $h_{2,1}=1$ one finds,
\begin{equation}
    P^{(p)}_3(X,T) = 1 + a^{(p)}_1 T + a^{(p)}_2  T^2 +  a^{(p)}_1 p^3T^3 + p^6 T^4\,,
\end{equation}
with the bounds 
\begin{align}
    \vert a^{(p)}_1 \vert\leq 4p^\frac{3}{2}<p^3\,,\quad \vert a^{(p)}_2 \vert\leq 6p^3<p^4\,,
\end{align}
where for both coefficients the second inequality holds for $p\geq 7$. We thus see that for one-parameter threefolds it is enough to calculate the Frobenius traces $\tr(\Frob_p)$ and $\tr(\Frob^2_p)$ to $p$-adic precision $n_{\text{acc}}=4$.

\subsubsection{Special fibres of one-parameter Calabi-Yau 3-fold families}
\label{sec:specialfibres}
Here we recall  the properties of the Hodge structure in special fibres $X_{z_*}$ of one 
parameter Calabi-Yau 3-folds $X_z$ and how this is reflected in properties of the Euler factors.        

Let us  first note the fact the coefficients of the Euler factors have further divisibility properties, in the case at hand for example we can write
\begin{align}
  P^{(p)}_3(X_z/\mathbb{F}_p,T) \, = \,  \det(1-T\text{Fr}_p^* | H^3(\overline{X_z},\mathbb{Q}_\ell)) \, = 1+\alpha_pT+\beta_ppT^2+\alpha_pp^3T^3+p^6T^4\,  
\end{align}
%
with integer $\alpha_p, \beta_p$. The divisibility of $a^{(p)}_2=p\,\beta_p$, or more generally $a^{(p)}_i$ for $i\geq2$, has a subtle geometric origin. From the Hodge numbers one can construct, in a purely combinatorial way, the so-called \emph{Hodge polygon}. Following Katz \cite{AST_1979__63__113_0} and Mazur \cite{bams/1183533965}, this polygon provides a lower bound on  $p$-adic divisibility of the coefficients $a_i^{(p)}$. We will later directly write the Euler factors with such prime powers factored out, and denote the remaining parameters $\alpha_i^{(p)}$ to distinguish from the $a_i^{(p)}$.

\noindent
{\sl Motives for generic $z_*\in \mathbb{Q}$}: For fibres over points $z_*\in \mathbb{Q}$ in the  moduli space of one-parameter threefolds the $\alpha_p$ and $\beta_p$ correspond 
conjecturally \cite{Gross2016OnTL} for good primes to the eigenvalues $\lambda_p,\mu_p$ of the two Hecke operators acting on weight $3$ Siegel modular forms invariant under the paramodular group  via 
\begin{equation}
    \alpha_p=-\lambda_p \quad \text{and} \quad \beta_p=\mu_p +p^2 +1.
\end{equation}
Many identifications of this type can be found in \cite{Gegelia:2024vhd}\cite{Blessemaster} an overview article is planned \cite{DEGGKKPPRSTV}.

\noindent
{\sl Motives for generic rank two attractor points}: 
At generic rank two attractor points in the complex moduli space one expects conjecturally that the lattices $\Lambda_\mathbb{Q}$ and $\Lambda^\perp_\mathbb{Q}$ correspond to two dimensional motives which implies according to the Serre–Khare–Wintenberger theorem a factorisation of the 
Euler factors in the form~\cite{Candelas:2019llw}\cite{samol}\cite{Candelas:2021tqt}\cite{bonisch}\cite{Bonisch:2022mgw} 
\begin{align}
  P_3^{(p)}(X_{z_*}/\mathbb{F}_p,T) \, = \,  (1-a_pT+p^3T^2)(1-b_p(pT)+p(pT)^2)\ ,
\end{align}
where $a_p$ and $b_p$ are the Hecke eigenvalues of the holomorphic Hecke newforms $f_4\in S_k(\Gamma_0(N_4)$ and $f_2\in  S_2(\Gamma_0(N_2) $.

\noindent
{\sl Motives for generic conifold points}: 
At conifold points the Calabi-Yau family develops a nodal singularity and the Euler factor degenerates into a degree 
three polynomial in $T$, which factorises in the form \cite{MR2785550}\cite{Bonisch:2022mgw}  
\begin{align}
 P_3^{(p)}(X_{z_*}/\mathbb{F}_p,T) \, = \,  (1-\chi(p)pT)(1-a_pT+p^3T^2)\ .
  \label{eq:FactorizationConifold}
\end{align}
Here the numbers $a_p$ are the Hecke eigenvalues of a weight 4 newform $f_4 \in S_k(\Gamma_0(N_4)$ and  
$\chi(p)$ is a Dirichlet character. For the hyper geometric one-parameter families it is given by $(\frac{\kappa}{p})$.
The two dimensional motive is expected for rigid Calabi-Yau 3 folds to which the singular Calabi-Yau can 
be resolved and the Serre–Khare–Wintenberger theorem applies. 

\noindent
{\sl Motives for generic K-points}: 
At the $K$-points one expects again a two dimensional motive but of a rigid K3 surface, which is in turn 
expected to be related to a weight 3 newform in $S_3(\Gamma_0(N),\chi)$. Note that weight 3 forms  necessarily 
have a non trivial character. The Euler factor takes the form 
\begin{align}
 P_3^{(p)}(X_{z_*}/\mathbb{F}_p,T) \, = \,  1-a_pT+\chi(p) p^2T^2\ .
  \label{eq:EulerfactorKpoint}
\end{align}
The unique newform $f_3 =q+q^2 +O(q^3)\in S_3(\Gamma(15),\chi_{15}(14,\cdot))$ has 
been identified numerically at the K-point of the AESZ34 model~\cite{bonischmaster}.     

Methods to infer the values of the complete period matrix at  the attractor points from the arithmetic data 
have been provided at the  attractor points in \cite{Bonisch:2022mgw} and at conifolds 
in~\cite{Bonisch:2025cax} in terms of periods of the corresponding modular forms or their $L$-functions.    
At at those points the modular forms were not at points of complex multiplication.      

Note that in each case the above motives can split further over $\mathbb{Q}$, which  
happens for example if the motives are in addition at points of complex multiplication, and 
explains the term generic above. Another intriguing case is that the factorisation of the Euler factor occurs only over 
extensions of $\mathbb{Q}$.     
An interesting place to study both phenomena is the $z_*=\infty$ point in the hypergeometric 
models. Here generic smooth points, rank two attractors, conifolds K-points and maximal unipotent  
monodromy points occur with an additional $\mathbb{Z}/k \mathbb{Z}$ action on the moduli 
space, with a fixed point at $z_*=\infty$. An important hint comes from the fact that that at $z_*=\infty$ the full period 
matrix is known by the Barnes integral analytic continuation methods and expressible in terms 
of $\Gamma$-functions at rational values with denominator $k$ and $k$-roots of unity or 
algebraic extensions involving roots of $k$, see~\cite{Bonisch:2022mgw}.

\noindent 
{\sl  Attractor points in which the motive further splits}.
The first example of an attractor point is the Gepner point $\psi=0$ of the mirror sextic $X^6\subset \mathbb{P}_{21111}$ \cite{Moore:1998pn}. Over the rationals the Euler factors split into two quadratic factors, signalling the splitting of the rank four motive into two motives of rank two. However, these two motives split further over the field $K\equiv \mathbb{Q}(\zeta_6)=\mathbb{Q}(\sqrt{-3})$. From the Euler factors a hint comes from the fact that for half of the primes, more precisely for $p\equiv 2 \mod 3$, the quadratic factors are even polynomials in $T$. For primes $p\equiv 1 \mod 3$ on the other hand, the Euler factors split over $K$ into two linear factors. Note in particular that while the motive splits over $K$, it is only for half of the primes that one expects further factorizations. This is because only for those primes the Frobenius preserves the rank one subspaces, as explained in more generality in \cref{subsec:Arithmetic and analytic analysis of flux vacua}.  In terms of the modular forms associated to the attractor splitting, $f_2=\eta(3\tau)^2\eta(9\tau)^2=q - 2q^4+\mathcal{O}(q^7)\in  S_2(\Gamma_0(27))$ and $f_4=q + 17q^7 + 89q^{13} + 107q^{19}+\mathcal{O}(q^{25})\in  S_4(\Gamma_0(108))$, this is reflected in these forms being CM by $\mathbb{Q}(\sqrt{-3})$. They represent Gr\"o\ss encharacters, 
whose critical  $L$-function values representing the periods of the elliptic curves are expressible in $\Gamma$ functions at rational values 
in accordance with the Chowla-Selberg theorem and the fact that the transition matrix yielding the values of the Calabi-Yau period matrix at $z_*=\infty$ 
is likewise expressible in terms of these values as follows from the Branes integral analytic continuation.  
Other cases with attracter points where the rank two motives split further due to a cyclic symmetry are the 
points $z=\infty$ of $X^{6,4}\subset \mathbb{P}_{322111}$ and $X^{4,3}\subset \mathbb{P}_{211111}$.

\noindent 
{\sl Conifold and K-points for  which the motive further splits.} 
Similarly one finds at the conifolds and K-points with an additional $\mathbb{Z}/k\mathbb{Z}$ action at $z=\infty$ in the hypergeometric models modular forms of weight four and three, respectively, that are of CM-type and which reflect a further splitting of the associated motives. In the model $X^{3,2,2}\subset \mathbb{P}^6$, for example, there is a conifold at $z=\infty$ with local exponents $(1/3,1/2,1/2,2/3)$ and the Euler factors correspond to the weight four form $f_4(q)=\eta(3\tau)^8=q - 8q^4 +\mathcal{O}(q^7)\in S_4(\Gamma_0(9))$, which is CM by $\mathbb{Q}(\sqrt{-3})$. The model $X^{4,4}\subset \mathbb{P}_{221111}$ has a K-point with local exponents $(1/4,1/4,3/4,3/4)$, here the associated weight three form is $f_3=\eta(4\tau)^6=q - 6q^5+\mathcal{O}(q^9)\in S_3(\Gamma_0(16),\chi_{16}(15,\cdot))$ and it is CM by $\mathbb{Q}(\sqrt{-1})$. Analogous observations apply to the other models with Conifold or K-points at infinity. 

\noindent 
{\sl  Rank four motives that splits over extensions of $\mathbb{Q}$}. For the hypergeometric models with smooth points at infinity, that are not attractor points, the motive is irreducible over $\mathbb{Q}$ but splits into four rank one motives over a cyclotomic field. The simplest example is the mirror quintic $X^5\subset \mathbb{P}^4$, where the cohomology splits over $\mathbb{Q}(\zeta_5)$. Over this field the Euler factors split into four linear factors for $p\equiv 1 \mod 5$, two quadratic factors for $p\equiv 4\mod 5$ and stay irreducible for $p\equiv 2,3\mod 5$. The origin of this splitting pattern arises from the Galois group of the field extension, as will be discussed in \cref{subsec:Arithmetic and analytic analysis of flux vacua}. Similarly for the models $X^8\subset \mathbb{P}_{41111}$, $X^{10}\subset P_{52111}$ and $X^{12,2}\subset \mathbb{P}_{641111}$ the $\mathbb{Q}$-irreducible motives at infinity split into four rank one motives over cyclotomic fields corresponding to the respective $\mathbb{Z}/k\mathbb{Z}$ action.

\subsection{One-parameter threefolds and apparent singularities}
In the case of one-parameter Calabi-Yau threefolds, the discriminant of its Picard-Fuchs operator $\mathcal{L}^{(4)}(z)$ is of the form $\Delta(z)=\Delta_{\text{app}}(z)\Delta_{\text{geom}}(z)$. The zeros of $\Delta_{\text{geom}}(z)$ are fibres, where the geometry degenerates, whereas $\Delta_{\text{app}}(z)$ are so-called apparent singularities, that are regular singular points of the operator, at which the geometry is nonsingular. As we will see below the computation of the Frobenius traces at apparent singularities is more subtle. The appearance of apparent singularities persists also to the one-parameter operators studied later, which are obtained by restricting multi-parameter Picard-Fuchs systems of Calabi-Yau threefolds and fourfolds to linear subspaces. Since the situation there is a straightforward adaptation of the more familiar case of ordinary one-parameter Calabi-Yau threefold operators, we will first discuss the treatment of apparent singularities in the setting of such fourth order operators.

In the AESZ list of Calabi-Yau operators there are apparent singularities with local exponents $(0,1,3,4)$ and $(0,2,3,5)$.\footnote{The first operator in the AESZ list with an apparent singularity of type $(0,1,3,4)$ is the operator \href{https://cycluster.mpim-bonn.mpg.de/operator.html?nn=4.4.33}{AESZ 4.4.33} and for type $(0,2,3,5)$ the operator \href{https://cycluster.mpim-bonn.mpg.de/operator.html?nn=4.7.11}{AESZ 4.7.11} is the first example.} Apparent singularities are distinguished by the property that the Wronskian $\mathbf{W}_{ij}(z)=\theta^j\varpi_i(z)$ becomes non-invertible. In terms of the Wronskian we can rewrite the Picard-Fuchs equation as 
\begin{equation}
\theta  \mathbf{W}(z)^{\text{T}} = \mathbf{S}(z)\mathbf{W}(z)^{\text{T}},
\end{equation}
with the matrix $\mathbf{S}(z)$
\begin{align}
\mathbf{S}(z)=
    \begin{pmatrix}
        0&1&0& 0\\
        0 & 0& 1&0\\
        0&0&0&1\\
        -S_0(z)/\Delta(z) & -S_1(z)/\Delta(z) &-S_2(z)/\Delta(z)&-S_3(z)/\Delta(z)\\
    \end{pmatrix},
\end{align}
when writing the operator as $\mathcal{L}^{(4)}(z)=\sum_{i=0}^{4}S_i(z) \theta^i$ with $S_i\in \mathbb{Q}[z]$ and $S_4(z)=\Delta(z)$. It follows that the Wronskian satisfies the differential equation 
\begin{align}
\label{eq:detWdiffeqn}
    \theta\det \mathbf{W}(z) = \tr \mathbf{S}(z) \det \mathbf{W}(z),
\end{align}
From this we can compute the determinant up to normalization. In the computed cases it takes the form
\begin{align}
    \label{eq:order4detW}
    \det \mathbf{W}(z)\propto \exp\left(-\int\text{d}z\frac{S_3(z)}{z\,\Delta(z)}\right)=\frac{\big(\Delta^{(0,1,3,4)}_{\text{app}}(z)\big)^2\big(\Delta^{(0,2,3,5)}_{\text{app}}(z)\big)^4}{{\Delta_{\text{geom}}(z)}^2}\,,
\end{align}
where we wrote the operator as $\mathcal{L}^{(4)}(z)=\sum_{i=0}^{4}S_i(z) \theta^i$ with $S_i\in \mathbb{Q}[z]$, $S_4(z)=\Delta(z)$ and the apparent discriminant has the form $\Delta_{\text{app}}(z)=\big(\Delta^{(0,1,3,4)}_{\text{app}}(z)\big)^2\big(\Delta^{(0,2,3,5)}_{\text{app}}(z)\big)^3$. 

In one-parameter cases we will replace in \eqref{eq:frob} the period matrix $\mathbf{\Pi}(z)$ by the Wronskian $\mathbf{W}(z)$,
\begin{align}
      \Frob_p(z)= \mathbf{W}(z^p)^{-1}\mathbf{V}_0\mathbf{W}(z)\,,
\end{align}
which leaves the characteristic polynomial unchanged. The Frobenius converges precisely for the correct value of $\chi/c_{111}$ to any finite $p$-adic accuracy $n_{\text{acc}}$ to a rational expression. In all computed cases we find that we can then write 
\begin{align}
\label{eq:order4rationalFrob}
   \Frob_p(z)\equiv \frac{\Frob^{(n_{\text{acc}})}_p(z)}{\Delta_{\text{app}}(z^p){\Delta_{\text{geom}}(z^p)}^{\max\{n_{\text{acc}}-4,0\}}}\mod p^{n_{\text{acc}}}\,, 
\end{align}
where $\Frob^{(n_{\text{acc}})}_p\in \text{Mat}_{4\times 4}(\mathbb{Q}[z])$, with the degrees of the entries growing linearly with $p$, typically $\lesssim p\cdot \deg_z(\mathcal{L}^{(4)}(z))$. As we saw above we need only calculate to $n_{\text{acc}}=4$ to compute the characteristic polynomial $P_3^{(p)}(X,T)$ and hence the $\Delta_{\text{geom}}$-factor does not contribute. We can thus also compute the Euler factors for singular fibres when calculating $\mod p^4$. 

Since the apparent singularities contribute to the denominator in \eqref{eq:order4rationalFrob} we can a priori not evaluate the Frobenius at those points. However, one can resolve the apparent singularities by modifying the Wronskian\footnote{We thank Kilian Bönisch for helpful explanations regarding the resolution of apparent singularities using a rational transformation of the Wronskian.} according to $\mathbf{W}(z)\mapsto \mathbf{W}(z) \mathbf{U}(z)$ where $\mathbf{U}(z)\in \text{Mat}_{4\times 4}(\mathbb{Q}(\!(z)\!))$ is chosen with $\det \mathbf{U}(z)=1/\big(\Delta^{(0,1,3,4)}_{\text{app}}(z)\big)^2\big(\Delta^{(0,2,3,5)}_{\text{app}}(z)\big)^4$. Away from apparent singularities such a transformation does not affect the end result for $P^{(p)}_3(X,T)$, as it drops out upon inserting the Teichmüller lift. In general, however, such a $\mathbf{U}(z)$ introduces poles around $\Delta_{\text{app}}(z)=0$ in the entries of $\mathbf{W}(z)$ when analytically continued to the apparent singularities. In such cases $\Delta_{\text{app}}$ still contributes to the denominator of the Frobenius. Hence, we need to choose the rational entries of $\mathbf{U}(z)$ in such a way as to cancel such poles. In the two cases, $(0,1,3,4)$ and $(0,2,3,5)$, we make the following ansatz for the inverse of $\mathbf{U}(z)$:
\begin{align}
\label{eq:order4Umatrix}
    &{\mathbf{U}^{(0,1,3,4)}(z)}^{-1}=
\begin{pmatrix}
1 & 0 & p_1(z) & p_3(z) \\
0 & 1 & p_2(z) & p_4(z) \\
0 & 0 & \Delta_{\text{app}}^{(0,1,3,4)}(z) & p_5(z) \\
0 & 0 & 0 & \Delta_{\text{app}}^{(0,1,3,4)}(z) \\
\end{pmatrix},
\\
    &{\mathbf{U}^{(0,2,3,5)}(z)}^{-1}=
\begin{pmatrix}
1 & p_1(z) & p_2(z) & q_1(z) \\
0 & \Delta_{\text{app}}^{(0,2,3,5)}(z) & p_3(z) & q_2(z) \\
0 & 0 & \Delta_{\text{app}}^{(0,2,3,5)}(z) & q_3(z) \\
0 & 0 & 0 & \big(\Delta_{\text{app}}^{(0,2,3,5)}(z)\big)^2 \\
\end{pmatrix},
\end{align}
where the $p_i$ are polynomials of degree at most $\deg(\Delta_{\text{app}}^{(0,1,3,4)/(0,2,3,5)})-1$ and the $q_i$ are of polynomials of degree at most $2\deg(\Delta_{\text{app}}^{(0,2,3,5)})-1$. We can fix the coefficients of these polynomials by requiring the poles in the entries of the analytic continuation of $\widetilde{\mathbf{W}}(z)=\mathbf{W}(z)\mathbf{U}(z)$ to $\Delta_{\text{app}}=0$ to cancel. In the case of multiple irreducible components of the apparent discriminant, one can successively apply this procedure for each component. When the Frobenius is computed with the modified Wronskian,
\begin{align}
\label{eq:order4modifiedFrob}
    \widetilde{\Frob}_p(z)=\widetilde{\mathbf{W}}(z^p)^{-1}\mathbf{V}_0\widetilde{\mathbf{W}}(z)\,,
\end{align}
we have to finite $p$-adic accuracy
\begin{align}
\label{eq:order4rationalmodifiedFrob}
   \widetilde{\Frob}_p(z)\equiv \frac{\widetilde{\Frob}^{(n_{\text{acc}})}_p(z)}{{\Delta_{\text{geom}}(z^p)}^{\max\{n_{\text{acc}}-4,0\}}}\mod p^{n_{\text{acc}}}\,, 
\end{align}
where again $\widetilde{\Frob}^{(n_{\text{acc}})}_p\in \text{Mat}_{4\times 4}(\mathbb{Q}[z])$. The apparent discriminant no longer contributes to the denominator and the characteristic polynomial can be evaluated for all smooth fibres.

\subsection{Reduction to linear subspaces}
\label{subsec:DeformationMethod}
Since the evaluation of multi-parameter periods to high order is computationally expensive, we will instead follow the strategy of computing the local zeta function along one-dimensional rays through the origin\footnote{We thank Duco van Straten for bringing this idea to our attention.}. For the purpose of locating splittings of Hodge structure this is sufficient: For example in the case of a threefold family, where the moduli space contains a codimension one supersymmetric vacuum locus, we expect to see persistent factorizations of the Euler factors for any generic direction\footnote{We will comment below on the case where this would not be true, namely when the vacuum itself lies along a line through the origin.}, coming from the intersection of the vacuum locus with the one-dimensional ray in the moduli space. By calculating the local zeta function for enough primes the intersection point on the ray can be reconstructed using the CRT. Repeating this procedure for several directions yields the algebraic equation defining the splitting locus. In the following sections we will describe how to apply the deformation method directly to one-dimensional subspaces of threefold and fourfold moduli spaces.

\subsubsection{Threefolds}
Since we will only compute Euler factors for two-parameter models in this work, we will outline the method for the case $h_{2,1}=2$ and afterwords briefly discuss how the methods carry over to general threefolds. 

Along a generic line $(z_1,z_2)=(\lambda_1 z,\lambda_2 z)$ in the moduli space of a two-parameter model the differential ideal is reduced to a sixth-order differential operator $\mathcal{L}^{(6)}(z)$ in one variable. The degree of the operator is the same for every generic line in a given model. This operator is not self-adjoint and at $z=0$ the local exponents are not of MUM type, but instead are $(0,0,0,0,1,r)$ with $r=1$ in all cases that we studied except one case where $r=3$.\footnote{In the cases where $r\neq 1$ the indicials $1,r$ give a series solution starting as $z^r$ and a corresponding $\log$-solution.} The indicial structure arises as follows: By restricting the two-parameter triple-logarithmic period in the Frobenius basis \eqref{eq:frobbasis} to the line, we obtain a triple-log solution of the operator $\mathcal{L}^{(6)}(z)$, whose cubic log-term is multiplied by the restriction of the holomorphic period, corresponding to the $(0,0,0,0)$-indicial block. The holomorphic solution of the additional $(1,r)$-indicial block arises by taking the difference of the restricted two-parameter single-logarithmic periods. The second solution of this indicial block is then a single-logarithmic solution, obtained by taking the difference of the restricted two-parameter double-logarithmic periods, so that the double-log terms cancel.

Even though we do not have a MUM point we find that we can apply the deformation method nonetheless. Conjecturally the Frobenius then takes the form 
\begin{align}
    \label{eq:oneparamfrob}
    \Frob_p(z)=\mathbf{W}^{(1)}(z^p)^{-1}\mathbf{B}_0^{-1}\mathbf{V}_0\mathbf{B}_0\mathbf{W}^{(1)}(z)\,,
\end{align}
where $\mathbf{V}_0$ is the matrix \eqref{eq:V0matrix} for the two-parameter model, $\mathbf{W}^{(1)}_{ij}(z)=\theta^j\varpi^{(1)}_i(z)$ is the Wronskian of the Frobenius basis $\underline{\varpi}^{(1)}(z)$ of the one-parameter operator $\mathcal{L}^{(6)}(z)$ around $z=0$ and $\mathbf{B}_0$ is the constant transformation matrix from said Frobenius basis into the basis \eqref{eq:frobbasis} of the two-parameter model restricted to $(z_1,z_2)=(\lambda_1 z,\lambda_2 z)$. The matrix $\mathbf{B}_0$ contains $\log(\lambda_1),\log(\lambda_2)$ values which are understood to be replaced by the $p$-adic Iwasawa logarithms\footnote{For a compact introduction to $p$-adic numbers and the definitions of the $p$-adic logarithms and zeta function we refer to appendix A in \cite{Candelas:2024vzf}.} $\log_p(\lambda_1),\log_p(\lambda_2)$. 

The discriminant of $\mathcal{L}^{(6)}(z)$ is of the form $\Delta^{(1)}(z)=\Delta^{(1)}_{\text{app}}(z)\Delta^{(1)}_{\text{geom}}(z)$, where $\Delta^{(1)}_{\text{geom}}(z)=\Delta^{(2)}(\lambda_1 z,\lambda_2 z)$, with $\Delta^{(2)}(z_1,z_2)$ being the discriminant of the two-parameter model as it appears in the denominator of the determinant of the two-parameter model period matrix $\boldsymbol{\Pi}^{(2)}(z_1,z_2)$. The factor $\Delta^{(1)}_{\text{app}}(z)$ is the locus of apparent singularities defined by the numerator of the determinant of the Wronskian $\mathbf{W}^{(1)}(z)$. The latter takes in the computed cases the form
\begin{align}
    \label{eq:order6detW}
    \det \mathbf{W}^{(1)}(z)\propto \exp\left(-\int\text{d}z\frac{S_5(z)}{z\,\Delta^{(1)}(z)}\right)=\frac{z^{r+1}\Delta^{(1)}_{\text{app}}(z)}{\big(\Delta^{(1)}_{\text{geom}}(z)\big)^4}\,,
\end{align}
where we wrote the operator as $\mathcal{L}^{(6)}(z)=\sum_{i=0}^{6}S_i(z) \theta^i$ with $S_i\in \mathbb{Q}[z]$ and $S_6(z)=\Delta^{(1)}(z)$. 

Again, the Frobenius converges for the correct values of $\chi/c_{ijk}$, to any finite $p$-adic accuracy $n_{\text{acc}}$, to a rational expression. In all computed cases we find that we can then write for almost all directions
\begin{align}
\label{eq:order6rationalFrob}
   \Frob_p(z)\equiv \frac{\Frob^{(n_{\text{acc}})}_p(z)}{\Delta^{(1)}_{\text{app}}(z^p)\big(\Delta^{(1)}_{\text{geom}}(z^p)\big)^{n_{\text{acc}}-6}}\mod p^{n_{\text{acc}}}\,, 
\end{align}
where now $\Frob^{(n_{\text{acc}})}_p\in \text{Mat}_{6\times 6}(\mathbb{Q}[z])$ and again typically $\deg((\Frob^{(n_{\text{acc}})}_p)_{ij})\lesssim p\cdot \deg_z(\mathcal{L}^{(6)}(z))$. We will see below that we need only calculate to $n_{\text{acc}}=6$ to compute the two-parameter Euler factors and hence the $\Delta^{(2)}$-factor does not contribute. We can thus also compute the Euler factors for singular fibres when calculating $\mod p^6$. There are  exceptional directions where \eqref{eq:order6rationalFrob} fails to hold. For example, in the case of toric hypersurfaces with five vertices, these are directions including a point of intersection of the two irreducible components of the discriminant $\Delta^{(2)}$. In this case $\Delta^{(2)}$ contributes to the denominator with $\big(\Delta^{(1)}_{\text{geom}}(z^p)\big)^{\max\{n_{\text{acc}}-4,0\}}$ instead.\footnote{Note that if we were to apply the deformation method to the whole two-parameter moduli space, we would have a denominator contribution of $\Delta^{(2)}(z_1^p,z_2^p)^{\max\{n_{\text{acc}}-4,0\}}$ and would not be able to compute the Frobenius traces at the singular fibres to accuracy $n_{\text{acc}}=6$ as possible here for singular fibres lying on a generic direction.} 

The presence of the apparent singularities in the denominator of \eqref{eq:order6rationalFrob} is again an obstruction to compute the Euler factors at all smooth fibres. Also here we can resolve the apparent singularities by modifying the Wronskian via $\mathbf{W}^{(1)}(z)\mapsto \mathbf{W}^{(1)}(z) \mathbf{U}(z)$ where $\mathbf{U}(z)\in \text{Mat}_{6\times 6}(\mathbb{Q}(\!(z)\!))$ is chosen with $\det \mathbf{U}(z)=1/\Delta_{\text{app}}^{(1)}(z)$. To avoid poles around $\Delta_{\text{app}}^{(1)}(z)=0$ in the entries of $\mathbf{W}^{(1)}(z)$, analytically continued to $\Delta^{(1)}_{\text{app}}(z)=0$, and get rid of the $\Delta^{(1)}_{\text{app}}$-contribution to the denominator of the Frobenius, we need to choose the rational entries of $\mathbf{U}(z)$ in such a way as to cancel such poles. In the examples we study, the apparent singularities always have local exponents $(0,1,2,3,4,6)$. This motivates to use the following ansatz for the inverse of $\mathbf{U}(z)$:
\begin{align}
\label{eq:order6Umatrix}
    \mathbf{U}(z)^{-1}=
\begin{pmatrix}
1 & 0 & 0 & 0 & 0 & p_1(z) \\
0 & 1 & 0 & 0 & 0 & p_2(z) \\
0 & 0 & 1 & 0 & 0 & p_3(z) \\
0 & 0 & 0 & 1 & 0 & p_4(z) \\
0 & 0 & 0 & 0 & 1 & p_5(z) \\
0 & 0 & 0 & 0 & 0 & \Delta_{\text{app}}^{(1)}(z)
\end{pmatrix}
\end{align}
where the $p_i$ are polynomials of degree at most $\deg(\Delta_{\text{app}}^{(1)})-1$. They are again fixed by the cancellation of poles in the entries of the analytic continuation of $\widetilde{\mathbf{W}}^{(1)}(z)=\mathbf{W}^{(1)}(z)\mathbf{U}(z)$ to $\Delta_{\text{app}}^{(1)}=0$. With the modified Frobenius,
\begin{align}
\label{eq:modifiedFrob}
    \widetilde{\Frob}_p(z)=\widetilde{\mathbf{W}}^{(1)}(z^p)^{-1}\mathbf{B}_0^{-1}\mathbf{V}_0\mathbf{B}_0\widetilde{\mathbf{W}}^{(1)}(z)\,,
\end{align}
we have to finite $p$-adic accuracy
\begin{align}
\label{eq:rationalmodifiedFrob}
   \widetilde{\Frob}_p(z)\equiv \frac{\widetilde{\Frob}^{(n_{\text{acc}})}_p(z)}{\big(\Delta^{(1)}_{\text{geom}}(z^p)\big)^{\max\{n_{\text{acc}}-6,0\}}}\mod p^{n_{\text{acc}}}\,, 
\end{align}
with $\widetilde{\Frob}^{(n_{\text{acc}})}_p\in \text{Mat}_{6\times 6}(\mathbb{Q}[z])$. The apparent discriminant no longer contributes to the denominator.

Having resolved the apparent singularities we can obtain the Frobenius traces from the deformation method to compute the Euler factors for all smooth points. The Weil conjectures imply that they take the form 
\begin{align}
\begin{split}
       P^{(p)}_3(X,T)=1+\alpha^{(p)}_1T+\alpha^{(p)}_2pT^2+\alpha^{(p)}_3p^2T^3+\alpha^{(p)}_2p^{4}T^4+\alpha^{(p)}_1p^{6}T^5+p^9T^6 
\end{split}
\end{align}
and the coefficients satisfy the bounds 
\begin{align}
    \vert \alpha^{(p)}_1\vert\leq 6p^{\frac{3}{2}}<p^3\,,\quad \vert p\alpha^{(p)}_2\vert \leq 15p^{3}<p^5\,,\quad \vert p^2\alpha^{(p)}_3\vert\leq 20p^{\frac{9}{2}}<p^6\,,
\end{align}
with the second $<$ holding for $p\geq 11$. From this we see that it is enough to calculate $\mod p^{6}$. To this $p$-adic accuracy, for a generic direction, also the singular discriminant factor cancels in the rational expression of the Frobenius and we can consequently also compute the Euler factors for singular fibres when working to this precision. Of course, then the Euler factors take a different form, depending on the degeneration type of the singularity\footnote{Even though the Weil conjectures imply, that for the singular fibres, less $p$-adic accuracy is required, we find it nonetheless useful to compute the Frobenius traces to the precision $n_{\text{acc}}=6$ as a consistency check.}. As outlined in \cref{app:LMHS} this form is dictated by the limiting mixed Hodge structure (LMHS), which can be computed from the local monodromy. We will later compute the Euler factors also for conifold and strong coupling singularities.

\paragraph{Example: Bicubic in $\mathbb{P}^2\times \mathbb{P}^2$}

To clarify the above notation we consider as a simple example the mirror of the bi-degree $(3,3)$-hypersurface in $\mathbb{P}^2\times \mathbb{P}^2$ and the direction $(z_1,z_2)=(2z,z)$ in moduli space. The operator annihilating the restriction of the periods to this direction is given by 
\begin{align}
\begin{split}
    \mathcal{L}^{(6)}(z)&=
    \left(\theta -1\right)^2 \theta ^4+9\,z\,\theta ^2 \left(\theta  \left(3 \theta  \left(9 \left(\theta -8\right) \theta +29\right)-16\right)-2\right)\\
    &\qquad\quad-27\,z^2 \left(3 \theta +1\right) \left(3 \theta +2\right) \left(3 \theta  \left(3 \theta  \left(9 \theta  \left(7 \theta +5\right)+59\right)+190\right)+200\right)\\ 
    &\qquad\quad-729\, z^3\left(3 \theta +1\right) \left(3 \theta +2\right) \left(3 \theta +4\right) \left(3 \theta +5\right) \left(9 \theta  \left(19 \theta +24\right)+143\right)\\
    &\qquad\quad-354294\, z^4\left(3 \theta +1\right) \left(3 \theta +2\right) \left(3 \theta +4\right) \left(3 \theta +5\right) \left(3 \theta +7\right) \left(3 \theta +8\right).
   \end{split}
\end{align}
The discriminant $\Delta^{(1)}(z)$ of the operator is 
\begin{align}
    \Delta^{(1)}(z)=\left(1 + 486 z\right) \left(1 - 243 z \left(1 + 81 z \left(1 + 27 z\right)\right)\right).
\end{align}
Using equation \eqref{eq:order6detW} we obtain the one-parameter Wronskian determinant:
\begin{align}
\label{eq:detWbicubic}
    \det \mathbf{W}^{(1)}(z)\propto  \frac{z^2 \left(1+486 z\right)}{\left(1-243 z \left(1+81 z \left(1+27 z\right)\right)\right)^4}\,,
\end{align}
where the denominator is ${\Delta^{(2)}(2z,z)}^4$, with the discriminant of the bicubic given by
\begin{align}
\begin{split}
    \Delta^{(2)}(z_1,z_2)=1-81 z_1-81 z_2&+2187 z_1^2-15309 z_1z_2+2187 z_2^2\\&-19683 z_1^3-59049 z_1^2z_2 -59049  z_1z_2^2-19683 z_2^3\,.    
\end{split}
\end{align}
From the numerator of \eqref{eq:detWbicubic} we can read off the apparent singularity of the operator:
\begin{align}
    \Delta^{(1)}_{\text{app}}(z)=1+486 z\,.
\end{align}
The modification $\mathbf{W}^{(1)}(z)\mapsto \widetilde{\mathbf{W}}^{(1)}(z)=\mathbf{W}^{(1)}(z) \mathbf{U}(z)$ to cure the apparent singularities is found to be
\begin{align}
    \mathbf{U}(z)=\left(
\begin{array}{cccccc}
 1 & 0 & 0 & 0 & 0 & -\frac{1240}{74601 (1+486 z)} \\
 0 & 1 & 0 & 0 & 0 & -\frac{2858}{24867  (1+486 z)} \\
 0 & 0 & 1 & 0 & 0 & -\frac{1969}{8289  (1+486 z)} \\
 0 & 0 & 0 & 1 & 0 & -\frac{59}{2763  (1+486 z)} \\
 0 & 0 & 0 & 0 & 1 & -\frac{553}{921  (1+486 z)} \\
 0 & 0 & 0 & 0 & 0 & \frac{1}{1+486 z} \\
\end{array}
\right).
\end{align}
Finally, we need the matrix $\mathbf{B}_0$ transforming the one-parameter Frobenius basis into the restricted two-parameter Frobenius basis, i.e. $\underline{\omega}^{(2)}(2z,z)=\mathbf{B}_0\,\underline{\omega}^{(1)}(z)$. We find
\begin{align}
    \mathbf{B}_0=\left(
\begin{array}{cccccc}
 1 & 0 & 0 & 0 & 0 & 0 \\
 \log (2) & 1 & 0 & 0 & 0 & 0 \\
 0 & 0 & 1 & 0 & 0 & 0 \\
 0 & 0 & 3 \log (2) & 1 & 0 & 0 \\
 \frac{3 \log ^2(2)}{2} & 3 \log (2) & 3 \log (2) & 0 & 1 & 0 \\
 0 & 0 & \frac{3 \log ^2(2)}{2} & \log (2) & 0 & 1 \\
\end{array}
\right)\left(
\begin{array}{cccccc}
 1 & 0 & 0 & 0 & 0 & 0 \\
 0 & 1 & 0 & 0 & -9 & 0 \\
 0 & 1 & 0 & 0 & 9 & 0 \\
 0 & 0 & 9 & 0 & 0 & 27 \\
 0 & 0 & 9 & 0 & 0 & -27 \\
 0 & 0 & 0 & 18 & 0 & 0 \\
\end{array}
\right).
\end{align}
The Frobenius can then be computed from \eqref{eq:modifiedFrob},
where $\mathbf{B}_0$ is as above with $\log(2)$ replaced by its $p$-adic analogue $\log_p(2)$ and $\mathbf{V}_0$ as in \eqref{eq:4foldV0matrix} with the Euler characteristic $\chi=-168$. The maximum degree of the entries of $\widetilde{\Frob}^{(6)}_p$ in \eqref{eq:rationalmodifiedFrob} determines the necessary expansion order of the periods. For this model we find that an expansion order of $5p$ suffices. This model is exceptional in the sense that the $z$-degree of the operator is low compared to most other $h^{2,1}=2$ models in the Kreuzer-Skarke list. The typical degrees range between $10-30$, the highest degree we found is $52$.

\paragraph{Generalization to $h_{2,1}>2$}
For the case of general $h_{2,1}$ we can apply the same reasoning. For a generic direction $\uz=\underline{\lambda}z$, the periods are annihilated by an operator $\mathcal{L}^{(b_3)}(z)$ of order $b_3=2+2h_{2,1}$, which has local exponents $(0,0,0,0,1,1,2,2,\dots)$. The indicial structure arises as follows: The appearance of four zeros again follows from restricting the triple-logarithmic multi-parameter period to the line. We can further build $h_{2,1}-1$ independent pairs of solutions of $\mathcal{L}^{(b_3)}(z)$ by taking differences of the $h_{2,1}$ double-logarithmic periods, normalized so that the quadratic log-terms cancel. This yields $h_{2,1}-1$ single-logarithmic solutions of $\mathcal{L}^{(b_3)}(z)$, whose leading log-term is multiplied by the series solution obtained by subtracting the holomorphic parts of the multi-parameter single-logarithmic periods. The apparent singularities have in cases that we tested indicials $(0,1,2,\dots,b_3-2,b_3)$ and they are resolved by modifying the Wronskian by a matrix $\mathbf{U}(z)\in  \text{Mat}_{b_3\times b_3}(\mathbb{Q}(\!(z)\!))$ analogous to \eqref{eq:order6Umatrix}.

\subsubsection{Fourfolds}
For fourfolds the above technique of reducing multi-parameter models to one-dimensional linear subspaces may be applied in a similar way to calculate the characteristic polynomial $P^{(p)}_{4,\text{hor}}(X,T)$ of the Frobenius action on the horizontal subspace of $H^{4}(X)$. We outline the method first for the case $h_{3,1}=h^{\text{hor}}_{2,2}=2$ and again briefly comment afterwards on how to treat the general case.

The Frobenius is expressed through the periods again via \eqref{eq:oneparamfrob}, where now $V_0$ is \eqref{eq:4foldV0matrix}. Since the Hodge numbers of the middle cohomology are $(1,2,2,2,1)$ the operators are then of order eight with local exponents at $z=0$ again not of MUM type, but $(0,0,0,0,0,1,r_1,r_2)$ where $r_1=r_2=1$ for all computed models. Using the same notation as for the threefold case the determinant of the Wronskian is given by
\begin{align}
    \label{eq:detWfourfold}
    \det \mathbf{W}^{(1)}(z)\propto \exp\left(-\int\text{d}z\frac{S_7(z)}{z\,\Delta^{(1)}(z)}\right)=\frac{z^{1+r_1+r_2}\Delta^{(1)}_{\text{app}}(z)}{\big(\Delta^{(1)}_{\text{geom}}(z)\big)^{11/2}}\,.
\end{align} 
The apparent singularities are resolved in the same way as before. The analogue of \eqref{eq:rationalmodifiedFrob} then reads
\begin{align}
\label{eq:rationalmodifiedFrob4fold}
   \widetilde{\Frob}_p(z)\equiv \frac{\widetilde{\Frob}^{(n_{\text{acc}})}_p(z)}{\big(\Delta^{(1)}_{\text{geom}}(z^p)\big)^{\max\{n_{\text{acc}}-9,0\}}}\mod p^{n_{\text{acc}}}\,.
\end{align}

According to the Weil conjectures the characteristic polynomial $P^{(p)}_4(X,T)$ at smooth fibres has zeros $\lambda_{4,j}^{-1}$ of absolute value $p^{-2}$. The factor $P^{(p)}_{4,\text{hor}}(X,T)$ corresponding to the horizontal part should satisfy the functional equation
\begin{align}
\label{eq:4foldfunctionalequation}
    P^{(p)}_{4,\text{hor}}(X,T)=\epsilon_p p^{2b_4}T^{b_4}P^{(p)}_{4,\text{hor}}(X,p^{-4}T^{-1})\,,
\end{align}
with $\epsilon_p=\pm1$.
From this form it follows that we can write for $b_4=8$ 
\begin{align}
\begin{split}
       P^{(p)}_{4,\text{hor}}(X,T)=1&+\alpha^{(p)}_1T+\alpha^{(p)}_2pT^2+\alpha^{(p)}_3p^2T^3+\alpha^{(p)}_4p^4T^4\\
    &+\epsilon_p(\alpha^{(p)}_3p^{6}T^5+\alpha^{(p)}_2p^{9}T^6+\alpha^{(p)}_1p^{12}T^7+p^{16}T^8)\,. 
\end{split}
\end{align}
The Weil conjectures imply the bounds 
\begin{align}
    \vert\alpha^{(p)}_1 \vert\leq 8p^2<p^3\,,\quad  \vert p\alpha^{(p)}_2 \vert\leq 28p^4<p^6\,,\quad  \vert p^2\alpha^{(p)}_3 \vert\leq 56p^6<p^8\,,\quad   \vert p^4\alpha^{(p)}_4 \vert\leq 70p^8<p^{10}\,,
\end{align}
where the second $<$ hold for $p\geq 11$. For these primes it is hence enough to calculate modulo $p^{10}$ and take the centered lift. 

\paragraph{Generalization to $h_{3,1}>2$ and $h_{2,2}>2$}
For general fourfolds we obtain along a generic line $\uz=\underline{\lambda}z$ an operator $\mathcal{L}^{(b^{\text{hor}}_4)}(z)$ of order $b^{\text{hor}}_4=2+2h_{3,1}+h^{\text{hor}}_{2,2}$. The local exponents at $z=0$ generically consist of one $(0,0,0,0,0)$-block, $h_{3,1}-1$ blocks of multiplicity three and $h^{\text{hor}}_{2,2}-h_{3,1}$ blocks of multiplicity one. The double-logarithmic solutions of the multiplicity three blocks arise by taking differences of the restricted multi-parameter triple-logarithmic periods to cancel their triple-log terms. The  $h^{\text{hor}}_{2,2}-h_{3,1}$ additional holomorphic solutions are obtained because the set of differences of the $h^{\text{hor}}_{2,2}$ restricted double-logarithmic periods contain $h^{\text{hor}}_{2,2}-h_{3,1}$ pairs of single-logarithmic solutions, where the series mulitiplying the single-log terms are identical. Hence taking the difference of those single-logarithmic solutions gives $h^{\text{hor}}_{2,2}-h_{3,1}$ additional pure series solutions. In the cases we tested the apparent singularities have indicials $(0,1,2,\dots,b^{\text{hor}}_4-2,b^{\text{hor}}_4)$ and they are resolved in the same way as above.

\subsection{Arithmetic and analytic analysis of flux vacua}
\label{subsec:Arithmetic and analytic analysis of flux vacua}
\paragraph{The general strategy}
Given the Euler factors for smooth fibres along a line through the moduli space, we will now describe the approach to search for type IIB flux vacua in codimension one will be as follows. Since we will perform the scan over two-parameter models, we will assume for concreteness $h^{2,1}=2$, however the general case may be studied analogously. We look for persistent factorizations of the local zeta function into one factor of degree two and one factor of degree four induced by the splitting of the $(1,2,2,1)$ Hodge structure into a $(0,1,1,0)$ and a $(1,1,1,1)$ Hodge substructure over $\mathbb{Q}$. As long as such a locus is not located itself along a line through the origin, we would expect there to be persistent factorizations along every generic line. However, we can test for this possibility separately by a simple criterion: For the sixth-order operator to reduce along a line to a fourth order operator it is necessary that the local solution around $z=0$ corresponding to the indicial $1$ does not appear. But the indicial $1$ solution is given by the restriction of the difference of the series parts of the single-logarithmic periods of the two-parameter model. This yields a simple necessary condition to find or exclude such cases. In practice we find that for toric models such cases can be identified simply by symmetries between the $l$-vectors.

\paragraph{Factorization patterns and the Chebotarëv density theorem}
An important question is what information can be extracted from the factorization patterns of Euler factors over a finite range of primes. To make statements about possible splittings of Hodge structure, particularly the field $K$ over which the splitting happens and the algebraic equation $\mathcal{V}(\uz)=0$ giving the locus of fibres over which it occurs, the so-called \textit{Chebotarëv density theorem} yields helpful information. We consider first the case where the splitting field is $\mathbb{Q}$.

Let $q\in \mathbb{Q}[z]$ be a monic polynomial, which in our context will be the polynomial $q(z)=\mathcal{V}(\underline{\lambda}z)$ whose roots give the intersection of the splitting locus with the line $\uz=\underline{\lambda}z$ along which we compute the Euler factors. For the moment we assume $q$ to be irreducible. Further, let $K_q$ denote the associated field extension of $\mathbb{Q}$. The elements of the Galois group $\text{Gal}(K_q/\mathbb{Q})$ act by permuting the roots of $q$, thus giving an embedding of the Galois group into the permutation group $S_{\text{deg}(q)}$. For a given prime $p$ we can reduce the polynomial modulo $p$ and consider its factorization into irreducible polynomials, $q(z)=\prod_{i=1}^{k}q_i(z)\mod p$. The list $\underline{m}\equiv (\deg(q_1),\dots,\deg(q_k))$ gives a partition of $\text{deg}(q)$. The \textit{Frobenius-Chebotarëv theorem} asserts that the density of primes for which a factorization pattern $\underline{m}$ occurs is equal to the fraction of elements of $\text{Gal}(K_q/\mathbb{Q})$, that when viewed as permutations, have cycle type equal to $\underline{m}$.

One consequence is that we get a lower bound on the frequency of primes for which $q$ completely factorizes, i.e. has $\text{deg}(q)$ zeros in $\mathbb{F}_p$. Since there is only the element $\mathds{1}\in \text{Gal}(K_f/\mathbb{Q})$ with cycle type $\underline{m}=(1,\dots,1)$, the above tells us that this frequency should be $1/\abs{\text{Gal}(K_f/\mathbb{Q})}$ and hence at least $1/(\text{deg}(q)!)$. Consider for example the situation of a supersymmetric vacuum along a cubic equation $\mathcal{V}(\uz)=0$. Then, for every generic direction, we should expect that the frequency of primes, where over three or more points of $\mathbb{F}_p$ the $P_n^{(p)}(X,T)$ factorize, to be at least $1/6$. If $q$ is reducible, one can apply the theorem to each irreducible component to make predictions about the factorization patterns and compare those to the Euler factor data. 

The number of irreducible components of $q$, can be inferred from an important corollary\footnote{For a compact proof of this corollary we refer to appendix A in \cite{Candelas:2021tqt}.} of the Frobenius-Chebotarëv theorem in combination with Burnside's lemma: If $q$ has $r$ irreducible components, the number of zeros of $q$, averaged over $p$, is equal to $r$. In practice this will be our main diagnostic when searching for supersymmetric vacua.

Aside from supersymmetric vacua, we will also find examples for splittings of Hodge structure over imaginary-quadratic field extensions $K/\mathbb{Q}$. The Galois group is here of order two and we write $\text{Gal}(K/\mathbb{Q})=\{\mathbb{1},c\}$, with the nontrivial element $c$ corresponding to complex conjugation. Given a Hodge splitting $V\otimes_{\mathbb{Q}}K\cong W\oplus \overline{W}$ we then have to distinguish between two possible cases for a prime $p$. The first one is the case where $p$ is \emph{split} over the ring of integers $\mathcal{O}_K$, i.e. $(p)=\mathfrak{p}_{+}\mathfrak{p}_{-}$ for two distinct prime ideals $\mathfrak{p}_{\pm}$ and the image of the Frobenius in $\text{Gal}(\overline{\mathbb{Q}}/\mathbb{Q}) / \text{Gal}(\overline{\mathbb{Q}}/K)\cong \text{Gal}(K/\mathbb{Q})=\{\mathbb{1},c\}$ is the identity. In other words the Frobenius is $K$-linear and preserves $W$ and $\overline{W}$. For the Euler factors coming from reductions of the splitting locus we then expect a factorization into two factors of equal degree, $P^{(p)}_3(X,T)=Q_1(T)Q_2(T)$, where the factors are complex conjugates of each other, $Q_1(T)=\overline{Q_2}(T)$. The second case, where $(p)$ is prime over $\mathcal{O}_K$, is called an \emph{inert} prime. Here the Frobenius instead maps to the nontrivial element of $\text{Gal}(K/\mathbb{Q})$. As a consequence the Frobenius exchanges the two spaces $W$ and $\overline{W}$ and hence takes a block-antidiagonal form in a suitable basis. The Euler factors for inert primes are thus even polynomials in $T$. 

For a Hodge splitting over a general Galois field extension $K/\mathbb{Q}$, more general splitting types of the ideal $(p)$ can occur. These splitting types correspond to the image of $\text{Frob}_p$ in $\text{Gal}(K/\mathbb{Q})$ having a particular cycle type, where we again view $\text{Gal}(K/\mathbb{Q})$ as a subgroup of the symmetric group permuting the roots of the polynomial defining $K$. The general Chebotarëv density theorem then states that the density of primes for which a given splitting type occurs is equal to the proportion of elements of $\text{Gal}(K/\mathbb{Q})$ with this cycle type. In particular, the density of \textit{fully split} primes, where the Frobenius maps to the identity in $\text{Gal}(K/\mathbb{Q})$, is $1/|\text{Gal}(K/\mathbb{Q})|$.

For a finite set of primes the above statements are of course only statistical in nature and one needs to compute Euler factors for larger and larger amounts of primes to increase confidence that a Hodge splitting has been identified or conversely excluded. In the case, that we will most commonly encounter, namely $K=\mathbb{Q}$ and a linear $\mathcal{V}(\uz)$, we expect for the Euler factors, for all except a finite number of primes, to see at least one that factorizes. The finite set of exceptions is, if we write $q(z)=z-z_*$ with $z_*\in\mathbb{Q}$, the primes of bad reduction, where $z_*=0 \mod p$, $1/z_*=0\mod p$ or $\Delta^{(1)}(z_*)=0\mod p$. In appendix B of \cite{Candelas:2019llw} the authors outline how arguments along these lines may be used to constrain the possibility of an attractor point for the quintic or further unknown attractor points of \href{https://cycluster.mpim-bonn.mpg.de/operator.html?nn=4.3.1}{AESZ 34}. 

Once we have identified a candidate splitting field $K$ and locus $\mathcal{V}(\uz)=0$ using the factorization analysis and Chinese remaindering, we can perform an analytic continuation of the periods to the candidate locus and check whether there are indeed vanishing $K$-linear combinations of the periods corresponding to the Hodge splitting.  

\paragraph{Modularity of flux vacua}
Focussing now on supersymmetric type IIB flux vacua, the quadratic and quartic pieces of the Euler factors along a vacuum $\mathcal{V}(\uz)=0$ carry a lot of information, that fits nicely with the analytic information of a global integral basis of periods at the vacuum locus. The quadratic factors correspond conjecturally to the Euler factors of a family of elliptic curves over $\mathbb{Q}$ \cite{Kachru:2020abh}. Following the arguments of \cite{Kachru:2020abh} one can choose a basis of the lattice of fluxes $f,h$ stabilizing to the vacuum configuration, so that a basis of its periods is given by the derivatives transversal to the vacuum locus, 
\begin{align}
    \left(f^{\text{T}}\Sigma \partial_{\mathcal{V}}\underline{\Pi},\,h^{\text{T}}\Sigma \partial_{\mathcal{V}}\underline{\Pi}\right)\big\vert_{\mathcal{V}(\uz)=0}\,.
\end{align}
Moreover the $j$-function of the axio-dilaton of the vacuum configuration 
\begin{align}
    \tau=\frac{f^{\text{T}}\Sigma\partial_{\mathcal{V}}\underline{\Pi}}{h^{\text{T}}\Sigma \partial_{\mathcal{V}}\underline{\Pi}}\bigg\vert_{\mathcal{V}(\uz)=0}\,,
\end{align}
is the $j$-invariant of the family of rational elliptic curves. In particular it is a rational function. The modularity theorem of elliptic curves states that to each rational isogeny class of elliptic curves over $\mathbb{Q}$ one can associate a weight two modular form. Concretely we can write the quadratic part of the Euler factor as
\begin{align}
    E^{(2)}_p=1-a_p \left(pT\right)+p\left(pT\right)^2\,,
\end{align}
where $a_p$ is the Hecke eigenvalue of a weight two modular form. The form of this factor is up to $T\mapsto pT$ the one of an elliptic curve. This reflects a so-called \emph{Tate twist}. Given values of $a_p$ for enough primes one can search up the modular form, for example on the \href{https://www.lmfdb.org}{LMFDB} \cite{lmfdb}. In the rational isogeny class of elliptic curves, associated to this modular form via the modularity theorem, one should then find a rational model, which has the same $j$-invariant as the $j$-function of the axio-dilaton at that fibre on the vacuum locus. 
 
Note that the choice of fluxes only determines a $\text{GL}(2,\mathbb{Z})$-orbit of the axio-dilaton and it is only for some choices of fluxes within this orbit that the associated $j$-invariants will agree with those of a rational model in the isogeny class of elliptic curves over $\mathbb{Q}$ corresponding to the weight two modular form. Acting with a $\text{GL}(2,\mathbb{Z})$-transformation on the fluxes $(f,h)^{T}$, and hence $\tau$, induces an isogeny in general over $\mathbb{C}$. However, in some cases rational rescalings of $\tau$ do give rise to isogenies between elliptic curves over $\mathbb{Q}$. In this case there exist multiple choices for the fluxes, which then yield isogenous families of elliptic curves over $\mathbb{Q}$ (isogenous curves have the same number of points over finite fields and in particular the same Euler factors). For a more detailed discussion of isogenies and rational models of elliptic curves in this context we refer to \cite{Candelas:2023yrg}.

The quartic factor is interpreted as the Euler factor of a $(1,1,1,1)$-Hodge substructure, whose periods are given by the restricting the two-parameter model periods to the vacuum locus. In explicit examples we find the corresponding motive to be given either by a one-parameter Calabi--Yau threefold or to be reducible, as for example in the case of the $\psi=0$ locus of the model $X_{22211}$, where the $(1,1,1,1)$-Hodge substructure splits for all points on $\psi=0$ over $\mathbb{Q}(i)$. 

In the case of attractor points in one-parameter models, where the Euler factor factors into a weight two and a weight four modular part, it was shown in \cite{Bonisch:2022mgw} that the entries of the period matrix at the attractor are rational linear combinations of the periods and quasiperiods of the modular forms\footnote{For the mixed motives associated to singular fibres of Calabi-Yau families similar statements hold, as shown in \cite{Bonisch:2025cax}.}. While we did not verify this in the vacua that we find, we expect that similarly for a fibre on the vacuum locus the period matrix to be expressible in terms of periods and quasiperiods of the weight two form and further transcendental numbers coming from the period matrix of the $(1,1,1,1)$-motive.

\paragraph{Symmetries and vacua}
A way of generating candidate loci for supersymmetric vacua is to consider models where an involution symmetry is acting on the complex structure moduli space, so that any given fibre and its image are isomorphic. At the fixed point locus the automorphism acting on the Calabi-Yau induces via pullback an involutory automorphism of the middle cohomology. If the decomposition of the latter into eigenspaces of positive and negative eigenvalues is respectively of Hodge types $(3,0)+(0,3)$ and $(2,1)+(1,2)$ one obtains a supersymmetric vacuum.

An example is the bicubic in $\mathbb{P}^2\times\mathbb{P}^2$. The complex structure parameters $z_1$ and $z_2$ are symmetric and as discussed in \cite{Candelas:2023yrg} along the diagonal $z_1=z_2$ one finds a vacuum. In one-parameter cases supersymmetric vacua are equivalent to attractor points. Examples of attractor points can be found in \cite{Elmi2020}\cite{McGovern:2023eop}\cite{Bonisch:2022slo}. In \Cref{tab:attractor_list} we list additional examples we found and their associated weight two and four modular forms.
\begin{table}[H]
    \centering
    \begin{tabular}{c|c|c|c|c|c}
         $\mathcal{L}^{(4)}$ & $z_\ast$ & $N_2$ & $N_4$ &$f_2$ & $f_4$\\
         \hline
         \href{https://cycluster.mpim-bonn.mpg.de/operator.html?nn=4.5.93}{5.93} & $-1/2^3 3^2$ & $72$ & $144$ & \href{https://www.lmfdb.org/ModularForm/GL2/Q/holomorphic/72/2/a/a/}{72.2.a.a} & \href{https://www.lmfdb.org/ModularForm/GL2/Q/holomorphic/144/4/a/g/}{144.4.a.g} \\
         \href{https://cycluster.mpim-bonn.mpg.de/operator.html?nn=4.5.111}{5.111} & $-1/2^2 5$  & $50$ & $10$ & \href{https://www.lmfdb.org/ModularForm/GL2/Q/holomorphic/50/2/a/b/}{50.2.a.b} & \href{https://www.lmfdb.org/ModularForm/GL2/Q/holomorphic/10/4/a/a/}{10.4.a.a} \\
         \href{https://cycluster.mpim-bonn.mpg.de/operator.html?nn=4.5.132}{5.132} & $1/2^6 3^4$ & $90$ & $630$ & \href{https://www.lmfdb.org/ModularForm/GL2/Q/holomorphic/90/2/a/c/}{90.2.a.c} & \href{https://www.lmfdb.org/ModularForm/GL2/Q/holomorphic/630/4/a/t/}{630.4.a.t} \\
         \href{https://cycluster.mpim-bonn.mpg.de/operator.html?nn=4.6.15}{6.15} & $1/2^7$ & $72$ & $8$ & \href{https://www.lmfdb.org/ModularForm/GL2/Q/holomorphic/72/2/a/a/}{72.2.a.a} & \href{https://www.lmfdb.org/ModularForm/GL2/Q/holomorphic/8/4/a/a/}{8.4.a.a}
    \end{tabular}
    \caption{List of new rank two attractor points. The operators $\cL^{(4)}$ are listed in the CYCluster. The weight $2$ and $4$ modular forms $f_2$ and $f_4$ are referenced by their LMFDB label and their respective level $N_2$ and $N_4$.}
    \label{tab:attractor_list}
\end{table}
Most examples arise by taking the Hadamard product of elliptic curve families, that have an involution symmetry \cite{Elmi2020}. The Hadamard product of two families of elliptic curves $\mathcal{E}\rightarrow \mathbb{P}^1$,  $\tilde{\mathcal{E}}\rightarrow \mathbb{P}^1$ is defined as the fibre product $X_z\equiv \mathcal{E}\times_{\mathbb{P}^1} \widetilde{\mathcal{E}}_{t(z)}$, where $\widetilde{\mathcal{E}}_{t(z)}$ denotes the elliptic family $\widetilde{\mathcal{E}}$ twisted by the map $t(z)\!:\mathbb{P}^1 \rightarrow \mathbb{P}^1,\;\; s\mapsto z/s$, depending on a parameter $z$. From the holomorphic periods of the operators $\mathcal{L}^{(2)}_{\mathcal{E}}(s)$ and $\mathcal{L}^{(2)}_{\widetilde{\mathcal{E}}}(s)$,
\begin{align}
     \varpi_0(s)=\sum_{i=0}^{\infty}a_n s^n\,,\quad \widetilde{\varpi}_0(s)=\sum_{i=0}^{\infty}b_n s^n\,,
\end{align} 
one can write down the holomorphic period for the fourth order operator $\mathcal{L}^{(4)}_{X}(z)$ of the Hadamard product:
\begin{align}
     (\varpi_0\ast \widetilde{\varpi}_0)(z)=\sum_{i=0}^{\infty}a_nb_n z^n\,.
\end{align}
As discussed in \cite{Elmi:2023hof} the middle homology can be explicity constructed by taking unions of appropriate two-cycles in the fibres over contours of the base. 

Many of the degree two Calabi-Yau operators are constructed in this way \cite{2021arXiv210308651A}. Integral monodromy bases of Hadamard products using a modified Gamma class conjecture can be found in \cite{arXiv:2505.07685}. If the families of elliptic curves now possess involution symmetries, i.e. $\mathcal{E}_s\cong \mathcal{E}_{1/(\alpha s)}$ and $\widetilde{\mathcal{E}}_s\cong \widetilde{\mathcal{E}}_{1/(\beta s)}$ we also have $X_z\cong X_{1/(\alpha \beta z)}$ \cite{Elmi2020}. Under the involution symmetries the periods transform, up to a Kähler gauge transformation, into each other, while preserving the integral structure.

Following \cite{Elmi2020} we consider as an example the self-Hadamard product $a\ast a$, where $a$ is the operator 
\begin{align}
    \mathcal{L}^{(2)}_a(s)=\theta^2-s\,(7 \theta^2+7\theta +2)-8s^2\,(\theta +1)^2\,.
\end{align}
It possesses an involution symmetry under $s\mapsto -\frac{1}{8s}$. Its self-Hadamard product gives a fourth order operator,
\begin{align}
\begin{split}
        \mathcal{L}^{(4)}_{a\ast a}(z)=\theta^{4}
&-
z \left(4 + \theta \left(28 + \theta \left(77 + \theta \left(98 + 73\theta\right)\right)\right)\right)\\
&+
8 z^{2}\left(-60 + \theta \left(-256 + \theta \left(-363 + 65(-2+\theta)\theta\right)\right)\right)\\
&+
64 z^{3}\left(28 + \theta \left(180 + \theta \left(417 + 65\theta(6+\theta)\right)\right)\right)\\
&-
512 z^{4}\left(28 + \theta \left(124 + \theta \left(221 + \theta \left(194 + 73\theta\right)\right)\right)\right)
+
32768 z^{5}\left(1+\theta\right)^{4},
\end{split}
\end{align}
equivalent via an algebraic coordinate transformation to operator \href{https://cycluster.mpim-bonn.mpg.de/operator.html?nn=4.3.8}{AESZ 4.3.8}. Its Riemann symbol is 
\begin{equation}
\mathcal{P}_{\mathcal{L}^{(4)}_{a\ast a}}\left\{
\begin{array}{cccccc}
-\frac{1}{8} & 0 & \frac{1}{64} & \frac{1}{8} & 1 & \infty \\ \hline
0 & 0 & 0 & 0 & 0 & 1 \\
1 & 0 & 1 & 1 & 1 & 1 \\
1 & 0 & 1 & 3 & 1 & 1 \\
2 & 0 & 2 & 4 & 2 & 1
\end{array}
,\ z\right\}.
\end{equation}
Under a combined coordinate and Kähler gauge transformation a global integral basis of periods is transformed into itself up to a rational basis change:
\begin{align}
   \underline{\Pi}(z)=\mathfrak{M}\frac{1}{64z}\underline{\Pi}\bigg(\frac{1}{64z}\bigg)\,, 
\end{align}
with some $\mathfrak{M}\in  \text{Mat}_{4\times 4}(\mathbb{Q})$ that is up to rescaling integral symplectic. There are two symmetry fixed points $z_\ast = \pm 1/8$, of which $-1/8$ is a conifold point, while $z=1/8$ is an apparent singularity and in fact an attractor point. The Euler factors for smooth fibres $p=11$ are shown in \Cref{tab:Hadaaeulerfactors11}. At the reduction of the attractor point, $1/8=7 \mod 11$, the Euler factor splits into a weight two modular factor, corresponding to the form $f_2$ with LMFDB label \href{https://www.lmfdb.org/ModularForm/GL2/Q/holomorphic/14/2/a/a/}{14.2.a.a}, and a weight four modular factor, corresponding to the form $f_4$ with LMFDB label \href{https://www.lmfdb.org/ModularForm/GL2/Q/holomorphic/14/4/a/b/}{14.4.a.b}.

   \begin{table}[]
\centering
\renewcommand{\arraystretch}{1.4}
\resizebox{0.5\textwidth}{!}{%
\begin{tabular}{|c|c|}
\hline
$z \mod p$ & $P^{(p)}_3(X,T)$ \\
\hline \hline
$2$ & $p^6T^4 + 12p^3T^3 -118pT^2 + 12T + 1$ \\
\hline
$3$ & $p^6T^4 + 36p^3T^3 + 98pT^2 + 36T + 1$ \\
\hline
$6$ & $p^6T^4 + 24p^3T^3 + 134pT^2 + 24T + 1$ \\
\hline
$7$ & $\left(p^3T^2+1\right)\left(p^3T^2 - 48T + 1\right)$ \\
\hline
$8$ & $p^6T^4 + 12p^3T^3 -118pT^2 + 12T + 1$ \\
\hline
$9$ & $p^6T^4 + 36p^3T^3 + 98pT^2 + 36T + 1$ \\
\hline
$10$ & $p^6T^4 + 24p^3T^3 + 134pT^2 + 24T + 1$ \\
\hline
\end{tabular}
}
\caption{Euler factors at smooth fibres of the Hadamard product $a\ast a$ for $p = 11$.}
\label{tab:Hadaaeulerfactors11}
\end{table}

However there are also numerous examples of attractor points, that do not admit an interpretation as fixed points of symmetries, such as the attractor points $z\in \{-1/7,33\pm 8\sqrt{17}\}$ of \href{https://cycluster.mpim-bonn.mpg.de/operator.html?nn=4.3.1}{AESZ 34} \cite{Candelas:2019llw} or the attractor point $z=-1/5832$ of the model $X^{3,3}\subset \mathbb{P}^5$ (\href{https://cycluster.mpim-bonn.mpg.de/operator.html?nn=4.1.4}{AESZ 4}). We will later find examples in two-parameter models that similarly lack an explanation in terms of symmetry fixed points. To check whether the moduli space of a model is $\mathbb{Z}_2$-symmetric the arithmetic approach offers a useful diagnostic: The Euler factors at fibres connected via symmetry should be equal, up to a possible twist $T\mapsto -T$, depending on whether the action of $\text{Frob}_p$ on $X$ commutes with the involution or not. For the above example of the Hadamard product $a \ast a$, this can be seen in \Cref{tab:Hadaaeulerfactors11}. Hence, observing almost all\footnote{Some finite fibres may be mapped to infinity under symmetry action and the symmetry images of their Euler factors would then be invisible when computing only Euler factors in the interior of the moduli space.} Euler factors at a given prime to come in pairs, strongly suggests the presence of a symmetry. Note that a given direction might not be mapped to itself under a supposed symmetry action, so we need to compute all Euler factors for that prime if we are to analyze possible symmetries. In a two-parameter model this means calculating Euler factors along $p$ directions. For a fixed fibre $z_\ast$, its supposed symmetry image can then be obtained by determining for a number of primes the point in $\mathbb{F}_p$ with the same Euler factor as its reduction $z_\ast \mod p$ and applying the CRT. By making an ansatz for the symmetry action on the moduli and repeating this procedure for enough points in the moduli space one can then determine the precise form of the symmetry.

\section{Threefolds}
In this section, we apply in parallel the analytic and arithmetic methods discussed in \cref{sec:periodgeometry,sec:arithmeticmethods} to special loci in the moduli spaces of Calabi--Yau threefolds.
We distinguish between singular loci giving rise to a limiting mixed Hodge structure and smooth loci with a splitting into pure variations of Hodge structures.
The former will be the topic of \Cref{sec:CY3trans}, where we analyse and classify one-parameter models arising at base degenerations in fibred families.
The local zeta functions obtained via the deformation method are in agreement with the predictions of the monodromy weight filtration of the LMHS.
In \cref{sec:fluxvacuaCY3}, we perform a systematic scan of the Kreuzer-Skarke list of toric hypersurfaces with $h^{2,1}=2$ for supersymmetric flux vacua, as well as a search within a selection of complete intersections in toric ambient spaces. 

\paragraph{}

\subsection{Special singular loci}\label{sec:CY3trans}
The toric description of threefolds allows for discriminant factors originating from faces of dimensions one and two~\cite{MR1264417}. 
Edges, i.e.\ one-dimensional faces, lead to strong-coupling\footnote{The term comes from the $N=2$ heterotic/type II duality in 4d. 
In this case the base of the K3 fibrations in the type II description shrinks to zero size if the heteoric string coupling becomes large~\cite{Klemm:1995tj}\cite{Kachru:1995wm}.} transitions with a shrinking two-cycle in the mirror geometry whose desingularisation is well-understood.
Loci in the complex-structure moduli spaces where a discriminant factor of a two-face vanishes also furnish a relative period geometry, which we interpret as coming from a candidate desingularisation.
\Cref{sec:sctrans,sec:contrans} discuss detailed examples of the limiting period geometry in the two cases.
We will extend the discussion to more general bases in \cref{sec:gentrans} and end the section with remarks on compatibility of transitions due to dependencies in distinct faces in \cref{sec:remarks}.

\subsubsection{Generalized strong-coupling transition}
\label{sec:sctrans}
A shrinking two-cycle in a threefold family $\mathcal{\hat{X}}$ is associated with a vanishing\footnote{Models with a $\Z_n$ B-field twist can have rational Kähler moduli at the transition, see for example~\cite{Katz:2022lyl}\cite{Katz:2023zan}\cite{Schimannek:2025cok}.} Kähler modulus $t\rightarrow 0$\,.
Transitions appear for suitable values of the moduli corresponding to points on edges (one-dimensional faces) of the polytope~\cite{Klemm:1996kv}\cite{Morrison:1996pp}.
We review the well-known transition between the degree-eight hypersurfaces in $X_{22211}$ to the complete intersection of bi-degree $(4,2)$ in $\P^5$ and extend the analysis by a precise analysis of the period geometry and the arithmetic of the singular fibres.
We explain how the transition can be seen from the perspective of the toric polyhedron and how the one-parameter model can be recovered analytically and arithmetically.
\paragraph{}
The toric and topological data is listed in \Cref{tab:22211} in \cref{app:ttdata}. 
The polytope describing $\P_{22211}$ includes an inner point
\begin{equation}
    \nu_6 = (-1,-1,-1,0) = \frac{\nu_4+\nu_5}{2}\,,\hspace{1cm} 
    \begin{tikzpicture}
            \draw[step=1cm,gray,very thin] (-1,0) grid (1,0);
            \draw[thick, black, fill=blue!20] (-1,0) -- (1,0);
            \foreach \x in {-1,0,1}
                \fill (\x,0) circle (2pt);
            \node[above=0.1cm] at (0,0) {$\nu_6$};
            \node[above=0.1cm] at (-1,0) {$\nu_4$};
            \node[above=0.1cm] at (1,0) {$\nu_5$};
    \end{tikzpicture},
\end{equation}
where we depicted the face on which $\nu_6$ lies.
The defining polynomial restricted to the face spanned by $\nu_4$ and $\nu_5$ reads
\begin{equation}
    a_4 t_4+a_5\frac{1}{t_1^2 t_2^2 t_3^2 t_4}+a_6\frac{1}{t_1 t_2 t_3}\,.
\end{equation}
Since we assign the vertex-deformations the value $a_4=a_5=1$, it follows that the deformation parameter corresponding to $\nu_6$ should be set to $a_6=2$ for this polynomial to factorise
\begin{equation}
    \frac{\left(t_1 t_2 t_3 t_4-1\right)^2}{t_1^2 t_2^2 t_3^2 t_4}\,.
\end{equation}
In the Batyrev coordinates, this one-dimensional locus is described by
\begin{equation}
    z_1 = \frac{1}{4}\,,\quad z_2 = 2 z\,,\label{eq:X42translocus}
\end{equation}
where the coordinate $z$ parametrises the family $X^{4,2}$.
From the $\hat{\Gamma}$-class formalism, we obtain the integral period vector of $X_{22211}$
\begin{equation}
    \vec{\Pi}(\vec{t}) = \begin{pmatrix}
    1\\t_1\\t_2\\4 t_2 \left(t_1+t_2\right)-4t_2 -\frac{7}{3}\\2 t_2^2-1\\-\frac{2}{3} t_2^2 \left(3 t_1+2 t_2\right)-t_1-\frac{7 t_2}{3}+\frac{21 i \zeta(3)}{\pi ^3}
    \end{pmatrix}+\mathcal{O}(\vec{q})
\end{equation}
with the two MUM-monodromy representations of $\mathfrak{M}_i:\ \vec{t}\mapsto\vec{t}+e_i$
\begin{equation}
\label{eq:mum_x22211}
    \mathfrak{M}_1 = \left(
\begin{array}{cccccc}
 1 & 0 & 0 & 0 & 0 & 0 \\
 1 & 1 & 0 & 0 & 0 & 0 \\
 0 & 0 & 1 & 0 & 0 & 0 \\
 0 & 0 & 4 & 1 & 0 & 0 \\
 0 & 0 & 0 & 0 & 1 & 0 \\
 -2 & 0 & 0 & 0 & -1 & 1 \\
\end{array}
\right)\quad\text{and}\quad \mathfrak{M}_2\left(
\begin{array}{cccccc}
 1 & 0 & 0 & 0 & 0 & 0 \\
 0 & 1 & 0 & 0 & 0 & 0 \\
 1 & 0 & 1 & 0 & 0 & 0 \\
 0 & 4 & 8 & 1 & 0 & 0 \\
 2 & 0 & 4 & 0 & 1 & 0 \\
 -6 & -2 & -8 & -1 & 0 & 1 \\
\end{array}
\right).
\end{equation}
The two discriminant components can be obtained, for example, by toric methods and read
\begin{equation}
    \Delta_1 = 1-4z_1\quad\text{and}\quad \Delta_2 = 1-512 z_2 \left(1-128 z_2\left(1-4 z_1\right) \right).
\end{equation}
The locus of transition \eqref{eq:X42translocus} is given by $\Delta_1=0$.
A local basis around the point $(\Delta_1,z_2)=0$ is given by
\begin{equation}
\label{eq:scfrobbasis}
    \vec{\varpi}_{\text{sc}}=\left(
\begin{array}{c}
 \varpi_0 \\
 \log(z_2)\varpi_0+\sigma_1 \\
 \frac{1}{2} \log(z_2) ^2\varpi_0+\log(z_2)\sigma_1+\sigma_2 \\
 \frac{1}{6} \log ^3\left(z_2\right)\varpi_0+\frac{1}{2} \log(z_2)^2\sigma_1+\log(z_2)\,\sigma_2  +\sigma_3\\
 \sqrt{\Delta _1}\,\sigma_4\\
 \sqrt{\Delta _1} \left(\log (\Delta _1)+\log (z_2)\right)\sigma_4 +\sqrt{\Delta _1}\,\sigma_5 \\
\end{array}
\right),
\end{equation}
where $\sigma_i$ and $\varpi_0$ are formal power series in $\Delta_1$ and $z_2$.
The rotation into the integral basis $\vec{\Pi}=T\vec{\varpi}_\text{sc}$ uses the transition matrix found numerically
\begin{equation}
    {\footnotesize T_{\Delta_1,z_2}=\left(
\begin{array}{cccccc}
 1 & 0 & 0 & 0 & 0 & 0 \\
 0 & 0 & 0 & 0 & \frac{2}{\pi } & 0 \\
 \frac{i \log (2)}{2 \pi } & -\frac{i}{2 \pi } & 0 & 0 & -\frac{1}{\pi } & 0 \\
 -\frac{7}{3}-\frac{\log ^2(2)}{\pi ^2}-\frac{2 i \log (2)}{\pi } & \frac{\log (4)+2 i \pi }{\pi ^2} & -\frac{2}{\pi ^2} & 0 & \frac{4}{\pi } & 0 \\
 \frac{1}{6} \left(-7-\frac{3 \log ^2(2)}{\pi ^2}\right) & \frac{\log (2)}{\pi ^2} & -\frac{1}{\pi ^2} & 0 & \frac{2 (\pi +i (\log (8)-2))}{\pi ^2} & \frac{2 i}{\pi ^2} \\
 -\frac{i \left(-132 \zeta(3)-\log ^3(2)+7 \pi ^2 \log (2)\right)}{6 \pi ^3} & \frac{i \left(7 \pi ^2-3 \log ^2(2)\right)}{6 \pi ^3} & \frac{i \log (2)}{\pi ^3} & -\frac{i}{\pi ^3} & 0 & 0 \\
\end{array}
\right)}.
\end{equation}
We find the two fluxes\footnote{To be precise, the fluxes seen as elements in the homology are given by $g\Sigma^{-1}$. For better readability we absorb the intersection form in the fluxes.}
\begin{equation}
    f^T\Sigma = (0,1,0,0,0,0)\,,\quad \text{and}\quad h^ T\Sigma=(0,0,4,1,-2,0)
\end{equation}
satisfying $g^T\Sigma \vec{\Pi}(\Delta_1=0,z_2)=0$.
The axio-dilaton on $\Delta_1$ is given by
\begin{equation}
    \tau = \frac{f^T\Sigma\, \partial_{\Delta_1}\vec{\Pi}}{h^T\Sigma\, \partial_{\Delta_1}\vec{\Pi}}\Bigg|_{\Delta_1=0}=-\frac{2\ii \log(8z_2)}{\pi}-2\,.
\end{equation}
The expressions $g^T\Sigma\, \partial_{\Delta_1}\vec{\Pi}\sim \sqrt{\Delta_1}^{-1}$ are divergent\footnote{This divergence is not avoided in a covering space parametrised by $\psi=\Delta_1^2$. Instead, the expression $f^T\Sigma\, \partial_{\psi}\vec{\Pi}$ becomes finite ($2/\pi$), while $h^T\Sigma\, \partial_{\psi}\vec{\Pi}$ diverges  as logarithmically as $\psi\rightarrow 0$.} along $\Delta_1=0$, so unlike the case of a supersymmetric vacuum, one cannot give a family of elliptic curves with complex structure modulus $\tau$. 

\paragraph{}
Note that the vanishing flux $f$ corresponds to $t_1\rightarrow 0$.
This implies that one can obtain the pre-potential of the one-parameter model $X^{4,2}$ from that of $X_{22211}$ by setting $t_1=0$ and thus $q_1=1$. 
This limit is exact, meaning that not only do the polynomial terms in the mirror coordinates $\vec{t}$ in this limit coincide, but also the instanton corrections.
Therefore, one can read-off all the topological data of the complete intersection from that of the two-parameter model.
Furthermore, the period vector $\Pi_{\Delta_1=0,z_2}$ is given by the integral periods of the model $X^{4,2}$ again obtained via the $\hat{\Gamma}$-class formalism. 
We use the fluxes to omit the periods $X^1$ and $F_1$ corresponding to $t_1$ and $\partial_{t_1}\mathcal{F}$, respectively.
Note, however, that one must first perform the change of variables to $z=\frac{1}{2}z_2$ to obtain this identification, cf.\ \cref{eq:X42translocus}.

On the arithmetic side, we first use the LMHS to constrain the form of the local zeta function at the strong coupling locus. 
Following \cref{app:LMHS} it is enough to know the Jordan decomposition of the local monodromy.
In  \eqref{eq:scfrobbasis} we have one solution of indicial $1/2$ and a corresponding logarithmic solution.
Assembling the ranks of the associated pure motives in a vector, with entries corresponding to the possible weights $j=0,\dots,6$ of their Hodge structure, one arrives at \eqref{eq:dWinvsc}, so in our case $(0,0,0,4,0,0,0)$.
We infer that the form of the Euler factors at the strong coupling locus matches that of a one-parameter threefold, i.e. a single pure motive of rank four and weight three:
\begin{align}
    P^{(p)}_\text{sc}(X,T)=1+\alpha^{(p)}_1T+\alpha^{(p)}_2 pT^2+\alpha^{(p)}_1p^3T^3+p^6T^4\,.
\end{align}
As an example we give the $p=11$ Euler factors for the diagonal $z_1=z_2$ in \Cref{tab:X8eulerfactors11}. Indeed the strong coupling Euler factor takes the expected form with the fourth order polynomial being the Euler factor of the corresponding fibre of the one-parameter family $X^{4,2}$.

    \begin{table}[]
\centering
\renewcommand{\arraystretch}{1.6}
\resizebox{\textwidth}{!}{%
\begin{tabular}{|c|c|c|}
\hline
\multicolumn{3}{|c|}{$p=11$ for direction $(z_1,z_2)=(z,z)$} \\
\hline
$z \mod p$ & smooth/sing. & $P^{(p)}_3(X,T)$ \\
\hline \hline
$1$ & smooth & $p^9T^6 - 28p^6T^5 - 5p^4T^4 + 344p^2T^3 - 5pT^2 - 28T + 1$ \\
\hline
$2$ & smooth & $\left(p^3T^2 - 4pT + 1\right)\left(p^6T^4 + 96p^3T^3 + 446pT^2 + 96T + 1\right)$ \\
\hline
$3$ & strong coupling & $p^6T^4 - 4p^3T^3 - 78pT^2 - 4T + 1$ \\
\hline
$4$ & smooth & $\left(p^3T^2 + 1\right)\left(p^6T^4 + 12p^3T^3 - 38pT^2 + 12T + 1\right)$ \\
\hline
$5$ & smooth & $p^9T^6 - 4p^7T^5 + 107p^4T^4 - 72p^2T^3 + 107pT^2 - 4pT + 1 $ \\
\hline
$6$ & smooth & $p^9T^6 - 8p^6T^5 + 119p^4T^4 + 72p^2T^3 + 119pT^2 - 8T + 1$ \\
\hline
$7$ & smooth & $p^9T^6 + 20p^6T^5 + 23p^4T^4 - 288p^2T^3 + 23pT^2 + 20T + 1$ \\
\hline
$8$ & smooth & $p^9T^6 + 28p^6T^5 + 63p^4T^4 + 288p^2T^3 + 63pT^2 + 28T + 1$ \\
\hline
$9$ & smooth & $p^9T^6 + 26p^6T^5 + 59p^4T^4 + 156p^2T^3 + 59pT^2 + 26T + 1$ \\
\hline
$10$ & smooth & $p^9T^6 - 42p^6T^5 + 103p^4T^4 - 140p^2T^3 + 103pT^2 - 42T + 1$ \\
\hline
\end{tabular}
}
\caption{Euler factors of the model $X_{22211}$ along the direction $(z_1,z_2)=(z,z)$ for $p = 11$.}
\label{tab:X8eulerfactors11}
\end{table}

As was noted under \eqref{eq:order6rationalFrob}, when choosing the direction containing the intersection between the conifold and the strong coupling locus, the discriminant appears in the denominator of the Frobenius with power\footnote{The phenomenon of an increase in the power of the discrmininant in the denominator of the Frobenius also occurs frequently in one-parameter Calabi-Yau operators for singularities that contribute with quadratic or higher multiplicity to the discriminant.} $n_{\text{acc}}-4$ instead of $n_{\text{acc}}-6$. In particular we can only determine the Frobenius traces at the points of bad reduction up to $p$-adic precision $n_{\text{acc}}=4$. However, according to eq. \eqref{eq:dWinvsccon}, the LMHS tells us that the Euler factor at the intersection should contain a linear factor satisfying $n=2$ Weil conjectures and a quadratic factor satisfying $n=3$ Weil conjectures, i.e. in accordance with the presence of the transition the same as for the conifold of a one-parameter threefold:
\begin{align}
\begin{split}
       P^{(p)}_3(X,T)=\left(1-p\chi(p)T\right)\big(1+\alpha^{(p)}T+p^3T^2\big).
\end{split}
\end{align}
It is then enough to calculate the linear and quadratic Frobenius traces modulo $p^3$ to determine $\alpha^{(p)}$ and $\chi(p)$ and we can further calculate the cubic Frobenius trace modulo $p^4$ and check that the coefficient of $T^3$ vanishes to this precision. We checked that this factor agrees as expected with the conifold Euler factors of the one-parameter model $X^{4,2}$, for which $\chi(p)=\big(\frac{2}{p}\big)$ and $-\alpha^{(p)}$ are the Hecke eigenvalues of the weight four form $f_4$ with LMFDB label \href{https://www.lmfdb.org/ModularForm/GL2/Q/holomorphic/16/4/a/a/}{16.4.a.a}, for primes up to $p=149$. 

Also note that so far we only computed the Euler factors at finite fibres.
However, the same methods can be used to compute the Euler factors at $z_1=\infty$ or $z_2=\infty$ after an algebraic coordinate transformation that maps these loci to the interior of the moduli space while keeping the MUM point at the origin.
Since the $z_1=\infty$ (equivalently $\psi=0$) locus furnishes a supersymmetric vacuum \cite{DeWolfe:2005gy,Kachru:2020sio} we compute also here the Euler factors. We transform the periods according to
\begin{align}
    \Pi(z_1,z_2)\mapsto \Pi^\prime(z^\prime_1,z^\prime_2)= \frac{1}{1-256z^\prime_1}\Pi\left(\frac{z^{\prime 4}_1}{(1 - 256z^\prime_1)^4},z^\prime_2\right).
\end{align}
The coordinate transformation has the effect that $z_1=\infty$ is mapped to $z^\prime_1=\frac{1}{256}$ and we go to a covering space where the local solutions are regular power series instead of having fourth root indicials in $z_1$. The Kähler gauge transformation just has the effect of shifting the indicials so that one solution starts with a constant term as for a generic regular point in the moduli space. Since the transformation scales $t^1\mapsto t^1/4$ we have to change the classical intersection numbers appearing in the chosen Frobenius basis around the MUM point to 
\begin{align}
    Y^\prime_{111}=4^3\cdot 8,\quad  Y^\prime_{112}=4^2\cdot 4,\quad Y^\prime_{122}=Y^\prime_{222}=0.
\end{align}
Then, we can again apply the deformation method as before along linear subspaces and compute Euler factors also for fibres corresponding to $z_1=\infty$.
As an example we give the Euler factors for $z_1=\infty, z_2=1$ for primes $p=29$ to $p=149$ in \Cref{tab:X8eulerfactorsinfinity}. As expected the Euler factors factorize into a quadratic and quartic factor, with the quadratic factor belonging to the weight two form \href{https://www.lmfdb.org/ModularForm/GL2/Q/holomorphic/48/2/a/a/}{48.2.a.a}. 

\begin{table}
\centering
\renewcommand{\arraystretch}{1.2}
\resizebox{0.5\textwidth}{!}{%
\begin{tabular}{|c|c|}
\hline
\multicolumn{2}{|c|}{$(z_1,z_2)=(\infty,1)$} \\
\hline
$p$ & $P^{(p)}_3(X,T)$ \\
\hline \hline
$29$ &  $\left(p^3T^2 - 6pT + 1\right)\left(p^6T^4 + 82p^3T^3 + 1738pT^2 + 82T + 1\right)$ \\
\hline
$31$ &  $\left(p^3T^2 - 4pT + 1\right)\left(p^3T^2 + 4pT + 1\right)\left(  p^3T^2 + 8pT + 1\right)$ \\
\hline
$37$ & $ \left(p^3T^2 - 6pT + 1\right)\left(p^6T^4 - 420p^3T^3 + 3494pT^2 - 420T + 1\right)$ \\
\hline
$41$ & $\left(p^3T^2 + 6pT + 1\right)\left( p^6T^4 - 434p^3T^3 + 2578pT^2 - 434T + 1\right)$ \\
\hline
$43$ &  $ \left(p^3T^2 - 8pT + 1\right)\left(p^3T^2 + 4pT + 1\right)\left(p^3T^2 + 8pT + 1\right)$ \\
\hline
$47$ &  $ \left(p^3T^2 + 1\right)\left(p^6T^4 + 62p^2T^2 + 1\right)$ \\
\hline
$53$ &  $ \left(p^3T^2 + 2pT + 1\right)\left(p^6T^4 - 102p^3T^3 - 4086pT^2 - 102T + 1\right)$ \\
\hline
$59$ &  $ \left(p^3T^2 + 4pT + 1\right)\left( p^6T^4 - 10p^2T^2 + 1\right)$ \\
\hline
$61$ &  $ \left(p^3T^2 + 2pT + 1\right)\left(p^6T^4 + 44p^3T^3 - 6714pT^2 + 44T + 1\right)$ \\
\hline
$67$ &  $ \left(p^3T^2 - 8pT + 1\right)\left(p^3T^2 - 4pT + 1\right)\left( p^3T^2 + 8pT + 1\right)$ \\
\hline
$71$ &  $ \left(p^3T^2 + 8pT + 1\right)\left(p^6T^4 + 110p^2T^2 + 1\right)$ \\
\hline
$73$ &  $ \left(p^3T^2 - 10pT + 1\right)\left(p^6T^4 - 1442pT^2 + 1\right)$ \\
\hline
$79$ &  $ \left(p^3T^2 - 8pT + 1\right)\left(p^3T^2 - 4pT + 1\right)\left( p^3T^2 + 4pT + 1\right)$ \\ 
\hline
$83$ &  $ \left(p^3T^2 - 4pT + 1\right)\left(p^6T^4 + 134p^2T^2 + 1\right)$ \\ 
\hline
$89$ &  $ \left(p^3T^2 + 1670T + 1\right)\left( p^3T^2 - 16pT + 1\right)\left( p^3T^2 + 6pT + 1\right)$ \\ 
\hline
$97$ &  $ \left(p^3T^2 - 594T + 1\right)\left(p^3T^2 - 2pT + 1\right)\left( p^3T^2 + 18pT + 1\right)$ \\ 
\hline
$101$ &  $ \left(p^3T^2 + 18pT + 1\right)\left(p^6T^4 - 2054p^3T^3 + 26218pT^2 - 2054T + 1\right)$ \\ 
\hline
$103$ &  $ \left(p^3T^2 - 4pT + 1\right)\left(p^3T^2 + 4pT + 1\right)\left( p^3T^2 + 16pT + 1\right)$ \\ 
\hline 
$107$ &  $ \left(p^3T^2 + 1\right)^2\left(p^3T^2 - 12pT + 1\right)$ \\ 
\hline
$109$ &  $ \left(p^3T^2 + 2pT + 1\right)\left(p^6T^4 + 1440p^3T^3 + 6334pT^2 + 1440T + 1\right)$ \\ 
\hline
$113$ &  $ \left(p^3T^2 - 1328T + 1\right)\left(p^3T^2 - 18pT + 1\right)\left(p^3T^2 + 14pT + 1\right)$ \\ 
\hline
$127$ &  $ \left(p^3T^2 - 12pT + 1\right)\left(p^3T^2 - 8pT + 1\right)\left( p^3T^2 + 12pT + 1\right)$ \\ 
\hline
$131$ &  $ \left(p^3T^2 - 4pT + 1\right)\left(p^6T^4 + 134p^2T^2 + 1\right)$ \\ 
\hline
$137$ &  $ \left(p^3T^2 + 6pT + 1\right)\left(p^6T^4 - 34p^3T^3 - 36686pT^2 - 34T + 1\right)$ \\ 
\hline
$139$ &  $ \left(p^3T^2 - 12pT + 1\right)\left(p^3T^2 - 8pT + 1\right)\left( p^3T^2 + 8pT + 1\right)$ \\ 
\hline
$149$ &  $ \left(p^3T^2 - 14pT + 1\right)\left(p^6T^4 + 178p^3T^3 - 28262pT^2 + 178T + 1\right)$ \\ 
\hline
\end{tabular}
}
\caption{Euler factors of the model $X_{22211}$ for $(z_1,z_2)=(\infty,1)$ for primes $p=29,\dots,149$.}
\label{tab:X8eulerfactorsinfinity}
\end{table}

Generators of the flux lattice $\Lambda$ realizing the $\psi=0$ vacuum locus can be chosen as 
\begin{align}
    &f^{\text{T}}=(2, 2, -1, 0, 0, 0)\,,\\
    &h^{\text{T}}=(0, 1, 0, 2, 0, 2)\,.
\end{align}
The derivatives transversal to the vacuum locus, 
\begin{align}
    \left(f^{\text{T}}\Sigma \partial_{\psi}\underline{\Pi},\,h^{\text{T}}\Sigma \partial_{\psi}\underline{\Pi}\right)\big\vert_{\psi=0}\,,
\end{align}
are periods of the family of elliptic curves with Picard--Fuchs operator 
\begin{align}
    \mathcal{L}^{(2)}_{\text{el}}=\theta^2-4z \left(4 \theta+1\right) \left(4 \theta+3\right),
\end{align}
where $z=z_2/16$. This hypergeometric operator is referred to as C in the literature, see \Cref{tab:elliptic1}. The axio-dilaton is given by 
\begin{align}
    \tau(z_2)=\frac{f^{\text{T}}\Sigma \partial_{\psi}\underline{\Pi}}{h^{\text{T}}\Sigma \partial_{\psi}\underline{\Pi}}\bigg\vert_{\psi=0}\,.
\end{align}
The corresponding $j$-invariant of this family of elliptic curves takes the rational form 
\begin{align}
    j(\tau(z_2))=\frac{256(1-3z_2)^3}{z_2^2(1-4z_2)}\,.
\end{align}
At the example point $z_2=1$ discussed above the $j$-invariant is $j(\tau(1))=2048/3$, which agrees with the $j$-invariant of the elliptic curve class \href{https://www.lmfdb.org/EllipticCurve/Q/48/a/5}{48.a5} in the isogeny class associated to the weight two modular form \href{https://www.lmfdb.org/ModularForm/GL2/Q/holomorphic/48/2/a/a/}{48.2.a.a} of the quadratic factor at that fibre. 

On the locus $\psi=0$ there are (after multiplying the periods by $\psi^{-1}$) two surviving periods annihilated by the second order operator 
\begin{align}
    \mathcal{L}^{(2)}=\theta^2-4z \left(8 \theta+1\right) \left(8 \theta+5\right),
\end{align}
with $z=z_2/64$.
One might wonder how the vanishing of four periods at the $\psi=0$ locus leaves room for a $(1,1,1,1)$ Hodge structure. A hint comes from the fact that as shown in \Cref{tab:X8CMsplitting} for primes $p\equiv 1 \mod 4 $ the fourth order factors in \Cref{tab:X8eulerfactorsinfinity} factorize further into two quadratic factors over $\mathbb{Q}(i)$ and for $p\equiv 3\mod 4$ the polynomials are even. As we will argue in the following this reflects an additional splitting over $\mathbb{Q}(i)$ of the $(1,1,1,1)$ Hodge structure into Hodge structures of type $(1,1,0,0)$ and $(0,0,1,1)$ at any value of $z_2$ and the above $\mathcal{L}^{(2)}$ describes the variation of Hodge structure of the latter piece. 

We choose for the generators of the orthogonal lattice $\Lambda^{\perp}$ the vectors 
\begin{align}
    g_i^{\text{T}}=(\delta_{i, 1}, \delta_{i, 2}, \delta_{i, 3}, -4\delta_{i, 3} - 4\delta_{i, 1} + 2\delta_{i, 4}, -2\delta_{i, 1} - 2\delta_{i, 3}, \delta_{i, 4})\,, \quad i=1,\dots,4
\end{align}
and define the corresponding periods of $\Lambda^{\perp}\otimes_{\mathbb{Q}} \mathbb{C}$,
\begin{align}
    \Pi^{\perp}_i\coloneqq g_i^{\text{T}}\Sigma \underline{\Pi}\,.
\end{align}
By analytic continuation of the periods to the $\psi=0$ locus we then find  
\begin{align}
   \gamma^{\text{T}}\underline{\Pi}^{\perp}= \gamma^{\text{T}}\partial_{z_2}\underline{\Pi}^{\perp}=0
\end{align}
with 
\begin{align}
    \gamma^{\text{T}}\in\text{Span}_{\mathbb{Q}(i)}\left\{\left(1, 1, -\frac{1 - \ii}{2}, 0\right),\; \left(0, \frac{1+\ii}{2}, 0,  -1\right)\right\}.
\end{align}
Denoting $K=\mathbb{Q}(i)$ this corresponds to a splitting 
\begin{align}
\label{eq:octicCMsplitting}
    \Lambda^{\perp}\otimes_{\mathbb{Q}} K \cong W\oplus \overline{W}\,,
\end{align}
 where $W$ and $\overline{W}$ are $K$-vector spaces of Hodge type $(1,1,0,0)$ and $(0,0,1,1)$, respectively. In light of the general theory, sketched in \Cref{subsec:Arithmetic and analytic analysis of flux vacua}, the form of the Euler factors in \Cref{tab:X8CMsplitting} then arises as follows. For $p\equiv 1\mod 4$ the primes are \emph{split} over the ring of integers $\mathcal{O}_K$ and the Frobenius preserves $W$ and $\overline{W}$, so that the quartic factors are products $Q_1(T)Q_2(T)$ of complex conjugate factors $Q_1(T)=\overline{Q_2}(T)$. In the case $p\equiv 3\mod 4$ the prime is an \emph{inert} prime and the Frobenius exchanges the two spaces $W$ and $\overline{W}$, implying the Euler factors to be even polynomials in $T$. 
\begin{table}[H]
\centering
\renewcommand{\arraystretch}{1.2}
\resizebox{0.65\textwidth}{!}{%
\begin{tabular}{|c|c|c|}
\hline
\multicolumn{3}{|c|}{$(z_1,z_2)=(\infty,1)$} \\
\hline
$p$ & split/inert over $K$& $P^{(p)}_3(X,T)$ \\
\hline \hline
$29$ & split & $p\left(840 + 41\sqrt{-1}\right)T^2 + \left(41 + \sqrt{-1}\right)T + 1$  \\
\hline
$31$ & inert & $p^{6}T^{4} + 46\,p^{2}T^{2} + 1$  \\
\hline
$37$ & split & $p\left(1081 -840\sqrt{-1}\right)T^2 + \left(-210 + 72\sqrt{-1}\right)T + 1$  \\
\hline
$41$ & split & $p\left(-720 -1519\sqrt{-1}\right)T^2 + \left(-217 + 343\sqrt{-1}\right)T + 1$  \\
\hline
$43$ & inert & $p^{6}T^{4} + 22\,p^{2}T^{2} + 1$  \\
\hline
$47$ & inert & $p^{6}T^{4} + 62\,p^{2}T^{2} + 1$  \\
\hline
$53$ & split & $p\left(-2520 -1241\sqrt{-1}\right)T^2 + \left(-51 + 219\sqrt{-1}\right)T + 1$  \\
\hline
$59$ & inert & $p^{6}T^{4} - 10\,p^{2}T^{2} + 1$  \\
\hline
$61$ & split & $p\left(-3479 + 1320\sqrt{-1}\right)T^2 + \left(22 + 120\sqrt{-1}\right)T + 1$  \\
\hline
$67$ & inert & $p^{6}T^{4} + 70\,p^{2}T^{2} + 1$  \\
\hline
$71$ & inert & $p^{6}T^{4} + 110\,p^{2}T^{2} + 1$  \\
\hline
$73$ & split & $p\left(-721 + 5280\sqrt{-1}\right)T^2 + 1$  \\
\hline
$79$ & inert & $p^{6}T^{4} + 142\,p^{2}T^{2} + 1$  \\
\hline
$83$ & inert & $p^{6}T^{4} + 134\,p^{2}T^{2} + 1$  \\
\hline
$89$ & split & $\left(p\left(-8 + 5\sqrt{-1}\right)T + 1\right)\left(\left(835 + 88\sqrt{-1}\right)T + 1\right)$  \\
\hline
$97$ & split & $\left(\left(-297 + 908\sqrt{-1}\right)T + 1\right)\left(p\left(9 + 4\sqrt{-1}\right)T + 1\right)$  \\
\hline
$101$ & split & $p\left(-3960 -9401\sqrt{-1}\right)T^2 + \left(-1027 + 1547\sqrt{-1}\right)T + 1$  \\
\hline
$103$ & inert & $p^{6}T^{4} + 190\,p^{2}T^{2} + 1$  \\
\hline
$107$ & inert & $p^{6}T^{4} + 2\,p^{3}T^{2} + 1$  \\
\hline
$109$ & split & $p\left(-4681 + 10920\sqrt{-1}\right)T^2 + \left(720 + 1092\sqrt{-1}\right)T + 1$  \\
\hline
$113$ & split & $\left(\left(-664 + 1001\sqrt{-1}\right)T + 1\right)\left(p\left(7 + 8\sqrt{-1}\right)T + 1\right)$  \\
\hline
$127$ & inert & $p^{6}T^{4} + 110\,p^{2}T^{2} + 1$  \\
\hline
$131$ & inert & $p^{6}T^{4} + 134\,p^{2}T^{2} + 1$  \\
\hline
$137$ & split & $p\left(-18480 -3281\sqrt{-1}\right)T^2 + \left(-17 + 193\sqrt{-1}\right)T + 1$  \\
\hline
$139$ & inert & $p^{6}T^{4} + 214\,p^{2}T^{2} + 1$  \\
\hline
$149$ & split & $p\left(-14280 + 16999\sqrt{-1}\right)T^2 + \left(89 + 191\sqrt{-1}\right)T + 1$  \\
\hline
\end{tabular}
}
\caption{Factorizations over $K=\mathbb{Q}(i)$ of Euler factors associated to the $(1,1,1,1)$ Hodge structure $\Lambda^{\perp}\otimes_{\mathbb{Q}} \mathbb{C}$ along $\psi=0$ in the model $X_{22211}$ for primes $p=29,\dots,149$ at $z_2=1$. For split primes we give one quadratic factor (or the product of two linear factors with positive imaginary parts); the remaining factor(s) are obtained by complex conjugation of the coefficients.}
\label{tab:X8CMsplitting}
\end{table}

\subsubsection{Nodal singularities}\label{sec:contrans}
    We study the period geometry of the elliptically fibred threefold family $X_{96111}$ at the degeneration locus, whose periods are described by a one-parameter family with degree two Picard--Fuchs operator.
    The degeneration allows us to obtain an integral period basis of this one-parameter model and identify its topological data.
    We suspect that these special loci have similar physical aspects as the strong couplings transitions for K3-fibrations. 
    As a further example, we discuss the elliptic fibration over $\P^1\times \P^1$ in \cref{app:P1P1el}.
    \paragraph{}
    The topological and toric data is contained in \Cref{tab:96111} in \cref{app:ttdata}.
    The polytope possesses an inner point
    \begin{equation}
        \nu_6 = (-3,-2,0,0) = \frac{\nu_3+\nu_4+\nu_5}{3}\,,\hspace{2cm}
        \raisebox{-1.5cm}{\begin{tikzpicture}
            \draw[step=1cm,gray,very thin] (-1,-1) grid (1,1);
            \draw[thick, blue, fill=blue!20] (1,0) -- (0,1) -- (-1,-1) -- cycle;
            \foreach \x in {-1,0,1}
                \foreach \y in {-1,0,1}
                    \fill (\x,\y) circle (2pt);
            \node[pin={[pin distance=0mm,font=\small]above:{$\nu_6$}}] at (0,0) {};
            \node[pin={[pin distance=0mm,font=\small] right:{$\nu_3$}}] at (1,0) {};
            \node[pin={[pin distance=0mm,font=\small]above :{$\nu_4$}}] at (0,1) {};
            \node[pin={[pin distance=0mm,font=\small]below left:{$\nu_5$}}] at (-1,-1) {};
        \end{tikzpicture}}
    \end{equation}
    inside a two-dimensional face and the conifold locus lies along $\Delta_1=1+27z_1=0$. 
    Similarly to the strong coupling transition discussed above, this value for $z_1$ can be obtain from a factorization of the polynomial
    \begin{equation}
        x^3+y^3+z^3+a_6 x yz \xrightarrow{a_6\rightarrow -3}(x+y+z) (x+\xi\,y+\xi^2z)(x+\xi^2y+\xi\,z)
    \end{equation}
    with $\xi^3=1$\,.
    Here, we used the abbreviations $x=t_3^{1/3}$, $y=t_4^{1/3}$ and $z=t_1^{-3}t_2^{-2}t_3^{-1/3}t_4^{-1/3}$.
    Note that $a_6\rightarrow -3$ corresponds to $\Delta_1\rightarrow 0$.
    \paragraph{}
    We obtain the integral period basis of the two-parameter model at $\vec{z}=0$ from the $\hat{\Gamma}$-class, see \cref{eq:frobbasis,eq:Tmum3}.
    An analytical continuation to the discriminant locus yields $\vec{\Pi}(\Delta_1,z_2)$. 
    Note that $\Delta_1=0$ and $z_2=0$ have normal crossing, making them suitable coordinates in that neighbourhood.
    We use the local Frobenius basis
    \begin{align}\
        \varpi(\Delta_1,z_2)={\footnotesize\left(
\begin{array}{c}
 \sigma_1 \\
 \Delta _1\sigma_2 \\
 \sigma_1 \log\! \left(z_2\right)+\sigma_3 \\
 \Delta _1\sigma_2 \log \!\left(\Delta _1\right)+\sigma_4 \\
 \sigma_1 \log\!\left(z_2\right)^2+2 \left(\sigma_3-9 \Delta _1\sigma_2\right) \log\! \left(z_2\right)+\sigma_5 \\
 \sigma_1 \log\!\left(z_2\right)^3+3 \left(\sigma_3-9 \Delta _1\sigma_2\right) \log\!\left(z_2\right)^2+3\sigma_5\log\! \left(z_2\right)+\sigma_6 \\
\end{array}
\right)}\,,
    \end{align}
    with formal power series $\sigma_i$ in $\Delta_1$ and $z_2$. This basis is rotated into the integral basis by the transition matrix
    \begin{equation}\label{eq:tmatrixX18}
        T_{\Delta_1,z_2}=\scalebox{0.6}{$\left(
\begin{array}{cccccc}
 1 & 0 & 0 & 0 & 0 & 0 \\
 \frac{1}{2}+\frac{27 i \sqrt{3} L}{8 \pi ^2} & \frac{27 i \sqrt{3}}{4 \pi ^2} & 0 & -\frac{27 i \sqrt{3}}{4 \pi ^2} & 0 & 0 \\
 \frac{i \left(\pi  \log (81)-9 \sqrt{3} L\right)}{8 \pi ^2} & -\frac{9 i \left(\sqrt{3}-2 \pi \right)}{4 \pi ^2} & -\frac{i}{2 \pi } & \frac{9 i \sqrt{3}}{4 \pi ^2} & 0 & 0 \\
 -\frac{-27 i \sqrt{3} L+35 \pi ^2+9 \log ^2(3)+12 i \pi  \log (3)}{8 \pi ^2} & \frac{27 i \left(\sqrt{3}-2 \pi +3 i \log (3)\right)}{4 \pi ^2} & \frac{3 (\log (27)+2 i \pi )}{4 \pi ^2} & -\frac{27 i \sqrt{3}}{4 \pi ^2} & -\frac{9}{8 \pi ^2} & 0 \\
 -\frac{35}{24}-\frac{3 \log ^2(3)}{8 \pi ^2} & \frac{3 \left(2 \sqrt{3} \pi -9 \log (3)\right)}{4 \pi ^2} & \frac{3 \log (3)}{4 \pi ^2} & 0 & -\frac{3}{8 \pi ^2} & 0 \\
 \frac{9 i \left(356 \zeta(3)+\log ^3(3)\right)-37 \pi ^3+9 \pi  \log ^2(3)-105 i \pi ^2 \log (3)}{48 \pi ^3} & \frac{81 i \left(\log ^2(3)-3 L\right)-3 \pi ^2 \left(4 \sqrt{3}+105 i\right)+54 \pi  \log (3)}{16 \pi ^3} & \frac{i \left(35 \pi ^2-9 \log ^2(3)+6 i \pi  \log (3)\right)}{16 \pi ^3} & 0 & \frac{3 (\pi +3 i \log (3))}{16 \pi ^3} & -\frac{3 i}{16 \pi ^3} \\
\end{array}
\right)$}.
    \end{equation}
    The constant $L$ is the Dirichlet $L$-function value
    \begin{equation}
        L \equiv L_{-3}(2)= \sum_{n=1}^\infty \frac{\chi_3(n)}{n^2}=0.781302412896486\cdots =\frac{4}{3 \sqrt{3}} \Im\text{Li}_2\left(\exp \left(\frac{\pi  i}{3}\right)\right)
    \end{equation}
    with the Dirichlet character \href{https://www.lmfdb.org/Character/Dirichlet/3/b}{3.b}
    \begin{equation}\label{eq:chi3n}
        \chi_3(n)=\left(\frac{-3}{n}\right) = \begin{cases}
        0 & n \equiv 0 \text{ mod } 3\\
        1 & n \equiv 1 \text{ mod } 3\\
        -1 & n \equiv 2 \text{ mod } 3\\
        \end{cases}\,.
    \end{equation}
    Its decimal expansion is listed in the OEIS \cite{OEIS} as \href{https://oeis.org/A086724}{A086724}.
    The appearance seems to be related to the value of the mirror map of the local model $\mathcal{O}(-3)\rightarrow \P^2$~\cite{kerr2008algebraic,Bonisch:2022mgw}.
    Along $\Delta_1=0$, we obtain one flux $g$ with $g^T\Sigma\vec{\Pi}(\Delta_1=0,z_2)=0$, where 
    \begin{equation}\label{eq:96111flux}
        g= (0,3,-1,3,0,0)\,
    \end{equation}
    and the period vector on this locus has the asymptotic form 
    \begin{equation}
        \vec{\Pi}(\Delta_1=0,z) \sim \left(
\begin{array}{c}
 1 \\
 \frac{1}{2}+\frac{27 i \sqrt{3} L}{8 \pi ^2} \\
 -\frac{9 i \sqrt{3} L}{8 \pi ^2}+t+\frac{1}{2} \\
 \frac{1}{8} \left(36 t^2+12 t-38+\frac{27 i \sqrt{3} L}{\pi ^2}\right) \\
 \frac{1}{12} \left(18 t^2+18 t-13\right) \\
 \frac{1}{12} \left(-18 t^3-36 t^2-75 t+\frac{801 i \zeta(3)}{\pi ^3}-40\right) \\
\end{array}
\right)
    \end{equation}
    where we already used the one-parameter coordinate $z=-\frac{1}{3} z_2$ together with $t=\frac{\log z}{2\pi\ii}$\,.
    We introduce the operator labelled  \href{https://cycluster.mpim-bonn.mpg.de/operator.html?nn=4.2.23}{4.2.23}
    \begin{align}\label{eq:AESZ223}
    \begin{split}
        \mathcal{L}^{(4)}_{\text{2.23}}(z)&=\theta ^4+3888 z^2(6 \theta +1) (6 \theta +5) (6 \theta +7) (6 \theta +11) \\
        &\quad +36z (6 \theta +1) (6 \theta +5) (3 \theta  (\theta +1)+1)
    \end{split}
    \end{align}
    with Riemann symbol
     \begin{equation}
        \mathcal{P}_{\mathcal{L}^{(4)}_{\text{2.23}}}\left\{\begin{array}{cccc}
            0 & a & \bar{a} &  \infty\\\hline
            0 & 0 & 0 & \frac{1}{6}\\
            0 & 1 & 1 & \frac{5}{6}\\
            0 & 1 & 1 & \frac{7}{6}\\
            0 & 2 & 2 & \frac{11}{6}
    \end{array},\ z\right\},
    \end{equation}
    where $a=\frac{-3+i \sqrt{3}}{7776}$ and $\bar{a}$ are the roots of $5038848 z^2+3888 z+1$.
    This operator describes the PF ideal of the Hadamard product $D*f$.
    The application of the operator $\mathcal{L}^{(4)}_{\text{2.23}}$ on the vector $\vec{\Pi}(\Delta_1=0,z)$ yields
    \begin{equation}
        \mathcal{L}^{(4)}_{\text{2.23}}(z) \vec{\Pi}(\Delta_1=0,z)=g\frac{135\ii\sqrt{3}z}{\pi^2}\,.
    \end{equation}
    Note that the inhomogeneity follows from $ \mathcal{L}^{(4)}_{\text{2.23}}(z)\Omega(z) = \dd \beta$ and
    \begin{equation}
        \mathcal{L}^{(4)}_{\text{2.23}}(z) \Pi_i = \int_{\Gamma_i}\dd \beta = \int_{\partial\Gamma_i} \beta = \underbrace{\langle \Gamma_i, \Phi \rangle}_{g_i} \int_{\partial \Phi^\vee} \beta\,.
    \end{equation}
    Here we denoted the vanishing cycle by $\Phi$, its dual by $\Phi^\vee$ and identify $\int_{\partial \Phi^\vee} \beta=\frac{135\ii\sqrt{3}z}{\pi^2}$\,.
    It follows that all elements are solutions to the fifth-order operator $(\theta-1)\mathcal{L}^{(4)}_{\text{2.23}}$\,.
    We define the one-parameter basis of \emph{relative} periods 
    \begin{equation}\label{eq:96111oneparam}
        \vec{\Pi}_1(z) =\left(
            \begin{array}{ccccc}
             1 & 0 & 0 & 0 & 0\\
             -2 & 1 & 0 & 0 & 0\\
             0 & 0 & 1 & 0 & 0\\
             0 & 0 & 2 & 1 & 0\\
             0 & 0 & 0 & 0 & 1\\
            \end{array}
            \right)\cdot \left(
            \begin{array}{cccccc}
             1 & 0 & 0 & 0 & 0 & 0 \\
             0 & 1 & 3 & 0 & 0 & 0 \\
             0 & 0 & 0 & 0 & 1 & 0 \\
             0 & 0 & 0 & 0 & 0 & 1 \\
             0 & 0 & 0 & 1 & 0 & 0 \\
            \end{array}
            \right)\vec{\Pi}(\Delta_1=0,z).
    \end{equation}
    The first four rows in the $5\times 6$-matrix pick out the integral linear combinations of $\vec{\Pi}$ that are solutions to a one-parameter operator modulo the relation of the flux in \cref{eq:96111flux}.
    The last row corresponds to the chain integral over $\Phi$, which intersects the vanishing $S^3$ exactly once, cf.\ \cref{eq:96111flux}. 
    The square matrix removes the quadratic terms of the triple-logarithmic period to allow for a pre-potential description of the homogeneous periods. 
    Note that the latter leaves the intersection form 
    \begin{equation}\label{eq:Sigmaoneparam}
        \Sigma = \left(
            \begin{array}{cccc}
             0 & 0 & 0 & 1 \\
             0 & 0 & 1 & 0 \\
             0 & -1 & 0 & 0 \\
             -1 & 0 & 0 & 0 \\
            \end{array}
            \right)
    \end{equation}
    invariant.
    The period vector $\vec{\Pi}_1$ has the following asymptotic behaviour in $t=\frac{\log z}{2\pi\ii}$
    \begin{equation}
        \vec{\Pi}_1\sim \left(1,3t,\frac{1}{3}\partial_{t}\mathcal{F} , 2\mathcal{F}-t\partial_t \mathcal{F}, \Pi^{\text{asy}}_{\text{rel}}(t)\right),
    \end{equation}
    where the closed periods are described by the pre-potential
    \begin{equation}\label{eq:prep223}
        \mathcal{F}(t) = \frac{9}{3!}t^3 + \frac{9/2}{2}t^2 - \frac{78}{24}t - \frac{\left(-534\right) \zeta(3)}{2 (2 \pi  i)^3}-\frac{11}{4}\,.
    \end{equation}
    The entry $\Pi^{\text{asy}}_{\text{rel}}(t)$ denotes the polynomial terms of the relative period
    \begin{equation}\label{eq:relperiod}
        \Pi_{\text{rel}}(t)=\frac{9t^2}{2}+\frac{3t}{2}-\frac{1}{(2\pi\ii)^2}\sum_{k\in\N_0}\sum_{j=0}^2 n_{k,j}\text{Li}_2\!\left( e^{2\pi\ii(k\,t-j)/3}\right)
    \end{equation}
    with coefficients $n_{k,j}$ given in \Cref{tab:openinstantons}. We interprete this relative period as chain integral that emerges when the dual cycle to 
    the vanishing cycle splits open at the conifold singularity as in the Kuga-Sato variety considered in \cite{Bonisch:2025cax}. In the resolution the singular boundaries 
    are eventually replaced by $\mathbb{P}^1$.          
    \begin{table}
    \centering
        \begin{tabular}{|c|ccccccc|}\hline
            $k$ & 0 & 1 & 2 & 3 & 4 & 5 &$\cdots$\\  \hline
            $n_{k,0}$ & $-114$ & $1620$ & $280260$ & $-589871052$ & $267770671560$ & $11501604396000$ &$\cdots$\\
            $n_{k,1}$ & $27/2$ & $1620$ & $135270$ & $22352760$ & $68625557100$ & $-6990503553000$ &$\cdots$\\ \hline 
        \end{tabular}
        \caption{The first integers appearing in the corrections to the relative period dual to the shrinking $S^3$ at the conifold locus of $X_{96111}$\,, cf. \cref{eq:relperiod}.
        The coefficients satisfy $n_{k,2}=-n_{k,1}$\,.
        Integrality was verified for $k\leq 80$.
        Denoting the instanton corrections of the one-parameter model \eqref{eq:prep223} by $n_k$, one finds $n_k = k\, n_{k,0}$ for $k>0$.}
        \label{tab:openinstantons}
    \end{table}
    We observe that the summation at $k=0$ yields the constant contribution
    \begin{equation}
        \frac{19}{4}-\frac{27/2}{(2\pi\ii)^2}\left(\text{Li}_2\!\left(e^{-2\pi\ii/3}\right)-\text{Li}_2\!\left(e^{-2\cdot 2\pi\ii/3}\right)\right) = \frac{19}{4}-\frac{27 i \sqrt{3} L_{-3}(2)}{8 \pi ^2},
    \end{equation}
    explaining the appearance of the Dirichlet character in the periods.
    We identify a limit of an open-string sector with an open-string parameter set to $1/3$\,. The interpretations of the chain integral as disk 
    instanton generating function is similar as in \cite{Walcher:2006rs}. It is an interesting question whether one can deform the  open string parameter and give an A-model description of the special Lagrangian 
    brane on which these disk instantons end.  
    
    The basis has integral monodromies and its closed part respects the canonical intersection form $\Sigma$ \eqref{eq:Sigmaoneparam}.
    We give the monodromy representations
    \begin{gather}
        \mathfrak{M}_0=\left(
\begin{array}{ccccc}
 1 & 0 & 0 & 0 & 0 \\
 3 & 1 & 0 & 0 & 0 \\
 3 & 1 & 1 & 0 & 0 \\
 -8 & 0 & -3 & 1 & 0 \\
 6 & 3 & 0 & 0 & 1 \\
\end{array}
\right)\ \mathfrak{M}_a=\left(
\begin{array}{ccccc}
 9 & -1 & 1 & 1 & 0\\
 8 & 0 & 1 & 1 & 0\\
 8 & -1 & 2 & 1 & 0\\
 -64 & 8 & -8 & -7 & 0\\
 0 & 0 & 0 & 0 & 1\\
\end{array}
\right)\\
\mathfrak{M}_{\bar{a}}=\left(
\begin{array}{ccccc}
 4 & 0 & -1 & 1 & 0\\
 -3 & 1 & 1 & -1 & 0\\
 0 & 0 & 1 & 0 & 0\\
 -9 & 0 & 3 & -2 & 0\\
  0 & 0 & 0 & 0 & 1\\
\end{array}
\right) \ \mathfrak{M}_\infty=\left(
\begin{array}{ccccc}
 6 & 0 & 0 & 1 & 0 \\
 0 & 0 & 1 & 0 & 0 \\
 0 & -1 & 1 & 0 & 0 \\
 -31 & 0 & 0 & -5 & 0 \\
 3 & -3 & 0 & 0 & 1 \\
\end{array}
\right).
    \end{gather}
    These satisfy the relation $\mathfrak{M}_a\cdot \mathfrak{M}_0\cdot \mathfrak{M}_{\bar{a}}\cdot\mathfrak{M}_\infty=\mathbbm{1}$\,.
    Note that the closed part is an invariant subspace of the monodromy group.
    
The Hadamard product is also reflected in the Euler factors at the conifold.
The Euler factors for the diagonal direction $z_1=z_2$ at $p=11$ are listed in \Cref{tab:X18eulerfactors11}.
At the conifold the Euler factor splits into a linear factor corresponding to a quadratic character and a fourth order factor satisfying threefold Weil conjectures (cf. eq. \eqref{eq:dWinvcon}):
\begin{align}
    P^{(p)}_\text{con}(X,T)=\left(1-p\chi(p)T\right)\big(1+\alpha^{(p)}_1T+\alpha^{(p)}_2 pT^2+\alpha^{(p)}_1p^3T^3+p^6T^4\big).
\end{align}
For the reductions of $\Delta_1=0$ the character is  $\chi(p)=\left(\frac{-3}{p}\right)$, introduced in \cref{eq:chi3n}. 
The critical value at $s=2$ of its $L$-function appeared in the transition matrix $T_{\Delta_1,z_2}$ in \cref{eq:tmatrixX18}.
The fourth order factor is as expected the Euler factor of the operator \href{https://cycluster.mpim-bonn.mpg.de/operator.html?nn=4.2.23}{4.2.23} at the fibre $z=-\frac{1}{3}\cdot(-\frac{1}{27})=\frac{1}{81}$. 
    \begin{table}[H]
\centering
\renewcommand{\arraystretch}{1.6}
\resizebox{\textwidth}{!}{%
\begin{tabular}{|c|c|c|}
\hline
\multicolumn{3}{|c|}{$p=11$ for direction $(z_1,z_2)=(z,z)$} \\
\hline
$z \mod p$ & smooth/sing. & $P^{(p)}_3(X,T)$ \\
\hline \hline
$1$ & conifold & $-\left(pT-1\right)\left(p^6T^4 + 27p^3T^3 + 123pT^2 + 27T + 1\right)$ \\
\hline
$2$ & conifold & $\left(pT+1\right)\left(p^6T^4 - 47pT^2 + 1\right)$ \\
\hline
$3$ & smooth & $p^9T^6 - 27p^6T^5 + 104p^4T^4 - 69p^3T^3 + 104pT^2 - 27T + 1 $ \\
\hline
$4$ & smooth & $p^9T^6 + p^6T^5 + p^5T^4 - 346p^2T^3 + p^2T^2 + T + 1$ \\
\hline
$5$ & smooth & $p^9T^6 - 4p^6T^5 + 172p^4T^4 - 309p^2T^3 + 172pT^2 - 4T + 1 $ \\
\hline
$6$ & smooth & $p^9T^6 - 9p^6T^5 + 23p^4T^4 - 390p^2T^3 + 23pT^2 - 9T + 1$ \\
\hline
$7$ & smooth & $p^9T^6 + 41p^6T^5 - 98p^4T^4 - 822p^2T^3 - 98pT^2 + 41T + 1$ \\
\hline
$8$ & smooth & $p^9T^6 - 19p^6T^5 - 20p^4T^4 + 426p^2T^3 - 20pT^2 - 19T + 1$ \\
\hline
$9$ & smooth & $p^9T^6 - 90p^6T^5 + 461p^4T^4 - 1668p^2T^3 + 461pT^2 - 90T + 1$ \\
\hline
$10$ & smooth & $p^9T^6 - 18p^6T^5 + 113p^4T^4 + 108p^2T^3 + 113pT^2 - 18T + 1$ \\
\hline
\end{tabular}
}
\caption{Euler factors of the model $X_{96111}$ along the direction $(z_1,z_2)=(z,z)$ for $p = 11$.}
\label{tab:X18eulerfactors11}
\end{table}
    \paragraph{}
    The exact same methods work for the genus-one fibration $X_{33111}$, whose data we list in \Cref{tab:33111} in \cref{app:ttdata}.
    The one-parameter model is described by \href{https://cycluster.mpim-bonn.mpg.de/operator.html?id=35}{4.2.21} and the pre-potential reads
    \begin{equation}
        \mathcal{F}(t) = \frac{9}{3!}t^3 + \frac{9/2}{2}t^2 - \frac{6}{24}t - \frac{\left(-66\right) \zeta(3)}{2 (2 \pi  i)^3}-\frac{3}{4}\,.
    \end{equation}
    The Picard--Fuchs operator arises from the Hadamard product $B*f$.
    \paragraph{}
    A further realisation of a one-parameter model exists in the hypersurfaces $X_{43311}$ with data listed in \Cref{tab:43311} of \cref{app:ttdata}.
    In this two-parameter\footnote{Note that the family has $h^{2,1}=5$ with three non-polynomial deformations.} model, we find a singular locus very similar to the ones discussed above.
    While the mondromies are only rational due to non-polynomial deformations,
    we obtain the operator \href{https://cycluster.mpim-bonn.mpg.de/operator.html?nn=4.4.74}{4.4.74}.
\subsubsection{Transitions in more general fibrations}\label{sec:gentrans}

\begin{figure}[ht!]
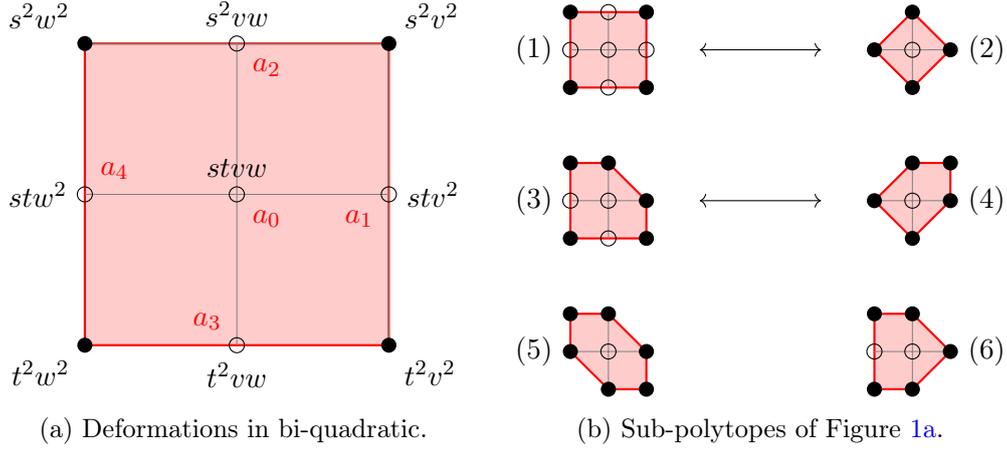

        \centering
        \begin{subfigure}{0.45\textwidth}
            \centering
            %
            \caption{Deformations in bi-quadratic.}
            \label{fig:defs22}
        \end{subfigure}
        \begin{subfigure}{0.45\textwidth}
            \centering
            %
            \caption{Sub-polytopes of \Cref{fig:defs22}.}
            \label{fig:subp22}
        \end{subfigure}

        \caption{Elliptic families embedded in the bi-quadratic in $\P^1\times \P^1$. Polytopes (1,2) and (3,4) form reflexive pairs while 5 and 6 are self-dual. Polytopes (2,4,6) can also be embedded in the quartic family in \Cref{fig:ellembed4} and (2-6) in the cubic family of \Cref{fig:ellembed3}.}
        \label{fig:ellembed22}
    \end{figure}

\begin{figure}[ht]
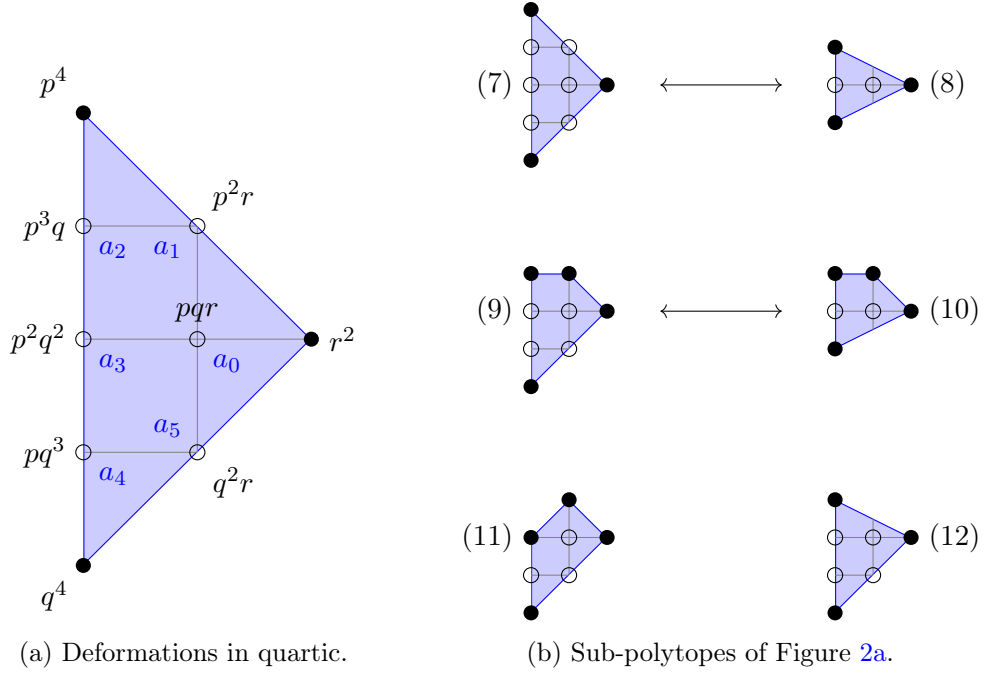

        \centering
        \begin{subfigure}[b]{0.45\textwidth}
            \centering
            %
            \caption{Deformations in quartic.}
            \label{fig:defs4}
        \end{subfigure}
        \begin{subfigure}[b]{0.45\textwidth}
            \centering
            %
            \caption{Sub-polytopes of \Cref{fig:defs4}.}
            \label{fig:subp4}
        \end{subfigure}

        \caption{Elliptic families embedded in the quartic in $\P_{211}$. Polytopes (7,8) and (9,10) form reflexive pairs while 11 and 12 are self-dual, where (8) and (10) can also be embedded in both the bi-quadratic in \Cref{fig:ellembed22} and the cubic in \Cref{fig:ellembed3}.}
        \label{fig:ellembed4}
    \end{figure}

\begin{figure}[ht]
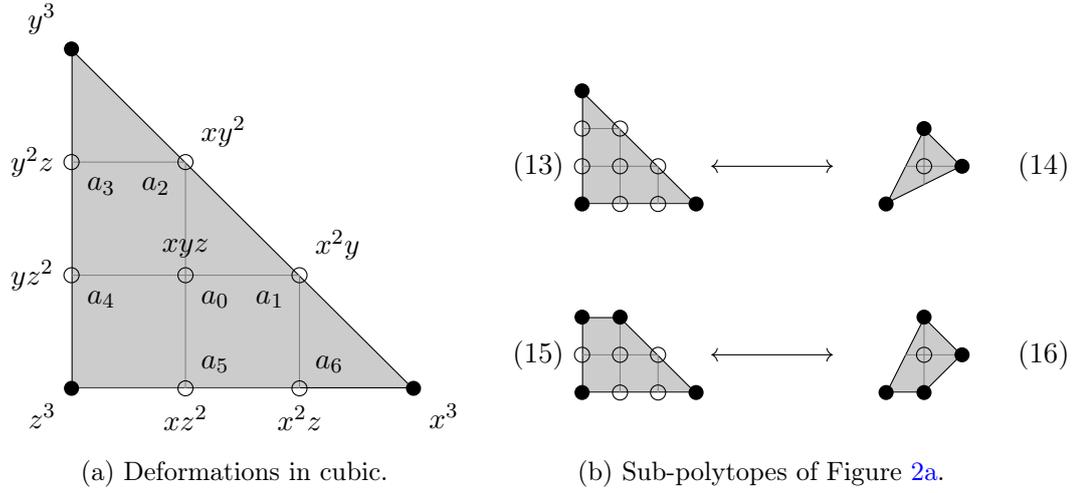

        \centering
        \begin{subfigure}[b]{0.45\textwidth}
            \centering
            %
            \caption{Deformations in cubic.}
            \label{fig:defs3}
        \end{subfigure}
        \begin{subfigure}[b]{0.45\textwidth}
            \centering
            %
            
            \caption{Sub-polytopes of \Cref{fig:defs4}.}
            \label{fig:subp3}
        \end{subfigure}

        \caption{Elliptic families embedded in the cubic in $\P^2$. Polytopes (13,14) and (15,16) form reflexive pairs, where (14) and (16) can also be realised in both the bi-quadratic and the quartic of \Cref{fig:ellembed22,fig:ellembed4}, respectively.}
        \label{fig:ellembed3}
    \end{figure}

\begin{table}[]
        \centering
        \begin{tabular}{|c|c|c|c|}\hline
            new number & AZ label & geometry & operator \\ \hline
            2.1.1 & A & $X^{2,2}\subset \P^1\times\P^1$ & $\theta^ 2-4z(2\theta+1)^2$\\
            2.1.2 & B & $X^{3}\subset \P^2$ & $\theta^2-3z(3\theta+1)(3\theta+2)$\\
            2.1.3 & C & $X^{4}\subset \P_{211}$ & $\theta^2-4z(4\theta+1)(4\theta+3)$\\
            2.1.4 & D & $X^{6}\subset \P_{321}$ & $\theta^2-12z(6\theta+1)(6\theta+5)$\\ \hline
        \end{tabular}
        \caption{Elliptic operators of degree one.}
        \label{tab:elliptic1}
    \end{table}
    \begin{table}[]
        \centering
        \begin{tabular}{|c|c|c|}\hline
            AZ label & info & operator \\ \hline
            f & BZ: B &$\theta ^2-3z\left(3 \theta ^2+3 \theta +1\right)+27 z^2 (\theta +1)^2$\\
            b & BZ: D &$\theta ^2-z\left(11 \theta ^2+11 \theta +3\right)- z^2(\theta +1)^2$\\
            d & BZ: E &$\theta ^2-4 z\left(3 \theta ^2+3 \theta +1\right)+32 z^2 (\theta +1)^2$\\
            a & BZ: A &$\theta ^2-z\left(7 \theta ^2+7 \theta +2\right)-8  z^2(\theta +1)^2$\\
            c & BZ: C &$\theta ^2-z\left(10 \theta ^2+10 \theta +3\right)+9 z^2(\theta +1)^2 $\\
            g & BZ: F &$\theta ^2-z\left(17 \theta ^2+17 \theta +6\right)+72 z^2 (\theta +1)^2$\\
             --- & --- &$(z-1) \left(\theta ^2-z(\theta +1)^2\right)$\\
            e & $\mu(A)$ &$\theta ^2-4 z\left(8 \theta ^2+8 \theta +3\right)+256 z^2 (\theta +1)^2$\\
            h & $\mu(B)$ &$\theta ^2-3 z\left(18 \theta ^2+18 \theta +7\right)+729 z^2 (\theta +1)^2$\\
            i & $\mu(C)$ &$\theta ^2-4 z\left(32 \theta ^2+32 \theta +13\right)+4096  z^2(\theta +1)^2$\\
            j & $\mu(D)$ &$\theta ^2-12z \left(72 \theta ^2+72 \theta +31\right)+186624z^2 (\theta +1)^2 $\\
            k & --- &$\theta ^2-3z (2 \theta +1)-81 z^2(\theta +1)^2 $\\
            l & --- &$\theta ^2-4 z(2 \theta +1)-64z^2 (\theta +1)^2$\\
            m & --- &$\theta ^2-24 z(2 \theta +1)-1296 z^2 (\theta +1)^2$\\ \hline
        \end{tabular}
        \caption{Elliptic operators of degree two.}
        \label{tab:elliptic2}
    \end{table}
The strong coupling transitions of \cref{sec:sctrans} describe a K3-fibration over a rational normal curve of degree $k$. 
There, we discussed quartic-fibrations for $k=2$ and $k=5$ yielding $X^{4,2}$ and $X^5$, respectively.
The fundamental periods of both hypergeometric threefold families can be written as a Hadamard product of those of the quartic and a contour integral depending on $k$.
Below, we will also review the equivalent formulation as an iterated integral found in \cite{Doran:2015xjb}.
The conifold degenerations on the other hand, stem from elliptic fibrations over two dimensional bases.
In \cref{sec:contrans}, we considered $X_{321}$-fibrations over $\P^2$ and $F_0$, which gave rise to $D*f$ and $D*d$, respectively, where $D$ describes the PF system of the fibre (cf.\ \Cref{tab:elliptic1}).
In this section, we catalogue the one-parameter models arising in both strong coupling transitions and along conifold loci as Hadamard products of the fibre and a function depending solely on the geometry of the base.
Parts of the following have already appeared in \cite{Piribauer2025}.
\paragraph{}
We begin with the relatively well-known structure of quartic fibrations. 
Let us fix the fibre to be the mirror quartic in $\P^3$ and consider the base described by a rational normal curve of degree $k$.
The latter can be embedded into $\P^k$ via
\begin{align}
    \begin{split}
        \nu:\quad \P^1&\longrightarrow \P^k\,,\\
        (x:y)&\longmapsto (X_0:\ldots : X_k)=(x^k:x^{k-1}y:\ldots :y^k)\,,
    \end{split}
\end{align}
where we will be using coordinates $s=x^k$ and $t=x/y$ for the $\P^1$.
In their toric description, the fibrations have an edge describing this normal curve that looks like
\begin{center}
    \begin{tikzpicture}
            \draw[thick, black, fill=blue!20] (-3,0) -- (-1,0) -- (-0.7,0);
            \draw[thick, black, fill=blue!20] (3,0) -- (1,0) -- (0.7,0);
            \foreach \x in {-3,-1,1,3}
                \fill (\x,0) circle (2pt);
            \node[above=0.1cm] at (-3,0) {$s$};
            \node[above=0.1cm] at (-1,0) {$s\,t$};
            \node[above=0.1cm] at (1,0) {$s\,t^{k-1}$};
            \node[above=0.1cm] at (3,0) {$s\,t^k$};
            \node at (0,0) {$\cdots$};
\end{tikzpicture}
\end{center}
with the labels corresponding to the polynomial deformations of the curve.
Setting $a_i$ to be the deformation parameter of the monomial $st^i$, the face polynomial is given by $P_k=\sum_{i=0}^k a_i s\,t^i$. 
Over the locus $a_i=\binom{k}{i}$, the polynomial decomposes into linear factors $P_k=s(1+t)^k$.
This is precisely where the generalized strong coupling transition appears, the fundamental period of the fibration becomes a Hadamard product of the fibre period and 
\begin{align}
\begin{split}\label{eq:varpisc}
    \varpi_0^{(k)}(z) &= \frac{1}{(2\pi\ii)^2}\oint_{\abs{s}=\abs{t}=1} \frac{1}{1+s(1+t)^k}\frac{\dd s\,\dd t}{s\,t+z}= \frac{1}{2\pi\ii}\oint_{\abs{t}=1}\frac{\dd t}{t-z(1+t)^k}\\
    &= \sum_{i=0}^\infty \binom{i\,k}{i}z^i\,.
\end{split}
\end{align}
Note that the Hadamard product of $\varpi_0^{(k)}(z)$ with a function $f(z)=\sum_{j=0}^\infty c_j z^j$ can be as well be written as a contour integral
\begin{equation}
\label{eq:hadamardquintic}
    \varpi_0^{(k)}(z) * f(z) = \frac{1}{2\pi\ii}\oint_{\abs{t}=1} \frac{\dd t}{t}f\!\left(\frac{z(1+t)^k}{t}\right),
\end{equation}
which can be seen by expanding the integrand as
\begin{equation}\label{eq:integrandHP}
    f\!\left(\frac{z(1+t)^k}{t}\right) = \sum_{j=0}^\infty c_j z^j(1+t)^{k\,j}t^{-j}.
\end{equation}
The contour integral picks out the coefficient in \cref{eq:integrandHP} independent of $t$, given by the Hadamard product $\sum_{j}c_j\binom{j\,k}{j}$.
As mentioned above, the function $f(z)$ is the fundamental period of the K3-fibre and the generalized strong coupling transitions reproduce many of the 
results found in \cite{Doran:2015xjb}. E.g. the precise relation between the fibering out formula from \cite{Doran:2015xjb}, used in section 
3 of~\cite{Bonisch:2025cax} to prove the modularity and determining the period matrix of the quintic in terms of integrals of modular forms, 
reproduces the formula \eqref{eq:hadamardquintic} upon the simple substitution $t\rightarrow \frac{t}{t+1}$.   
\paragraph{}
A similar analysis can be performed for conifold loci, where we restrict ourselves to bases given by the reflexive polygons in two dimensions.
We saw in \cref{sec:sctrans,sec:contrans} how the limiting mixed Hodge structure reflect the transitions as we tune the complex-structure moduli to suitable values.
Focusing on the holomorphic period and assuming a projection on closed cycles, the one-parameter models arising along conifold loci can again be expressed as Hadamard products.
As for the generalized strong coupling transitions above, the fibre and the base each contribute one factor in this product.
Thus, we can fix a fibre and study the transitions that arise when varying the deformation parameters in the base.
We find that, whenever the two-dimensional base is described torically by a reflexive polygon, there exists at least one one-parameter model in the sense that its holomorphic period is annihilated by a order four Calabi--Yau operator.
\paragraph{}
Following \cite{Huang:2013yta}, we embed the 16 polygons as depicted in \Cref{fig:ellembed22,fig:ellembed4,fig:ellembed3}. 
Note that the assignment is not unique and several polygons have multiple possible embeddings.
Using these prescriptions, we find that their Hadamard factors along the special loci are described by the periods
\begin{align}\label{eq:varpi22}
        \varpi_0^{(2,2)}(z) &= \frac{1}{(2\pi\ii)^4}\oint_{\mathcal{C}} \frac{1}{1+P(s,t,v,w)}\frac{\dd s\,\dd t\,\dd v\,\dd w}{stvw+z}\,,\\
        \varpi_0^{(4)}(z) &= \frac{1}{(2\pi\ii)^3}\oint_{\mathcal{C}} \frac{1}{1+P(p,q,r)}\frac{\dd p\,\dd q\,\dd r}{pqr+z}\,,\label{eq:varpi4}\\
        \varpi_0^{(3)}(\tilde{z}) &= \frac{1}{(2\pi\ii)^3}\oint_{\mathcal{C}} \frac{1}{1+P(x,y,z)}\frac{\dd x\,\dd y\,\dd z}{xyz+\tilde{z}}\,,\label{eq:varpi3}
    \end{align}
where $P$ is the face polynomial and the contour $\mathcal{C}$ is given by the absolute values of all integration variables set to one.
The variable $\tilde{z}$ of $\varpi_0^{(3)}$ is used simply to allow us to follow the notation of \cite{Huang:2013yta} and has no further meaning.
Note that \cref{eq:varpi22,eq:varpi4,eq:varpi3} are direct generalisations of the strong coupling Hadamard factor in \cref{eq:varpisc}.
We collect our findings of special loci in \Cref{tab:ellident22,tab:ellident4,tab:ellident3}.
Again, we find that the holomorphic periods can be given by iterated integrals.
For example,
\begin{equation}
    f(z)*\varpi_0^{\P^2} = \frac{1}{(2\pi\ii)^2}\oint_{|s|=|t|=1}\frac{\dd s\,\dd t}{s\,t}\,f\!\left(z\,\frac{1+s^3+t^3-3\,s\,t}{s\,t}\right),
\end{equation}
for the $\P^2$ base described by (14) in \Cref{fig:ellembed3}.
In these tables, we always set the deformations parametrising vertices to unity, depicted by bold ${\bf 1}$'s.
Bold ${\bf 0}$'s indicate that the deformation lies outside the polygon.
While leaving at first the parameter of the inner point $a_0$ open, we search for possible finite values of the edge parameters that allow for one-parameter models.
With the parametrisation of the Weierstraß coefficients $f$ and $g$ in \cite{Huang:2013yta}, we find the $j$-invariant as a function of $a_0$.
We observe that restrictions to one-parameter models lie at rational degeneration points of the family.
As a verification, we identify the Kodaira classification~\cite{Kodaira1963} of singular fibres with those corresponding to the pencils describing the elliptic operators~\cite{zagier2009integral,almkvist2023calabi} (cf.\ \Cref{tab:elliptic1,tab:elliptic2}).
Based on their classification, it comes as no surprise that the hypergeometric operators and their $\mu$-transform arise in the same embeddings.
The same holds for the operators $(a,c,g)$, which also differ only by a Möbius transformation.
As an exception, the only restriction in bases given by polygons (4) and (10) seems to yield a (self-dual) degree four operator with five singular fibres
\begin{align}
    \begin{split}
        \mathcal{L}^{(2)}_{5a}(z) &= 7 \theta ^2-z\left(4 \theta ^2+24 \theta +7\right)-z^2\left(139 \theta ^2+238 \theta +105\right)\\
        &\quad-5 z^3\left(69 \theta ^2+139 \theta +72\right)-250 z^4(\theta +1)^2\,.
    \end{split}
    \end{align}
    Three of those fibres are at the roots of $\Delta=1-2 z-17 z^2-25 z^3$ and the Riemann symbol reads
    \begin{equation}\renewcommand{\arraystretch}{1.3}
        \mathcal{P}_{\mathcal{L}^{(2)}_{5a}}\left\{\begin{array}{cccc}  -7/10 & 0 & \Delta=0 &   \infty\\\hline
         0 & 0 & 0 & 1\\
         2 & 0 & 0 & 1
       \end{array},\ z\right\}\,.
    \end{equation}
    Note that $z=-7/10$ is an apparent singularity with trivial monodromy. 
\paragraph{}
Lastly, we note that only special loci with non-vanishing deformations were considered.
While leaving the edge deformations $a_i$, $i>0$, untouched as in \Cref{tab:ellident22,tab:ellident4,tab:ellident3}, some discriminants have further zeros at $a_0=0$.
Since the situation is the same for all configurations yielding a certain operator, we can summarise the $a_0=0$ cases as follows.
The operator $d$ becomes $\mathcal{L}_A^{(2)}(z^2)$ and $f$ becomes $\mathcal{L}_B^{(2)}(z^3)$.
Instead of obtaining $B$ (or $\mu(B)=h$) one obtains a reparametrised operator $B$ (still with three singular fibres, where the orbifold is over $z=-1/6$).
    Lastly, polygon number 16 has a one-parameter model at $a_0=0$, where the base-factor of the Hadamard product is annihilated by the operator
    \begin{align}
    \begin{split}
        \mathcal{L}^{(2)}_{5b}(z) &= 8 \theta ^2+z\,\theta\,  (17 \theta -1)- z^2\left(55 \theta ^2+128 \theta +64\right)-12 z^3\left(30 \theta ^2+78 \theta +47\right)\\
        &\quad -4 z^4\left(103 \theta ^2+250 \theta +147\right)  -99  z^5\left(\theta ^2+3 \theta +2\right).
    \end{split}
    \end{align}
    This self-adjoint operator has now four singular points at $\Delta=1+z-8 z^2-36 z^3-11 z^4$ and the full Riemann symbol is given by
    \begin{equation}\renewcommand{\arraystretch}{1.3}
        \mathcal{P}_{\mathcal{L}^{(2)}_{5b}}\left\{
\begin{array}{cccc}  -8/9 & 0 & \Delta=0 &   \infty\\\hline
         0 & 0 & 0 & 1\\
         2 & 0 & 0 & 2
       \end{array},\ z\right\}.
    \end{equation}
    Here, an apparent singularity lies at $z=-8/9$ while infinity is a regular point.
    We have no reason to believe that the $z$-degrees of these operators are minimal.
    However, the one-parameter models as obtained from the multi-parameter ambient fibrations are described by these parametrisations of the elliptic curves.
    \newpage
    
    \begin{table}[ht]
        \centering
        \begin{tabular}{|c||c|c|c|c|c|c|c|c|}
        \hline
        \multirow{2}{*}{no.}
          & \multicolumn{4}{c|}{deformations}
          & \multirow{2}{*}{$j(a_{0})$}
          & \multicolumn{3}{c|}{special loci}\\
        \cline{2-5} \cline{7-9}
          & $a_1$ & $a_2$ & $a_3$ & $a_4$ &  & $a_0$ & $\mathcal{L}^{(2)}$ & sing.\ fibres\\
        \hline \hline
        \multirow{3}{*}{1} & \multicolumn{3}{c|}{\multirow{3}{*}{2}}  & \multirow{2}{*}{2} & \multirow{2}{*}{$\frac{\left(a_0^2+8 a_0-32\right){}^3}{1728 \left(a_0-4\right) \left(a_0+12\right)}$} & 4 & $A$ & \multirow{2}{*}{$I_1$, $I_1^*$, $I_4$}\\ \cline{7-8}
        & \multicolumn{3}{c|}{} & & & -12 & $e=\mu(A)$ &\\ \cline{5-9}
        & \multicolumn{3}{c|}{} & -2& $\frac{\left(a_0^4-16 a_0^2+256\right){}^3}{1728 a_0^4 \left(a_0-4\right){}^2 \left(a_0+4\right){}^2}$ & $\pm 4$ & $d$ & $I_2$, $I_2$, $I_4$, $I_4$\\ \hline
        2 & \multicolumn{4}{c|}{\bf{1}}  & $\frac{\left(a_0^4-16 a_0^2+16\right){}^3}{1728 a_0^2 \left(a_0+4\right)\left(a_0-4\right)}$ & $\pm 4$ & $d$ & $I_1$, $I_1$, $I_2$, $I_8$\\ \hline
        3 & \multicolumn{2}{c|}{\bf{1}} & \multicolumn{2}{c|}{2} & $-\frac{\left(a_0^4-40 a_0^2+120 a_0-80\right){}^3}{1728 \left(a_0-3\right){}^5 \left(a_0^2+5 a_0-25\right)}$ & $3$ & $b$ & $I_1$, $I_1$, $I_5$, $I_5$\\ \hline
        4 & \multicolumn{4}{c|}{\bf{1}} & $-\frac{\left(a_0^4-16 a_0^2+24 a_0+16\right){}^3}{1728 \left(a_0-1\right){}^2 \left(a_0^3-a_0^2-18 a_0+43\right)}$ & 1 & $\mathcal{L}^{(2)}_{5a}$ & $I_1$, $I_1$, $I_1$, $I_2$, $I_7$\\
\hline
        \end{tabular}
        \caption{Special loci for bi-quadratic bases depicted in \Cref{fig:ellembed22} \cite{Piribauer2025}. The singular fibres follow the Kodaira classification and were obtained from the polynomials $f$ and $g$ of the Weierstraß form.}
        \label{tab:ellident22}
    \end{table}
    
        \begin{table}[ht]
        \centering
        \begin{tabular}{|c||c|c|c|c|c|c|c|c|c|}
        \hline
        \multirow{2}{*}{no.}
          & \multicolumn{5}{c|}{deformations}
          & \multirow{2}{*}{$j(a_{0})$}
          & \multicolumn{3}{c|}{special loci}\\
        \cline{2-6} \cline{8-10}
          & $a_1$ & $a_2$ & $a_3$ & $a_4$ & $a_5$ &  & $a_0$ & $\mathcal{L}^{(2)}$ & sing.\ fibres\\
        \hline \hline
        
        \multirow{3}{*}{7} & \multirow{2}{*}{2} & \multirow{3}{*}{4} & \multirow{3}{*}{6} & \multirow{3}{*}{4} & \multirow{3}{*}{2} & \multirow{2}{*}{$\frac{\left(a_0^2+8 a_0-32\right){}^3}{1728 \left(a_0-4\right) \left(a_0+12\right)}$} & 4 & $A$ & \multirow{2}{*}{$I_1$, $I_1^*$, $I_4$}\\ \cline{8-9}
        
        & \multirow{2}{*}{} & \multirow{3}{*}{} & \multirow{3}{*}{} & \multirow{3}{*}{} & \multirow{3}{*}{} & & $-12$ & $e=\mu(A)$ &\\ \cline{2-2} \cline{7-10}
        
         & -2 & \multirow{3}{*}{} & \multirow{3}{*}{} & \multirow{3}{*}{} & \multirow{3}{*}{} & $\frac{\left(a_0^4-16 a_0^2+256\right){}^3}{1728 a_0^4 \left(a_0^2-16\right){}^2}$ & $\pm 4$ & $d$ & $I_2$, $I_2$, $I_4$, $I_4$\\ \hline
        
        8 & \bf{0} & \bf{1} & 2 & \bf{1} & \bf{0}  & $\frac{\left(a_0^4-16 a_0^2+16\right){}^3}{1728 a_0^2 \left(a_0^2-16\right)}$ & $\pm 4$ & $d$ & $I_2$, $I_2$, $I_4$, $I_4$\\ \hline
        
        9 & \multicolumn{2}{c|}{\bf{1}} & $3$ & $3$ & 2 & $\frac{-\left(a_0^4-40 a_0^2+120 a_0-80\right){}^3}{1728 \left(a_0-3\right){}^5 \left(a_0^2+5 a_0-25\right)}$ & $3$ & $b$ & $I_1$, $I_1$, $I_5$, $I_5$\\ \hline

        10 & \multicolumn{2}{c|}{\bf{1}} & $2$ & \bf{1} & \bf{0} & \scalebox{0.8}{$\frac{-\left(a_0^4-16 a_0^2+24 a_0+16\right){}^3}{1728 \left(a_0-1\right){}^2 \left(a_0^3-a_0^2-18 a_0+43\right)}$} & 1 & $\mathcal{L}^{(2)}_{5a}$ & $I_1$, $I_1$, $I_1$, $I_2$, $I_7$\\
\hline
        
        \end{tabular}
        \caption{Special loci for quartic bases depicted in \Cref{fig:ellembed4} \cite{Piribauer2025}. The singular fibres follow the Kodaira classification and were obtained from the polynomials $f$ and $g$ of the Weierstraß form.}
        \label{tab:ellident4}
    \end{table}

    \newpage
    \begin{table}[ht]
        \centering
        \scalebox{0.92}{
        \begin{tabular}{|c||c|c|c|c|c|c|c|c|c|c|}
        \hline
        \multirow{2}{*}{no.}
          & \multicolumn{6}{c|}{deformations}
          & \multirow{2}{*}{$j(a_{0})$}
          & \multicolumn{3}{c|}{special loci}\\
        \cline{2-7} \cline{9-11}
          & $a_1$ & $a_2$ & $a_3$ & $a_4$ & $a_5$ & $a_6$ &  & $a_0$ & $\mathcal{L}^{(2)}$ & sing.\ fibres\\
        \hline \hline
        \multirow{6}{*}{13} & \multicolumn{6}{c|}{\multirow{2}{*}{3}} & \multirow{2}{*}{$-\frac{\left(a_0-6\right) \left(a_0+18\right){}^3}{1728 \left(a_0+21\right)}$} & 6 & $B$ & \multirow{2}{*}{$I_1$, $I_3$, $IV^*$}\\ \cline{9-10}
        & \multicolumn{6}{c|}{} & & -21 & $h=\mu(B)$ &\\ \cline{2-11}
        & \multicolumn{6}{c|}{\multirow{3}{*}{-1}} & \multirow{3}{*}{$-\frac{\left(\frac{a_0^4}{12}-2 a_0^2-16 a_0+60\right){}^3}{\left(a_0-3\right) \left(a_0-2\right){}^6 \left(a_0+6\right){}^2}$} & 2 & $a$ & \multirow{3}{*}{$I_1$, $I_2$, $I_3$, $I_6$}\\ \cline{9-10}
        & \multicolumn{6}{c|}{} & & 3 & $c$ &\\ \cline{9-10}
        & \multicolumn{6}{c|}{} & & -6 & $g$ &\\ \cline{2-11}
        
        & \multicolumn{6}{c|}{\multirow{2}{*}{0}} & \multirow{2}{*}{$-\frac{a_0^3 \left(a_0^3-216\right){}^3}{1728 \left(a_0^3+27\right){}^3}$} & 6 & \multirow{2}{*}{$f$} & \multirow{2}{*}{$I_1$, $I_1$, $I_1$, $I_9$}\\
\hline
        \end{tabular}}
        \caption{Special loci for cubic bases depicted in \Cref{fig:ellembed3} \cite{Piribauer2025}. The singular fibres follow the Kodaira classification and were obtained from the polynomials $f$ and $g$ of the Weierstraß form.}
        \label{tab:ellident3}
    \end{table}
    \footnotetext[1]{As we explain in the text, there exists a degeneration for $a_0=0$.}
\newpage
\clearpage
\subsubsection{Further remarks}\label{sec:remarks}
\paragraph{Two points inside two-dimensional face}
There exist further examples with more than one point inside a single two-face.
One such model is $X_{55311}$, whose toric and topological data we listed in \Cref{tab:55311} in \cref{app:ttdata}.
The discriminant factor of the face $\langle\nu_3,\nu_4,\nu_5\rangle$ reads
\begin{equation}
    \Delta_{345}=(1 - 4 z_3)^2 + 3125 z_2^2 z_3^3 + z_2 (27 + 25 z_3 (-9 + 20 z_3))
\end{equation}
The one-parameter model lies along $a_6=5$, $a_7=-5$, which is a zero of $\Delta_{345}$.
The one-parameter model is described by the operator
\href{https://cycluster.mpim-bonn.mpg.de/operator.html?nn=4.5.103}{4.5.103}.
\paragraph{Combining restrictions}
We mostly identified special loci in models where all deformations besides the origin lie in a single two-face.
In the examples we considered, it was possible to treat faces with inner points independently as long as their set of vertices were disjoint.
One minimal example are the toric hypersurfaces described in \Cref{tab:9611toric} in \cref{app:ttdata}.
There exists one point on a two-face and one on an edge, which, following the discussion above, give rise to $C*f$ when set to $-3$ and $2$, respectively.

The above methods fail in generalisations to multiple faces when the faces share vertices.
This can be seen most easily on the model described torically in \Cref{tab:7312toric} in \cref{app:ttdata}.
Both triangulations yield discriminant components for the relevant one- and two-dimensional faces, which do not intersect.

\paragraph{Reverse engineering and fibering out}
It is often possible to find various multi-parameter models in which the low parameter model is embedded by an analysis of the holomorphic
period at the MUM point. To illustrate the point let us consider the degree-eight hypersurfaces in $X_{22211}$ with $h_{11}=2$, $h_{21}=86$, 
which according to \cite{Candelas:1993dm} specialises to the degree $d_1=4$, $d_2=2$ complete intersection in $\mathbb{P}^5$ with $h_{11}=1$, $h_{21}=89$. 
The change of the Hodge numbers have been explained by the physical $N=2$ Higgs transition in \cite{Klemm:1996kv}\cite{Katz:1996ht}. 
The fundamental period of the complete intersection is according to \eqref{eq:fundamentalpiclosed} given by 
\be 
\varpi_0(z)=\sum_{n=0}^\infty \frac{(4n)! (2n)!}{(n!)^6} z^n=\sum_{n=0}^\infty\frac{(4n)!}{(n!)^3} \sum_{k=0}^{\lfloor \frac{n}{2}\rfloor } \frac{1}{4^k(n-2k)! (k!)^2} (2 z)^n
\label{eq:fpX42}
\ee
Here we used an identity for $(2n)!/(n!)^3$ of the kind of binomial identities, whose study and proofs have a long history \cite{DixonA}, see \cite{MR3574517} 
for a modern account.  Noting that the $X_{22211}$ mirror geometry has Mori cone vectors $l^{(1)}=(-4;1,1,1,1,0,0)$ and $l^{(2)}=(0;0,0,0,-2,1,1)$ \cite{MR1316509},
we see that again by comparing with \eqref{eq:fundamentalpiclosed} that the period \eqref{eq:fpX42} is a specialisation of the two parameter 
fundamental period of $X_{22211}$ for $z_2=\frac{1}{4}$. Note that the AESZ list~\cite{cycluster} contains many expressions for the fundamental 
periods in terms of binomials, which already contain finite sums like the right hand side of \eqref{eq:fpX42}, for which  multi-parameter 
Gelfand-Kapranov-Zelevinsky (GKZ) systems can be easily determined that specialise to the one-parameter system. Because of the identities studied 
in~\cite{MR3574517} the procedure is of course not unique and a given one-parameter model can be embedded in many ways in higher dimensional GKZ systems. 
Moreover given a toric specialisation of any reflexive polyhedron to a one-parameter family one can turn the techniques of the proofs \cite{MR3574517} around and
perform the residue calculations for the fundamental period in not unique terms of binomial sums. 

We have seen that our approach produces a simple reduction of the number of moduli in multi-parameter GKZ systems associated to 
Calabi-Yau $n$-folds defined in toric ambient spaces defined by reflexive polyhedra  with fibration structures. 
Simple combinatorial criteria for such fibration structures to occur  are given in~\cite{Avram:1996pj} and it is also pointed out there  
that (multi) fibrations are prevalent among the 4d reflexive polyhedra classified in~\cite{Kreuzer:2000xy}. This guarantees a wide range of 
applications of the formalism among the known Calabi-Yau differential systems and it is an interesting question whether all examples in the AESZ list~\cite{cycluster}
are obtained by specializing resonant multi-parameter GKZ systems in the way described above. Many of the associated polyhedra  have successive levels of fibration structures
making the fibering out formalism multiple times applicable and reducing the emerging differential equations into simplest Hadamard blocks, very much in 
the spirit of~\cite{MR4069107}.

\subsection{Type IIB flux vacua}\label{sec:fluxvacuaCY3}
In contrast to the singular loci before, here, we will use the Dwork deformation method on generic fibres.
We perform a systematical search for supersymmetric vacua in codimension one along one-dimensional rays in two-parameter moduli spaces.
In \cref{subsubsec:Toric Hypersurfaces} we present the results of scanning the $h^{2,1}=2$ toric hypersurface models in the Kreuzer--Skarke list. We find two models with supersymmetric vacua that do not lie along a line through the origin. In \cref{subsubsec:ToricBlowups} we discuss further examples given by CICYs, constructed as toric blowups of hypergeometric one-parameter models.

\subsubsection{Toric hypersurfaces}
\label{subsubsec:Toric Hypersurfaces}
Applying the strategy outlined in \cref{subsec:DeformationMethod} we searched for supersymmetric vacua in codimension one for all the $h^{2,1}=2$ models in the Kreuzer-Skarke list.
For each model we picked at least two directions to search for persistent factorizations, which we chose to be $(z_1,z_2)=(2z,z)$ and $(z_1,z_2)=(3z,z)$.
A list of the models together with the average number of splittings along the direction $(z_1,z_2)=(2z,z)$ for primes $p=49,\dots,249$ is given in \Cref{tab:fiveverticesresults,tab:sixverticesresults}.
We found two models with nontrivial vacua, by which me mean to exclude cases as described at the end of \cref{subsec:Arithmetic and analytic analysis of flux vacua}, where as a consequence of a symmetry between the two moduli there is a vacuum along a line through the origin
\footnote{The cases where this occurs are the bicubic in $\mathbb{P}^2\times\mathbb{P}^2$ and its free $\mathbb{Z}_3$-quotient, corresponding respectively to the third and first entry of \Cref{tab:sixverticesresults}. 
These models share the same Picard-Fuchs ideal and the two A-model geometries correspond to different integral monodromy structures of the Picard-Fuchs system. The vacuum along the diagonal $z_1=z_2=z$ was discussed in \cite{Candelas:2023yrg}. The Hodge substructure is described by the operator $\mathcal{L}^{(4)}_{\text{AESZ 15}}(z)$}
. 
Both models have six vertices and are discussed in the following.

\begin{table}[H]
\centering
\renewcommand{\arraystretch}{1.8}
\resizebox{\textwidth}{!}{%
\begin{tabular}{|c|c|c|c|c|}
\hline
$\chi$ & $l$-vectors & intersection ring & geometry & $\overline{N}_{\text{fact}}$ \\
\hline \hline
$-72$  & $(-3; 1, 1, 0, 0, 0, 1),\,(0; 0, 0, 1, 1, 1, -3)$ &  $9J_1^3+3J_1^2J_2+J_1J_2^2$ & $X\subset \mathbb{P}(\mathcal{O}_{\mathbb{P}^2}(-3)\oplus \mathcal{O}_{\mathbb{P}^2}\oplus \mathcal{O}_{\mathbb{P}^2})\big/\mathbb{Z}_3$ & $12\%$ \\ \hline
$-144$ & $(0; 0, 1, 0, 0, 1, -2),\,(-6; 2, 0, 1, 1, -1, 3)$ & $3J_1^3+3J_1^2J_2+3J_1J_2^2+2J_2^3$ & $X_{43221}$ & $2\%$ \\ \hline
$-168$ & $(-4; 0, 1, 1, 1, 0, 1),\,(0; 1, 0, 0, 0, 1, -2)$ & $8J_1^3+4J_1^2J_2$ & $X_{22211}$ & $6\%$ \\ \hline
$-208$ & $(0; 1, 0, 0, 1, 0, -2),\,(-8; 0, 1, 1, -3, 1, 8)$ & $36J_1^3+12J_1^2J_2+4J_1J_2^2+J_2^3$ & $X_{32111}$ &  $10\%$ \\ \hline
$-240$ & $(0; 1, 0, 0, 0, 1, -2),\,(-7; 0, 1, 1, 1, -3, 7)$ & $63J_1^3+21J_1^2J_2+7J_1J_2^2+2J_2^3$ & $X_{72221}$ & $6\%$ \\ \hline
$-252$ & $(-6; 3, 1, 1, 0, 0, 1),\,(0; 0, 0, 0, 1, 1, -2)$ & $4J_1^3+2J_1^2J_2$ & $X_{62211}$ & $16\%$ \\ \hline
$-540$ & $(-6; 3, 2, 0, 0, 0, 1),\,(0; 0, 0, 1, 1, 1, -3)$ & $9J_1^3+3J_1^2J_2+J_1J_2^2$ & $X_{96111}$ & $4\%$\\ \hline
\end{tabular}
}
\caption{Overview of the factorization results for the $h^{2,1}=2$ toric hypersurfaces from the Kreuzer-Skarke list with five vertices. For each model we give the Euler characteristic $\chi$ as well as the Mori-cone generators and the intersection ring. We further give the representation of the geometry in terms of a resolved hypersurface in a weighted projective space or in the case of the first model as a $\mathbb{Z}_3$ quotient of a hypersurface in a projectivization of a rank three vector bundle over $\mathbb{P}^2$. The last column lists the average number of factorizations of the local zeta function numerator $P^{(p)}_3(X,T)$ for the direction $(z_1,z_2)=(2z,z)$ over $50$ primes $p=41,\dots,293$.}
\label{tab:fiveverticesresults}
\end{table}

\begin{table}[H]
\centering
\renewcommand{\arraystretch}{1.8}
\resizebox{\textwidth}{!}{%
\begin{tabular}{|c|c|c|c|c|}
\hline
$\chi$ & $l$-vectors & intersection ring & geometry & $\overline{N}_{\text{fact}}$ \\
\hline \hline 
$-54$ & $(-3; 1, 0, 0, 0, 1, 1),\,(-3; 0, 1, 1, 1, 0, 0)$ & $J_1^2J_2+J_1J_2^2$ & $\left[
\renewcommand{\arraystretch}{0.7}%
\begin{array}{c|c}
\mathbb{P}^2 & 3 \\
\mathbb{P}^2 & 3
\end{array}
\right]\big/ \mathbb{Z}_3$ & $28\%$\\ \hline
$-144$  & $(0; 1, -1, 0, 0, -1, 1),\,(-6; 0, 3, 1, 1, 2, -1)$ &  $3J_1^3+3J_1^2J_2+3J_1J_2^2+2J_2^3$ & $\text{Bl}^{(1)}(X_{21111})\equiv X^{\text{III}(1)}_{43221}$ & $4\%$\\ \hline
$-162$  & $(-3; 1, 0, 0, 0, 1, 1),\,(-3; 0, 1, 1, 1, 0, 0)$ & $3J_1^2J_2+3J_1J_2^2$ & $\left[
\renewcommand{\arraystretch}{0.7}%
\begin{array}{c|c}
\mathbb{P}^2 & 3 \\
\mathbb{P}^2 & 3
\end{array}
\right]$ & $28\%$\\ \hline
$-164$ & $(-1; 1, -1, 0, 0, 0, 1),\,(-4; 0, 2, 1, 1, 1, -1)$ & $5J_1^3+5J_1^2J_2+5J_1J_2^2+3J_2^3$ & $\text{Bl}^{(1)}(X^5)$ & $2\%$\\ \hline
$-168$ & $(-4; 1, 1, 0, 1, 1, 0),\,(0; 0, 0, 1, -1, -1, 1)$ & $8J_1^3+4J_1^2J_2$ & $\left[
\renewcommand{\arraystretch}{0.7}%
\begin{array}{c|ccc}
\mathbb{P}^5 & 4 & 1 & 1\\
\mathbb{P}^1 & 0 & 1 & 1
\end{array}
\right]$ & $34\%$ \\ \hline
$-168$ & $(-2; 1, 1, 0, 0, 0, 0),\,(-4; 0, 0, 1, 1, 1, 1)$ & $4J_1J_2^2+2J_2^3$ & $\left[
\renewcommand{\arraystretch}{0.7}%
\begin{array}{c|c}
\mathbb{P}^1 & 2 \\
\mathbb{P}^3 & 4
\end{array}
\right]$ & $28\%$ \\ \hline
$-168$ & $(-3; 1, 1, 1, 0, 0, 0),\,(-2; -1, 0, 0, 1, 1, 1)$ & $5J_1^3+5J_1^2J_2+3J_1J_2^2$ & $X\subset \mathbb{P}(\mathcal{O}_{\mathbb{P}^2}(-1)\oplus \mathcal{O}_{\mathbb{P}^2}\oplus \mathcal{O}_{\mathbb{P}^2})$ & $6\%$ \\ \hline
$-168$ & $(-3; 1, 1, 1, 0, 0, 0),\,(-1; -1, -1, 0, 1, 1, 1)$ & $11J_1^3+7J_1^2J_2+3J_1J_2^2$ & $X\subset \mathbb{P}(\mathcal{O}_{\mathbb{P}^2}(-1)\oplus \mathcal{O}_{\mathbb{P}^2}(-1)\oplus \mathcal{O}_{\mathbb{P}^2})$ & $2\%$\\ \hline
$-168$ & $(-2; 1, -1, 0, 0, 1, 1),\,(-2; 0, 2, 1, 1, -1, -1)$ & $14J_1^3+10J_1^2J_2+6J_1J_2^2+3J_2^3$ & $\text{Bl}^{(2)}(X_{21111})$ & $6\%$ \\ \hline
$-168$ & $(-4; 1, 1, 1, 1, 0, 0),\,(-1; -1, 0, 0, 0, 1, 1)$ & $5J_1^3+4J_1^2J_2$ & $\left[
\renewcommand{\arraystretch}{0.7}%
\begin{array}{c|cc}
\mathbb{P}^4 & 4 & 1\\
\mathbb{P}^1 & 1 & 1
\end{array}
\right]$ & $14\%$\\ \hline
$-168$ & $(-1; 1, -2, 0, 0, 1, 1),\,(-3; 0, 3, 1, 1, -1, -1)$& $17J_1^3+13J_1^2J_2+9J_1J_2^2+6J_2^3$ & $X^{\text{III}(1)}_{32211}$ & $4\%$ \\ \hline
$-168$ & $(-2; 1, 1, 0, 0, 0, 0),\,(-4; -1, 0, 2, 1, 1, 1)$& $3J_1^3+3J_1^2J_2+3J_1J_2^2+J_2^3$ &  $\text{Bl}^{(3)}(X_{21111})$ & $10\%$ \\ \hline
$-176$ & $(-2; 1, 1, 0, 0, 0, 0),\,(-3; -1, 0, 1, 1, 1, 1)$& $5J_1^3+5J_1^2J_2+5J_1J_2^2+2J_2^3$ & $\text{Bl}^{(2)}(X^5)$ & $10\%$ \\ \hline
$-180$ & $(-3; 1, 1, 1, 0, 0, 0),\,(0; 0, -2, -1, 1, 1, 1)$& $21J_1^3+9J_1^2J_2+3J_1J_2^2$ &  $\text{Bl}^{(4)}(X_{21111})\equiv X\subset \mathbb{P}(\mathcal{O}_{\mathbb{P}^2}(-2)\oplus \mathcal{O}_{\mathbb{P}^2}(-1)\oplus \mathcal{O}_{\mathbb{P}^2})$ & $176\%$ \\ \hline
$-186$ & $(-3; 1, 1, 1, 0, 0, 0),\,(-1; -2, 0, 0, 1, 1, 1)$& $14J_1^3+7J_1^2J_2+3J_1J_2^2$ & $X\subset \mathbb{P}(\mathcal{O}_{\mathbb{P}^2}(-2)\oplus \mathcal{O}_{\mathbb{P}^2}\oplus \mathcal{O}_{\mathbb{P}^2})$  & $4\%$ \\ \hline
$-200$ & $(-2; 1, 1, 0, 0, 0, 0),\,(-2; -2, 0, 1, 1, 1, 1)$& $24J_1^3+12J_1^2J_2+6J_1J_2^2+2J_2^3$ & $\text{Bl}^{(5)}(X_{21111})$  & $18\%$ \\ \hline
$-208$ & $(-1; 1, -1, 1, 0, 0, 0),\,(-5; 0, 5, -3, 1, 1, 1)$& $36J_1^3+12J_1^2J_2+4J_1J_2^2+J_2^3$ & $\text{Bl}^{(1)}(X_{52111})\equiv X^{\text{III}(1)}_{32111}$ & $106\%$ \\ \hline
$-208$ & $(-2; 1, 1, 0, 0, 0, 0),\,(-2; -3, 0, 2, 1, 1, 1)$& $36J_1^3+12J_1^2J_2+4J_1J_2^2+J_2^3$ & $X^{\text{III}(2)}_{32111}$  & $6\%$\\ \hline
$-236$ & $(-2; 1, 1, -1, 0, 0, 1),\,(0; 0, -3, 5, 1, 1, -4)$& $101J_1^3+23J_1^2J_2+5J_1J_2^2+J_2^3$ & $\text{Bl}^{(1)}(X_{41111})$  & $2\%$\\  \hline
$-240$ & $(-1; 1, -1, 0, 0, 0, 1),\,(-4; 0, 4, 1, 1, 1, -3)$& $63J_1^3+21J_1^2J_2+7J_1J_2^2+2J_2^3$ & $\text{Bl}^{(2)}(X_{41111})\equiv X^{\text{III}(1)}_{31111}$  & $10\%$ \\ \hline
$-240$ & $(-2; 1, 1, 0, 0, 0, 0),\,(-1; -3, 0, 1, 1, 1, 1)$& $63J_1^3+21J_1^2J_2+7J_1J_2^2+2J_2^3$ & $X^{\text{III}(2)}_{31111}$  &$2\%$ \\ \hline
$-252$ & $(-6; 3, 1, 0, 1, 0, 1),\,(0; 0, 0, 1, -1, 1, -1)$& $4J_1^3+2J_1^2J_2$ & $X\subset\P_{3111}(\mathcal{O}_{\P^1}\oplus\mathcal{O}_{\P^1}\oplus\mathcal{O}_{\P^1}(-1)\oplus\mathcal{O}_{\P^1}(-1))$  & $8\%$ \\ \hline
$-252$ & $(-2; 1, 0, 2, 1, -1, -1),\,(0; 0, 1, -5, -2, 3, 3)$& $109J_1^3+42J_1^2J_2+16J_1J_2^2+6J_2^3$ & $\text{Bl}^{(2)}(X_{52111})$  & $8\%$\\ \hline
$-252$ & $(-2; 1, -1, 0, 0, 1, 1),\,(0; 0, 4, 1, 1, -3, -3)$& $108J_1^3+30J_1^2J_2+8J_1J_2^2+2J_2^3$ &  $\text{Bl}^{(3)}(X_{41111})$ & $6\%$\\ \hline
$-260$ & $(-4; 2, -1, 0, 1, 1, 1),\,(0; 0, 2, 1, -1, -1, -1)$& $14J_1^3+8J_1^2J_2+4J_1J_2^2+2J_2^3$ & $\text{Bl}^{(4)}(X_{41111})\equiv X\subset \mathbb{P}_{211}(\mathcal{O}_{\mathbb{P}^2}(-2)\oplus \mathcal{O}_{\mathbb{P}^2}(-1)\oplus \mathcal{O}_{\mathbb{P}^2})$ & $8\%$\\ \hline
$-260$ & $(-2; 1, -3, 0, 2, 1, 1),\,(0; 0, 7, 1, -4, -2, -2)$& $114J_1^3+49J_1^2J_2+21J_1J_2^2+9J_2^3$ & $X^{\text{III}(1)}_{73211}$  & $2\%$\\ \hline
$-284$ & $(-2; 1, -2, 0, 1, 1, 1),\,(0; 0, 5, 1, -2, -2, -2)$& $124J_1^3+50J_1^2J_2+20J_1J_2^2+8J_2^3$ & $X\subset \mathbb{P}_{211}(\mathcal{O}_{\mathbb{P}^2}(-3)\oplus \mathcal{O}_{\mathbb{P}^2}(-1)\oplus \mathcal{O}_{\mathbb{P}^2}(-1))$ & $14\%$ \\ \hline
\end{tabular}
}
\caption{Overview of the factorization results for the $h^{2,1}=2$ toric hypersurfaces from the Kreuzer-Skarke list with six vertices.
For models with multiple triangulations we list only one phase, corresponding to the given Mori-cone generators and intersection ring.
If available, we further give the associated geometry in terms of (i) a complete intersection in a product of projective spaces (or a free quotient thereof), (ii) type III desingularizations~\cite{Hosono:1996jv} of non-Fermat hypersurfaces (different such desingularizations are distinguished using a superscript) or (iii) for models with a genus-one/K$3$-fibred phase as a hypersurface in a projectivization of a rank three/four vector bundle over $\mathbb{P}^2/\mathbb{P}^1$.
We also introduced the notation $\text{Bl}^{(i)}(\dots)$ for models that can be realized as toric blowups of hypergeometric one-parameter models, where the superscript distinguishes different blowups. The last column lists the average number of factorizations of the local zeta function numerator $P^{(p)}_3(X,T)$ for the direction $(z_1,z_2)=(2z,z)$ over $50$ primes $p=41,\dots,293$.}
\label{tab:sixverticesresults}
\end{table}

\paragraph{$\text{Bl}^{(4)}(X_{21111})$}
The first model is defined by the polytope given as the convex hull of the points
\begin{equation}
    \label{eq:example1}
    \begin{alignedat}{4}
    \nu_1 &= (1, 0, 0, 0)\,,\quad &&\nu_2 = (0, 1, 0, 0)\,,&&\nu_3 = (-1, -1, 0, 0)\,,\\
    \nu_4 &= (0, 0, 1, 0)\,,\quad &&\nu_5 = (0, 0, 0, 1)\,,\quad &&\nu_6 = (-1, 1, -1, -1)\,.&&\\
    \end{alignedat}
    \end{equation}
This polytope has two fine star triangulations corresponding to birationally equivalent Calabi--Yau spaces. The results we find simply carry over from one model to the other by a simple transformation of the Batyrev coordinates, $(z_1,z_2)\mapsto(z_1z_2,\frac{1}{z_2})$, so we just focus on one of the triangulations. This triangulation has the Mori-cone generators
\begin{align}
 l^{(1)}&=(-3; 1, 1, 1, 0, 0, 0)\,,\\
 l^{(2)}&=(0; 0, -2, -1, 1, 1, 1)\,.
\end{align}
This model can be realized as a hypersurface in $\mathbb{P}(\mathcal{O}_{\mathbb{P}^2}(-2)\oplus \mathcal{O}_{\mathbb{P}^2}(-1)\oplus \mathcal{O}_{\mathbb{P}^2})$. It is a genus one fibration with 3-section \cite{Pioline:2025uov}.
Further it can be viewed as a toric blowup of the one-parameter model $X_{21111}$, whose Mori-cone generator is obtained as 
\begin{align}
\label{eq:X6blowuplvectors}
 2l^{(1)}+l^{(2)}=(-6; 2, 0, 1, 1, 1, 1)\,. 
\end{align}
Since it is the fourth $X_{21111}$-blowup in \Cref{tab:sixverticesresults} we refer to this model as $\text{Bl}^{(4)}(X_{21111})$. The Picard-Fuchs ideal and the discriminant are given in \Cref{tab:Bl4X6} in \cref{app:ttdata}.

\begin{figure}[H]
  \centering
  \begin{subfigure}[b]{\textwidth}
    \centering
    \includegraphics[width=\textwidth]{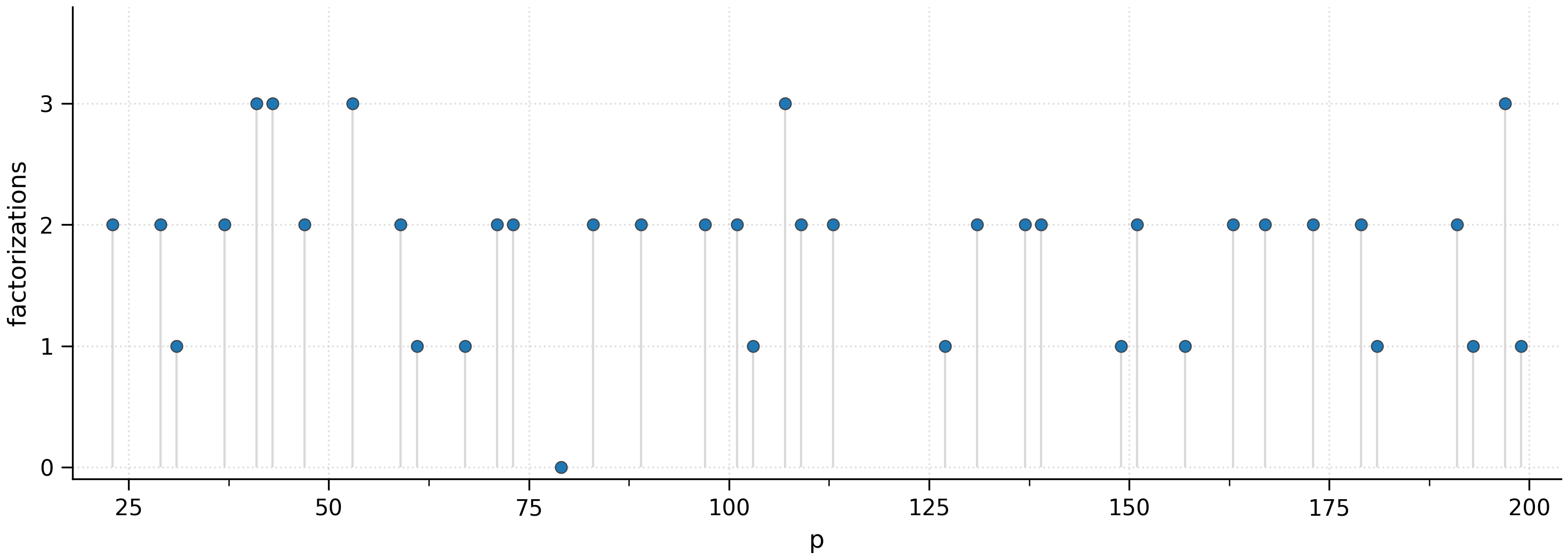}
    \caption{Direction $(z_1,z_2)=(z,z)$}
    \label{fig:factorizations-14triang2-z-z}
  \end{subfigure}

  \begin{subfigure}[b]{\textwidth}
    \centering
    \includegraphics[width=\textwidth]{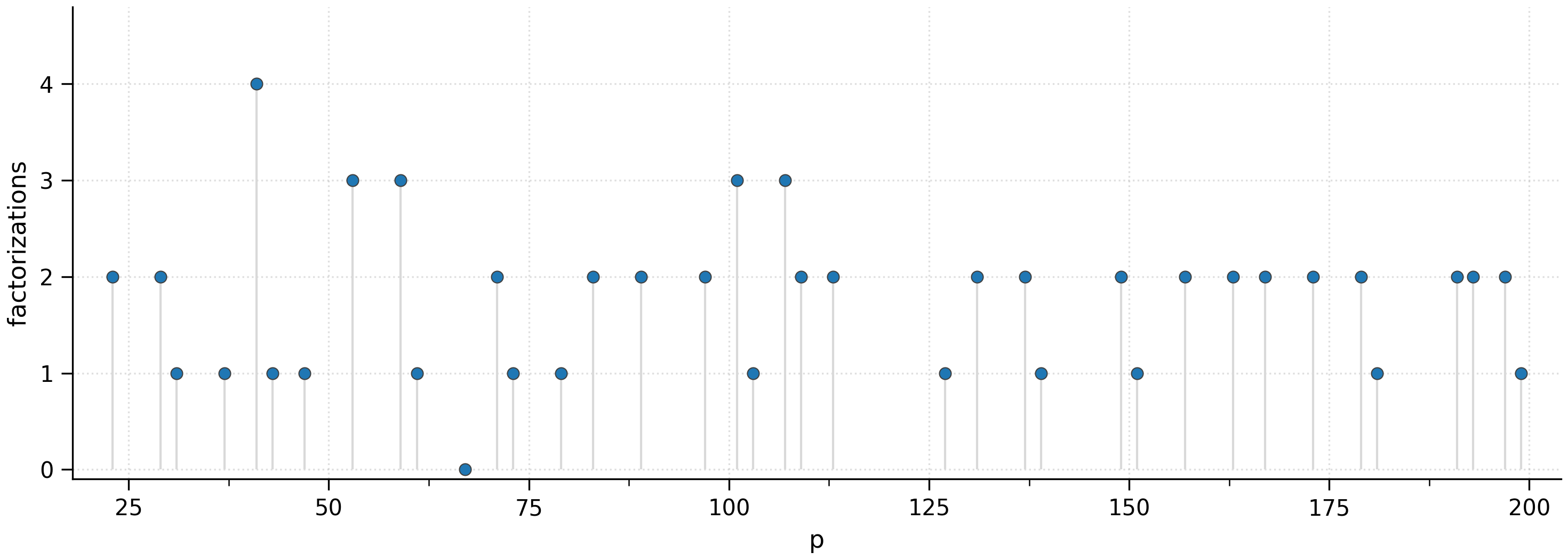}
    \caption{Direction $(z_1,z_2)=(2z,z)$}
    \label{fig:factorizations-14triang2-2z-z}
  \end{subfigure}

  \begin{subfigure}[b]{\textwidth}
    \centering
    \includegraphics[width=\textwidth]{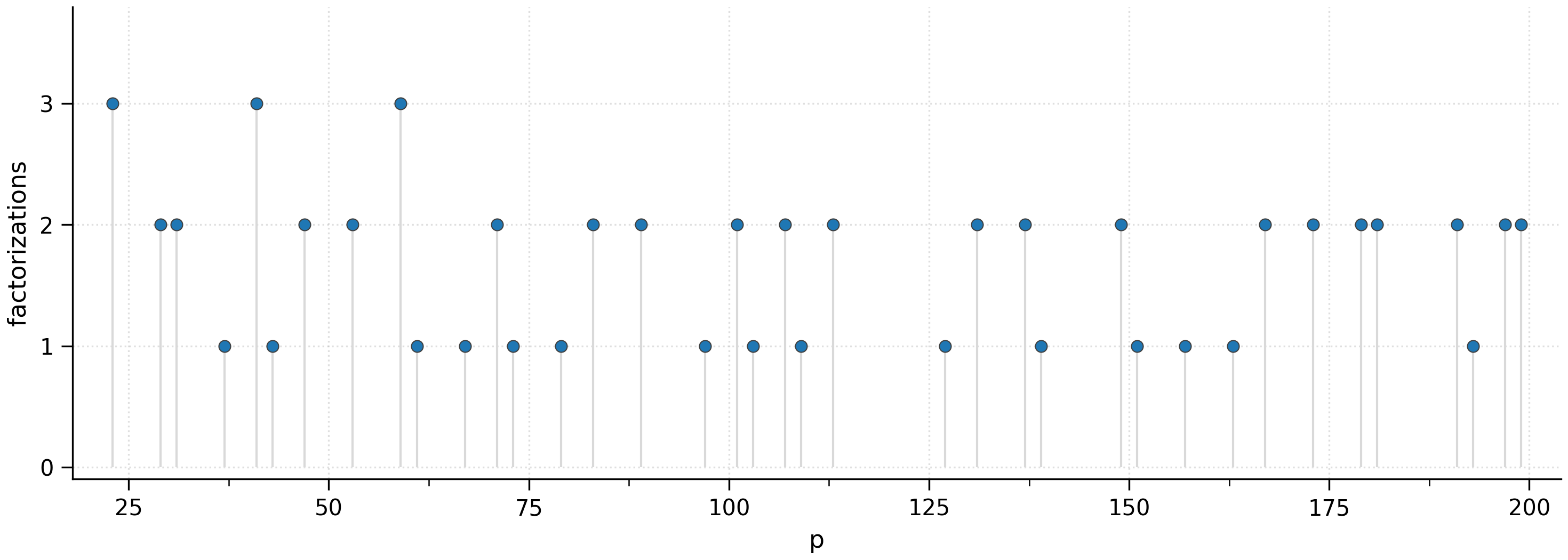}
    \caption{Direction $(z_1,z_2)=(3z,z)$}
    \label{fig:factorizations-14triang2-3z-z}
  \end{subfigure}

  \caption{Factorizations for the model $\text{Bl}^{(4)}(X_{21111})$ in various directions $(z_1,z_2)=(nz,z)$. For each direction the average number of factorizations is above one and, apart from points of bad reduction, for each prime there is at least one factorization at the reduction modulo $p$ of $z_1=1/27$.}
  \label{fig:factorizations-14triang2}
\end{figure}

We display the factorization plots for different directions in this model below in \Cref{fig:factorizations-14triang2-z-z,fig:factorizations-14triang2-2z-z,fig:factorizations-14triang2-3z-z}. One can see that for every direction for all but a few primes there is at least one factorization  indicating that they come from an equation linear in $z_1,z_2$. Indeed we find that the factorizations come from the fibres $1-27z_1=0$. On this locus the periods in $z_2$ are the ones of the hypergeometric model $X_6\subset \mathbb{P}_{21111}$. The corresponding operator is \href{https://cycluster.mpim-bonn.mpg.de/operator.html?nn=4.1.8}{AESZ 8} and its coordinate $\tilde{z}$ is related via $z_2=-3^9\tilde{z}$. As an example \Cref{tab:example1eulerfactors11} contains the Euler factors along the diagonal $z_1=z_2=z$. At $z=\frac{1}{27}\equiv 9\mod 11$ the Euler factor splits into a quadratic factor associated to the weight two modular form \href{https://www.lmfdb.org/ModularForm/GL2/Q/holomorphic/27/2/a/a/}{27.2.a.a} and the corresponding Euler factor of the model $X_{21111}$ (up to $T\mapsto-T$).
\newpage

    \begin{table}[H]
\centering
\renewcommand{\arraystretch}{1.6}
\resizebox{\textwidth}{!}{%
\begin{tabular}{|c|c|c|}
\hline
\multicolumn{3}{|c|}{$p=11$ for direction $(z_1,z_2)=(z,z)$} \\
\hline
$z \mod p$ & smooth/sing. & $P^{(p)}_3(X,T)$ \\
\hline \hline
$1$ & smooth & $p^9T^6 + 18p^6T^5 + 108p^4T^4 + 531p^2T^3 + 108pT^2 + 18T + 1$ \\
\hline
$2$ & smooth & $p^9T^6 + 2p^7T^5 - 12p^4T^4 - 245p^2T^3 - 12pT^2 + 2pT + 1$ \\
\hline
$3$ & smooth & $\left(p^3T^2 + 1\right)\left(p^6T^4 - 37pT^2 + 1\right)$ \\
\hline
$4$ & smooth & $p^9T^6 - 17p^6T^5 + 12p^5T^4 - 266p^2T^3 + 12p^2T^2 - 17T + 1$ \\
\hline
$5$ & smooth & $p^9T^6 - 2p^7T^5 - 12p^4T^4 + 245p^2T^3 - 12pT^2 - 2pT + 1 $ \\
\hline
$6$ & smooth & $p^9T^6 + 56p^6T^5 - 523p^2T^3 + 56T + 1$ \\
\hline
$7$ & smooth & $p^9T^6 + 3p^7T^5 + 171p^4T^4 + 456p^2T^3 + 171pT^2 + 3pT + 1$ \\
\hline
$8$ & smooth & $p^9T^6 + 42p^6T^5 + 126p^4T^4 + 384p^2T^3 + 126pT^2 + 42T + 1$ \\
\hline
$9$ & smooth & $\left(p^3T^2 + 1\right)\left(p^6T^4 + 60p^3T^3 + 188pT^2 + 60T + 1\right)$ \\
\hline
$10$ & smooth & $p^9T^6 - 59p^6T^5 + 204p^4T^4 - 488p^2T^3 + 204pT^2 - 59T + 1$ \\
\hline
\end{tabular}
}
\caption{Euler factors of the model $\text{Bl}^{(4)}(X_{21111})$ along the direction $(z_1,z_2)=(z,z)$ for $p = 11$.}
\label{tab:example1eulerfactors11}
\end{table}
By analytically continuing the periods from the MUM point
to the region around the point $z_1=\frac{1}{27},z_2=\infty$, we compute fluxes that realise $1-27z_1=0$ as a supersymmetric vacuum. We find as a possible choice
\begin{align}
\label{eq:X6vacuumfluxes}
    &f^{\text{T}}=(-266, -133, 513, 1368, -418, 0)\,,\\
    &h^{\text{T}}=(-190, -95, 437, 1083, -722, 0)\,.
\end{align}
The derivatives in the direction orthogonal to the vacuum locus
\begin{align}
    \left(f^{\text{T}}\Sigma \partial_{z_{1}}\underline{\Pi},\,h^{\text{T}}\Sigma \partial_{z_{1}}\underline{\Pi}\right)\big\vert_{z_1=\frac{1}{27}}
\end{align}
are then proportional to $z_2^{-\frac{1}{3}}$. Consequently, the axio-dilaton is constant instead of a nontrivial function of $z_2$. More precisely for the above choice of fluxes
\begin{align}
\label{eq:X6axiodilaton}
    \tau(z_2)=\frac{1+\ii\sqrt{3}}{2}\,.
\end{align}
This corresponds to a vanishing $j$-invariant. \Cref{tab:model14vacuumfactorizations} shows the Euler factors for a fibre on the vacuum locus. The quadratic factor corresponds to the weight two form \href{https://www.lmfdb.org/ModularForm/GL2/Q/holomorphic/27/2/a/a/}{27.2.a.a} and the fourth order factors are the Euler factors of $X_{21111}$ at the corresponding fibre $z_2=-3^9\tilde{z}$ (up to $T\mapsto -T$). 
\paragraph{}
Modularity associates the isogeny class \href{https://www.lmfdb.org/EllipticCurve/Q/27/a/}{27.a} of elliptic curves to the modular form.
This class contains elements with vanishing $j$-invariant and is therefore consistent with the value of the axio-dilaton \eqref{eq:X6axiodilaton}.
We checked for a few values of $z_2$ that the weight two modular factor always corresponds to an isogeny class containing at least one rational model with vanishing $j$-invariant.

\begin{table}[H]
\centering
\renewcommand{\arraystretch}{1.2}
\resizebox{0.65\textwidth}{!}{%
\begin{tabular}{|c|c|}
\hline
\multicolumn{2}{|c|}{$(z_1,z_2)=(\frac{1}{27},1)$} \\
\hline
$p$ & $P^{(p)}_3(X,T)$ \\
\hline \hline
$29$ &  $\left(p^3T^2 + 1\right)\left(p^6T^4 + 144p^3T^3 + 1196pT^2 + 144T + 1\right)$ \\
\hline
$31$ &  $\left(p^3T^2 + 4pT + 1\right)\left(p^6T^4 + 23p^3T^3 + 438pT^2 + 23T + 1\right)$ \\
\hline
$37$ & $ \left(p^3T^2 - 11pT + 1\right)\left(p^6T^4 + 169p^3T^3 + 2126pT^2 + 169T + 1\right)$ \\
\hline
$41$ & $\left(p^3T^2 + 1\right)\left(p^6T^4 + 52p^3T^3 - 535pT^2 + 52T + 1\right)$ \\
\hline
$47$ &  $ \left(p^3T^2 + 1\right)\left(p^6T^4 + 164p^3T^3 - 586pT^2 + 164T + 1\right)$ \\
\hline
$53$ &  $ \left(p^3T^2 + 1\right)
\left(p^6T^4 - 256p^3T^3 + 5114pT^2 - 256T + 1\right)$ \\
\hline
$59$ &  $ \left(p^3T^2 + 1\right)\left(p^6T^4 - 228p^3T^3 + 5756pT^2 - 228T + 1\right)$ \\
\hline
$61$ &  $ \left(p^3T^2 + pT + 1\right)\left(p^6T^4 + 425p^3T^3 + 1650pT^2 + 425T + 1\right)$ \\
\hline
$67$ &  $ \left(p^3T^2 - 5pT + 1\right)\left(p^6T^4 - 389p^3T^3 + 1634pT^2 - 389T + 1\right)$ \\
\hline
$71$ &  $ \left(p^3T^2 + 1\right)\left( p^6T^4 + 1144p^3T^3 + 12278pT^2 + 1144T + 1\right)$ \\
\hline
$73$ &  $ \left(p^3T^2 + 7pT + 1\right)\left(p^6T^4 + 779p^3T^3 + 5082pT^2 + 779T + 1\right)$ \\
\hline
$79$ &  $ \left(p^3T^2 - 17pT + 1\right)\left( p^6T^4 - 234p^3T^3 + 5266pT^2 - 234T + 1\right)$ \\ 
\hline
$83$ &  $ \left(p^3T^2 + 1\right)\left( p^6T^4 - 672p^3T^3 + 6560pT^2 - 672T + 1\right)$ \\ 
\hline
$89$ &  $ \left(p^3T^2 + 1\right)\left( p^6T^4 - 596p^3T^3 + 9866pT^2 - 596T + 1\right)$ \\ 
\hline
$97$ &  $ \left(p^3T^2 + 19pT + 1\right)\left( p^6T^4 + 492p^3T^3 + 10726pT^2 + 492T + 1\right)$ \\ 
\hline
$101$ &  $ \left(p^3T^2 + 1\right)\left(p^6T^4 + 1260p^3T^3 + 8549pT^2 + 1260T + 1\right)$ \\ 
\hline
$103$ &  $ \left(p^3T^2 + 13pT + 1\right)\left( p^6T^4 + 327p^3T^3 - 8624pT^2 + 327T + 1\right)$ \\ 
\hline 
$107$ &  $ \left(p^3T^2 + 1\right)\left(p^6T^4 - 408p^3T^3 - 10582pT^2 - 408T + 1\right)$ \\ 
\hline
$109$ &  $ \left(p^3T^2 - 2pT + 1\right)\left( p^6T^4 - 18p^3T^3 - 14222pT^2 - 18T + 1\right)$ \\ 
\hline
$113$ &  $ \left(p^3T^2 + 1\right)\left( p^6T^4 + 1148p^3T^3 + 13172pT^2 + 1148T + 1\right)$ \\ 
\hline
$127$ &  $ \left(p^3T^2 - 20pT + 1\right)\left( p^6T^4 - 1509p^3T^3 + 16876pT^2 - 1509T + 1\right)$ \\ 
\hline
$131$ &  $ \left(p^3T^2 + 1\right)\left(p^6T^4 + 316p^3T^3 - 1471pT^2 + 316T + 1\right)$ \\ 
\hline
$137$ &  $ \left(p^3T^2 + 1\right)\left(p^6T^4 + 3540p^3T^3 + 58058pT^2 + 3540T + 1\right)$ \\ 
\hline
$139$ &  $ \left(p^3T^2 - 23pT + 1\right)\left( p^6T^4 + 2385p^3T^3 + 35158pT^2 + 2385T + 1\right)$ \\ 
\hline
$149$ &  $ \left(p^3T^2 + 1\right)\left(p^6T^4 - 2904p^3T^3 + 47264pT^2 - 2904T + 1\right)$ \\ 
\hline
\end{tabular}
}
\caption{Euler factors of the model $\text{Bl}^{(4)}(X_{21111})$ for $(z_1,z_2)=(\frac{1}{27},1)$ for good primes $p=29,\dots,149$. We exclude the prime of bad reduction $p=43$. The quadratic factors correspond to the weight two form \href{https://www.lmfdb.org/ModularForm/GL2/Q/holomorphic/27/2/a/a/}{27.2.a.a} and the fourth order factors are Euler factors of $X_{21111}$ at the fibre $\tilde{z}=-1/19683$.}
\label{tab:model14vacuumfactorizations}
\end{table}

An interesting question is whether this vacuum is the fixed point locus of some symmetry.
When computing the Euler factors for all reductions at a given prime one observes that except from reductions of $z_1=-\frac{1}{216}$ they come in pairs, equal up to $T\mapsto -T$.
This doubling of Euler factors points to an involution symmetry of the moduli space, however with fixed point locus not along the vacuum but rather $z_1=-\frac{1}{216}$.
Determining for given values of the moduli for several primes the reductions of the supposed images of this involution symmetry using Chinese remaindering, one finds that the following involution reproduces the found values:
\begin{align}
\label{eq:BlX6involution}
    \mathcal{I}:\; (z_1,z_2)\mapsto \left(\frac{\big(1-3z_1^{\frac{1}{3}}\big)^3}{27\big(1+6z_1^{\frac{1}{3}}\big)^3},-\frac{3^9z_1^2z_2}{\big(1-3z_1^{\frac{1}{3}}\big)^6}\right).
\end{align}
We indeed have $\mathcal{I}^2=\mathbb{1}$ and $\mathcal{I}(-\frac{1}{216},z_2)=(-\frac{1}{216},z_2)$.
As expected of a symmetry the discriminant locus $\Delta^{(2)}=0$ is mapped to itself under $\mathcal{I}$. The vacuum locus $z_1=\frac{1}{27}$ is not fixed but rather mapped to the point $(z_1,z_2)=(0,\infty)$. We comment on a possible geometric origin of this symmetry below.
Its presence offers a potential explanation for the existence of the vacuum: As mentioned above the model at hand is a toric blowup of the hypergeometric model $X_{21111}$.
We want to argue that the supersymmetric vacuum of the two-parameter model is the image under the involution $\mathcal{I}$ of the locus in the moduli space where the blowup deformation vanishes. 

To make this more precise we consider how the hypersurface constraint of $\text{Bl}^{(4)}(X_{21111})$ reduces to that of $X_{21111}$ in a certain limit. The dual polytope $\Delta^\circ$ of \eqref{eq:example1} has seven vertices,
\begin{equation}
    \label{eq:example1dual}
    \begin{alignedat}{4}
    \nu^\circ_1 &= (2, -1, -1, -1)\,,\quad &&\nu^\circ_2 = (-1, -1, 2, -1)\,,&&\nu^\circ_3 = (-1, 2, 5, -1)\,,\\
    \nu^\circ_4 &= (-1, 2, -1, 5)\,,\quad &&\nu^\circ_5 = (-1, 2, -1, -1)\,,\quad &&\nu^\circ_6 = (-1, -1, -1, 2)\,,\\
    \nu^\circ_7 &= (-1, -1, -1, -1)\,.
    \end{alignedat}
\end{equation}
The hypersurface constraint $\{P_\Delta=0\} \subset \mathbb{P}_{\Delta^\circ}$ defining $\text{Bl}^{(4)}(X_{21111})$ in terms of the homogeneous coordinates associated to the $\nu^\circ_i$ is
\begin{align}
P_\Delta(\underline{a},\underline{x})=  a_{1} x_{1}^{3}
+ a_{2} x_{3}^{3} x_{4}^{3} x_{5}^{3}
+ a_{3} x_{2}^{3} x_{6}^{3} x_{7}^{3}
+ a_{4} x_{2}^{3} x_{3}^{6}
+ a_{5} x_{4}^{6} x_{6}^{3}
+ a_{6} x_{5}^{6} x_{7}^{3}+ a_{0} x_{1} x_{2} x_{3} x_{4} x_{5} x_{6} x_{7}\,,
\end{align}
with the following scaling relations acting on the $x_i$,
\begin{align}
    Q_1=(2, 2, 0, 1, 1, 0, 0)\,,\;\;
    Q_2=(2, 0, 1, 0, 1, 2, 0)\,,\;\;
    Q_3=(3, 1, 1, 1, 1, 1, 1)\,,
\end{align}
where the entries give the exponents under the respective $\mathbb{C}^\ast$-scaling. Using $Q_2$ and $Q_3$ to set $x_6=x_7=1$ and introducing new coordinates 
\begin{align}
    x_1 = X_1\,,\;\; x_2 = X_2^2\,,\;\; x_3 = \frac{X_3}{X_2}\,, \;\;x_4 = X_4\,,\;\; x_5 = X_5\,,
\end{align}
we are left with 
\begin{align}
P_\Delta(\underline{a},\underline{X})= a_{1} X_{1}^{3}
+ a_{3} X_{2}^{6}
+ a_{4} X_{3}^{6}
+ a_{5} X_{4}^{6}
+ a_{6} X_{5}^{6}+ a_{0} X_{1} X_{2} X_{3} X_{4} X_{5}+ a_2\frac{X_{3}^{3} X_{4}^{3} X_{5}^{3}}{X_{2}^{3}}\,,
\end{align}
and the scaling relation $Q^X= (2, 1, 1, 1, 1)$ acting on the $X_i$. We can use coordinate rescalings to set $a_1=a_3=a_4=a_5=a_6=1$. We see that in this patch the hypersurface is that of the sextic plus a blowup deformation controlled by the modulus $a_2$. In the Batyrev coordinates,
\begin{align}
    z_1=\frac{a_1a_2a_3}{a_0^3}=\frac{a_2}{a_0^3}\,,\quad z_2=\frac{a_4a_5a_6}{a_2^2a_3}=\frac{1}{a_2^2}\,,
\end{align}
setting the blowup deformation $a_2$ to zero thus corresponds for finite $a_0$ to $z_1=0$, ${z_2=\infty}$, with the $X_{21111}$ moduli space coordinate $\tilde{z}=\frac{1}{a_0^6}$ embedded as $\tilde{z}=z_1^2z_2$. Note that the identification of the blowup mode and the one-parameter coordinate can already be anticipated from the $l$-vector relation \eqref{eq:X6blowuplvectors}. The vacuum at $z_1=\frac{1}{27}$ is mapped under $\mathcal{I}$ to $z_1=0$, $z_2=\infty$ and from \eqref{eq:BlX6involution} we can deduce which $\tilde{z}$ coordinate the $\mathcal{I}$-images of $z_1=\frac{1}{27}$ correspond to:
\begin{align}
    \tilde{z}\left(\mathcal{I}\left(\frac{1}{27},z_2\right)\right)=-\frac{27z_1^2z_2}{\big(1+6z_1^{\frac{1}{3}}\big)^6}\Bigg\vert_{z_1=\frac{1}{27}}=-\frac{z_2}{3^9}\,,
\end{align}
which as explained earlier is precisely the normalization in which the periods of $X_{21111}$ appear along $z_1=\frac{1}{27}$.

\begin{figure}[H]
  \centering
  \includegraphics[width=\textwidth]{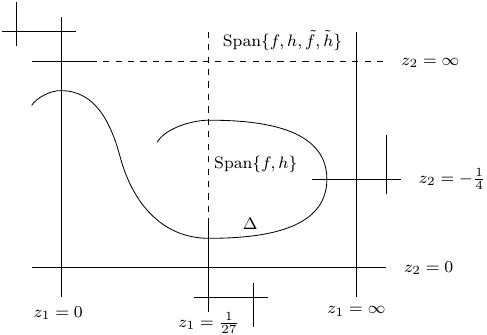}
  \caption{Schematic picture of the moduli space of the model $\text{Bl}^{(4)}(X_{21111})$. Shown are the discriminant components and exceptional divisors resolving tangencies. The dashed lines show the location of the supersymmetric vacua discussed in the text and we indicate the flux lattices they support.}
  \label{fig:BlX6modulispace}
\end{figure}

Note that since on the vacuum locus $z_1=\frac{1}{27}$ the quartic parts of the Euler factors are those of the sextic and the latter has an attractor splitting at the orbifold point, the Euler factor further factorizes at $z_2=\infty$. At this point we expect to find two more independent fluxes than those in \eqref{eq:X6vacuumfluxes}. In fact analytic continuation of the periods shows that there are four independent fluxes for all fibres at $z_2=\infty$, with a basis for the flux lattice given by the fluxes $f,h$ that are present also at $z_1=\frac{1}{27}$\,.
Additionally, one finds
\begin{align}
\label{eq:z2inftyfluxes}
    &\tilde{f}^{\text{T}}=(2, 0, 0, 2, -2, 4)\,,\\
    &\tilde{h}^{\text{T}}=(0, 0, 0, 1, 0, 2)\,,
\end{align} 
such that 
\begin{align}
\tilde{f}^T\Sigma\underline{\Pi}\big\vert_{z_2=\infty}=\tilde{f}^T\Sigma\partial_{z_1}\underline{\Pi}\big\vert_{z_2=\infty}=0\,,
\end{align}
and the same for the other three fluxes. 
Picking the fluxes $\tilde{f}$ and $\tilde{h}$ as a background, one stabilizes to $z_2=\infty$. The axio-dilaton expanded in $u=z_1-\frac{1}{27}$ is then 
\begin{align}
\label{eq:axiodilatonz2infty}
\tau(u)=\frac{1 + \sqrt{3}\, i}{2}\Big(
1
- \frac{9}{2} u
+ \frac{243}{4} u^2
- \frac{8505}{8} u^3
+ \mathcal{O}(u^4)
\Big).
\end{align}
With fluxes $f$ and $h$ as before, the moduli are stabilizes to ${\{z_1=\frac{1}{27}\}\cup\{z_2=\infty\}}$. The axio-dilaton then takes the constant value $\eqref{eq:X6axiodilaton}$. While we did not compute the Euler factors at $z_2=\infty$ we expect them to factorize for all values of $z_1$ into three quadratic factors. Based on our earlier discussion two of those factors should be the weight two and weight four modular factors of the sextic orbifold point (\href{https://www.lmfdb.org/ModularForm/GL2/Q/holomorphic/27/2/a/a/}{27.2.a.a} and \href{https://www.lmfdb.org/ModularForm/GL2/Q/holomorphic/108/4/a/c/}{108.4.a.c}): For finite $z_1$ and $z_2=\infty$ we have $a_0=a_2=0$ and the hypersurface constraint reduces to the Fermat point sextic. The remaining quadratic factor should be modular of weight two and the $j$-invariants should vary with $z_1$ according to the axio-dilaton profile \eqref{eq:axiodilatonz2infty}. The two periods surviving at $z_2=\infty$ are linearly dependent and proportional to $z_1^{-\frac{1}{3}}$. The fact that only one independent period remains is due an additional splitting of the  $(1,0,0,1)$ Hodge structure into $(1,0,0,0)\oplus(0,0,0,1)$ over $\mathbb{Q}(\sqrt{-3})$. The structure of the moduli space together with the location of the found supersymmetric vacua is shown schematically in figure \Cref{fig:BlX6modulispace}.

For the discussion\footnote{We thank Thorsten Schimannek for his detailed explanations regarding the following paragraph, in particular the A-model interpretation of the second MUM point and the analogy to the model AESZ 3 with regards to the involution symmetry.} of the origin of the symmetry it is important to note that there is a second MUM point\footnote{In fact there is a third MUM point at $(z_1,z_2)=(0,\infty)$ in the coordinates $(z_1z_2,\frac{1}{z_2})$, which corresponds to the other triangulation of the toric polytope underlying this model.} at $(z_1,z_2)=(\frac{1}{27},0)$. As indicated in \Cref{fig:BlX6modulispace}, at this point the discriminant intersects the divisor $z_2=0$ tangentially and hence has to be resolved with blowups. The local coordinates of the second MUM point are then
\begin{align}
\label{eq:blowupcoord}
    (z_1,z_2)\mapsto \Big(-z_1 + \frac{1}{27}, -\frac{z_1^3z_2}{(1-27z_1)^3}\Big).
\end{align}
The A-model geometry whose large volume limit is mirror to this MUM point is that of a Jacobian fibration with nodal singularities, stabilized by a nontrivial $B$-field \cite{Schimannek:2021pau}\cite{Katz:2022lyl}. The intersection ring is
\begin{align}
    9J_1^3+3J_1^2J_2+J_1J_2^2\,.
\end{align}
The associated integral monodromy struture is genuinely different from the one of the MUM point at the origin:
The matrix transforming the integral basis of the MUM point at $(0,0)$, analytically continued to $(\frac{1}{27},0)$, and the integral basis associated to the Jacobian fibration, is irrational.
The change of basis transformation maps the torus fibre mirror coordinate according to the Fricke involution $\tau=t_1\mapsto -\frac{1}{3\tau}$, whose action on the A-model topological string partition function is discussed in \cite{Schimannek:2021pau}. 
Note that this situation is different from the example of the elliptically fibred model $X_{96111}$, where both MUM points correspond to the same A-model geometry and the transformation between the integral structures of the two MUM points is integral symplectic.
The elliptic fibre is mapped in that case according to $\tau\mapsto -\frac{1}{\tau}$, which from the string theory point of view corresponds to applying T-duality to the $A$- and $B$-cycle of the fibre torus \cite{Candelas:1994hw,Klemm:2012sx,Andreas:2001ve,Alim:2012ss}. Hence, in the case of $X_{96111}$, the moduli space symmetry, that sends the coordinates $z_i$ to the blowup coordinates, preserves the integral monodromy structure and reflects a physical symmetry. It was derived in \cite{Candelas:1994hw} from the residual freedom after using the root automorphisms to set certain polynomial deformations to zero. In the model $\text{Bl}^{(4)}(X_{21111})$ on the other hand, the involution is seemingly an artefact of an additional integral structure present in this model when going to the covering space $z_1\mapsto z_1^3$. The situation is analogous to the one-parameter model $X^{2,2,2,2}\subset \mathbb{P}^7$ (\href{https://cycluster.mpim-bonn.mpg.de/operator.html?nn=4.1.3}{AESZ 3}), which has a second MUM point at $z=\infty$ corresponding on the A-side to a singular double cover of $\mathbb{P}^3$ with nontrivial $B$-field \cite{Katz:2022lyl} and whose integral structure is not related to the one at $z=0$ by an integral symplectic basis change. When transforming the operator $\mathcal{L}_{2,2,2,2}$ according to $z\mapsto z^2$ however, both MUM points correspond on the A-side to the same geometry (a different singular double cover of $\mathbb{P}^3$ with nontrivial $B$-field) and the integral structure is preserved under the moduli space symmetry \cite{Katz:2023zan}.

In our case, we can make the symmetry manifest, at least on the level of the rational monodromy structure (which is enough to explain the observed symmetry of the Euler factors), when we introduce coordinates $(\tilde{z}_1,\tilde{z}_2)=(z_1^{\frac{1}{3}},z_2)$ around the MUM point at the origin and for the MUM point at $z_1=\frac{1}{27},z_2=0$ the coordinates following from the involution:
\begin{align}
 u_1=\frac{1-3\tilde{z}_1}{3(1+6\tilde{z}_1)}\,,\quad u_2=-\frac{3^9\tilde{z}_1^6\tilde{z}_2}{(1-3\tilde{z}_1)^6}\,.
\end{align}
In the $\underline{u}$ coordinates the triple intersection numbers following from the leading logarithms of the triple-logarithmic solution are equal up to an overall factor to those of the $\underline{\tilde{z}}$ coordinates. The rational monodromy bases around the two MUM points following from these intersection numbers differ up to a rescaling by a rational symplectic transformation. Furthermore the local periods are equal up to a Kähler gauge transformation: 
\begin{align}
    \underline{\omega}^u(\underline{u})=(1+6u_1)\, \underline{\omega}^{\tilde{z}}(\underline{u})\,.
\end{align}
In particular it follows that the instanton expansions around both MUM points are the same. Note that this also shows that the periods of the Jacobian fibration are up to an algebraic coordinate transformation (given by the transformation between the blowup coordinates and the coordinates defined by the involution) and a Kähler gauge transformation equal to the periods of the model $\text{Bl}^{(4)}(X_{21111})$. 

Before closing the discussion of this model we study the arithmetic properties of the fixed point locus of the involution symmetry.
At the fixed point locus the Euler factors factorize frequently over the field $K=\mathbb{Q}(\sqrt{-3})$ into two cubic factors.
More precisely, they factorize when $p\equiv 1\mod 3$, i.e. for those primes that are split over the ring of integers $\mathcal{O}_K$. 
For inert primes, $p\equiv 2\mod 3$, the Euler factors are even polynomials\footnote{We observe that the sign flip of $T$ in the Euler factors related by symmetry is given by the character $\chi_3(p)=\big(\frac{-3}{p}\big)$.}. 
This is shown for the fibre $(z_1,z_2)=(-\frac{1}{216},1)$ in \Cref{tab:model14triang2fixedpoint}.  Analytic continuation of the periods to the fixed point locus shows that there is a rank three lattice $\Lambda$ of vectors $g$ satisfying
\begin{align}
\label{eq:fixedpointsplitting}
     g^{\text{T}}\Sigma \underline{\Pi}\big\vert_{z_1=-\frac{1}{216}}=g^{\text{T}}\Sigma \partial_{z_2}\underline{\Pi}\big\vert_{z_1=-\frac{1}{216}}=0\,,
\end{align}
with 
\begin{align}
\begin{aligned}
g^{T}\in \Lambda
&= \operatorname{Span}_K\Big\{
\Big(1,\tfrac{1+\sqrt{-3}}{2},-\tfrac{5+8\sqrt{-3}}{6},-\tfrac{5+13\sqrt{-3}}{4},0,0\Big),\\[-2pt]
&\qquad\Big(0,0,\tfrac{1+\sqrt{-3}}{12},-\tfrac{1-\sqrt{-3}}{4},-1,0\Big),\;
\Big(0,0,\tfrac{1}{6},\tfrac{1+\sqrt{-3}}{4},0,-1\Big)
\Big\}\,.
\end{aligned}
\end{align}
The implies a splitting of cohomology into Hodge types $(1,1,1,0)\oplus (0,1,1,1)$ over $K$. Analogous to the discussion under \eqref{eq:octicCMsplitting} this explains the behaviour of the Euler factors for split and inert primes in \Cref{tab:model14triang2fixedpoint}. The nonvanishing periods at $z_1=-\frac{1}{216}$ are annihilated by the operator
\begin{align}
    \mathcal{L}^{(3)}=\theta ^3-18 z\,(6 \theta +1)(3 \theta +1) (3 \theta +2)\,,
\end{align}
with $z=z_2/3^8$.

\begin{table}[H]
\centering
\renewcommand{\arraystretch}{1.5}
\resizebox{\textwidth}{!}{%
\begin{tabular}{|c|c|c|}
\hline
\multicolumn{3}{|c|}{$(z_1,z_2)=(-\frac{1}{216},1)$} \\
\hline
$p$ & split/inert over $K$ & $P^{(p)}_3(X,T)$ \\
\hline \hline
$29$ & inert & $p^{9}T^{6} + 114\,p^{4}T^{4} + 114\,pT^{2} + 1$  \\
\hline
$31$ & split & $p^{3}\left(-154 -45\sqrt{-3}\right)T^3 + \frac{p}{2}\left(-807 + 45\sqrt{-3}\right)T^2 + \frac{1}{2}\left(123 + 45\sqrt{-3}\right)T + 1$  \\
\hline
$37$ & split & $p^{3}\left(55 -126\sqrt{-3}\right)T^3 + p\left(413 + 846\sqrt{-3}\right)T^2 + \left(-217 -72\sqrt{-3}\right)T + 1$  \\
\hline
$41$ & inert & $p^{9}T^{6} + 3426\,p^{4}T^{4} + 3426\,pT^{2} + 1$  \\
\hline
$43$ & split & $\frac{p^{3}}{2}\left(449 + 197\sqrt{-3}\right)T^3 + \frac{p}{2}\left(-101 + 1397\sqrt{-3}\right)T^2 + \frac{1}{2}\left(211 -175\sqrt{-3}\right)T + 1$  \\
\hline
$47$ & inert & $p^{9}T^{6} + 474\,p^{4}T^{4} + 474\,pT^{2} + 1$  \\
\hline
$53$ & inert & $p^{9}T^{6} + 5529\,p^{4}T^{4} + 5529\,pT^{2} + 1$  \\
\hline
$59$ & inert & $p^{9}T^{6} + 7311\,p^{4}T^{4} + 7311\,pT^{2} + 1$  \\
\hline
$61$ & split & $p^{3}\left(91 + 270\sqrt{-3}\right)T^3 + p\left(-292 + 483\sqrt{-3}\right)T^2 + \left(98 -33\sqrt{-3}\right)T + 1$  \\
\hline
$67$ & split & $\frac{p^{3}}{2}\left(-127 -629\sqrt{-3}\right)T^3 + p^{2}\left(-64 + 8\sqrt{-3}\right)T^2 + \left(-52 + 308\sqrt{-3}\right)T + 1$  \\
\hline
$71$ & inert & $p^{9}T^{6} + 6621\,p^{4}T^{4} + 6621\,pT^{2} + 1$  \\
\hline
$73$ & split & $\frac{p^{3}}{2}\left(271 + 703\sqrt{-3}\right)T^3 + p\left(-3237 + 1553\sqrt{-3}\right)T^2 + \left(225 -253\sqrt{-3}\right)T + 1$  \\
\hline
$79$ & split & $\frac{p^{3}}{2}\left(503 + 757\sqrt{-3}\right)T^3 + p\left(-1549 + 1937\sqrt{-3}\right)T^2 + \left(290 -172\sqrt{-3}\right)T + 1$  \\
\hline
$83$ & inert & $p^{9}T^{6} + 5556\,p^{4}T^{4} + 5556\,pT^{2} + 1$  \\
\hline
$89$ & inert & $p^{9}T^{6} + 19911\,p^{4}T^{4} + 19911\,pT^{2} + 1$  \\
\hline
$97$ & split & $\frac{p^{3}}{2}\left(1853 + 269\sqrt{-3}\right)T^3 + p\left(-6967 + 5026\sqrt{-3}\right)T^2 + \frac{1}{2}\left(-941 -1189\sqrt{-3}\right)T + 1$  \\
\hline
$101$ & inert & $p^{9}T^{6} + 11703\,p^{4}T^{4} + 11703\,pT^{2} + 1$  \\
\hline
$103$ & split & $p^{3}\left(910 -297\sqrt{-3}\right)T^3 + \frac{p}{2}\left(-9311 + 20841\sqrt{-3}\right)T^2 + \frac{1}{2}\left(-2549 -1527\sqrt{-3}\right)T + 1$  \\
\hline
$107$ & inert & $p^{9}T^{6} - 6486\,p^{4}T^{4} - 6486\,pT^{2} + 1$  \\
\hline
$109$ & split & $p^{3}\left(-323 -630\sqrt{-3}\right)T^3 + p\left(-1594 + 3843\sqrt{-3}\right)T^2 + \left(-568 + 189\sqrt{-3}\right)T + 1$  \\
\hline
$113$ & inert & $p^{9}T^{6} - 12093\,p^{4}T^{4} - 12093\,pT^{2} + 1$  \\
\hline
$127$ & split & $\frac{p^{3}}{2}\left(-2267 -1009\sqrt{-3}\right)T^3 + \frac{p}{2}\left(-10821 + 719\sqrt{-3}\right)T^2 + \frac{1}{2}\left(693 + 389\sqrt{-3}\right)T + 1$  \\
\hline
$131$ & inert & $p^{9}T^{6} + 42306\,p^{4}T^{4} + 42306\,pT^{2} + 1$  \\
\hline
$137$ & inert & $p^{9}T^{6} + 23079\,p^{4}T^{4} + 23079\,pT^{2} + 1$  \\
\hline
$139$ & split & $p^{3}\left(-1288 + 585\sqrt{-3}\right)T^3 + \frac{p}{2}\left(-9231 + 30249\sqrt{-3}\right)T^2 + \frac{1}{2}\left(3363 + 1737\sqrt{-3}\right)T + 1$  \\
\hline
$149$ & inert & $p^{9}T^{6} + 18795\,p^{4}T^{4} + 18795\,pT^{2} + 1$  \\
\hline
\end{tabular}
}
\caption{Factorizations over $K=\mathbb{Q}(\sqrt{-3})$ of Euler factors of model $\text{Bl}^{(4)}(X_{21111})$ at involution fixed point $(z_1,z_2)=(-\frac{1}{216},1)$ for primes $p=29,\dots,149$. For split primes ($p\equiv1\mod3$) we display one cubic factor (the other is its complex conjugate).}
\label{tab:model14triang2fixedpoint}
\end{table}

\paragraph{$\text{Bl}^{(1)}(X_{52111})$}
The vertices of the polytope describing the second model are 
\begin{equation}
    \label{eq:example2}
    \begin{alignedat}{4}
    \nu_1 &= (1, 0, 0, 0)\,,\quad &&\nu_2 = (0, 1, 0, 0)\,,&&\nu_3 = (-1, 1, 0, 0)\,,\\
    \nu_4 &= (0, 0, 1, 0)\,,\quad &&\nu_5 = (0, 0, 0, 1)\,,\quad &&\nu_6 = (-3, -2, -1, -1)\,.&&\\
    \end{alignedat}
    \end{equation}
It has one fine star triangulation, for which the Mori-cone generators are 
\begin{align}
 l^{(1)}&=(-1; 1, -1, 1, 0, 0, 0)\,,\\
 l^{(2)}&=(-5; 0, 5, -3, 1, 1, 1)\,.
\end{align}
Since 
\begin{align}
\label{eq:X10blowuplvectors}
    5l^{(1)}+l^{(2)}=(-10;5,0,2,1,1,1)\,,
\end{align}
we refer to this model as $\text{Bl}^{(1)}(X_{52111})$ (the superscript again just refers to it being the first of the two $X_{52111}$-blowups in \Cref{tab:sixverticesresults}). The Picard--Fuchs differential ideal and the discriminant are given in \Cref{tab:Bl1X10}.

\begin{figure}[H]
  \centering
  \begin{subfigure}[b]{\textwidth}
    \centering
    \includegraphics[width=\textwidth]{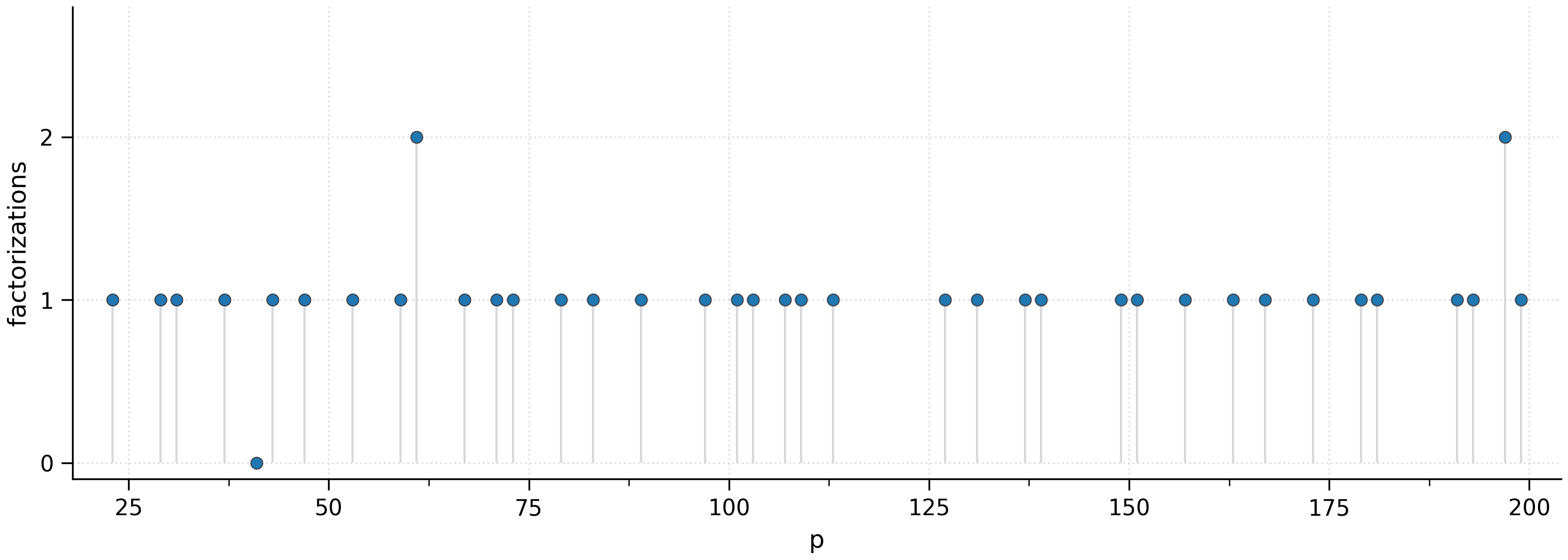}
    \caption{Direction $(z_1,z_2)=(z,z)$}
    \label{fig:factorizations-17-z-z}
  \end{subfigure}

  \begin{subfigure}[b]{\textwidth}
    \centering
    \includegraphics[width=\textwidth]{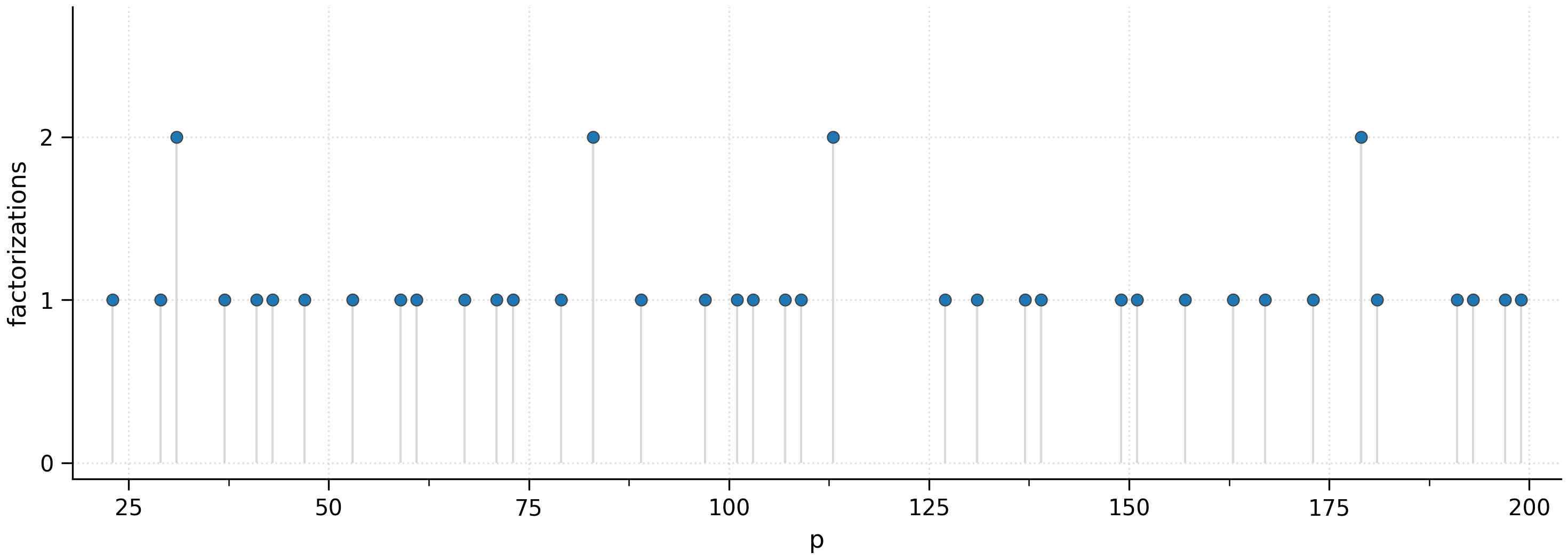}
    \caption{Direction $(z_1,z_2)=(2z,z)$}
    \label{fig:factorizations-17-2z-z}
  \end{subfigure}

  \begin{subfigure}[b]{\textwidth}
    \centering
    \includegraphics[width=\textwidth]{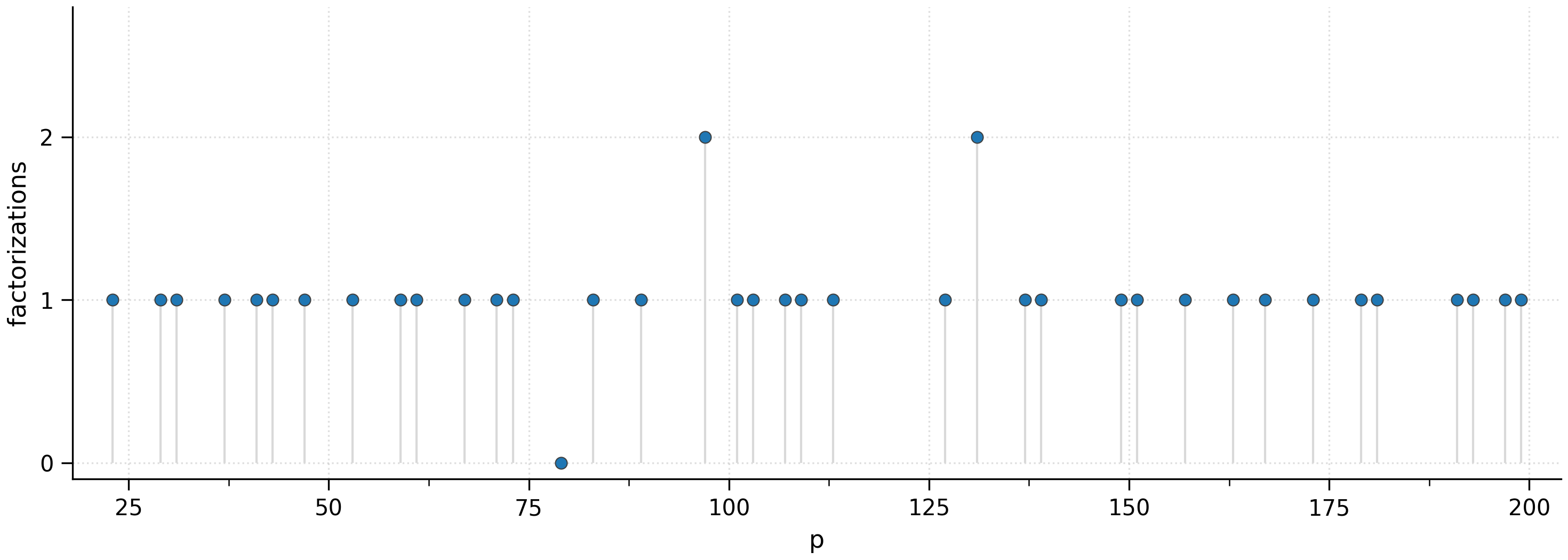}
    \caption{Direction $(z_1,z_2)=(3z,z)$}
    \label{fig:factorizations-17-3z-z}
  \end{subfigure}

  \caption{Factorizations for the model $\text{Bl}^{(1)}(X_{52111})$ in various directions $(z_1,z_2)=(nz,z)$. There are persistent factorizations in each direction, which come from the reductions of $z_1=-1/2$.}
  \label{fig:factorizations-17}
\end{figure}

The factorization data is shown in \Cref{fig:factorizations-17-z-z,fig:factorizations-17-2z-z,fig:factorizations-17-3z-z}. The persistent factorizations again stem from a linear equation, which we determine to be $1+2z_1=0$. The one-parameter model along this locus is the hypergeometric operator corresponding to $X_{41111}$. We compute fluxes leading to the vacuum configuration via analytic continuation of the periods. 
We then find that the fluxes can be chosen as
\begin{align}
    &f^{\text{T}}=(0,-1, 4, 0, 0,  -1)\,,\\
    &h^{\text{T}}=(0, 0, 0, 0, 2, -1)\,.
\end{align}
The derivatives transversal to the vacuum locus, 
\begin{align}
    \left(f^{\text{T}}\Sigma \partial_{z_{1}}\underline{\Pi},\,h^{\text{T}}\Sigma \partial_{z_{1}}\underline{\Pi}\right)\big\vert_{z_1=-\frac{1}{2}}\,,
\end{align}
are periods of the family of elliptic curves with Picard--Fuchs operator 
\begin{align}
    \mathcal{L}^{(2)}_{\text{el}}=\theta^2-4z \left(4 \theta+1\right) \left(4 \theta+3\right),
\end{align}
where $z=z_2/32$. This hypergeometric operator is referred to as C in the literature, cf.\ \Cref{tab:elliptic1}. The Riemann symbol is
\begin{align}
    \mathcal{P}_{\mathcal{L}^{(2)}_{\text{el}}}\left\{\begin{array}{ccc}  0 & \frac{1}{64} & \infty\\\hline
     0 & 0 & \frac{1}{4}\\
     0 & 0 & \frac{3}{4}
   \end{array},\ z\right\}.
\end{align}
The axio-dilaton is given by 
\begin{align}
    \tau(z_2)=\frac{f^{\text{T}}\Sigma \partial_{z_{1}}\underline{\Pi}}{h^{\text{T}}\Sigma \partial_{z_{1}}\underline{\Pi}}\bigg\vert_{z_1=-\frac{1}{2}}\,.
\end{align}
The corresponding $j$-invariant of this family of elliptic curves takes the rational form 
\begin{align}
    j(\tau(z_2))=\frac{(1-1536z_2)^3}{1024z_2^2(1-2048z_2)}\,.
\end{align}
We checked for several values of $z_2$ that the values of the $j$-invariants  agree with the ones of the elliptic curves associated to the quadratic factors of the Euler factors.  

Also for this model there is a doubling of the Euler factors and applying the same strategy as before one finds that the fibres are paired according to the involution 
\begin{align}
\label{eq:BlX10symmetry}
    \mathcal{I}:\; (z_1,z_2)\mapsto \left(-z_1 -1,-\frac{z_1^3z_2}{(z_1+1)^3}\right).
\end{align}
The MUM point at $z_1=-1,z_2=0$ has the same local periods and associated A-model geometry as the MUM point $z_1=z_2=0$ and the symmetry of the Euler factors originates from the integral structure having this symmetry (the analytic continuation of the integral basis from the origin differs from the integral basis at $z_1=-1,z_2=0$ by an integral symplectic transformation). We see that in this case the vacuum $z_1=-\frac{1}{2}$ \emph{does} coincide with the fixed point locus. The vacuum of this model is thus a consequence of the splitting of the middle cohomology into eigenspaces of the $\mathbb{Z}_2$-symmetry $\mathcal{I}$ happening over the rationals. 

Since this model has an involution symmetry and is also a blowup of a one-parameter model, one might have expected, analogous to the model $\text{Bl}^{(4)}(X_{21111})$, to find a supersymmetric vacuum with Hodge substructure given by $X_{52111}$ at the locus $z_1=-1$, which maps under $\mathcal{I}$ to $z_1=0,z_2=\infty$. However, while \eqref{eq:X10blowuplvectors} tells us that the blowup deformation corresponds also here to $a_2=0$, in the Batyrev coordinates,
\begin{align}
    z_1=\frac{1}{a_0a_2}\,,\quad z_2=\frac{a_2^5}{a_0^5}\,,
\end{align}
the locus where the one-parameter model is recovered is $z_1=\infty, z_2=0$ (for finite $a_0$) with the moduli space coordinate of $X_{52111}$ given by $1/a_0^{10}=z_1^5z_2$. Equation \eqref{eq:BlX10symmetry} shows that this is not mapped under $\mathcal{I}$ to the interior of the moduli space, explaining the absence of a vacuum arising as the $\mathcal{I}$-image of the one-parameter model.

\subsubsection{Blowups of $h^{2,1}=1$ models}
\label{subsubsec:ToricBlowups}
Motivated by the example $\text{Bl}^{(4)}(X_{21111})$ above, where the vacuum arises as the image under a $\mathbb{Z}_2$-symmetry of a locus in the moduli space that corresponds to blowing down to a one-parameter model, we want to study whether one may generalise this mechanism to other toric blowups of hypergeometric $h^{2,1}=1$ models. For this we constructed for all hypergeometric one-parameter models toric blowups. Since the blowups of the three hypergeometric hypersurface models would already appear in \Cref{tab:sixverticesresults}, we can focus on the complete intersection cases. The blowups are not unique and we picked just one choice for a positive linear combination of the vertices to add to the polyhedron describing the ambient space of each model. An overview of the models is shown in \Cref{tab:3foldblowupsdata}. The convention for the vertices of the ambient weighted projective space  $\P_{w_1\ldots w_{n+r+1}}$ (with $w_{n+r+1}=1$) of a hypergeometric one-parameter model $X^{k_1,\ldots,k_r}$ is 
\begin{equation}
\begin{split}
\begin{gathered}
    \nu_1=(1,0,0,\dots)\,,\qquad   \nu_2=(0,1,0,\dots)\,,\qquad  \dots\,,\qquad  \nu_{n+r}=(0,\dots,0,1)\,,\qquad\\
    \nu_{n+r+1}=(-w_1,-w_2,\dots,-w_{n+r})\,.
\end{gathered}
\end{split}
\end{equation}
The table lists the extra vertex $\nu_{n+r+2}$ that is added to define $\text{Bl}(X^{k_1,\ldots,k_r})$ as well as the NEF partition specifying the complete intersection and the topological data of the resulting model. The last column again shows the average number of factorizations along one direction for fifty primes $p=41,\dots,293$.

\begin{table}[H]
\centering
\renewcommand{\arraystretch}{1.8}
\resizebox{\textwidth}{!}{%
\begin{tabular}{|c|c|c|c|c|c|c|}
\hline
model & blowup vertex & NEF partition & $\chi$  &$l$-vectors & intersection ring & $\overline{N}_{\text{fact}}$ \\
\hline \hline 
$X^{2,2,2,2}\subset \mathbb{P}^7$ & $\nu_{9}=(-1,1,0,0,0,0,0)$ & $\{\nu_1,\nu_2,\nu_{9}\}\cup\{\nu_3,\nu_4\}\cup\{\nu_5,\nu_6\}\cup\{\nu_7,\nu_8\}$&$-112$ & $(-1, 0, 0, 0;1, -1, 0, 0, 0, 0, 0, 0, 1),\,(-1, -2, -2, -2;0, 2, 1, 1, 1, 1, 1, 1, -1)$& $16J_1^3+16J_1^2J_2+16J_1J_2^2+12J_2^3$ & $72\%$\\ \hline
$X^{3,3}\subset \mathbb{P}^5$ & $\nu_7=(1,1,1,0,0)$ & $\{\nu_1,\nu_2,\nu_3,\nu_7\} \cup \{\nu_4,\nu_5,\nu_6\}$ &  $-108$ &$(-1, -3;0, 0, 0, 1, 1, 1, 1),\,(-2, 0;1, 1, 1, 0, 0, 0, -1)$ & $9J_1^3+9J_1^2J_2+3J_1J_2^2$ & $20\%$\\ \hline 
$X^{3,2,2}\subset \mathbb{P}^6$ & $\nu_8=(0,1,1,1,0,0)$ & $\{\nu_1,\nu_2,\nu_3\}\cup\{\nu_4,\nu_5,\nu_8\}\cup\{\nu_6,\nu_7\}$& $-96$ & $(-1, -2, -2;1, 0, 0, 0, 1, 1, 1, 1),\,(-2, 0, 0;0, 1, 1, 1, 0, 0, 0, -1)$& $12J_1^3+12J_1^2J_2+4J_1J_2^2$ &$48\%$ \\ \hline
$X^{4,2}\subset \mathbb{P}^5$ & $\nu_7=(1,1,1,0,0)$ & $\{\nu_1,\nu_2,\nu_3,\nu_4,\nu_7\}\cup\{\nu_5,\nu_6\}$ &$-132$  & $(-2, -2;0, 0, 0, 1, 1, 1, 1),\,(-2, 0;1, 1, 1, 0, 0, 0, -1)$& $8J_1^3+8J_1^2J_2+4J_1J_2^2$ & $26\%$\\  \hline
$X^{12,2}\subset \mathbb{P}_{641111}$ & $\nu_7=(3,2,1,1,1)$ & $\{\nu_1,\nu_2,\nu_3,\nu_4,\nu_7\}\cup\{\nu_5,\nu_6\}$&$-424$ & $(-6, -1;3, 2, 0, 0, 0, 1, 1),\,(0, 0;0, 0, 1, 1, 1, -1, -2)$ & $7J_1^3+3J_1^2J_2+J_1J_2^2$ & $0\%$ \\ 
\hline
$X^{4,4}\subset \mathbb{P}_{221111}$ & $\nu_7=(1,1,1,1,1)$ & $\{\nu_1,\nu_2\}\cup\{\nu_3,\nu_4,\nu_5,\nu_6,\nu_7\}$& $-128$ & $(-2, -2;1, 1, 0, 0, 0, 1, 1),\,(0, 0;0, 0, 1, 1, 1, -1, -2)$& $28J_1^3+12J_1^2J_2+4J_1J_2^2$ & $22\%$ \\ \hline
$X^{6,4}\subset \mathbb{P}_{322111}$ & $\nu_7=(-1,-1,0,0,0)$ & $\{\nu_1,\nu_4,\nu_5,\nu_6,\nu_7\}\cup\{\nu_2,\nu_3\}$ & $-152$&  $(-2, -1;1, 1, 0, 0, 0, 0, 1),(0, -1;0, -1, 2, 1, 1, 1, -3)$& $53J_1^3+17J_1^2J_2+5J_1J_2^2+J_2^3$ & $0\%$\\ \hline
$X^{4,3}\subset \mathbb{P}_{211111}$ & $\nu_7=(-1,-1,0,0,0)$& $\{\nu_1,\nu_2,\nu_3\}\cup\{\nu_4,\nu_5,\nu_6,\nu_7\}$& $-140$& $(-2, -1;1, 1, 0, 0, 0, 0, 1),\,(0, -1;0, -1, 1, 1, 1, 1, -2)$ & $44J_1^3+20J_1^2J_2+8J_1J_2^2+2J_2^3$ & $14\%$ \\ \hline
$X^{6,6}\subset \mathbb{P}_{332211}$ & $\nu_7=(1,1,1,0,0)$ & $\{\nu_1,\nu_2\}\cup\{\nu_3,\nu_4,\nu_5,\nu_6,\nu_7\}$& $-84$& $(-2, 0;1, 1, 1, 0, 0, 0, -1),\,(0, -6;0, 0, -1, 2, 1, 1, 3)$ & $3J_1^3+3J_1^2J_2+3J_1J_2^2+J_2^3$ & $100\%$ \\ \hline
$X^{6,2}\subset \mathbb{P}_{311111}$ & $\nu_7=(1,1,0,0,1)$ &$\{\nu_1,\nu_2,\nu_3,\nu_4\}\cup\{\nu_5,\nu_6,\nu_7\}$ & $-196$ & $(-2, 0;1, 1, 0, 0, 1, 0, -1),\,(0, -2;0, -2, 1, 1, -2, 1, 3)$& $80J_1^3+32J_1^2J_2+12J_1J_2^2+4J_2^3$ & $208\%$ \\ \hline
\end{tabular}
}
\caption{Data for the blowups of the hypergeometric $h^{2,1}=1$ complete intersections. We list the vertex added to the toric polyhedron of the ambient weighted projective space and the chosen NEF partition of the blowup model. We further list topological data and the $l$-vectors. The last column lists the average number of factorizations of the local zeta function numerator $P^{(p)}_3(X,T)$ for the direction $(z_1,z_2)=(2z,z)$ over $50$ primes $p=41,\dots,293$.}
\label{tab:3foldblowupsdata}
\end{table}
Two of the above models feature vacuum loci, namely the models $\text{Bl}(X^{6,6})$ and $\text{Bl}(X^{6,2})$, which we discuss in the following.
Notably, we do not observe for either of these models evidence for a symmetry acting on the moduli space, as almost all Euler factor are pairwise different.
In particular, the vacua seem to arise neither as a fixed point locus of a $\mathbb{Z}_2$-symmetry nor via the blowup mechanism found in the model $\text{Bl}^{(4)}(X_{21111})$.
The case $\text{Bl}(X^{6,2})$ stands out in that the $(1,1,1,1)$ Hodge substructure does not have a MUM point, i.e. the associated Calabi--Yau operator is a so-called orphan operator. 

\begin{figure}[H]
  \centering
  \begin{subfigure}[b]{\textwidth}
    \includegraphics[width=\textwidth]{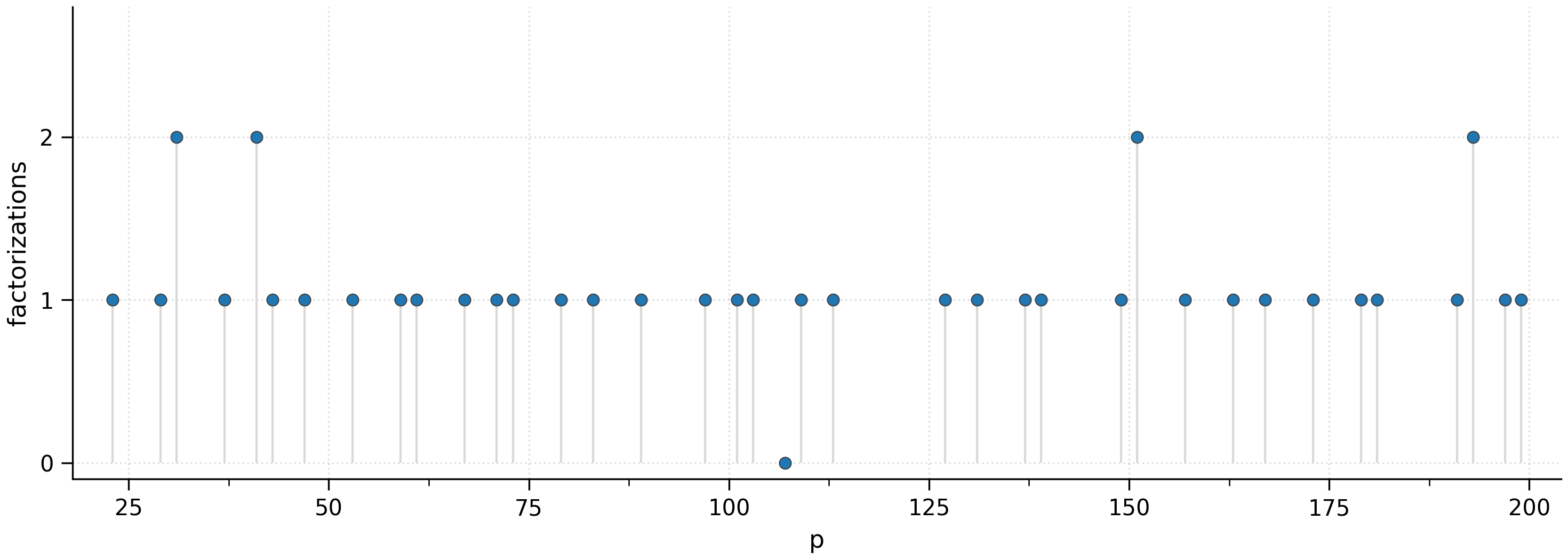}
    \caption{Direction $(z_1,z_2)=(z,z)$}
    \label{fig:factorizations-BlX66-z-z}
  \end{subfigure}
  \hfill
  \begin{subfigure}[b]{\textwidth}
    \includegraphics[width=\textwidth]{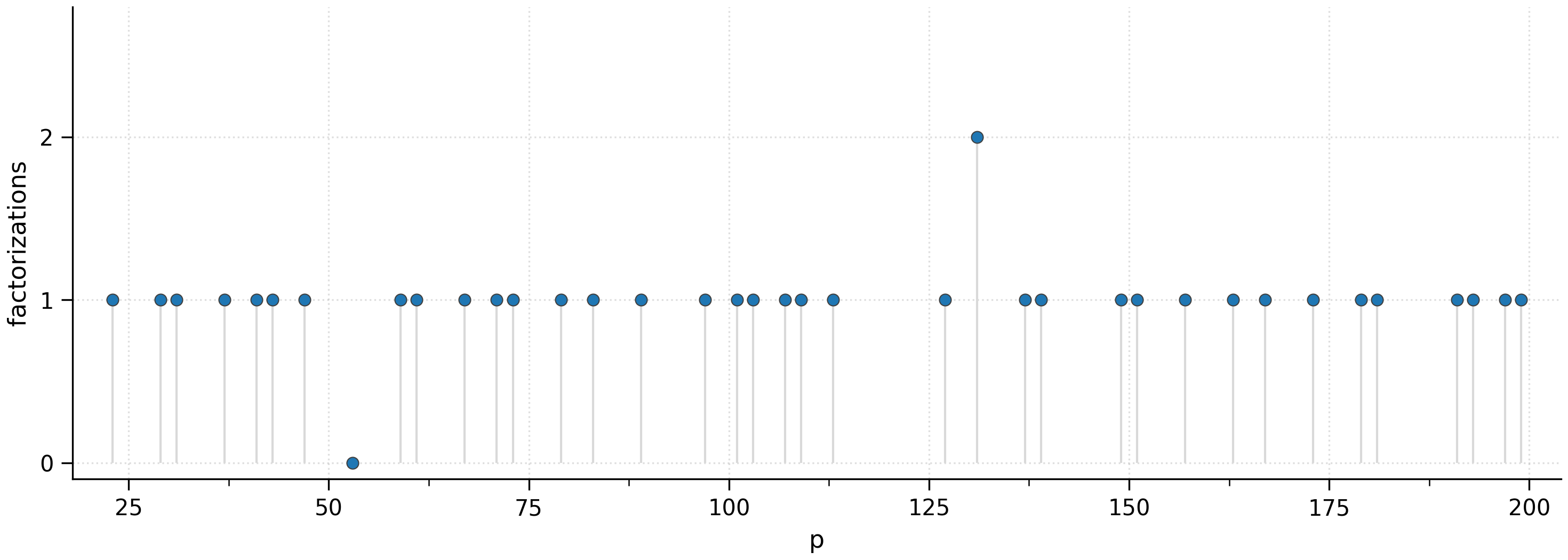}
    \caption{Direction $(z_1,z_2)=(2z,z)$}
    \label{fig:factorizations-BlX66-2z-z}
  \end{subfigure}
  \hfill
  \begin{subfigure}[b]{\textwidth}
    \includegraphics[width=\textwidth]{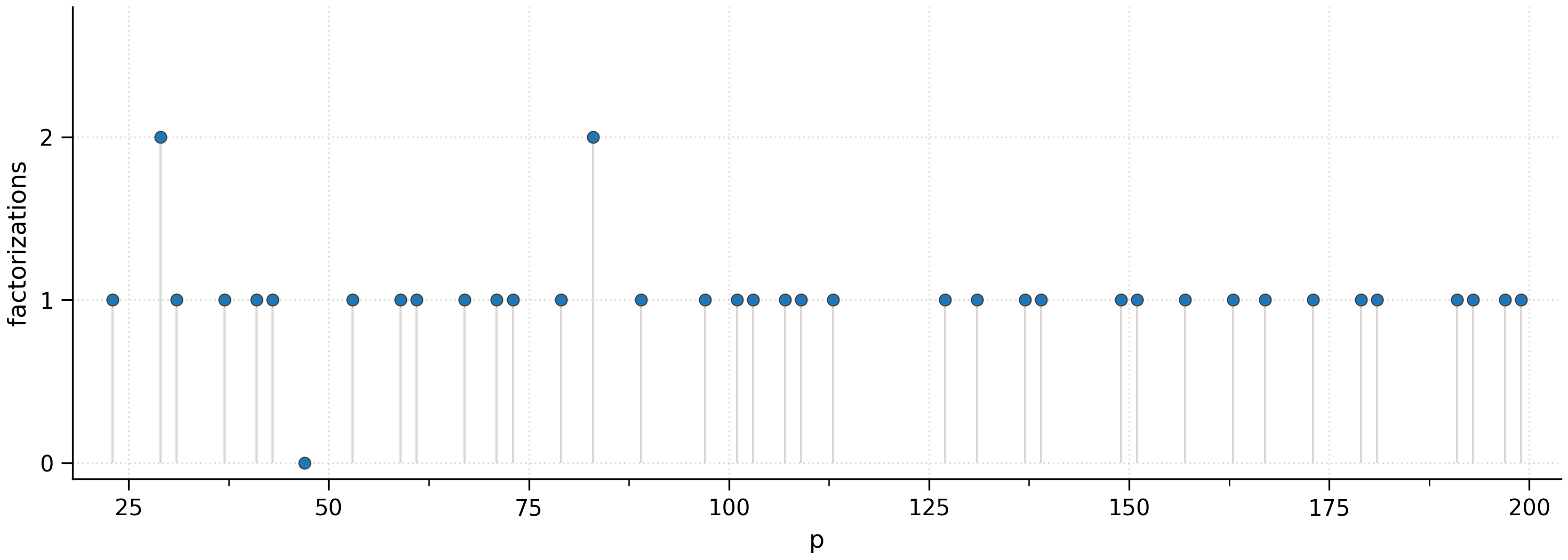}
    \caption{Direction $(z_1,z_2)=(3z,z)$}
    \label{fig:factorizations-BlX66-3z-z}
  \end{subfigure}
  \caption{Factorizations for the model $\text{Bl}(X_{6,6})$ in various directions $(z_1,z_2)=(n z,z)$. There are persistent factorizations in each direction, which come from the reductions of $z_1=-1/3$.}
  \label{fig:factorizations-BlX66}
\end{figure}

\paragraph{$\text{Bl}(X^{6,6})$}
The Picard--Fuchs ideal and the discriminant of this model are given in \Cref{tab:BlX66} of the appendix \ref{app:ttdata}.
The factorizations for this model for different directions are shown in \Cref{fig:factorizations-BlX66-z-z,fig:factorizations-BlX66-2z-z,fig:factorizations-BlX66-3z-z}. For almost all primes there is at least one factorizations. We find that these come from the good reductions of $z_1=-\frac{1}{3}$. We again confirm this via analytic continuation of the periods to the point $z_1=-\frac{1}{3}, z_2=0$. 
We find fluxes 
\begin{align}
    &f^{\text{T}}=(0,-1,3,1,8,-4)\,,\\
    &h^{\text{T}}=(0,0,0,0,-6,2)\,.
\end{align}
The $(1,1,1,1)$ Hodge structure on this line is that of $X_{21111}$ with the coordinate $\tilde{z}$ of the corresponding operator \href{https://cycluster.mpim-bonn.mpg.de/operator.html?nn=4.1.8}{AESZ 8} related via $z_2 = -27\tilde{z}$.
The derivatives transversal to the vacuum locus, 
\begin{align}
    \left(f^{\text{T}}\Sigma \partial_{z_{1}}\underline{\Pi},\,h^{\text{T}}\Sigma \partial_{z_{1}}\underline{\Pi}\right)\big\vert_{z_1=-\frac{1}{3}}
\end{align}
are periods of the family of elliptic curves with Picard--Fuchs operator 
\begin{align}
    \mathcal{L}^{(2)}_{\text{el}}=\theta^2-12z \left(6 \theta+1\right) \left(6 \theta+5\right)
\end{align}
where $z=-z_2$. This is the hypergeometric elliptic operator D, cf.\ \Cref{tab:elliptic1}. The Riemann symbol is
\begin{align}
    \mathcal{P}_{\mathcal{L}^{(2)}_{\text{el}}}\left\{\begin{array}{ccc}  0 & \frac{1}{432} & \infty\\\hline
     0 & 0 & \frac{1}{6}\\
     0 & 0 & \frac{5}{6}
   \end{array},\ z\right\}.
\end{align}
The $j$-invariant of the axio-dilaton profile,
\begin{align}
    \tau(z_2)=\frac{f^{\text{T}}\Sigma \partial_{z_{1}}\underline{\Pi}}{h^{\text{T}}\Sigma \partial_{z_{1}}\underline{\Pi}}\bigg\vert_{z_1=-\frac{1}{3}}\,,
\end{align}
is given by 
\begin{align}
    j(\tau(z_2))=\frac{1}{z_2(1-432z_2)}\,.
\end{align}

\begin{figure}[H]
  \centering
  \begin{subfigure}[b]{\textwidth}
    \includegraphics[width=\textwidth]{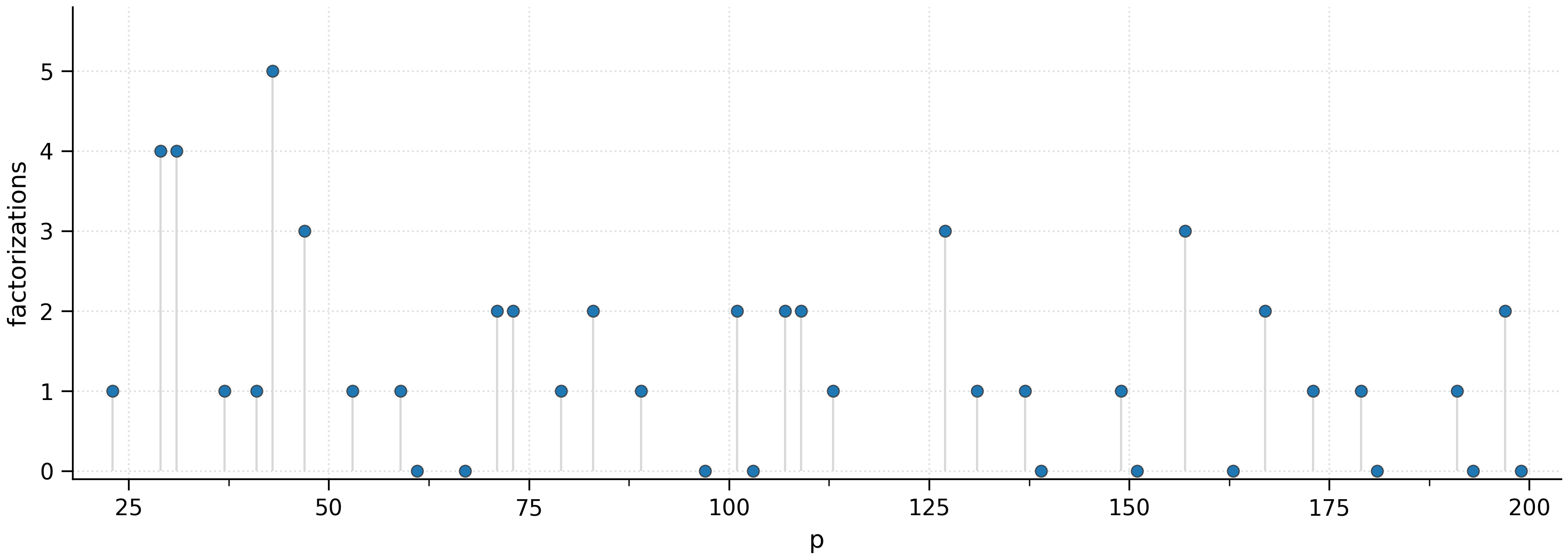}
    \caption{Direction $(z_1,z_2)=(z,z)$}
    \label{fig:factorizations-BlX62-z-z}
  \end{subfigure}
  \hfill
  \begin{subfigure}[b]{\textwidth}
    \includegraphics[width=\textwidth]{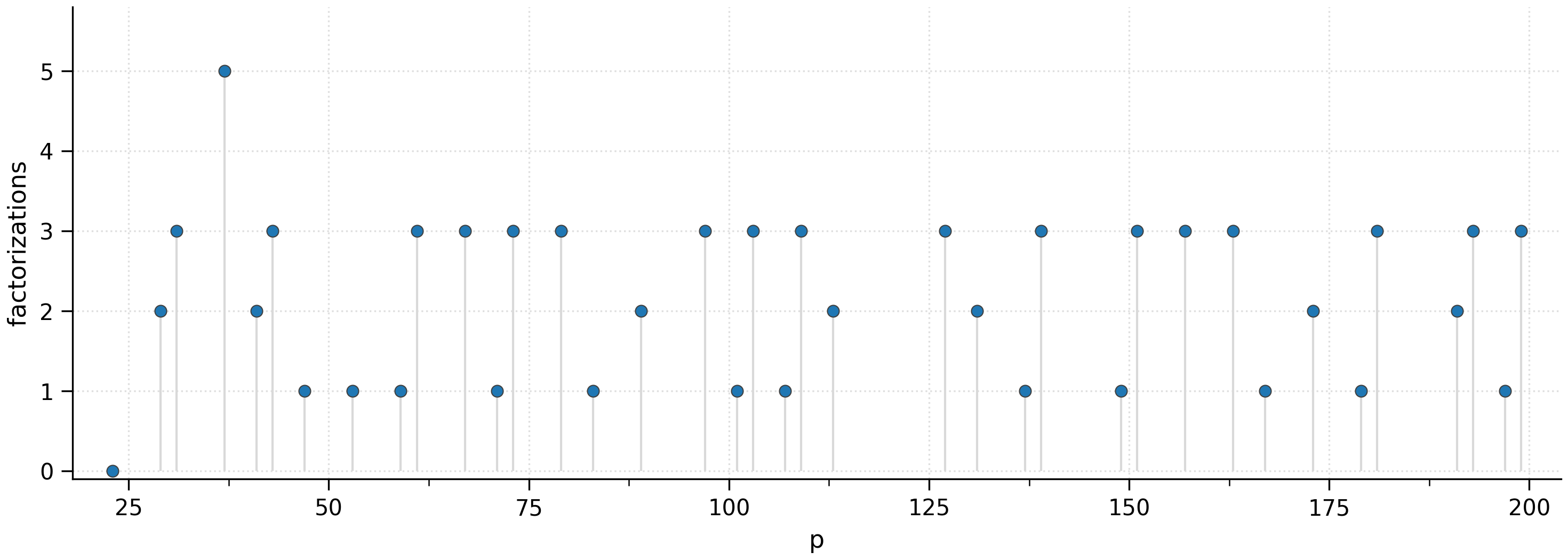}
    \caption{Direction $(z_1,z_2)=(2z,z)$}
    \label{fig:factorizations-BlX62-2z-z}
  \end{subfigure}
  \hfill
  \begin{subfigure}[b]{\textwidth}
    \includegraphics[width=\textwidth]{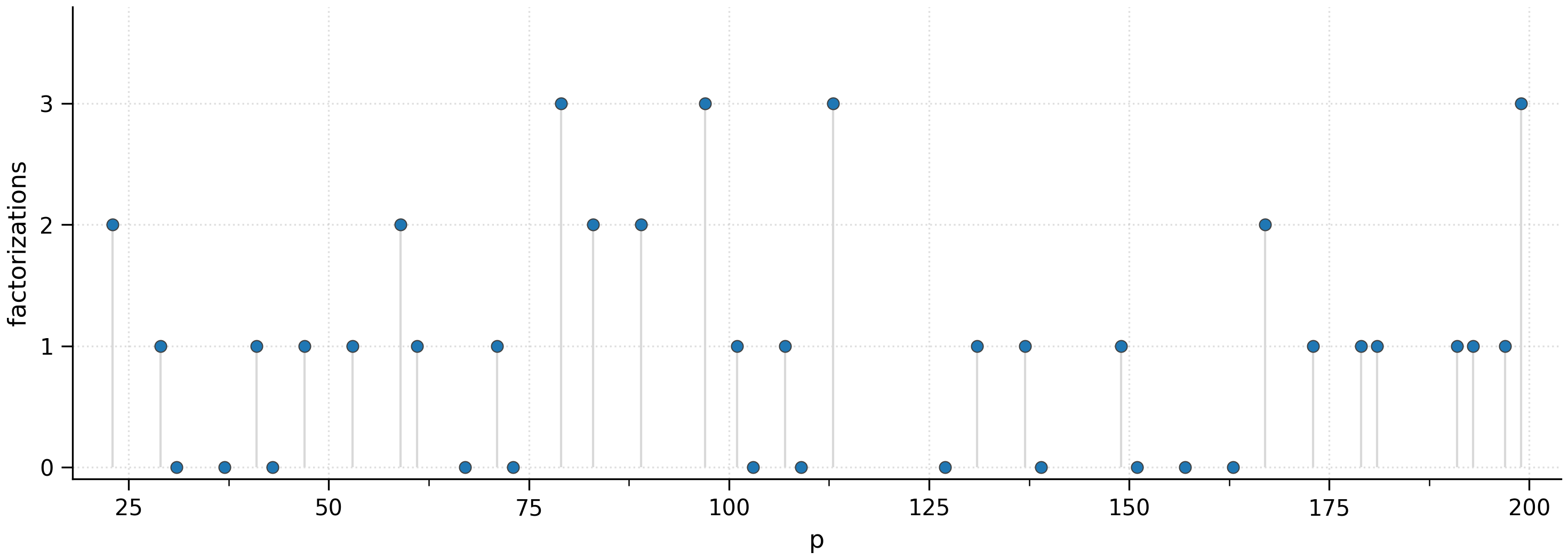}
    \caption{Direction $(z_1,z_2)=(3z,z)$}
    \label{fig:factorizations-BlX62-3z-z}
  \end{subfigure}
  \caption{Factorizations for the model $\text{Bl}(X^{6,2})$ in various directions $(z_1,z_2)=(nz,z)$. For all directions the average number of factorizations is above one. In the direction $(z_1,z_2)=(2z,z)$ the average is above two and there is for large enough primes always at least one factorization. We find the persistent factorizations to come from good reductions of $1+1024z_1^2z_2=0$, which factorizes into a linear and a quadratic factor along $(z_1,z_2)=(2z,z)$.}
  \label{fig:factorizations-BlX62}
\end{figure}

\paragraph{$\text{Bl}(X^{6,2})$}
The Picard-Fuchs ideal and the discriminant of this model are given in \Cref{tab:BlX62} of the appendix \ref{app:ttdata}. The factorization plots are shown in \Cref{fig:factorizations-BlX62-z-z,fig:factorizations-BlX62-2z-z,fig:factorizations-BlX62-3z-z}. We observe for directions $z_1=z_2$ and $z_1=3z_2$ that while the occurence of primes with zero factorizations persists also to higher and higher primes, the average number of factorizations does not go below one. This is the behaviour we find for a generic direction in this model. For the direction $z_1=2z_2$ for large enough primes one has at least one factorization for every prime and the average stays above two. The frequency of primes for which there are at least three factorizations along a generic direction suggests a cubic equation for the vacuum. Indeed we find the factorizations to come from the reductions of the locus $1+1024z_1^2z_2=0$. The behaviour along the direction $z_1=2z_2$ comes from the fact that the equation factorizes then into two irreducible factors, one linear and one quadratic factor, explaining the observed factorization pattern.

In order to compute the fluxes it is convenient instead of analytically continuing from the MUM point at $z_1=z_2=0$ to start from the integral basis at the MUM point at $z_1=0,z_2=\infty$ in the coordinates\footnote{Note that one can of course also directly work with the $\underline{u}$ coordinates and then find again a linear equation for the vacuum locus.}  $u_1=z_1^2z_2,u_2=\frac{1}{z_1z_2}$. 
In this way we find the fluxes 
\begin{align}
    &f^{\text{T}}=(2,1,0,4,6,-8)\,,\\
    &h^{\text{T}}=(0,0,-1,0,-2,4)\,.
\end{align}
The surviving periods on the vacuum locus are those of the operator 
\begin{align}
\label{eq:orphan}
    \mathcal{L}^{(4)}_{\text{orphan}}(z)=97200 \left(\theta-1\right)^2 \theta^2-144z \left(52 \theta^2+3\right) \theta^2 +z^2\left(2 \theta+1\right)^2 \left(6 \theta+1\right) \left(6 \theta+5\right),
\end{align}
where we introduced $z=-108z_1$. Its Riemann symbol reads
\begin{align}
    \mathcal{P}_{\mathcal{L}^{(4)}_{\text{orphan}}}\left\{\begin{array}{cccc}  0 & 25 & 27 &  \infty\\\hline
     0 & 0 & 0 & \frac{1}{6}\\
     0 & 1 & 1 & \frac{1}{2}\\
     1 & 1 & 1 & \frac{1}{2}\\
     1 & 2 & 2 & \frac{5}{6}
   \end{array},\ z\right\}.
\end{align}
Due to the absence of a MUM point such operators are called \emph{orphans} in the literature \cite{2017arXiv170909752C}. Our case has one K-point and three conifold points. The transversal derivatives, 
\begin{align}
    \left(f^{\text{T}}\Sigma \partial_{u_1}\underline{\Pi},\,h^{\text{T}}\Sigma \partial_{u_1}\underline{\Pi}\right)\big\vert_{u_1=-\frac{1}{1024}}\,,
\end{align}
are periods of the Legendre curve, whose Picard--Fuchs operator 
\begin{align}
    \mathcal{L}^{(2)}_{\text{el}}=\theta^2-4z \left(2\theta+1\right)^2,
\end{align}
where $z=-\frac{z_1}{4}$, is also known as the elliptic operator A, cf.\ \Cref{tab:elliptic1}. The Riemann symbol is
\begin{align}
    \mathcal{P}_{\mathcal{L}^{(2)}_{\text{el}}}\left\{\begin{array}{ccc}  0 & \frac{1}{16} & \infty\\\hline
     0 & 0 & \frac{1}{2}\\
     0 & 0 & \frac{1}{2}
   \end{array},\ z\right\}.
\end{align}
The $j$-invariant of the axio-dilaton profile,
\begin{align}
    \tau(z_1)=\frac{f^{\text{T}}\Sigma \partial_{u_1}\underline{\Pi}}{h^{\text{T}}\Sigma \partial_{u_1}\underline{\Pi}}\bigg\vert_{u_1=-\frac{1}{1024}}\,,
\end{align}
is given by 
\begin{align}
\label{eq:BlX62jinv}
    j(\tau(z_1))=\frac{16(1+ 4z_1+16z_1^2)^3}{z_1^2(1 + 4z_1)^2}\,.
\end{align}
The Euler factors at the fibre $(z_1,z_2)=(-\frac{1}{108},-\frac{729}{64})$ are given in \Cref{tab:orphaneulerfactors}. The quartic factors are then those of the orphan at $z=1$ and the quadratic factor corresponds to the weight two modular form \href{https://www.lmfdb.org/ModularForm/GL2/Q/holomorphic/3744/2/a/o/}{3744.2.a.o}. The isogeny class corresponding to the latter contains the rational model \href{https://www.lmfdb.org/EllipticCurve/Q/3744/o/3}{3744.o3}, which has $j$-invariant $22235451328/123201$, in accordance with the value of \eqref{eq:BlX62jinv} at $z_1=-\frac{1}{108}$.
\newpage

\begin{table}[H]
\centering
\renewcommand{\arraystretch}{1.2}
\resizebox{0.5\textwidth}{!}{%
\begin{tabular}{|c|c|}
\hline
\multicolumn{2}{|c|}{$(z_1,z_2)=(-\frac{1}{108},-\frac{729}{64})$} \\
\hline
$p$ & $P^{(p)}_3(X,T)$ \\
\hline \hline
$29$ &  $\left(p^{3} T^{2} + 2 p T + 1\right) \, \left(p^{6} T^{4} + p^{3} T^{3} - 452 p T^{2} + T + 1\right)$ \\ 
\hline
$31$ &  $\left(p^{3} T^{2} + 8 p T + 1\right) \, \left(p^{6} T^{4} - 88 p^{3} T^{3} + 714 p T^{2} - 88 T + 1\right)$ \\ 
\hline
$37$ &  $\left(p^{3} T^{2} + 10 p T + 1\right) \, \left(p^{6} T^{4} - 46 p^{3} T^{3} + 1938 p T^{2} - 46 T + 1\right)$ \\ 
\hline
$41$ &  $\left(p^{3} T^{2} + 6 p T + 1\right) \, \left(p^{6} T^{4} - 510 p^{3} T^{3} + 3546 p T^{2} - 510 T + 1\right)$ \\ 
\hline
$43$ &  $\left(p^{3} T^{2} - 4 p T + 1\right) \, \left(p^{6} T^{4} + 306 p^{3} T^{3} + 2586 p T^{2} + 306 T + 1\right)$ \\ 
\hline
$47$ &  $\left(p^{3} T^{2} + 1\right) \, \left(p^{6} T^{4} + 524 p^{3} T^{3} + 3418 p T^{2} + 524 T + 1\right)$ \\ 
\hline
$53$ &  $\left(p^{3} T^{2} - 14 p T + 1\right) \, \left(p^{6} T^{4} + 595 p^{3} T^{3} + 4480 p T^{2} + 595 T + 1\right)$ \\ 
\hline
$59$ &  $\left(p^{3} T^{2} + 12 p T + 1\right) \, \left(p^{6} T^{4} - 335 p^{3} T^{3} - 394 p T^{2} - 335 T + 1\right)$ \\ 
\hline
$61$ &  $\left(p^{3} T^{2} + 10 p T + 1\right) \, \left(p^{6} T^{4} - 155 p^{3} T^{3} + 3752 p T^{2} - 155 T + 1\right)$ \\ 
\hline
$67$ &  $\left(p^{3} T^{2} - 8 p T + 1\right) \, \left(p^{6} T^{4} - 396 p^{3} T^{3} + 8378 p T^{2} - 396 T + 1\right)$ \\ 
\hline
$71$ &  $\left(p^{3} T^{2} + 1\right) \, \left(p^{6} T^{4} + 191 p^{3} T^{3} + 3202 p T^{2} + 191 T + 1\right)$ \\ 
\hline
$73$ &  $\left(p^{3} T^{2} + 14 p T + 1\right) \, \left(p^{6} T^{4} + 130 p^{3} T^{3} - 2070 p T^{2} + 130 T + 1\right)$ \\ 
\hline
$79$ &  $\left(p^{3} T^{2} - 4 p T + 1\right) \, \left(p^{6} T^{4} + 194 p^{3} T^{3} + 7982 p T^{2} + 194 T + 1\right)$ \\ 
\hline
$83$ &  $\left(p^{3} T^{2} - 4 p T + 1\right) \, \left(p^{6} T^{4} + 461 p^{3} T^{3} + 4646 p T^{2} + 461 T + 1\right)$ \\ 
\hline
$89$ &  $\left(p^{3} T^{2} - 10 p T + 1\right) \, \left(p^{6} T^{4} - 1012 p^{3} T^{3} + 6862 p T^{2} - 1012 T + 1\right)$ \\ 
\hline
$97$ &  $\left(p^{3} T^{2} - 10 p T + 1\right) \, \left(p^{6} T^{4} - 494 p^{3} T^{3} + 4618 p T^{2} - 494 T + 1\right)$ \\ 
\hline
$101$ &  $\left(p^{3} T^{2} + 10 p T + 1\right) \, \left(p^{6} T^{4} - 719 p^{3} T^{3} - 2268 p T^{2} - 719 T + 1\right)$ \\ 
\hline
$103$ &  $\left(p^{3} T^{2} - 4 p T + 1\right) \, \left(p^{6} T^{4} - 1462 p^{3} T^{3} + 13858 p T^{2} - 1462 T + 1\right)$ \\ 
\hline
$107$ &  $\left(p^{3} T^{2} - 12 p T + 1\right) \, \left(p^{6} T^{4} + 566 p^{3} T^{3} + 12998 p T^{2} + 566 T + 1\right)$ \\ 
\hline
$109$ &  $\left(p^{3} T^{2} - 14 p T + 1\right) \, \left(p^{6} T^{4} + 46 p^{3} T^{3} + 18434 p T^{2} + 46 T + 1\right)$ \\ 
\hline
$113$ &  $\left(p^{3} T^{2} - 6 p T + 1\right) \, \left(p^{6} T^{4} - 1521 p^{3} T^{3} + 10076 p T^{2} - 1521 T + 1\right)$ \\ 
\hline
$127$ &  $\left(p^{3} T^{2} - 4 p T + 1\right) \, \left(p^{6} T^{4} - 862 p^{3} T^{3} + 15554 p T^{2} - 862 T + 1\right)$ \\ 
\hline
$131$ &  $\left(p^{3} T^{2} + 12 p T + 1\right) \, \left(p^{6} T^{4} - 2274 p^{3} T^{3} + 29530 p T^{2} - 2274 T + 1\right)$ \\ 
\hline
$137$ &  $\left(p^{3} T^{2} - 10 p T + 1\right) \, \left(p^{6} T^{4} - 240 p^{3} T^{3} - 12322 p T^{2} - 240 T + 1\right)$ \\ 
\hline
$139$ &  $\left(p^{3} T^{2} - 4 p T + 1\right) \, \left(p^{6} T^{4} + 1556 p^{3} T^{3} + 38174 p T^{2} + 1556 T + 1\right)$ \\ 
\hline
$149$ &  $\left(p^{3} T^{2} + 6 p T + 1\right) \, \left(p^{6} T^{4} - 1752 p^{3} T^{3} - 274 p T^{2} - 1752 T + 1\right)$ \\ 
\hline
\end{tabular}
}
\caption{Euler factors of the model $\text{Bl}(X^{6,2})$ at $(z_1,z_2)=(-\frac{1}{108},-\frac{729}{64})$. The quadratic factors correspond to the weight two modular form \href{https://www.lmfdb.org/ModularForm/GL2/Q/holomorphic/3744/2/a/o/}{3744.2.a.o} and the quartic factors are the Euler factors of $\mathcal{L}^{(4)}_\text{orphan}(z)$, given in \eqref{eq:orphan}, at $z=1$.}
\label{tab:orphaneulerfactors}
\end{table}

\section{Fourfolds}\label{sec:fourfolds}
Supersymmetric compactifications of M- or F-theory to three or four dimensions have Calabi--Yau fourfolds as internal spaces. 
Similarly to the compactification of type II theories on threefolds, we discuss special singular loci at suitable values of parameters originating from faces of the toric polytope.
\paragraph{}
The embeddings of one-parameter Hadamard products arise similarly as in K3- and genus-one-fibred threefolds.
The classification of bases in \cref{sec:CY3trans} can be combined with higher-dimensional fibres. 
We will exemplify the generalisation together with computations of the local zeta function in \cref{sec:CY4fibr12}. Since the discriminant components arise by tuning the deformations associated with interior points of edges and two-faces, respectively, we will name the corresponding singularity types accordingly. 
Genus-one fibred fourfolds specialise to Hadamard products of the fibre and a third-order Picard--Fuchs operator originating at suitable values for the deformations in the base face. 
A $G_4$ flux in the cohomology class dual to the $S^4$ shrinking at the degeneration of the base yields a supersymmetric scalar potential whose F-terms are satisfied along a one-parameter locus in the complex-structure moduli space.
This mechanism is discussed in \cref{sec:CY4fibr3}.

\subsection{Special loci in threefold- and K3-fibrations}\label{sec:CY4fibr12}
Since the special singular loci we investigate are dependent solely on the base of a fibration, the discussion of one- and two-dimensional bases with K3 and genus-one fibres follows that for threefolds in \cref{sec:CY3trans}.

\subsubsection{CY3-fibration}
In analogy to the threefold $X_{22211}$ given by the quartic K3-fibration over $\P^1$, we begin with the quintic CY3-fibration over the same $\P^1$-base.
This hypersurface family $X_{222211}$ has a horizontal middle cohomology of type $(1,2,2,2,1)$.
The situation is completely analogous as for the threefold hypersurface $X_{22211}$ discussed earlier, however, we find the example useful nonetheless to further illustrate the methods.
The toric and topological data are collected in \Cref{tab:222211}.
The polytope contains an edge with the inner point
    \begin{equation}
    \nu_7 = (-1, -1, -1, -1, 0) = \frac{\nu_5+\nu_6}{2}\,
    \end{equation}
and the discriminant reads
\begin{align}
    \Delta=\left(1-6250z_1+9765625z_1^2-39062500z_1^2z_2\right)\left(1-4z_2\right)^3\,.
\end{align}
The first factor of $\Delta$ gives a conifold locus whereas the second component is associated to the edge, which in the threefold case would correspond to a strong coupling locus.
As expected of the general pattern, we consider a restriction to the latter component, obtained by setting $a_7=2$.
In this limit, the periods are annihilated by the Picard--Fuchs operator of the complete intersection of bi-degree $(5,2)$ in $\mathbb{P}^6$:
\begin{gather}
     \mathcal{L}^{(5)}_{5,2}(z)=\theta ^5-10 z(2 \theta +1) (5 \theta +1) (5 \theta +2) (5 \theta +3) (5 \theta +4)\,,    \\
    \renewcommand{\arraystretch}{1.3}
        \mathcal{P}_{\mathcal{L}^{(5)}_{5,2}}\left\{\begin{array}{ccc}   0 & 1/12500 &   \infty\\\hline
         0 & 0 & \frac{1}{5}\\
         0 & 1 & \frac{2}{5}\\
         0 & \frac{3}{2} & \frac{1}{2}\\
         0 & 2 & \frac{3}{5}\\
         0 & 3 & \frac{4}{5}\\
       \end{array},\ z\right\}.
\end{gather}
where $z_1=2z$.
The holomorphic solution of this operator around $z=0$ is given by the Hadamard product of that of the quintic threefold and $\varpi_0^{(2)}(z)$ as in \cref{eq:varpisc}.
The genus zero invariants from corrections to the triple couplings on the CICY
\begin{center}
    \begin{tabular}{| c | ccccc|}\hline
        degree $k$ & 1 & 2 & 3 & 4  & $\cdots$ \\ \hline
        $n^{(0)}_k$ & 24500&  48263250 & 181688069500 & 905026660335000 & $\cdots$ \\ \hline
    \end{tabular}
\end{center}
follow immediately in the limit $t_2\rightarrow 0$ from that of the two-parameter coupling $C_{11\alpha}$ with $\alpha$ an $H^{2,2}$-index corresponding to the $A$-brane supported on $D_1\cap D_1$ :
\begin{center}
    \begin{tabular}{| c | ccccc |}\hline
 $n_{d_1d_2}^{\quad\ \ 1}$ &  $d_1=0$ & 1 & 2 & 3 & 4\\\hline
 $d_2=0$ & -- & 12250 & 6462250 & 5718284750 & 6349209995000 \\
 1 & 0 & 12250 & 35338750 & 85125750000 & 192339896968750 \\
 2 & 0 & 0 & 6462250 & 85125750000 & 507648446407500 \\
 3 & 0 & 0 & 0 & 5718284750 & 192339896968750 \\
 4 & 0 & 0 & 0 & 0 & 6349209995000\\\hline
    \end{tabular}
\end{center}

The monodromy around $z_2=1/4$ is generated by local solutions of indicial $1/2$ with multiplicity three, i.e. divergencies up to $\log^2 (1-4z_2)$.
According to the monodromy weight filtration \eqref{eq:dWinvCY3fibred}, we expect the Euler factor at these points to have the same form as a pure $(1,1,1,1,1)$-fourfold motive,
\begin{align}
     P^{(p)}_{4,\text{hor}}(X,T)=1+\alpha^{(p)}_1T+\alpha^{(p)}_2pT^2+\epsilon_p\,(\alpha^{(p)}_2p^3T^3+\alpha^{(p)}_1p^{6}T^4+p^{10}T^5)\,,
\end{align}
with the following bounds for $p\geq 11$:
\begin{align}
    \vert\alpha^{(p)}_1\vert\leq 5p^2<p^3\,,\quad \vert p\alpha^{(p)}_2\vert\leq 10p^4<p^5\,.
\end{align}
By calculating the cubic Frobenius trace modulo $p^7$ we can determine the coefficient of $T^3$ and thus $\epsilon_p$. 
Since the discriminant locus only contributes to the denominator of the Frobenius matrix, when the $p$-adic accuracy is $n_{\text{acc}}=10$ or higher (see \eqref{eq:rationalmodifiedFrob4fold}), we can do a consistency check by calculating also the quartic Frobenius trace modulo $p^9$ and check whether the coefficient of $T^4$ equals $\epsilon_pa_pp^{6}$.
In our computations the latter condition is always fulfilled. The smooth fibres and the conifold fibres are calculated as before. The results for the diagonal direction at $p=11$ are given in \Cref{tab:Y10eulerfactors11}.
The Euler factor at the reduction from the point $z_1=z_2=1/4$ (i.e.\ $z=3\mod 11$ in the table) is just the corresponding Euler factor of the hypergeometric one-parameter model $X^{5,2}$ at $z=\frac{1}{2}z_1=1/8$. 
\begin{table}[H]
\centering
\renewcommand{\arraystretch}{1.6}
\resizebox{\textwidth}{!}{%
\begin{tabular}{|c|c|c|}
\hline
\multicolumn{3}{|c|}{$p=11$ for direction $(z_1,z_2)=(z,z)$} \\
\hline
$z \mod p$ & smooth/sing. & $P^{(p)}_{4,\text{hor}}(X,T)$ \\
\hline \hline
$1$ & smooth & $ p^{16}T^8 - 135p^{12}T^7 + 526p^9T^6 + 5415p^6T^5 - 6514p^4T^4 + 5415p^2T^3 + 526pT^2 - 135T + 1$ \\
\hline
$2$ & smooth & $-\left(p^2T - 1\right)\left(p^2T + 1\right)\left(p^12T^6 + 14p^8T^5 + 700p^5T^4 - 1270p^2T^3 + 700pT^2 + 14T + 1\right)$ \\
\hline
$3$ & edge & $\left(p^2T + 1\right)\left(p^8T^4 - 259p^4T^3 + 3116pT^2 - 259T + 1\right)$ \\
\hline
$4$ & smooth & $p^{16}T^8 + 90p^{12}T^7 + 696p^9T^6 - 21150p^6T^5 - 22794p^4T^4 - 21150p^2T^3 + 696pT^2 + 90T + 1$ \\
\hline
$5$ & conifold & $-\left(p^2T - 1\right)\left(p^{12}T^6 + 59p^8T^5 + 95p^5T^4 + 850p^3T^3 + 95pT^2 + 59T + 1\right)$ \\
\hline
$6$ & smooth & $p^{16}T^8 - 100p^{12}T^7 - 1139p^9T^6 + 12450p^6T^5 - 3119p^4T^4 + 12450p^2T^3 - 1139pT^2 - 100T + 1$ \\
\hline
$7$ & smooth & $p^{16}T^8 + 30p^{12}T^7 + 96p^{10}T^6 - 9510p^6T^5 - 3774p^4T^4 - 9510p^2T^3 + 96p^2T^2 + 30T + 1$ \\
\hline
$8$ & smooth & $-\left(p^2T - 1\right)\left(p^2T + 1\right)\left(p^{12}T^6 - 102p^8T^5 - 381p^5T^4 + 14334p^2T^3 - 381pT^2 - 102T + 1\right)$ \\
\hline
$9$ & smooth & $-\left(p^2T - 1\right)\left(p^2T + 1\right)\left(p^{12}T^6 + 119p^8T^5 + 1325p^5T^4 + 19770p^2T^3 + 1325pT^2 + 119T + 1\right)$ \\
\hline
$10$ & smooth & $-\left(p^2T - 1\right)\left(p^2T + 1\right)\left(p^{12}T^6 + 54p^8T^5 - 95p^5T^4 - 16710p^2T^3 - 95pT^2 + 54T + 1\right)$ \\
\hline
\end{tabular}
}
\caption{Euler factors of the model $X_{222211}$ along the direction $(z_1,z_2)=(z,z)$ for $p = 11$.}
\label{tab:Y10eulerfactors11}
\end{table}

\subsubsection{K3-fibration}
The classification of degenerations of two-dimensional bases in \cref{sec:gentrans} extends to fibres of higher dimensions. 
As an example, we consider the quartic K3-fibration over $\P^2$ with horizontal Hodge structure (1,2,3,2,1).
We list the toric and topological data in \Cref{tab:333111}.

The toric polytope contains the base two-face (14) in \Cref{fig:ellembed3}, whose linear dependence is reflected in the second $l$-vector.
As the local model develops a conic singularity in the limit $z_1\rightarrow -3z$ and $z_2\rightarrow (-3)^{-3}$, the system is described by the operator 
\begin{align}
\begin{split}
    \mathcal{L}^{(6)}(z) &= (\theta -1)\, \theta ^5+24z\theta (2 \theta +1) (4 \theta +1) (4 \theta +3) (3 \theta  (\theta +1)+1) \\
    &\quad + 1728  z^2(2 \theta +1) (2 \theta +3) (4 \theta +1) (4 \theta +3) (4 \theta +5) (4 \theta +7)
\end{split}
\end{align}
\begin{equation}
\renewcommand{\arraystretch}{1.3}
\setlength{\arraycolsep}{6pt}
        \mathcal{P}_{\mathcal{L}^{(6)}}\left\{\begin{array}{cccc}   0 & a & \bar{a} & \infty\\\hline
         0 & 0 & 0 & \frac{1}{4}\\
         0 & 1 & 1 & \frac{1}{2}\\
         0 & \frac{3}{2} & \frac{3}{2} & \frac{3}{4}\\
         0 & 2 & 2 & \frac{5}{4}\\
         0 & 3 & 3 & \frac{3}{2}\\
         1 & 4 & 4 & \frac{7}{4}\\
       \end{array},\ z\right\}
\end{equation}
with $a=\frac{-3+i \sqrt{3}}{4608}$ and $\bar{a}$ being the roots of $1+2304 z+1769472 z^2$\,. 
This operator is the Hadamard product of the quartic fibre $X^4$ (C in \Cref{tab:elliptic1}) and the operator $f$ in \Cref{tab:elliptic2}.
This is a direct consequence of the classification of two-dimensional bases (cf.\ \Cref{tab:ellident3}, no.\ 14) and exemplifies the generalisation to higher-dimensional fibres.

At $z_2=-\frac{1}{27}$, the monodromy is generated by local solutions of indicial $1$, again with multiplicity three. From \eqref{eq:dWinvK3fibred} we then conclude that the Euler factor at these points have a sixth-order factor, which we identify with the above one-parameter model of Hodge type $(1,1,2,1,1)$, and a Dirichlet character factor fulfilling $n=2$ Weil conjectures:
\begin{multline}
P^{(p)}_{4,\text{hor}}(X,T)=\left(1-\chi(p) pT\right)\big(1+\alpha^{(p)}_1T+\alpha^{(p)}_2pT^2+\alpha^{(p)}_3p^3T^3\\
+\epsilon_p\big(\alpha^{(p)}_2p^5T^4+\alpha^{(p)}_1p^{8}T^5+p^{12}T^6\big)\big).
\end{multline}
The Euler factors along the diagonal $z_1=z_2$ are shown in \Cref{tab:Y333111eulerfactors17}. We checked up to $p=149$ that the character is $\chi(p)=\left(\frac{-3}{p}\right)$.
\begin{table}[H]
\centering
\renewcommand{\arraystretch}{1.4}
\resizebox{\textwidth}{!}{%
\begin{tabular}{|c|c|c|}
\hline
\multicolumn{3}{|c|}{$p=17$ for direction $(z_1,z_2)=(z,z)$} \\
\hline
$z \mod p$ & smooth/sing. & $P^{(p)}_{4,\text{hor}}(X,T)$ \\
\hline \hline
$1$ & smooth & $ p^{16}T^8 - 252p^{12}T^7 + 2170p^9T^6 - 63180p^6T^5 + 72666p^4T^4 - 63180p^2T^3 + 2170pT^2 - 252T + 1 $ \\
\hline
$2$ & smooth & $ p^{16}T^8 + 227p^{12}T^7 - 3597p^9T^6 - 28700p^6T^5 + 68266p^4T^4 - 28700p^2T^3 - 3597pT^2 + 227T + 1 $ \\
\hline
$3$ & smooth & $ p^{16}T^8 - 344p^{12}T^7 + 3955p^9T^6 - 33164p^6T^5 + 10928p^4T^4 - 33164p^2T^3 + 3955pT^2 - 344T + 1 $ \\
\hline
$4$ & smooth & $ p^{16}T^8 - 504p^{12}T^7 + 7906p^9T^6 - 110952p^6T^5 + 79962p^4T^4 - 110952p^2T^3 + 7906pT^2 - 504T + 1 $ \\
\hline 
$5$ & two-face & $\left(pT+1\right)\left(p^{12}T^6 + 426p^{8}T^5 - 161p^6T^4 - 10644p^3T^3 - 161p^2T^2 + 426T + 1\right)$\\
\hline
$6$ & smooth & $ p^{16}T^8 - 592p^{12}T^7 + 14417p^9T^6 - 316082p^6T^5 + 358592p^4T^4 - 316082p^2T^3 + 14417pT^2 - 592T + 1 $ \\
\hline
$7$ & smooth & $ p^{16}T^8 + 811p^{12}T^7 + 18147p^9T^6 + 268100p^6T^5 + 233914p^4T^4 + 268100p^2T^3 + 18147pT^2 + 811T + 1 $ \\
\hline
$8$ & smooth & $ p^{16}T^8 - 260p^{12}T^7 + 3076p^9T^6 - 49532p^6T^5 - 3850p^4T^4 - 49532p^2T^3 + 3076pT^2 - 260T + 1 $ \\
\hline
$9$ & smooth & $ p^{16}T^8 + 742p^{12}T^7 + 21568p^9T^6 + 507226p^6T^5 + 559358p^4T^4 + 507226p^2T^3 + 21568pT^2 + 742T + 1 $ \\
\hline
$10$ & smooth & $ p^{16}T^8 + 421p^{12}T^7 + 4096p^9T^6 + 146431p^6T^5 + 242438p^4T^4 + 146431p^2T^3 + 4096pT^2 + 421T + 1 $ \\
\hline
$11$ & smooth & $ p^{16}T^8 - 78p^{12}T^7 + 162p^9T^6 + 4110p^6T^5 + 30202p^4T^4 + 4110p^2T^3 + 162pT^2 - 78T + 1 $ \\
\hline
$12$ & smooth & $ p^{16}T^8 + 653p^{12}T^7 + 6856p^9T^6 - 91513p^6T^5 - 191338p^4T^4 - 91513p^2T^3 + 6856pT^2 + 653T + 1 $ \\
\hline
$13$ & smooth & $ p^{16}T^8 + 296p^{12}T^7 + 600p^9T^6 - 45992p^6T^5 - 10994p^4T^4 - 45992p^2T^3 + 600pT^2 + 296T + 1 $ \\
\hline
$14$ & smooth & $ p^{16}T^8 - 13p^{12}T^7 - 2026p^9T^6 + 16125p^6T^5 - 71118p^4T^4 + 16125p^2T^3 - 2026pT^2 - 13T + 1 $ \\
\hline
$15$ & smooth & $ p^{16}T^8 + 344p^{12}T^7 + 4001p^9T^6 - 27698p^6T^5 - 30640p^4T^4 - 27698p^2T^3 + 4001pT^2 + 344T + 1 $ \\
\hline
$16$ & smooth & $ p^{16}T^8 + 147p^{12}T^7 + 7099p^9T^6 + 122172p^6T^5 + 134346p^4T^4 + 122172p^2T^3 + 7099pT^2 + 147T + 1 $ \\
\hline
\end{tabular}
}
\caption{Euler factors of the model $X_{333111}$ along the direction $(z_1,z_2)=(z,z)$ for $p = 17$.}
\label{tab:Y333111eulerfactors17}
\end{table}

\subsection{Flux vacua from degenerating three-faces}\label{sec:CY4fibr3}
The toric and topological data of the two-parameter fourfold family $X_{12\,81111}$ are given in \Cref{tab:1281111}.
The polytope contains a point inside a three-face
    \begin{equation}
        \nu_7 = (-3, -2, 0, 0, 0) = \frac{\nu_3+\nu_4+\nu_5+\nu_6}{4}\,.
    \end{equation}
The discriminant of this model is 
\begin{align}
    \Delta=\big((1 - 432z_1)^4-8916100448256\,z_1^4z_2 \big)\left(1-256z_2\right).
\end{align}
We consider the locus $a_7=4$, which corresponds to the conifold component $1-256z_2=0$.
This locus was shown to be a supersymmetric F-theory flux vacuum in \cite{Cota:2017aal}. 
In this limit, the periods as a function of $z=z_1$ are annihilated by the operator
\begin{align}
\begin{split}
     \mathcal{L}^{(7)}(z)&=(\theta -2) (\theta -1)\, \theta ^5-12 z(\theta -1) (2 \theta +1) (6 \theta +1) (6 \theta +5) (2 \theta  (\theta +1)+1) \\&\quad +144z^2 \theta\,(6 \theta +1) (6
   \theta +5) (6 \theta +7) (6 \theta +11) (6 \theta  (\theta +2)+7)\\&\quad -3456z^3 \theta\,(2 \theta +3) (6 \theta +1) (6 \theta +5) (6 \theta +7) (6 \theta +11) (6 \theta +13) (6 \theta +17)  
   \,.    
\end{split}
\end{align}
\begin{equation}\label{eq:Riemannfffv}
    \renewcommand{\arraystretch}{1.3}
\setlength{\arraycolsep}{6pt}
        \mathcal{P}_{\mathcal{L}^{(7)}}\left\{\begin{array}{ccccc}   0 & a & b & c & \infty\\\hline
         0 & 0 & 0 & 0 & \frac{1}{6}\\
         0 & 1 & 1 & 1 & \frac{5}{6}\\
         0 & \frac{3}{2} & \frac{3}{2} & \frac{3}{2} & \frac{7}{6}\\
         0 & 2 & 2 & 2 & \frac{3}{2}\\
         0 & 3 & 3 & 3 & \frac{11}{6}\\
         1 & 4 & 4 & 4 &  \frac{13}{6}\\
         2 & 5 & 5 & 5 &  \frac{17}{6}\\
       \end{array},\ z\right\}
\end{equation}
with $a=1/864$, $b=(1+\ii)/864$ and $c=\bar{b}$ roots of the first discriminant factor restricted to $z_2=1/4^4$.
The base $\P^3$ gives a Hadamard factor 
\begin{align}
\begin{split}
    \mathcal{L}^{(3)}_{\P^3}(z)&=\theta ^3-z\,(2 \theta  (\theta  (2 \theta +3)+2)+1) +z^2(\theta +1) (6 \theta  (\theta +2)+7)\\
    &\quad -2 z^3(\theta +1) (\theta +2) (2 \theta +3)\,,
\end{split}
\end{align}
\begin{equation}\label{eq:RiemannfffvK3}
    \renewcommand{\arraystretch}{1.3}
\setlength{\arraycolsep}{6pt}
        \mathcal{P}_{\mathcal{L}^{(3)}_{\P^3}}\left\{\begin{array}{ccccc}   0 & 432a & 432b & 432c & \infty\\\hline
         0 & 0 & 0 & 0 & 1\\
         0 & \frac{1}{2} & \frac{1}{2} & \frac{1}{2} & \frac{3}{2}\\
         0 & 1 & 1 & 1 & 2\\
       \end{array},\ z\right\},
\end{equation}
which, together with the operator $D$ (cf. \Cref{tab:elliptic1}) yields $\mathcal{L}^{(7)}=D*\mathcal{L}^{(3)}_{\P^3}$.

From \eqref{eq:dWinvgenusonefibred} we expect the conifold Euler factor to be of order seven, satisfying the $n=4$ Weil conjectures:
\begin{multline}
P^{(p)}_{4,\text{hor}}(X,T)=1+\alpha^{(p)}_1T+\alpha^{(p)}_2pT^2+\alpha^{(p)}_3p^2T^3\\
+\epsilon_p\,\big(\alpha^{(p)}_3p^4T^4+\alpha^{(p)}_2p^7T^5+\alpha^{(p)}_1p^{10}T^6+p^{14}T^7\big).
\end{multline}
with the bounds
\begin{align}
    \vert \alpha^{(p)}_1\vert \leq 7p^2<p^3\,,\quad \vert p\alpha^{(p)}_2\vert \leq 21p^4<p^6\,,\quad \vert p^2\alpha^{(p)}_3\vert \leq 35p^6<p^8\,.
\end{align}
By computing the Frobenius traces modulo $p^8$ we can thus compute the coefficients up to order $T^3$. To determine $\epsilon_p=\pm 1$ we need to compute the coefficient of $T^4$, whose exact value from the Frobenius traces would require one to work $\mod p^{10}$. At this precision the discriminant would appear in the denominator of the Frobenius. However, it suffices to demand the $T^4$-coefficient to be equal to $\epsilon_pp^4c_p \mod p^8$ to infer $\epsilon_p$. Since one of the zeros of the Euler factor is necessarily $\pm 1/p^2$, it has in this case always a linear factor. 

The Euler factors (except at the MUM point) along the diagonal ${(z_1,z_2)=(z,z)}$ for $p=11$ are shown in \Cref{tab:Y24eulerfactors11}. We note that the degree seven of the Euler factors at the conifold is consistent with the order seven operator that we find at this locus. Further note that under the involution symmetry
\begin{align}
(z_1,z_2)\overset{\mathcal{I}}{\longleftrightarrow} \left(-z_1+\frac{1}{2^43^3},\frac{2^{16}3^{12}z_1^4z_2}{(1-2^43^3z_1)^4}\right)
\end{align}
of the moduli space the fibres $z=1,2,3,7,8\mod 11$ are mapped again to the diagonal direction modulo $p=11$. In particular $z=1,3\mod 11$ are images of each other under $\mathcal{I}$ modulo $p$ as well as $z=8,9\mod 11$.
Correspondingly, their Euler factors are equal up to $T\mapsto -T$. The fibre $z=2\mod 11$ is a fixed point modulo $p$ of this symmetry.

\begin{table}[H]
\centering
\renewcommand{\arraystretch}{1.6}
\resizebox{\textwidth}{!}{%
\begin{tabular}{|c|c|c|}
\hline
\multicolumn{3}{|c|}{$p=11$ for direction $(z_1,z_2)=(z,z)$} \\
\hline
$z \mod p$ & smooth/sing. & $P^{(p)}_{4,\text{hor}}(X,T)$ \\
\hline \hline
$1$ & smooth & $ p^{16}T^8 - 61p^{12}T^7 - 834p^9T^6 + 4309p^6T^5 - 4534p^4T^4 + 4309p^2T^3 - 834pT^2 - 61T + 1 $ \\
\hline
$2$ & smooth & $p^{16}T^8 - 480p^9T^6 + 2594p^4T^4 - 480pT^2 + 1 $ \\
\hline
$3$ & smooth & $p^{16}T^8 + 61p^{12}T^7 - 834p^9T^6 - 4309p^6T^5 - 4534p^4T^4 - 4309p^2T^3 - 834pT^2 + 61T + 1$ \\
\hline
$4$ & conifold & $\left(p^2T+1\right)\left(p^{12}T^6 - 197p^8T^5 + 2521p^5T^4 -35470p^2T^3 + 2521pT^2 - 197T + 1\right)$ \\
\hline
$5$ & smooth & $-\left(p^2T-1\right)\left(p^2T+1\right)\left(p^{12}T^6 - 169p^8T^5 + 2752p^5T^4 - 29083p^2T^3 + 2752pT^2 - 169T + 1\right) $ \\
\hline
$6$ & smooth & $-\left(p^2T-1\right)\left(p^2T+1\right)\left(p^{12}T^6 - 196p^8T^5 + 1036p^5T^4 - 2592p^2T^3 + 1036pT^2 - 196T + 1\right)$ \\
\hline
$7$ & smooth & $p^{16}T^8 + 140p^{12}T^7 + 138p^9T^6 - 7840p^6T^5 - 8493p^4T^4 - 7840p^2T^3 + 138pT^2 + 140T + 1$ \\
\hline
$8$ & smooth & $ p^{16}T^8 - 140p^{12}T^7 + 138p^9T^6 + 7840p^6T^5 - 8493p^4T^4 + 7840p^2T^3 + 138pT^2 - 140T + 1$ \\
\hline
$9$ & smooth & $ p^{16}T^8 + 20p^{12}T^7 + 204p^9T^6 + 5980p^6T^5 - 16714p^4T^4 + 5980p^2T^3 + 204pT^2 + 20T + 1$ \\
\hline
$10$ & smooth & $p^{16}T^8 - 40p^{12}T^7 + 1040p^9T^6 + 3424p^6T^5 + 3570p^4T^4 + 3424p^2T^3 + 1040pT^2 - 40T + 1$ \\
\hline
\end{tabular}
}
\caption{Euler factors of the model $X_{12\,81111}$ along the direction $(z_1,z_2)=(z,z)$ for $p = 11$.}
\label{tab:Y24eulerfactors11}
\end{table}
\paragraph{Further examples.}

We begin with another toric Fano base given by $\P^1\times \P^1\times\P^1$ with $l$-vectors
\begin{align}
    l^{(1)} &= (-6,3,2,1,0,0,0,0,0,0)\,,\\
    l^{(2)} &= (0;0,0,-2,1,1,0,0,0,0)\,,\\
    l^{(3)} &= (0;0,0,-2,0,0,1,1,0,0)\,,\\
    l^{(4)} &= (0;0,0,-2,0,0,0,0,1,1)\,.
\end{align}
We consider the family over the symmetric locus $z_2=z_3=z_4$.
Note that the system is not hypergeometric and the Picard--Fuchs ideal is generated by
\begin{align}
    \mathcal{L}^{(2)}(z_1,z_2) &= \theta _1 \left(\theta _1-2 \theta _2\right)-12z_1 \left(6 \theta _1+1\right) \left(6 \theta _1+5\right),\\
    \mathcal{L}^{(4)}(z_1,z_2) &= \theta _2^4+z_2\left(\theta _1-2 \theta _2\right) \left(20 \theta _2^3-10 \left(\theta _1-3\right) \theta _2^2+2 \left(8-5 \theta _1\right) \theta _2+3 \left(\theta _1-2\right) \theta _1^2+3\right) \nonumber\\
    & \quad +9 z_2 \left(z_2\left(\theta _1-2 \theta _2-3\right) \left(\theta _1-2 \theta _2-1\right) \left(\theta _1-2 \theta _2\right) \left(\theta _1-2 \left(\theta _2+1\right)\right) \right.\\
    &\quad \left. -4 z_1\theta _1^2 \left(6 \theta _1+1\right) \left(6 \theta _1+5\right) \right).\nonumber
\end{align}
To identify flux vacua, we can consider the discriminant factor of the three-parameter base restricted to the symmetric locus
\begin{equation}
    \Delta_{(\P^1)^3}^{\text{sym}} = (1-36 z_2)^2 (1-4 z_2)^6 z_2^8\,.
\end{equation}
While $z_2=0$ corresponds to a large volume limit in the base, we consider the locus $z_2\in\{\frac{1}{36},\frac{1}{4}\}$\,.
Since the one-parameter models are of a similar form, we give only the former one in $z=6z_1$
\begin{align}
\begin{split}
    \mathcal{L}^{(7)}(z) &= \left(\theta _1-2\right) \left(\theta _1-1\right) \theta _1^5-72 z_1\theta _1 \left(\theta _1-1\right) \left(2 \theta _1+1\right) \left(6 \theta _1+1\right) \left(6 \theta _1+5\right) \\
    &\quad \times \left(2 \theta _1 \left(\theta _1+1\right)+1\right)+576 z_1^2\theta _1\left(6 \theta _1+1\right) \left(6 \theta _1+5\right) \left(6 \theta _1+7\right) \left(6 \theta _1+11\right) \\
    &\quad \times \left(44 \theta _1 \left(\theta _1+2\right)+51\right)  -331776z_1^3 \left(2 \theta _1+3\right) \left(6 \theta _1+1\right) \left(6 \theta _1+5\right) \\
    &\quad \times \left(6 \theta _1+7\right)\left(6 \theta _1+11\right) \left(6 \theta _1+13\right) \left(6 \theta _1+17\right) \,.
\end{split}
\end{align}
Its Riemann symbol is of the general form given in \cref{eq:Riemannfffv} with $a=2^{-6}3^{-4}$\,, $b=2^{-7}3^{-3}$ and $c=2^{-6}3^{-3}$\,.
This Hadamard product is given by the fibre $D$ and an operator from the base given by
\begin{align}
\begin{split}
    \mathcal{L}^{(3)}_{(\P^1)^3,\frac{1}{36}}(z) &= \theta ^3-6z\, (2 \theta \, (\theta \,(2 \theta +3)+2)+1)+4 z^2(\theta +1) (44 \theta \, (\theta +2)+51) \\
    &\quad -192z^3 (\theta +1) (\theta +2) (2 \theta +3)
\end{split}
\end{align}
with Riemann symbol as in \cref{eq:RiemannfffvK3}.

As for threefolds, degenerations of faces in a single three-face can appear simultaneously.
We consider the base $\P_{3111}$ with an elliptic $X^6$-section and Hodge structure (1,3,4,3,1).
Note that the point $(-1,0,0)$ in codimension one (a two-face) of the base face yields a third complex-structure parameter.
The local model has $l$-vectors
\begin{equation}
    l^{(2)} = (-2;1,0,0,0,1)\,,\quad l^{(3)} = (0;0,1,1,1,-3)\,.
\end{equation}
The linear dependence in the face of $(-1,0,0)$ yields the discriminant factor $1-27z_3$, which intersects the second discriminant factor along $z_2=a\coloneqq \frac{1}{24} \left(3- \ii\sqrt{3}\right)$ and $z_2=\bar{a}$.
These are the roots of $1-12z+48 z^2$\,.
We list the toric and topological data of the global model in \Cref{tab:18123111}.
The one-parameter family over this locus is described by the operator

\begin{align}
    \begin{split}
        \mathcal{L}^{(7)}(z)&=4439 (\sqrt{3}+\ii) (\theta -2) (\theta -1) \theta ^5\\
        &\quad +53268\ii\,z\,\theta\, (\sqrt{3}+\ii) (\theta -1) (2 \theta +1)(6 \theta +1) (6 \theta +5) (1+(1+\ii) \theta ) ((1+\ii) \theta +i)\\ 
        &\quad -15341184\ii z^3 (2 \theta +3) (6 \theta +1) (6 \theta +5)(6 \theta +7)(6 \theta +11) (6 \theta +13) (6 \theta +17)\\
        &\quad +24z^2 \theta (9 \theta  (193 \theta  (2208 \theta  (\theta  (18(2 \sqrt{3}+3 \ii) \theta  (\theta +6)+780 i+517 \sqrt{3})+960 i+628 \sqrt{3}) \\
        &\quad +14989 (59 \sqrt{3}+92 \ii))+426328 (121 \sqrt{3}+193 \ii))+3418030 (13 \sqrt{3}+21 \ii))
    \end{split}
\end{align}
with Riemann symbol given by \cref{eq:Riemannfffv} with $a=\frac{1}{864}$, $b=\frac{1}{864} i \left((2-i)-\sqrt{3}\right)$ and $c=\bar{b}$.
This Hadamard product has the base factor 
\begin{align}
\begin{split}
    \mathcal{L}^{(3)}_{\P_{3111},\frac{1}{24} \left(3-\ii \sqrt{3}\right)}&=6 \theta ^3-6 z \left(4 \theta ^3+6 \theta ^2+4 \theta +1\right) \\
    &\quad +\ii\, z^2 (\theta +1) \left(3 \left(\sqrt{3}-9 \ii\right) \theta ^2+6 \left(\sqrt{3}-9 \ii\right) \theta -30 \ii+4 \sqrt{3}\right)\\
    &\quad -3z^3 \left(1+i \sqrt{3}\right) \left(2 \theta ^3+9 \theta ^2+13 \theta +6\right),
\end{split}
\end{align}
Its Riemann symbol follows again the general form presented in \cref{eq:RiemannfffvK3}.

As a last example, we consider the elliptic $X^6$-fibration over $\P_{2211}$, whose polytope contains a point on an edge. 
The toric and topological data are collected in \Cref{tab:18122211}.
The point $(-9,-6,-1,-1,0)$ lies on an edge and is a vertex of a polygon, which lies in the base three-face with the inner point $(-3,-2,0,0,0)$\,.
Along $z_2 = -1/54$ and $z_3=1/4$, the periods are described in $z=z_1$ via
\begin{align}
\begin{split}
    \mathcal{L}^{(7)}(z)&=(\theta -2) (\theta -1) \theta ^5\\
    &\quad -12z\,\theta\, (\theta -1) (2 \theta +1) (6 \theta +1) (6 \theta +5) (2 \theta  (\theta +1)+1) \\
    &\quad +144z^2\theta\, (6 \theta +1) (6 \theta +5) (6 \theta +7) (6 \theta +11) (6 \theta\,  (\theta +2)+7)   \\
    & \quad -2592z^3 (2 \theta +3) (6 \theta +1) (6 \theta +5) (6 \theta +7) (6 \theta +11)(6 \theta +13) (6 \theta +17)\,.
    \end{split}
\end{align}
The Hadamard factor of the base is given by
\begin{align}
    \begin{split}
        \mathcal{L}^{(3)}_{\P_{2211}}(z) &= 2 \theta ^3-6z\, (2 \theta  (\theta  (2 \theta +3)+2)+1)+18  z^2(\theta +1) (6 \theta  (\theta +2)+7)\\
        &\quad -81 z^3 (\theta +1) (\theta +2) (2 \theta +3)\,.
    \end{split}
\end{align}
The Riemann symbols again follow the general forms of \cref{eq:Riemannfffv,eq:RiemannfffvK3} with $a=\frac{1}{432}$, $b=\frac{3-\ii \sqrt{3}}{2592}$ and $c=\bar{b}$\,.
\section{Conclusions}

    We performed a scan for supersymmetric vacua for two-parameter models in the Kreuzer-Skarke list by looking for persistent factorizations in the local zeta function. To avoid computation of multi-parameter periods to high orders, we reduced the Dwork deformation method to one-parameter subloci in the moduli space. If persistent factorizations are found along several rays through the origin, one can reconstruct the location of a hypersurface in the moduli space, where a splitting of Hodge structure occurs. For this purpose, the strategy of looking only at a few selected directions, is more economical, as the number of points for which Euler factors have to be computed is linear in the prime number $p$, instead of scaling with $p^r$, where $r$ is the number of moduli. 

    The scan resulted in new examples of supersymmetric vacua, in particular an example, that does neither arise from setting symmetric moduli equal nor lies on an orbifold locus. This model, which we called $\text{Bl}^{(4)}(X_{21111})$, furnishes the model $X_{21111}$ as a Hodge substructure along the vacuum. By observing that the Euler factors come in pairs we derived a symmetry acting on the moduli space, that we interpreted to be a symmetry of an additional integral structure of the model, present when going to covering space coordinates. The vacuum can then be viewed as the symmetry image of the region in moduli space, where the toric model becomes birational to the model $X_{21111}$. In the scan we also encountered examples of complex splittings, namely $(1,1,1,0)\oplus(0,1,1,1)$ over an imaginary quadratic field extension of $\mathbb{Q}$.
    
    To see whether the vacuum of $\text{Bl}^{(4)}(X_{21111})$ can be generalized, we further analyzed two-parameter toric complete intersection models that are blowups of hypergeometric one-parameter models. For the vacua that we find in this class of examples, we do not observe a doubling of the Euler factors. In particular they seem to neither arise via the blowup mechanism observed in $\text{Bl}^{(4)}(X_{21111})$, nor as a fixed point locus of a symmetry. Notably one of the vacua features an orphan operator as a Hodge substructure.
    
    We also computed Euler factors for different types of singular loci of three- and fourfold moduli spaces. In particular, we studied how the one-parameter models are reflected in the Euler factors over these fibres. Finally, we applied for the first time the deformation method to multi-parameter fourfolds of Hodge types $(1,2,2,2,1)$ and $(1,2,3,2,1)$. We plan to perform also a systematic scan for Hodge splittings in fourfold toric hypersurfaces, e.g. splittings of type $(0,1,0,1,0)\oplus(1,h^{3,1}-1,h^{2,2},h^{3,1}-1,1)$. 
    
    Following the reverse engineering described shortly in Section \ref{sec:remarks} we have identified many more models with interesting subslices and chains of 
    potential transitions. In a preliminary investigation of the integral monodromy bases we found that the corresponding geometries have MUM points  with
    non-standard $\hat \Gamma$-classes and as mentioned shortly in the introduction to Section \ref{sec:periodgeometry} we expect them to have 
    A-model interpretations as geometries with torsion and non-commutative resolutions, which we plan to investigate in \cite{BDKPT}.  
    
    \paragraph{}
    We studied simple toric fibrations characterised by having the base polytope as one of its faces.
    For threefolds, one-dimensional bases yield generalized strong coupling transitions while two-dimensional bases allow for a new type of degeneration resulting in a system of relative periods. 
    We used the hypersurfaces $X_{22211}$ and $X_{96111}$ for a detailed analysis of the period geometry in both types of degeneration and classified the one-parameter models arising at the singular loci for general bases.
    In Section \ref{sec:contrans} we investigated flux that drives the theory to conifolds. This is a two parameter version of the example discussed in the paper 
    of Polchinski and Strominger \cite{Polchinski:1995sm}. It seems that  the nodal singularities of the model have no projective algebraic small resolution like the nodal singularities in \cite{Katz:2022lyl}. In 
    \cite{Polchinski:1995sm} and similarly in \cite{Candelas:1990rm} a possible non-K\"ahler resolution is proposed with a supersymmetry breaking parameter given by the size of the resolution.
    It would be important to understand this geometry and its mirror geometry with the dual special Lagrangian better in order to interprete the proposed open disc 
    instantons invariants, that seem to have a frozen open string parameter, in a direct enumerative way.            

    The results were generalised to fourfolds, where degenerations of the three-dimensional bases in elliptic fibrations allow for supersymmetric flux vacua in F-theory compactifications.
    We exemplified that, independent of the dimension of the manifold or the base, the one-parameter models can be written as Hadamard products (with inhomogeneous terms for relative parts of the cohomology).
    Furthermore, the face polynomial of the base can be used to give the holomorphic periods of the Hadamard products as contour integrals over the fibre's holomorphic period.

\section*{Acknowledgements}    
We would like to thank Kilian Bönisch for comments on the treatment of the apparent singularities and a note on the $L$-function values of Gr\"o\ss encharacters and the Chowla-Selberg formula,  Thorsten Schimannek for comments on singular geometries and 
the geometric interpretation of the transitions, Vasily Golyshev for discussions on the fibering out and restrictions of toric deformations, Duco van Straten for the hospitality in Mainz and the idea to use one-parameter restrictions and 
Pyry Kuusela for discussion on related topics. A.K. and P.B. would like to acknowledge the support of the Deutsche Forschungsgemeinschaft Projektnummer 508889767/FOR5582 “Modern Foundations of Scattering Amplitudes” and A.K. thanks the Leverhulme Trust 
for their support of the project Quantum Geometry and Arithmetic through an International Professorship  at Sheffield.          

\newpage
\appendix

\section{Mixed Hodge structure and Euler factors}
\label{app:LMHS}

The form of the local zeta function for singular fibres of a family of projective varieties is reflected by the limiting mixed Hodge structure associated to the singularity. The idea is that the limiting mixed Hodge structure is divided into pure pieces, each of which contributes a factor to the local zeta function, whose form is dictated by the standard Weil conjectures. In the following we will very briefly review the limiting mixed Hodge structure and monodromy weight filtration and the information they carry about the arithmetic structure. For an introduction to mixed Hodge structures we refer to \cite{Kulikov_1998}. A concise introduction with focus on applications to the landscape distance conjecture can be found in \cite{Grimm:2018ohb}.

To define the limiting mixed Hodge structure we pass from the Hodge filtration $\mathcal{F}^0\subset\ldots \subset \mathcal{F}^n=\mathcal{H}^n(X,\C)$, whose sections diverge when approaching a divisor component of the discriminant, given in some local coordinate by $u=0$, to
\begin{align}
    F^k_{\infty}\equiv \lim_{u\rightarrow 0} \exp \left[-\frac{1}{2\pi \ii }\log (u)N\right]F^k\,,
\end{align}
where we introduced the nilpotent operator $N\equiv\log(\mathfrak{M})$.
From this operator we construct the monodromy weight filtration of $V_{\mathbb{C}}\equiv F^0_\infty$,
\begin{equation}
    0= W_{-1}(N)\subset W_0(N)\subset W_1(N) \subset \dots \subset W_{2n}(N) = V_{\C},
\end{equation}
using the formulae 
\begin{align*}
    W_0(N) &= \mathrm{im} N^n \\
    W_1(N) &= \mathrm{im} N^{n-1} \cap \mathrm{ker} N \\
    W_2(N) &= \mathrm{im} N^{n-2} \cap \mathrm{ker} N \oplus \mathrm{im} N^{n-1} \cap \mathrm{ker} N^2\\
     &\vdots \\
    W_{2n-1}(N) &= \mathrm{ker}N^n\\
    W_{2n}(N) &= V_{\C}\,.
\end{align*}

Following \cite{gugiatti2024hypergeometriclocalsystemsmathbbq}, where the arithmetic of conifolds in one-parameter hypergeometric motives was considered, we use the following recipe to obtain the form of the Euler factors: We consider the weight filtration on the invariant subspace $\ker(N)\subset V_{\mathbb{C}}$ induced by the weight filtration of $V_{\mathbb{C}}$, 
\begin{align}
\label{eq:monodromyinv}
    W_j^{\text{inv}}(N)\equiv W_j(N) \cap \ker(N)\,.
\end{align}
Then each nontrivial graded piece $\Gr_j^{W_{\text{inv}}}=W^{\text{inv}}_j(N)/W^{\text{inv}}_{j-1}(N)$ from the filtration contributes an Euler factor of degree $\dim_{\mathbb{C}}(\Gr_j^{W_{\text{inv}}})$, whose form is that of an Euler factor of a pure Hodge structure of dimension $j$. 

For the Euler factor at an intersection of two discriminant components with monodromies $\mathfrak{M}_1$ and $\mathfrak{M}_2$, we instead consider the weight filtration on $\ker(N_1)\cap \ker(N_2)$ induced by the monodromy weight filtration of $\mathfrak{M}_1\mathfrak{M}_2$. For this, note \cite{Cattani1982} that the weight filtrations associated to $a_1N_1+a_2N_2$ are the same for any positive $a_1,a_2$ and we may write $W_j(\mathcal{C})$, where $\mathcal{C}\equiv\mathbb{R}_{>0}N_1+\mathbb{R}_{>0}N_2$. We then consider the induced weight filtration on $\ker(N)_1\cap\ker(N_2)$,
\begin{align}
\label{eq:monodromyinvcone}
    W_j^{\text{inv}}(\mathcal{C})\equiv W_j(\mathcal{C})\cap \ker(N_1)\cap \ker(N_2)\,.
\end{align}
From the graded pieces $\Gr_j^{W_{\text{inv}}(\mathcal{C})}=W^{\text{inv}}_j(\mathcal{C})/W^{\text{inv}}_{j-1}(\mathcal{C})$ we can then again deduce the form of the Euler factor as a product of factors of degree $\dim_{\mathbb{C}}(\Gr_j^{W_{\text{inv}}(\mathcal{C})})$ associated to pure Hodge structures of dimension $j$. 

We will now discuss the resulting predictions for the Euler factors for the toric examples that appear in the text. For threefolds we thus consider the degenerations coming in the K3-fibred case from the factorization of the face-polynomial of an edge in the base, which give rise to strong coupling singularities, and in the genus-one fibred case from a two-face in the base, yielding conifold singularities. As an application of \eqref{eq:monodromyinvcone} we also consider the intersection between a strong coupling and a conifold discriminant. For fourfolds we discuss analogously degenerations from face-polynomials of the base of CY3-, K3- and genus-one fibrations. In each case we give the Jordan decomposition of the local monodromy and the tuple $\underline{d}^{W_{\text{inv}}}\equiv (\dim_{\mathbb{C}}(\Gr_j^{W_{\text{inv}}}))_{j=0,\dots,2n}$ of vector space dimensions of the graded pieces. 

\paragraph{Threefolds}
\begin{itemize}
    \item discriminant associated to edge in base of K3-fibration (strong coupling):
    \begin{align}
       \mathfrak{M}&\sim \bigg(\begin{array}{cc}
-1 & 1 \\
 0 &-1 \\
\end{array}\bigg)\oplus \mathbb{1}_{(b_3-2)\times (b_3-2)}\,,\\
\label{eq:dWinvsc}
\underline{d}^{W_{\text{inv}}}&=(0,0,0,b_3-2,0,0,0)\,.
    \end{align}
    \item discriminant associated to two-face in base of genus-one fibration (conifold):
    \begin{align}
    \mathfrak{M}&\sim\bigg(\begin{array}{cc}
 1 & 1 \\
 0 & 1 \\
\end{array}\bigg)\oplus \mathbb{1}_{(b_3-2)\times (b_3-2)}\,,\\
    \label{eq:dWinvcon}
 \underline{d}^{W_{\text{inv}}}&=(0,0,1,b_3-1,0,0,0)\,.
    \end{align}
    \item intersection of strong coupling and conifold discrminant:
    \begin{align}
       \mathfrak{M}_{\text{sc}}\mathfrak{M}_{\text{con}} &\sim \bigg(\begin{array}{cc}
-1 & 1 \\
 0 &-1 \\
\end{array}\bigg)\oplus\bigg(\begin{array}{cc}
 1 & 1 \\
 0 & 1 \\
\end{array}\bigg)\oplus \mathbb{1}_{(b_3-4)\times (b_3-4)}\,,\\
    \label{eq:dWinvsccon}
\underline{d}^{W_{\text{inv}}(\mathcal{C})}&=(0,0,1,b_3-4,0,0,0)\,.
    \end{align}
\end{itemize}
    
\paragraph{Fourfolds}
\begin{itemize}
    \item discriminant associated to edge in base of CY3-fibration: 
    \begin{align}
        \mathfrak{M}&\sim\left(\begin{array}{ccc}
 -1& 1 & 0 \\
 0 &-1 & 1 \\
 0 & 0 &-1 \\
\end{array}\right)\oplus \mathbb{1}_{(b_4-3)\times (b_4-3)}\,,\\
    \label{eq:dWinvCY3fibred}
 \underline{d}^{W_{\text{inv}}}&=(0,0,0,0,b_4-3,0,0,0,0)\,.
    \end{align}
    \item discriminant associated to two-face in base of K3-fibration:\footnote{Note that, if we were not to restrict to the monodromy invariant part \eqref{eq:monodromyinv}, we would incorrectly obtain $\dim_{\mathbb{C}}(\Gr_4^{W})=b_3-2$ for the degree of the second factor.}
 \begin{align}
        \mathfrak{M}&\sim\left(\begin{array}{ccc}
 1 & 1 & 0 \\
 0 & 1 & 1 \\
 0 & 0 & 1 \\
\end{array}\right)\oplus \mathbb{1}_{(b_4-3)\times (b_4-3)}\,,\\
    \label{eq:dWinvK3fibred}
 \underline{d}^{W_{\text{inv}}}&=(0,0,1,0,b_4-3,0,0,0,0)\,.
    \end{align}
    \item discriminant associated to three-face in base of genus-one fibration (conifold):
\begin{align}
        \mathfrak{M}&\sim (\begin{array}{c}
 -1 \\
\end{array})\oplus \mathbb{1}_{(b_4-1)\times (b_4-1)}\,,\\
    \label{eq:dWinvgenusonefibred}
 \underline{d}^{W_{\text{inv}}}&=(0,0,0,0,b_4-1,0,0,0,0)\,.
    \end{align}
\end{itemize}

\section{Conifold locus in fibration over $\P^1\times \P^1$}\label{app:P1P1el}
The identification in \cref{sec:contrans} of an inhomogeneous solution arising at the conifold locus as a chain integral is exemplified on the elliptic $X_{321}$-fibration over $\P^1\times \P^1$. 
The results are similar to the ones for $\P^2$ discussed in \cref{sec:contrans}.
\paragraph{}
The toric and topological data are collected in \Cref{tab:P1P1el}.
We restrict the three-parameter model to the symmetry locus $z_3=z_2$\,, where the Picard--Fuchs system is generated by
\begin{align}
    \mathcal{L}^{(2)}(\vec{z}) &= \theta _1 \left(\theta _1-2 \theta _2\right) - 12 z_1\left(36 \theta _1^2+36 \theta _1+5\right),\\
    \mathcal{L}^{(3)}(\vec{z}) &=\theta _2^3-2z_2 \left(2 \theta _2+1\right) \left(\theta _1^2-\left(4 \theta _2+1\right) \theta _1+2 \theta _2 \left(2 \theta _2+1\right)\right) \,.
\end{align}
The integral two-parameter basis is obtained from the data in \Cref{tab:P1P1el} and the $\hat{\Gamma}$-class formalism after setting $t^3=t^2$ and removing the entries belonging to $t^3$ and $\partial_{3}\mathcal{F}$. 
The cycles yielding $t_2$ and $\partial_{2}\mathcal{F}$ then have intersection number two, which is reflected in the intersection form.
\paragraph{}
From \Cref{tab:ellident22} (2), we deduce that a one-parameter model arises along $0=\Delta_2=1-16z_2$ and that the homogeneous system will be described by the Hadamard product $D*d$\,.
We choose the local Frobenius solution to be of the form
\begin{align}\
        \varpi(z_1,\Delta_2)={\footnotesize\left(
\begin{array}{c}
 \sigma_1 \\
 \Delta _2\sigma_2 \\
 \sigma_1 \log\! \left(z_1\right)+\sigma_3 \\
 \Delta _2\sigma_2 \log \!\left(\Delta _2\right)+\sigma_4 \\
 \sigma_1 \log\!\left(z_1\right)^2+2 \left(\sigma_3+8 \Delta _2\sigma_2\right) \log\! \left(z_1\right)+\sigma_5 \\
 \sigma_1 \log\!\left(z_1\right)^3+3 \left(\sigma_3+8 \Delta _2\sigma_2\right) \log\!\left(z_1\right)^2-6\sigma_5\log\! \left(z_1\right)+\sigma_6 \\
\end{array}
\right)}\,
    \end{align}
    with formal power series $\sigma_i$ in $z_1$ and $\Delta_2$\,.
    The transition matrix to the integral basis is identified numerically with 
    \begin{equation}
        {\footnotesize T_{z_1,\Delta_2}=\left(
\begin{array}{cccccc}
 1 & 0 & 0 & 0 & 0 & 0 \\
 \frac{i (\pi  \log (2)-2 G)}{\pi ^2} & \frac{4 i-(4+4 i) \pi }{\pi ^2} & -\frac{i}{2 \pi } & -\frac{4 i}{\pi ^2} & 0 & 0 \\
 \frac{4 i G}{\pi ^2} & \frac{8 (\pi -i)}{\pi ^2} & 0 & \frac{8 i}{\pi ^2} & 0 & 0 \\
 -\frac{23}{24}-\frac{\log (2) (\log (2)+i \pi )-2 i G}{\pi ^2} & \frac{2 (-2 i+(1+2 i) \pi +\log (16))}{\pi ^2} & \frac{\log (4)+i \pi }{2 \pi ^2} & \frac{4 i}{\pi ^2} & -\frac{1}{4 \pi ^2} & 0 \\
 -\frac{23}{6}-\frac{4 i \log (2) (\pi -i \log (2))}{\pi ^2} & \frac{16 (\log (4)+i \pi )}{\pi ^2} & \frac{\log (16)+2 i \pi }{\pi ^2} & 0 & -\frac{1}{\pi ^2} & 0 \\
 \frac{i \left(705 \zeta (3)+16 \log ^3(2)-46 \pi ^2 \log (2)\right)}{12 \pi ^3} & \frac{2 i \left(24 G+23 \pi ^2-24 \log ^2(2)\right)}{3 \pi ^3} & \frac{i \left(23 \pi ^2-24 \log ^2(2)\right)}{12 \pi ^3} & 0 & \frac{i \log (2)}{\pi ^3} & -\frac{i}{6 \pi ^3} \\
\end{array}
\right)}.
    \end{equation}
    Here, we found Catalan's constant
    \begin{equation}
        G \equiv L_{-4}(2)= \sum_{n=1}^\infty \frac{\chi_4(n)}{n^2}=0.915965594177219\cdots =\frac{\ii}{2} \left(\text{Li}_2(-\ii)-\text{Li}_2(\ii)\right)
    \end{equation}
    with the quadratic Dirichlet character \href{https://www.lmfdb.org/Character/Dirichlet/4/b}{4.b}
    \begin{equation}
        \chi_4(n)=\left(\frac{-4}{n}\right) = \begin{cases}
        1 & n \equiv 1 \text{ mod } 4\\
        -1 & n \equiv 3 \text{ mod } 4\\
        0 & n \equiv 0 \text{ mod } 2\\
        \end{cases}\,.
    \end{equation}
    We again note the appearance of $G$ in the mirror map at the conifold of the local model  $\mathcal{O}(-2,-2)\rightarrow \P^1\times \P^1$~\cite{kerr2008algebraic,Bonisch:2022mgw}.
    The asymptotic period vector along $\Delta_2=0$ in $z=z_1/4$, $z\sim e^{2\pi\ii t}$, is given by
    \begin{equation}
        \vec{\Pi}(\Delta_1=0,z) \sim \begin{pmatrix}1\\t-\frac{2 i G}{\pi ^2}\\\frac{4 i G}{\pi ^2}\\\frac{2 i G}{\pi ^2}+(t-1) t-\frac{23}{24}\\4 (t-1) t-\frac{23}{6}\\-\frac{1}{6} t \left(8 t^2+23\right)+\frac{235 i \zeta (3)}{4 \pi ^3}\end{pmatrix}.
    \end{equation}
    The linear combination corresponding to the homology class of the shrinking $S^3$ is represented by the vector
    \begin{equation}
        f = (0, -1, 2, 1, 0, 0).
    \end{equation}
    
    As mentioned above, the homogeneous system is given by periods of \href{https://cycluster.mpim-bonn.mpg.de/operator.html?nn=4.2.16}{AESZ 2.16}
    \begin{align}\label{eq:AESZ216}
    \begin{split}
        \mathcal{L}^{(4)}_{\text{2.16}}(z)&=\theta ^4-48 z(6 \theta +1) (6 \theta +5) (3 \theta  (\theta +1)+1)\\
        &\quad +4608z^2 (6 \theta +1) (6 \theta +5) (6 \theta +7) (6 \theta +11) 
    \end{split}
    \end{align}
    with Riemann symbol
     \begin{equation}
        \mathcal{P}_{\mathcal{L}^{(4)}_{\text{2.16}}}\left\{\begin{array}{cccc}
            0 & 1/3456 & 1/1728 &  \infty\\\hline
            0 & 0 & 0 & \frac{1}{6}\\
            0 & 1 & 1 & \frac{5}{6}\\
            0 & 1 & 1 & \frac{7}{6}\\
            0 & 2 & 2 & \frac{11}{6}
    \end{array},\ z\right\}.
    \end{equation}
    Here, we find the following inhomogeneity of the periods
    \begin{equation}
        \mathcal{L}^{(4)}_{\text{2.16}}(z) \vec{\Pi}(\Delta_1=0,z)=f\,\frac{240\ii z}{\pi^2}\,.
    \end{equation}
    An integral one-parameter basis of the relative periods is given by
    \begin{equation}\label{eq:P1P1oneparam}
        \vec{\Pi}_1(z) =\left(
            \begin{array}{cccccc}
             1 & 0 & 0 & 0 & 0 & 0 \\
             0 & 2 & 1 & 0 & 0 & 0 \\
             0 & 1 & \frac{1}{2} & 0 & \frac{1}{4} & 0 \\
             0 & 0 & 0 & 0 & 0 & 1 \\
             0 & 0 & 0 & 1 & 0 & 0 \\
            \end{array}
            \right)\vec{\Pi}(\Delta_1=0,z)\,.
    \end{equation}
    We can express the periods as
    \begin{equation}
        \vec{\Pi}_1\sim \left(1,2t,\frac{1}{4}\partial_{t}\mathcal{F} , 2\mathcal{F}-t\partial_t \mathcal{F}, \Pi^{\text{asy}}_{\text{rel}}(t)\right),
    \end{equation}
    with a pre-potential
    \begin{equation}\label{eq:prep216}
        \mathcal{F}(t) = \frac{8 t^3}{3!}-\frac{92 t}{24}-\frac{(-470) \zeta (3)}{2 (2 \pi  \ii)^3}\,.
    \end{equation}
    The exact expression of the relative period is given by
    \begin{equation}\label{eq:relperiodP1P1}
        \Pi_{\text{rel}}(t)=t^2-t-\frac{1}{(2\pi\ii)^2}\sum_{k\in\N_0}\sum_{j=0}^3 n_{k,j}\text{Li}_2\!\left( e^{2\pi\ii(2k\,t-j)/4}\right).
    \end{equation}
    The coefficients $n_{k,j}$ are restricted to $n_{k,3}=-n_{k,1}$ and $n_{k,2}=0$\,.
    We list them for $k\leq 5$ in \Cref{tab:openinstantons2}.
    We verified their integrality up to $k= 80$\,.
    \begin{table}
    \centering
        \begin{tabular}{|c|ccccccc|}\hline
            $k$ & 0 & 1 & 2 & 3 & 4 & 5 &$\cdots$\\  \hline
            $n_{k,0}$ & $-23$ & $-480$ & $-228408$ & $-114635040$ & $-118481151480$ & $-100178206144800$ &$\cdots$\\
            $n_{k,1}$ & $-4$ &  $480$ &  $-53280$ &  $41798880$ &  $-2005514400$ &  $19861272633120$ &$\cdots$\\ \hline 
        \end{tabular}
        \caption{The first integers appearing in the corrections to the relative period dual to the shrinking $S^3$ at the conifold locus of the $X_{321}$-fibration over $\P^1\times\P^1$\,, cf. \cref{eq:relperiodP1P1}.
        The coefficients satisfy $n_{k,3}=-n_{k,1}$ and $n_{k,2}=0$\,.
        Integrality was verified for $k\leq 80$.
        Denoting the instanton corrections of the one-parameter model \eqref{eq:prep216} by $n_k$, one finds $n_k = 4 k\, n_{k,0}$ for $k>0$.}
        \label{tab:openinstantons2}
    \end{table}
    The summation at $k=0$ yields the constant contribution
    \begin{equation}
        -\frac{23}{24}+\frac{1}{\pi ^2} (\text{Li}_2(e^{-2\pi\ii/4})-\text{Li}_2(e^{-3\cdot 2\pi\ii/4})) = -\frac{23}{24}+\frac{2 \ii\, G}{\pi ^2}\,.
    \end{equation}
    We give the generators of the monodromy group of the relative periods obtained by analytical continuation
    \begin{gather}
        \mathfrak{M}_0=\left(
\begin{array}{ccccc}
 1 & 0 & 0 & 0 & 0 \\
 2 & 1 & 0 & 0 & 0 \\
 1 & 1 & 1 & 0 & 0 \\
 -9 & -2 & -4 & 1 & 0 \\
 0 & 1 & 0 & 0 & 1 \\
\end{array}
\right),\ \mathfrak{M}_{\frac{1}{3456}}=\left(
\begin{array}{ccccc}
 1 & 0 & 0 & 1 & 0 \\
 0 & 1 & 0 & 0 & 0 \\
 0 & 0 & 1 & 0 & 0 \\
 0 & 0 & 0 & 1 & 0 \\
 0 & 0 & 0 & 0 & 1 \\
\end{array}
\right),\\
\mathfrak{M}_{\frac{1}{1728}}=\left(
\begin{array}{ccccc}
 5 & 0 & 2 & 2 & 0 \\
 2 & 1 & 1 & 1 & 0 \\
 0 & 0 & 1 & 0 & 0 \\
 -8 & 0 & -4 & -3 & 0 \\
 0 & 0 & 0 & 0 & 1 \\
\end{array}
\right), \ \mathfrak{M}_\infty=\left(
\begin{array}{ccccc}
 4 & 0 & 2 & 1 & 0 \\
 4 & 0 & 3 & 1 & 0 \\
 1 & -1 & 1 & 0 & 0 \\
 -15 & 2 & -8 & -3 & 0 \\
 2 & -1 & 0 & 0 & 1 \\
\end{array}
\right).
    \end{gather}
    They satisfy the closed-contour relation $\mathfrak{M}_0\cdot \mathfrak{M}_\frac{1}{3456}\cdot \mathfrak{M}_{\frac{1}{1728}}\cdot\mathfrak{M}_\infty=\mathbbm{1}$\,.
    The closed part is an invariant subspace of the monodromy group and the intersection form of the homogeneous system is 
    \begin{equation}
        \Sigma = \left(
\begin{array}{cccc}
 0 & 0 & 0 & -1 \\
 0 & 0 & -2 & 0 \\
 0 & 2 & 0 & 0 \\
 1 & 0 & 0 & 0 \\
\end{array}
\right).
    \end{equation}
    \paragraph{}
    For completeness, we list the first couple of instanton numbers of this symmetric $(z_2=z_3)$ elliptic fibration that brings forth the relative one-parameter family above.
    \begin{center}
        \begin{tabular}{|c|ccccc|} \hline
         $n_{k_1,k_2}$ & $k_2=0$ & 1 & 2 & 3 & 4 \\ \hline
         $k_1=0$ & -- & $-4$ & $-4$ & $-12$ & $-48$ \\
         1 & 480 & 960 & 1440 & 4800 & 23520 \\
         2&480 & 565776 & $-225120$ & $-904320$ & $-5450400$ \\
         3&480 & 34117120 & 85633920 & 206750400 & 1134457600 \\
         4&480 & 955033560 & 125127827280 & $-31072029360$ & $-181077930240$ \\ \hline
        \end{tabular}
    \end{center}
    
\section{Supplementary data for discussed models}\label{app:ttdata}
The integrations of the Chern classes are listed in lexicographical order, e.g.\ for fourfolds we give ($c_2\cdot J_1^2,c_2\cdot J_1J_2,c_2\cdot J_2^2)$.

For three-parameter models, we give the discriminant factors $\bar{\Delta}$ omitting multiplicities and the factor from the top-dimensional face due to its length.
\begin{table}[H]
    \renewcommand{\arraystretch}{1.2}
    \centering
    \begin{tabular}{| c | c c c c|c c || c | l |}
        \hline \multicolumn{5}{|c}{points} & \multicolumn{2}{| c||}{$l$-vectors} & \multicolumn{2}{c|}{topological data} \\ \hline \hline
        (1 & 0 & 0 & 0 & 0) &  0 & -4 & $R$ & $4J_1J_2^2+8J_2^3$\\ \cline{1-7}
        (1 & 1 & 0 & 0 & 0) &  0 &  1 & $c_2$ & $(24,56)$\\
        (1 & 0 & 1 & 0 & 0) &  0 &  1 & $\chi$ & $-168$\\
        (1 & 0 & 0 & 1 & 0) &  0 &  1 &&\\
        (1 & 0 & 0 & 0 & 1) &  1 &  0 &&\\
        (1 &-2 &-2 &-2 &-1) &  1 &  0 &&\\
        (1 &-1 &-1 &-1 & 0) & -2 &  1 &&\\ \hline\hline
        \multicolumn{9}{|l|}{$\mathcal{L}^{(2)}= \theta _1^2-z_1\left(2 \theta _1-\theta _2\right) \left(2 \theta _1-\theta _2+1\right)$}\\
        \multicolumn{9}{|l|}{$\mathcal{L}^{(3)} = \left(2 \theta _1-\theta _2\right) \theta _2^2+8 z_2\left(2 \theta _2+1\right) \left(4 \theta _2+1\right) \left(4 \theta _2+3\right)$}\\\hline
    \multicolumn{9}{|l|}{$\Delta=\left(1-4z_1\right)^2\left(1-512 z_2 \left(1-128 z_2\left(1-4 z_1\right) \right)\right)$}\\ \hline
    \end{tabular}
    \caption{Toric and topological data for $X_{22211}$. This model is a K3 $X^4$-fibration over $\P^1$.}
    \label{tab:22211}
\end{table}
\begin{table}[H]
    \renewcommand{\arraystretch}{1.2}
    \centering
    \begin{tabular}{| c | c c c c|c c || c | l |}
        \hline \multicolumn{5}{|c}{points} & \multicolumn{2}{| c||}{$l$-vectors} & \multicolumn{2}{c|}{topological data} \\ \hline \hline
        (1 & 0 & 0 & 0 & 0) &  0 & -6 & $R$ & $J_1^2J_2+3J_1J_2^2+9J_2^3$\\ \cline{1-7}
        (1 & 1 & 0 & 0 & 0) &  0 &  3 & $c_2$ & $(36,102)$\\
        (1 & 0 & 1 & 0 & 0) &  0 &  2 & $\chi$ & $-540$\\
        (1 & 0 & 0 & 1 & 0) &  1 &  0 &&\\
        (1 & 0 & 0 & 0 & 1) &  1 &  0 &&\\
        (1 &-9 &-6 &-1 &-1) &  1 &  0 &&\\
        (1 &-3 &-2 & 0 & 0) & -3 &  1 &&\\ \hline\hline
        \multicolumn{9}{|l|}{$\mathcal{L}^{(2)}= \left(3 \theta _1-\theta _2\right) \theta _2+12 z_2 \left(6 \theta _2+1\right) \left(6 \theta _2+5\right) $}\\
        \multicolumn{9}{|l|}{$\mathcal{L}^{(3)} = \theta _1^3+z_1\left(3 \theta _1-\theta _2\right) \left(3 \theta _1-\theta _2+1\right) \left(3 \theta _1-\theta _2+2\right)$}\\\hline
        \multicolumn{9}{|l|}{$\Delta=\left(1 + 27z_1\right)\big(2176782336 z_1 z_2^3-\left(1-432 z_2\right)^3\big)$}\\ \hline
    \end{tabular}
    \caption{Toric and topological data for $X_{96111}$. This model is an elliptic $X^6$-fibration over $\P^2$.}
    \label{tab:96111}
\end{table}
\begin{table}[H]
    \renewcommand{\arraystretch}{1.2}
    \centering
    \begin{tabular}{| c | c c c c|c c || c | l |}
        \hline \multicolumn{5}{|c}{points} & \multicolumn{2}{| c||}{$l$-vectors} & \multicolumn{2}{c|}{topological data} \\ \hline \hline
        (1 & 0 & 0 & 0 & 0) &  0 & -3 & $R$ & $J_1^2J_2+3J_1J_2^2+9J_2^3$\\ \cline{1-7}
        (1 & 1 & 0 & 0 & 0) &  0 &  1 & $c_2$ & $(12,30)$\\
        (1 & 0 & 1 & 0 & 0) &  0 &  1 & $\chi$ & $-72$\\
        (1 & 0 & 0 & 1 & 0) &  1 &  0 &&\\
        (1 & 1 & 2 & 1 & 3) &  1 &  0 &&\\
        (1 &-4 &-5 &-2 &-3) &  1 &  0 &&\\
        (1 &-1 &-1 & 0 & 0) & -3 &  1 &&\\ \hline\hline
        \multicolumn{9}{|l|}{$\mathcal{L}^{(2)}=\theta _1 \left(\theta _1-3 \theta _2\right)- 3z_1 \left(9 \theta _1 \left(\theta _1+1\right)+2\right)$}\\
        \multicolumn{9}{|l|}{$\mathcal{L}^{(3)} = \theta _2^3-z_2\left(\theta _1-3 \theta _2-2\right) \left(\theta _1-3 \theta _2-1\right) \left(\theta _1-3 \theta _2\right)$}\\\hline
        \multicolumn{9}{|l|}{$\Delta=\left(1+27 z_1\right) \left(1-81 z_2+2187 z_2^2-19683 \left(1+27 z_1\right)z_2^3\right)$}\\ \hline
    \end{tabular}
    \caption{Toric and topological data for the $\mathbb{Z}_3$-quotient of $X_{33111}$. This model is an elliptic $X^3$-fibration over $\P^2$.}
    \label{tab:33111}
\end{table}
\begin{table}[H]
    \renewcommand{\arraystretch}{1.2}
    \centering
    \begin{tabular}{| c | c c c c|c c c|| c | l |}
        \hline \multicolumn{5}{|c}{points} & \multicolumn{3}{| c||}{$l$-vectors} & \multicolumn{2}{c|}{topological data} \\ \hline \hline
        (1 & 0 & 0 & 0 & 0) & -3 & 0 &  0 & $R$ & $25 J_1^3+30 J_2 J_1^2+15 J_3 J_1^2+36 J_2^2 J_1+9 J_3^2 J_1$\\ \cline{1-8}
        (1 & 1 & 0 & 0 & 0) &  1 & 0 &  0 & & $+18 J_2 J_3 J_1+42 J_2^3+9 J_2 J_3^2+21 J_2^2 J_3$\\
        (1 & 0 & 1 & 0 & 0) &  1 & 0 &  0 & $c_2$ & $(82,108,60)$\\
        (1 & 0 & 0 & 1 & 0) &  0 & 0 &  1 & $\chi$ & $-192$\\
        (1 & 0 & 0 & 0 & 1) & -1 & 1 &  0 &&\\
        (1 &-5 &-5 &-3 &-1) & -1 & 1 &  0 &&\\
        (1 &-2 &-2 &-1 & 0) &  3 &-3 &  1 &&\\
        (1 &-1 &-1 & 0 & 0) &  0 & 1 & -2 &&\\\hline\hline
        \multicolumn{10}{|l|}{$\mathcal{L}^{(2)}_1=\theta _2 \left(-3 \theta _1+\theta _2+3\theta _3\right)-z_2\left(\theta _1-2 \theta _2-1\right) \left(\theta _1-2 \theta _2\right)$}\\
        \multicolumn{10}{|l|}{$\mathcal{L}^{(2)}_2=\left(\theta _1-2 \theta _2\right) \theta _3-3 z_1 z_3 \left(9 \theta _3 \left(\theta _3+1\right)+2\right)$}\\
        \multicolumn{10}{|l|}{$\mathcal{L}^{(3)} = \left(\theta _1-2 \theta _2\right) \left(\theta _1-\theta _3\right){}^2$}\\
        \multicolumn{10}{|l|}{$\hspace{2cm} +z_1\left(3 \theta _1-\theta _2-3 \theta _3\right) \left(3 \theta _1-\theta _2-3 \theta _3+1\right) \left(3 \theta _1-\theta _2-3 \theta _3+2\right)$}\\\hline
        \multicolumn{10}{|l|}{$\bar{\Delta} = \left(3125 z_2^2 z_3^3+\left(1-4 z_3\right){}^2+z_2 \left(500 z_3^2-225 z_3+27\right)\right)$}\\\hline
    \end{tabular}
    \caption{Toric and topological data for $X_{55311}$. This model has a two-face with two inner points.}
    \label{tab:55311}
\end{table}
\begin{table}[H]
    \renewcommand{\arraystretch}{1.2}
    \centering
    \begin{tabular}{| c | c c c c|c c || c | l |}
        \hline \multicolumn{5}{|c}{points} & \multicolumn{2}{| c||}{$l$-vectors} & \multicolumn{2}{c|}{topological data} \\ \hline \hline
        (1 & 0 & 0 & 0 & 0) &  0 & -4 & $R$ & $20J_1^3+16J_1^2J_2+12J_1J_2^2+9J_2^3$\\ \cline{1-7}
        (1 & 1 & 0 & 0 & 0) &  1 &  0 & $c_2$ & $(80,54)$\\
        (1 & 0 & 1 & 0 & 0) &  0 &  1 & $\chi$ & $-168$\\
        (1 & 0 & 0 & 1 & 0) &  0 &  1 &&\\
        (1 & 0 & 0 & 0 & 1) &  1 &  -1 &&\\
        (1 &-4 &-3 &-3 &-1) &  1 &  -1 &&\\
        (1 &-1 &-1 &-1 & 0) & -3 &  4 &&\\ \hline\hline
        \multicolumn{9}{|l|}{$\mathcal{L}^{(2)}= \theta _2 \left(3 \theta _1-4 \theta _2\right)+4 z_2 \left(\theta _1-\theta _2\right){}^2+12 z_1 z_2\left(9 \theta _1^2+48 \theta _2^2+12 \left(1-3 \theta _1\right) \theta _2+2\right) $}\\
        \multicolumn{9}{|l|}{$\mathcal{L}^{(3)} = \theta _1 \left(\theta _1-\theta _2\right){}^2+z_1\left(3 \theta _1-4 \theta _2\right) \left(3 \theta _1-4 \theta _2+1\right) \left(3 \theta _1-4 \theta _2+2\right) $}\\\hline
        \multicolumn{9}{|l|}{$\Delta=\left(1+27 z_1\right) \left(1-z_2-576 z_1z_2 +512 z_1z_2^2 +96768  z_1^2z_2^2-65536 z_1^2z_2^3 \right. $}\\
        \multicolumn{9}{|l|}{$\hspace{8cm}\left.-3538944 z_1^3 z_2^3-47775744  z_1^4z_2^3\right) $}
        \\\hline
    \end{tabular}
    \caption{Toric and topological data for $X_{43311}$. Note that this model has three non-polynomial deformations.}
    \label{tab:43311}
\end{table}
\begin{table}[H]
    \renewcommand{\arraystretch}{1.2}
    \centering
    \begin{tabular}{| c | c c c c|c c c|| c | l |}
        \hline \multicolumn{5}{|c}{points} & \multicolumn{3}{| c||}{$l$-vectors} & \multicolumn{2}{c|}{topological data} \\ \hline \hline
        (1 & 0 & 0 & 0 & 0) &-4 & 0 &  0 & $R$ & $9 J_1^3+3 J_2 J_1^2+18 J_3 J_1^2+J_2^2 J_1+18 J_3^2 J_1$\\ \cline{1-8}
        (1 & 1 & 0 & 0 & 0) & 0 & 0 &  1 & &$+6 J_2 J_3 J_1+18 J_3^3+6 J_2 J_3^2+2 J_2^2 J_3 $\\
        (1 & 1 & 2 & 0 & 0) &-1 & 0 &  1 & $c_2$ & $(102,36,96)$\\
        (1 & 0 & 0 & 1 & 0) & 0 & 1 &  0 & $\chi$ & $-324$\\
        (1 & 0 & 0 & 0 & 1) & 0 & 1 &  0 &&\\
        (1 &-9 &-6 &-1 &-1) & 0 & 1 &  0 &&\\
        (1 &-3 &-2 & 0 & 0) & 1 &-3 &  0 &&\\
        (1 & 1 & 1 & 0 & 0) & 4 & 0 & -2 &&\\\hline\hline
        \multicolumn{10}{|l|}{$\mathcal{L}^{(2)}= \theta _3 \left(\theta _3-\theta _1\right)-2 z_3 \left(4 \theta _1-2 \theta _3-1\right) \left(2 \theta _1-\theta _3\right)$}\\
        \multicolumn{10}{|l|}{$\mathcal{L}^{(2)} = \left(\theta _1-3 \theta _2\right) \left(2 \theta _1-\theta _3\right)-z_1 \left(-2 \theta _1^2+\theta _1 \left(17 \theta _3-2\right)+\theta _3 \left(2-15 \theta _3\right) \right.$}\\
        \multicolumn{10}{|l|}{\qquad\qquad $\left.+64 \left(10 \theta _1^2-9 \theta _3 \theta _1+2 \theta _3^2\right) z_3^2+2 \left(144 \theta _1^2-100 \theta _3 \theta _1+14 \theta _3^2+19 \theta _3+6\right) z_3\right)$}\\
        \multicolumn{10}{|l|}{$\mathcal{L}^{(3)} = \theta _2^3-z_2\left(\theta _1-3 \theta _2-2\right) \left(\theta _1-3 \theta _2-1\right) \left(\theta _1-3 \theta _2\right)$}\\\hline
        \multicolumn{10}{|l|}{$\bar{\Delta}= \left(1+27z_2\right)\left(1-4z_3\right)$}\\\hline
    \end{tabular}
    \caption{Toric and topological data for toric hypersurfaces. This model has two inner points in a two-face and an edge with disjoint sets of vertices.}
    \label{tab:9611toric}
\end{table}
\begin{table}[H]
    \renewcommand{\arraystretch}{1.2}
    \centering
    \begin{tabular}{| c | c c c c|c c c|| c | l |}
        \hline \multicolumn{5}{|c}{points} & \multicolumn{3}{| c||}{$l$-vectors} & \multicolumn{2}{c|}{topological data} \\ \hline \hline
        (1 & 0 & 0 & 0 & 0) &-4 & 0 &  0 & $R$ & $9 J_1^3+21 J_2 J_1^2+18 J_3 J_1^2+43 J_2^2 J_1+30 J_3^2 J_1$\\ \cline{1-8}
        (1 & 1 & 0 & 0 & 0) & 0 & 0 &  1 & &$+36 J_2 J_3 J_1+86 J_2^3+50 J_3^3+60 J_2 J_3^2+72 J_2^2 J_3$\\
        (1 & 0 & 1 & 0 & 0) & 1 & 0 &  0 & $c_2$ & $(66,128,104)$\\
        (1 & 0 & 0 & 1 & 0) & 0 & 1 &  -1 & $\chi$ & $-204$\\
        (1 & 1 & 0 & 0 & 2) & -2 & 1 &  0 &&\\
        (1 &-7 &-3 &-1 &-2) & 0 & 1 &  -1 &&\\
        (1 &-2 &-1 & 0 & 0) & 1 &-3 &  3 &&\\
\hline
    \end{tabular}
    \caption{Toric and topological data for toric hypersurfaces. This model has two inner points in a two-face and an edge, which share one vertex.}
    \label{tab:7312toric}
\end{table}

\begin{table}[H]
    \renewcommand{\arraystretch}{1.2}
    \centering
    \begin{tabular}{| c | c c c c|c c || c | l |}
        \hline \multicolumn{5}{|c}{points} & \multicolumn{2}{| c||}{$l$-vectors} & \multicolumn{2}{c|}{topological data} \\ \hline \hline
        (1 & 0 & 0 & 0 & 0) &  0 & -3 & $R$ & $21J_1^3+9J_1^2J_2+3J_1J_2^2$\\ \cline{1-7}
        (1 & 1 & 0 & 0 & 0) &  0 &  1 & $c_2$ & $(78,36)$\\
        (1 & 0 & 1 & 0 & 0) & -2 &  1 & $\chi$ & $-180$\\
        (1 &-1 &-1 & 0 & 0) & -1 &  1 &&\\
        (1 & 0 & 0 & 1 & 0) &  1 &  0 &&\\
        (1 & 0 & 0 & 0 & 1) &  1 &  0 &&\\ 
        (1 &-1 & 1 &-1 &-1) &  1 &  0 &&\\ \hline\hline
        \multicolumn{9}{|l|}{$\mathcal{L}^{(2)}=\left(\theta _1-2 \theta _2\right) \left(\theta _1-\theta _2\right)-3z_1 \left(9 \theta _1 \left(\theta _1+1\right)+2\right) $}\\
        \multicolumn{9}{|l|}{$\mathcal{L}^{(3)}=-54 z_1 z_2 \theta_1^3 + \left(\left(27 z_1 + 4 z_2 - 1\right) \theta_2 - 54 z_1 z_2\right) \theta_1^2$}\\
        \multicolumn{9}{|l|}{$\quad\quad+ \left(\left(-12 z_2 + 3\right) \theta_2^2 + 27 z_1 \theta_2 - 12 z_1 z_2\right) \theta_1 + 8 z_2 \theta_2^3 + 6 z_1 \theta_2$}\\\hline
        \multicolumn{9}{|l|}{$\Delta=1- 81z_1 + 27z_1z_2+ 2187z_1^2- 14580z_1^2z_2 - 19683z_1^3- 157464z_1^3z_2 -314928z_1^3z_2^2$}\\ \hline
    \end{tabular}
    \caption{Toric and topological data for $\text{Bl}^{(4)}(X_{21111})$. This model is an elliptic $X^3$-fibration over $\P^2$.}
    \label{tab:Bl4X6}
\end{table}

\begin{table}[H]
    \renewcommand{\arraystretch}{1.2}
    \centering
    \begin{tabular}{| c | c c c c|c c || c | l |}
        \hline \multicolumn{5}{|c}{points} & \multicolumn{2}{| c||}{$l$-vectors} & \multicolumn{2}{c|}{topological data} \\ \hline \hline
        (1 & 0 & 0 & 0 & 0) & -1 & -5 & $R$ & $36J_1^3+12J_1^2J_2+4J_1J_2^2+J_2^3$\\ \cline{1-7}
        (1 & 1 & 0 & 0 & 0) &  1 &  0 & $c_2$ & $(96,34)$\\
        (1 & 0 & 1 & 0 & 0) & -1 &  5 & $\chi$ & $-208$\\
        (1 &-1 & 1 & 0 & 0) &  1 & -3 &&\\
        (1 & 0 & 0 & 1 & 0) &  0 &  1 &&\\
        (1 & 0 & 0 & 0 & 1) &  0 &  1 &&\\ 
        (1 &-3 &-2 &-1 &-1) &  0 &  1 &&\\
\hline
    \end{tabular}
    \caption{Toric and topological data for $\text{Bl}^{(1)}(X_{52111})$.}
    \label{tab:Bl1X10}
\end{table}

\begin{table}[H]
    \renewcommand{\arraystretch}{1.2}
    \centering
    \begin{tabular}{| c | c c c c c |c c || c | l |}
        \hline \multicolumn{6}{|c}{points} & \multicolumn{2}{| c||}{$l$-vectors} & \multicolumn{2}{c|}{topological data} \\ \hline \hline
        (1 & 0 & 0 & 0 & 0 & 0) & -2 &  0 & $R$ & $3J_1^3+3J_1^2J_2+3J_1J_2^2+J_2^3$\\
        (1 & 0 & 0 & 0 & 0 & 0) &  0 & -6 & $c_2$ & $(42,22)$\\\cline{1-8}
        (1 & 1 & 0 & 0 & 0 & 0) &  1 &  0 & $\chi$ & $-84$ \\
        (1 & 0 & 1 & 0 & 0 & 0) &  1 &  0 &  & \\
        (1 & 0 & 0 & 1 & 0 & 0) &  1 & -1 &&\\
        (1 & 0 & 0 & 0 & 1 & 0) &  0 &  2 &&\\
        (1 & 0 & 0 & 0 & 0 & 1) &  0 &  1 &&\\ 
        (1 &-3 &-3 &-2 &-2 &-1) &  0 &  1 &&\\
        (1 & 1 & 1 & 1 & 0 & 0) & -1 &  3 &&\\  \hline\hline
        \multicolumn{10}{|l|}{$\mathcal{L}^{(2)}=\theta _1 \left(\theta _1-\theta _2\right)+2 z_1\left(2 \theta _1+1\right) \left(\theta _1-3 \theta _2\right) $}\\
        \multicolumn{10}{|l|}{$\mathcal{L}^{(3)}=\theta _2^2 \left(3 \theta _2-\theta _1\right)-432 \left(2 \theta _1+5\right) \left(2 \theta _1+1\right){}^2 z_1^2 z_2-72 \left(\theta _1 \left(6 \theta _1+7\right){}^2+10\right) z_1 z_2$}\\
        \multicolumn{10}{|l|}{$\quad\quad-36 z_2\left(\theta _1-\theta _2\right) \left(36 \theta _2+12 \left(\theta _1^2+2 \theta _2 \theta _1+\theta _1+3 \theta _2^2\right)+5\right) $}\\\hline
        \multicolumn{10}{|l|}{$\Delta= (1+4 z_{1})^{2}\,(1+432 z_{2})\,
(1- 11664 z_{1} z_{2}- 93312 z_{1}^{2} z_{2}- 186624 z_{1}^{3} z_{2})$}\\\hline
    \end{tabular}
    \caption{Toric and topological data for $\text{Bl}(X^{6,6})$.}
    \label{tab:BlX66}
\end{table}

\begin{table}[H]
    \renewcommand{\arraystretch}{1.2}
    \centering
    \begin{tabular}{| c | c c c c c |c c || c | l |}
        \hline \multicolumn{6}{|c}{points} & \multicolumn{2}{| c||}{$l$-vectors} & \multicolumn{2}{c|}{topological data} \\ \hline \hline
        (1 & 0 & 0 & 0 & 0 & 0) & -2 &  0 & $R$ & $80J_1^3+32J_1^2J_2+12J_1J_2^2+4J_2^3$\\
        (1 & 0 & 0 & 0 & 0 & 0) &  0 & -2 & $c_2$ & $(128,52)$\\\cline{1-8}
        (1 & 1 & 0 & 0 & 0 & 0) &  1 &  0 & $\chi$ & $-196$ \\
        (1 & 0 & 1 & 0 & 0 & 0) &  1 & -2 &  & \\
        (1 & 0 & 0 & 1 & 0 & 0) &  0 &  1 &&\\
        (1 & 0 & 0 & 0 & 1 & 0) &  0 &  1 &&\\
        (1 & 0 & 0 & 0 & 0 & 1) &  1 & -2 &&\\ 
        (1 &-3 &-1 &-1 &-1 &-1) &  0 &  1 &&\\
        (1 & 1 & 1 & 0 & 0 & 1) & -1 &  3 &&\\  \hline\hline
        \multicolumn{10}{|l|}{$\mathcal{L}^{(2)}=\left(\theta _1-2 \theta _2\right){}^2+2z_1 \left(2 \theta _1+1\right) \left(\theta _1-3 \theta _2\right)  $}\\
        \multicolumn{10}{|l|}{$\mathcal{L}^{(3)}=\theta _1^3-20 \theta _2^3+2 z_1\left(2 \theta _1+1\right) \left(\theta _1 \left(4 \theta _1-11 \theta _2+1\right)-3 \theta _2\right) $}\\
        \multicolumn{10}{|l|}{$\quad\quad+12 z_1^2\left(\theta _1-3 \theta _2+8z_2 \left(2 \theta _2+1\right) \right) \left(4 \theta _1 \left(\theta _1+2\right)+3\right)  $}\\\hline
        \multicolumn{10}{|l|}{$\Delta= (1+4 z_{1})
\left(1+ 4 z_{1}- 512 z_{1}^{2} z_{2}- 9216 z_{1}^{3} z_{2}- 27648z_{1}^{4} z_{2}+ 65536 z_{1}^{4} z_{2}^{2}\right)$}\\\hline
    \end{tabular}
    \caption{Toric and topological data for $\text{Bl}(X^{6,2})$.}
    \label{tab:BlX62}
\end{table}

\begin{table}[H]
    \renewcommand{\arraystretch}{1.2}
    \centering
    \begin{tabular}{| c | c c c c|c c c|| c | l |}
        \hline \multicolumn{5}{|c}{points} & \multicolumn{3}{| c||}{$l$-vectors} & \multicolumn{2}{c|}{topological data} \\ \hline \hline
        (1 & 0 & 0 & 0 & 0) &-6 & 0 & 0 & $R$ & $8 J_1^3+2J_1^2 J_2 +2J_1^2 J_3 +J_1 J_2 J_3$\\ \cline{1-8}
        (1 & 0 & 0 & 1 & 0) & 3 & 0 & 0 &$c_2$ & $(92, 24, 24)$\\
        (1 & 0 & 0 & 0 & 1) & 2 & 0 & 0 & $\chi$ &$-480$ \\
        (1 & 0 & 0 &-3 &-2) & 1 &-2 &-2 &  & \\
        (1 & 1 & 0 &-3 &-2) & 0 & 1 & 0 &&\\
        (1 &-1 & 0 &-3 &-2) & 0 & 1 & 0 &&\\
        (1 & 0 & 1 &-3 &-2) & 0 & 0 & 1 &&\\
        (1 & 0 &-1 &-3 &-2) & 0 & 0 & 1 &&\\\hline\hline
        \multicolumn{10}{|l|}{$\mathcal{L}^{(2)}= \theta _1 \left(2 \left(\theta _2+\theta _3\right)-\theta _1\right)+12 z_1\left(36 \theta _1^2+36 \theta _1+5\right) $}\\
        \multicolumn{10}{|l|}{$\mathcal{L}^{(2)} = \theta _2^2-z_2\left(\theta _1^2-\left(4 \theta _2+4 \theta _3+1\right) \theta _1+2 \left(2 \theta _2^2+4 \theta _3 \theta _2+\theta _2+2 \theta _3^2+\theta _3\right)\right) $}\\
        \multicolumn{10}{|l|}{$\mathcal{L}^{(2)} = \theta _3^2 \left(1-4 z_2\right)-z_3\left(\theta _1^2-\left(4 \theta _2+4 \theta _3+1\right) \theta _1+2 \left(2 \theta _3^2+4 \theta _2 \theta _3+\theta _3+\theta _2\right)\right) $}\\ \hline
        \multicolumn{10}{|l|}{$\bar{\Delta}= 16 z_2^2-32 z_2 z_3-8 z_2+16 z_3^2-8 z_3+1$}\\\hline
    \end{tabular}
    \caption{Toric and topological data for $X_{321}$-fibration over $\P^1\times \P^1$\,.
    The discriminant locus along $z_3=z_2$ becomes $\bar{\Delta}|_{z_3\rightarrow z_2}=1-16z_2 $\,.}
    \label{tab:P1P1el}
\end{table}

\begin{table}[H]
    \renewcommand{\arraystretch}{1.2}
    \centering
    \begin{tabular}{| c | c c c c c |c c || c | l |}
        \hline \multicolumn{6}{|c}{points} & \multicolumn{2}{| c||}{$l$-vectors} & \multicolumn{2}{c|}{topological data} \\ \hline \hline
        (1 & 0 & 0 & 0 & 0 & 0) & -5 & 0 & $R$ & $10J_1^4+20J_1^3J_2^2$\\ \cline{1-8}
        (1 & 1 & 0 & 0 & 0 & 0) &  1 &  0 & $c_2$ & $(110,50,0)$\\
        (1 & 0 & 1 & 0 & 0 & 0) &  1 &  0 & $c_3$ & $(-410, -200)$\\
        (1 & 0 & 0 & 1 & 0 & 0) &  1 &  0 & $\chi$ & $2160$\\
        (1 & 0 & 0 & 0 & 1 & 0) &  1 &  0 &&\\
        (1 & 0 & 0 & 0 & 0 & 1) &  0 &  1 &&\\
        (1 &-2 &-2 &-2 &-2 &-1) &  0 &  1 &&\\
        (1 &-1 &-1 &-1 &-1 & 0) &  1 &  -2 &&\\ \hline\hline
        \multicolumn{10}{|l|}{$\mathcal{L}^{(2)}=\theta _2^2-z_2\left(\theta _1-2 \theta _2\right) \left(\theta _1-2 \theta _2-1\right) $}\\
        \multicolumn{10}{|l|}{$\mathcal{L}^{(3)} = \left(\theta _1-2 \theta _2\right) \theta _1^3-5 z_1\left(5 \theta _1+1\right) \left(5 \theta _1+2\right) \left(5 \theta _1+3\right) \left(5 \theta _1+4\right)$}\\\hline
        \multicolumn{10}{|l|}{$\Delta=\left(1-4z_2\right)^3\left(1-6250z_1+9765625z_1^2-39062500z_1^2z_2\right)$}\\\hline
    \end{tabular}
    \caption{Toric and topological data for $X_{222211}$. This model is a CY3 $X^5$-fibration over $\P^1$.}
    \label{tab:222211}
\end{table}

\begin{table}[H]
    \renewcommand{\arraystretch}{1.2}
    \centering
    \begin{tabular}{| c | c c c c c |c c || c | l |}
        \hline \multicolumn{6}{|c}{points} & \multicolumn{2}{| c||}{$l$-vectors} & \multicolumn{2}{c|}{topological data} \\ \hline \hline
        (1 & 0 & 0 & 0 & 0 & 0) &  -4 & 0 & $R$ & $36J_1^4+12J_1^3J_2+4J_1^2J_2^2$\\ \cline{1-8}
        (1 & 1 & 0 & 0 & 0 & 0) &  1 &  0 & $c_2$ & $(228, 84, 24)$\\
        (1 & 0 & 1 & 0 & 0 & 0) &  1 &  0 & $c_3$ & $(-720, -252)$\\
        (1 & 0 & 0 & 1 & 0 & 0) &  1 &  0 & $\chi$ & $2988$\\
        (1 & 0 & 0 & 0 & 1 & 0) &  0 &  1 &&\\
        (1 & 0 & 0 & 0 & 0 & 1) &  0 &  1 &&\\
        (1 &-3 &-3 &-3 &-1 &-1) &  0 &  1 &&\\
        (1 &-1 &-1 &-1 & 0 & 0) &  1 & -3 &&\\ \hline\hline
        \multicolumn{10}{|l|}{$\mathcal{L}^{(3)}=\theta _1^2 \left(\theta _1-3 \theta _2\right)-8 z_1\left(2 \theta _1+1\right) \left(4 \theta _1+1\right) \left(4 \theta _1+3\right)$}\\
        \multicolumn{10}{|l|}{$\mathcal{L}^{(3)}=\theta _2^3-z_2\left(\theta _1-3 \theta _2-2\right) \left(\theta _1-3 \theta _2-1\right) \left(\theta _1-3 \theta _2\right)$}\\\hline
        \multicolumn{10}{|l|}{$\Delta=\left(1+27 z_2\right)^2 \left( 1-768 z_1+196608 z_1^2-16777216 z_1^3-452984832  z_1^3z_2\right)$}\\\hline
    \end{tabular}
    \caption{Toric and topological data for $X_{333111}$. This model is a K3 $X^4$-fibration over $\P^2$.}
    \label{tab:333111}
\end{table}

\begin{table}[H]
    \renewcommand{\arraystretch}{1.2}
    \centering
    \begin{tabular}{| c | c c c c c |c c || c | l |}
        \hline \multicolumn{6}{|c}{points} & \multicolumn{2}{| c||}{$l$-vectors} & \multicolumn{2}{c|}{topological data} \\ \hline \hline
        (1 & 0 & 0 & 0 & 0 & 0) &  -6 & 0 & $R$ & $64J_1^4+16J_1^3J_2+4J_1^2J_2^2+J_1J_2^3$\\ \cline{1-8}
        (1 & 1 & 0 & 0 & 0 & 0) &  3 &  0 & $c_2$ & $(728, 182, 48)$\\
        (1 & 0 & 1 & 0 & 0 & 0) &  2 &  0 & $c_3$ & $(-3860, -960)$\\
        (1 & 0 & 0 & 1 & 0 & 0) &  0 &  1 & $\chi$ & $23328$\\
        (1 & 0 & 0 & 0 & 1 & 0) &  0 &  1 &&\\
        (1 & 0 & 0 & 0 & 0 & 1) &  0 &  1 &&\\
        (1 &-12&-8 &-1 &-1 &-1) &  0 &  1 &&\\
        (1 &-3 &-2 & 0 & 0 & 0) &  1 & -4 &&\\ \hline\hline
        \multicolumn{10}{|l|}{$\mathcal{L}^{(2)}=\theta _1 \left(\theta _1-4 \theta _2\right)-12z_1 \left(6 \theta _1+1\right) \left(6 \theta _1+5\right)$}\\
        \multicolumn{10}{|l|}{$\mathcal{L}^{(4)}=\theta _2^4-z_2\left(\theta _1-4 \theta _2-3\right) \left(\theta _1-4 \theta _2-2\right) \left(\theta _1-4 \theta _2-1\right) \left(\theta _1-4 \theta _2\right)$}\\\hline
        \multicolumn{10}{|l|}{$\Delta=\left(1-256z_2\right)\big((1 - 432z_1)^4-8916100448256\,z_1^4z_2 \big)$}\\\hline
    \end{tabular}
    \caption{Toric and topological data for $X_{12\,81111}$. This model is an elliptic $X^6$-fibration over $\P^3$.}
    \label{tab:1281111}
\end{table}

\begin{table}[H]
    \renewcommand{\arraystretch}{1.2}
    \centering
    \begin{tabular}{| c | c c c c c |c c c|| c | l |}
        \hline \multicolumn{6}{|c}{points} & \multicolumn{3}{| c||}{$l$-vectors} & \multicolumn{2}{c|}{topological data} \\ \hline \hline
        (1 & 0 & 0 & 0 & 0 & 0) &  -6 & 0 & 0 & $R$ & $72 J_1^4+36 J_2 J_1^3+12 J_3 J_1^3+18 J_2^2 J_1^2$\\ \cline{1-9}
        (1 & 1 & 0 & 0 & 0 & 0) &  3 &  0 & 0 &  & $+2 J_3^2 J_1^2+6 J_2 J_3 J_1^2+9 J_2^3 J_1+J_2 J_3^2 J_1$\\
        (1 & 0 & 1 & 0 & 0 & 0) &  2 &  0 & 0 && $+3 J_2^2 J_3 J_1$\\
        (1 & 0 & 0 & 1 & 0 & 0) &  0 &  1 & 0 & $c_2$ & $(816, 408, 138, 216, 72, 24)$\\
        (1 & 0 & 0 & 0 & 1 & 0) &  0 &  0 & 1 & $c_3$ & $(-4338, -2160, -720)$\\
        (1 & 0 & 0 & 0 & 0 & 1) &  0 &  0 & 1 & $\chi$& $26208$\\
        (1 &-18&-12&-3 &-1 &-1) &  0 &  0 & 1 &&\\
        (1 &-6 &-4 &-1 & 0 & 0) &  0 &  1 &-3 &&\\
        (1 &-3 &-2 & 0 & 0 & 0) &  1 & -2 & 0 &&\\\hline\hline
        \multicolumn{11}{|l|}{$\mathcal{L}^{(2)}= \theta _1 \left(\theta _1-2 \theta _2\right)-12z_1 \left(6 \theta _1+1\right) \left(6 \theta _1+5\right)$}\\
        \multicolumn{11}{|l|}{$\mathcal{L}^{(2)} = \theta _2 \left(\theta _2-3 \theta _3\right)- z_2\left(\theta _1-2 \theta _2-1\right) \left(\theta _1-2 \theta _2\right)$}\\
        \multicolumn{11}{|l|}{$\mathcal{L}^{(3)} = \theta _3^3-z_3\left(\theta _2-3 \theta _3-2\right) \left(\theta _2-3 \theta _3-1\right) \left(\theta _2-3 \theta _3\right)$}\\\hline
        \multicolumn{11}{|l|}{$\bar{\Delta} = \left(1+27z_3\right)
        \left(1-12 z_2+48 z_2^2-64z_2^3 \left(1+27 z_3\right) \right)$}\\\hline
    \end{tabular}
    \caption{Toric and topological data for $X_{18\,12\,3111}$. This model is an elliptic $X^6$-fibration over $\P_{3111}$.}
    \label{tab:18123111}
\end{table}

\begin{table}[H]
    \renewcommand{\arraystretch}{1.2}
    \centering
    \begin{tabular}{| c | c c c c c |c c c|| c | l |}
        \hline \multicolumn{6}{|c}{points} & \multicolumn{3}{| c||}{$l$-vectors} & \multicolumn{2}{c|}{topological data} \\ \hline \hline
        (1 & 0 & 0 & 0 & 0 & 0) &  -6 & 0 & 0 & $R$ & $54 J_1^4+18 J_2 J_1^3+9 J_3 J_1^3+6 J_2^2 J_1^2$\\ \cline{1-9}
        (1 & 1 & 0 & 0 & 0 & 0) &  3 &  0 & 0 &  & $+3 J_2 J_3 J_1^2+2 J_2^3 J_1+J_2^2 J_3 J_1$\\
        (1 & 0 & 1 & 0 & 0 & 0) &  2 &  0 & 0 & $c_2$ & $(618, 206, 102, 72, 36, 0)$\\
        (1 & 0 & 0 & 1 & 0 & 0) &  0 &  1 & 0 & $c_3$ & $(-3258, -1080, -540)$\\
        (1 & 0 & 0 & 0 & 1 & 0) &  0 &  1 & 0 & $\chi$& $19728$\\
        (1 & 0 & 0 & 0 & 0 & 1) &  0 &  0 & 1 &&\\
        (1 &-18&-12&-2 &-2 &-1) &  0 &  0 & 1 &&\\
        (1 &-9 &-6 &-1 &-1 & 0) &  0 &  1 &-2 &&\\
        (1 &-3 &-2 & 0 & 0 & 0) &  1 & -3 & 0 &&\\\hline\hline
        \multicolumn{11}{|l|}{$\mathcal{L}^{(2)}= \theta _1 \left(\theta _1-3 \theta _2\right)-12z_1 \left(6 \theta _1+1\right) \left(6 \theta _1+5\right)$}\\
        \multicolumn{11}{|l|}{$\mathcal{L}^{(2)} = \theta _3^2-z_3\left(\theta _2-2 \theta _3-1\right) \left(\theta _2-2 \theta _3\right)$}\\
        \multicolumn{11}{|l|}{$\mathcal{L}^{(3)} = \theta _2^2 \left(\theta _2-2 \theta _3\right)-z_2\left(\theta _1-3 \theta _2-2\right) \left(\theta _1-3 \theta _2-1\right) \left(\theta _1-3 \theta _2\right)$}\\\hline
        \multicolumn{11}{|l|}{$\bar{\Delta} = \left(1-4z_3\right)
        \left(1+54 z_2+729z_2^2 \left(1-4 z_3\right) \right)$}\\\hline
    \end{tabular}
    \caption{Toric and topological data for $X_{18\,12\,2211}$. This model is an elliptic $X^6$-fibration over $\P_{2211}$.}
    \label{tab:18122211}
\end{table}

\bibliography{refs.bib}
\end{document}